\PassOptionsToPackage{unicode}{hyperref}
\PassOptionsToPackage{naturalnames}{hyperref}
\documentclass[twocolumn]{aastex631}
\extrafloats{100}

\usepackage{CJK}
\usepackage{amsmath}
\usepackage{subfigure}
\shorttitle{Spatial Distribution of FRBs}
\shortauthors{Gordon et al.}

\begin{document}
\begin{CJK*}{UTF8}{gbsn}

\title{Mapping the Spatial Distribution of Fast Radio Bursts within their Host Galaxies}

\correspondingauthor{Alexa C. Gordon}
\email{alexagordon2026@u.northwestern.edu}

\newcommand{\NU}{\affiliation{Center for Interdisciplinary Exploration and Research in Astrophysics (CIERA) and Department of Physics and Astronomy, Northwestern University, Evanston, IL 60208, USA}}

\newcommand{\Swin}{\affiliation{Centre for Astrophysics and Supercomputing, Swinburne University of Technology, Hawthorn, VIC 3122, Australia}}

\newcommand{\macqa}{\affiliation{School of Mathematical and Physical Sciences, Macquarie University, NSW 2109, Australia}}

\newcommand{\macqb}{\affiliation{Astrophysics and Space Technologies Research Centre, Macquarie University, Sydney, NSW 2109, Australia}}

\newcommand{\ATNF}{\affiliation{Australia Telescope National Facility, CSIRO Space \& Astronomy, Box 76 Epping, NSW 1710, Australia}}

\newcommand{\Yonsei}{\affiliation{Department of Astronomy, Yonsei University, 50 Yonsei-ro, Seodaemun-gu, Seoul 03722, Republic of Korea}}

\newcommand{\NOIRLab}{\affiliation{NSF's NOIRLab, 950 N. Cherry Avenue, Tucson, AZ 85719, USA}}

\newcommand{\Syd}{\affiliation{Sydney Institute for Astronomy, School of Physics, The University of Sydney, NSW 2006, Australia}}

\newcommand{\ICRAR}{\affiliation{International Centre for Radio Astronomy Research, Curtin University, Bentley, WA 6102, Australia}}

\newcommand{\Edin}{\affiliation{Institute for Astronomy, University of Edinburgh, Royal Observatory, Edinburgh, EH9 3HJ, United Kingdom}}

\newcommand{\Cape}{\affiliation{Inter-University Institute for Data Intensive Astronomy, Department of Astronomy, University of Cape Town, Cape Town, South Africa}}

\newcommand{\UChi}{\affiliation{Dept. of Astronomy and Astrophysics, University of Chicago}}

\newcommand{\PUCV}{\affiliation{Instituto de F\'isica, Pontificia Universidad Cat\'olica de Valpara\'iso, Casilla 4059, Valpara\'iso, Chile}}

\newcommand{\NAOJ}{\affiliation{Division of Science, National Astronomical Observatory of Japan,2-21-1 Osawa, Mitaka, Tokyo 181-8588, Japan}}

\newcommand{\IPMU}{\affiliation{Kavli Institute for the Physics and Mathematics of the Universe (Kavli IPMU), 5-1-5 Kashiwanoha, Kashiwa, 277-8583, Japan}}

\newcommand{\UCSC}{\affiliation{Department of Astronomy and Astrophysics, University of California, Santa Cruz, CA 95064, USA}}
\author[0000-0002-5025-4645]{Alexa~C.~Gordon}
\NU

\author[0000-0002-7374-935X]{Wen-fai~Fong}
\NU

\author[0000-0001-9434-3837]{Adam~T.~Deller}
\Swin

\author[0000-0003-1483-0147]{Lachlan Marnoch}
\macqa
\macqb
\ATNF

\author[0000-0002-5049-4390]{Sungsoon Lim}
\Yonsei

\author[0000-0002-2073-2781]{Eric~W.~Peng}
\NOIRLab


\author[0000-0003-2149-0363]{Keith~W.~Bannister}
\ATNF
\Syd

\author[0000-0002-2864-4110]{Apurba Bera}
\ICRAR

\author[0000-0002-8383-5059]{N.~D.~R.~Bhat}
\ICRAR

\author[0009-0004-1205-8805]{Tyson Dial}
\Swin

\author[0000-0002-9363-8606]{Yuxin~Dong~(董雨欣)}
\NU

\author[0000-0003-0307-9984]{Tarraneh Eftekhari}
\NU

\author[0000-0002-5067-8894]{Marcin Glowacki}
\Edin
\Cape

\author[0000-0002-0152-1129]{Kelly Gourdji}
\ATNF

\author[0000-0001-9817-4938]{Vivek Gupta}
\ATNF

\author[0000-0003-4193-6158]{Joscha~N.~Jahns-Schindler}
\Swin

\author[0000-0002-8987-1544]{Akhil Jaini}
\Swin

\author[0000-0002-5740-7747]{Charles~D.~Kilpatrick}
\NU

\author[0000-0002-7866-4531]{Chang Liu}
\NU

\author[0000-0002-7738-6875]{J.~Xavier~Prochaska}
\UCSC
\IPMU
\NAOJ

\author[0000-0003-4501-8100]{Stuart~D.~Ryder}
\macqa
\macqb

\author[0000-0002-7285-6348]{Ryan~M.~Shannon}
\Swin

\author[0000-0003-3801-1496]{Sunil Simha}
\NU
\UChi

\author[0000-0002-1883-4252]{Nicolas Tejos}
\PUCV

\author[0000-0003-0203-1196]{Yuanming Wang}
\Swin 

\author[0000-0002-2066-9823]{Ziteng Wang}
\ICRAR 



\begin{abstract} 
We present deep optical and near-infrared observations of the host galaxies of 34 fast radio bursts (FRBs) detected by the Commensal Real-time ASKAP Fast Transient (CRAFT) survey on the Australian SKA Pathfinder (ASKAP) to compare the locations of FRBs relative to their host light distributions. Incorporating three additional FRBs from the literature, for a total of four repeating and 33 apparently non-repeating FRBs, we determine their projected galactocentric offsets and find a median of $ 4.2^{+5.7}_{-2.5}$ kpc ($1.0^{+1.5}_{-0.6}r_e$). We model their host surface brightness profiles and develop synthetic spatial distributions of their globular clusters based on host properties. We calculate the likelihood the observed location of each FRB is consistent with the smooth light of its host galaxy, residual (primarily spiral) substructure, or globular cluster distributions. The majority of FRBs favor locations within the disks of their galaxies, while only 11$\pm$5\% favor a globular cluster origin, primarily those with galactocentric offsets $\gtrsim3r_e$. At $z<0.15$, where spiral structure is apparent in 86\% of our sample of FRB hosts, we find $\approx 20-46\%$ of FRBs favor an association with spiral arms. Assuming FRBs derive from magnetars, our results support multiple formation channels with the majority of progenitors associated with massive stars and a minority formed through dynamical channels. However, the moderate fraction of FRBs associated with spiral structure indicates that high star formation efficiency of the youngest and most massive stars is not a predominant driver in the production of FRB progenitors.
\end{abstract}



\section{Introduction} \label{sec:intro}

Despite the detection of hundreds of fast radio bursts (FRBs) to date (e.g., \citealt{CHIME-catalog,Petroff+22}), their precise astrophysical origins remain uncertain. Historically, FRBs have been detected by a wide range of facilities such as the Australian SKA Pathfinder (ASKAP; \citealt{ASKAP}), Canadian Hydrogen Intensity Mapping Experiment (CHIME/FRB; \citealt{CHIME+18}), Deep Synoptic Array (DSA; \citealt{Ravi+19}), More (meer) Karoo Array Telescope (MeerKAT; \citealt{Rajwade+22}), and Five hundred meter Aperture Spherical radio Telescope (FAST; \citealt{FAST}). These experiments have revealed them to be energetic ($\approx10$~mJy--$100$~Jy; \citealt{Ravi19,Hashimoto22,Zhang23}), millisecond-duration radio transients, with high inferred brightness temperatures ($\gtrsim10^{30}$K) implying coherent non-thermal emission (e.g., \citealt{Nimmo+22}), short duration implying a small emitting region \citep{Nimmo+25}, and typically high linear polarization degree \citep{Day+20,Pandhi+24} implying strong, ordered  magnetic fields. The discovery of the repeating FRB\,20200428A from the Galactic magnetar SGR\,1935+2154 \citep{Bochenek20,CHIME-magnetar,Mereghetti+20} at the geometric center of a known supernova remnant \citep{Sun+2011,Zhou20} led to strong consensus that at least some FRBs originate from magnetars produced in core-collapse supernova (CCSN) explosions. The Galactic event, coupled with the observed characteristics of FRBs, strongly reinforces a magnetar progenitor for cosmological events (e.g., \citealt{Zhang23}). 

Alongside studies of FRB properties, $\approx$100 FRBs have been localized with a precision of $\sim$few arcsec or better which allow for robust association to their host galaxies (e.g., \citealt{Chatterjee+17,Marcote_etal_2020,Bannister+19,Prochaska+19,Ravi+19,Bhandari+20,Heintz+20,Bhardwaj21,Fong+21,Niu+22,Bhandari+23,Gordon+23,LeeWaddell+23,Panther+23,Ravi+23,Ryder23,Bhardwaj+24,Hewitt+24,Law+24,Shannon+25,Sharma+24,Rajwade+24,CHIME+25,Shah+25}). At the most basic level, this has enabled constraints on their global host stellar populations and thus inference on both the conditions under which FRB progenitors form and how they trace galaxy properties. Commensurate with expectations for young magnetar progenitors born from CCSNe, the majority of FRBs occur in star-forming galaxies that lie along the galaxy `star-forming main sequence' \citep{Heintz+20,Bhandari+20,Gordon+23,Bhardwaj+24,Sharma+24}, with only a small fraction occurring in quiescent hosts \citep{Ravi+19,Sharma_etal_2023,Gordon+23,Law+24,Shah+25,Eftekhari+25}. A comparison of their stellar population properties (for instance, stellar mass and present-date star formation rate) to field galaxies suggests that the FRB progenitor rate either traces star formation alone \citep{Sharma+24,Loudas+25}, consistent with CCSNe \citep{Bochenek+21}, or a combination of star formation and stellar mass \citep{Horowicz-Margalit2025}. The latter scenario would indicate stellar progenitors with longer delay times relative to star formation rather than young magnetars born from CCSNe.

Studies of the local environments of FRBs have added significant value to this picture by providing unambiguous detail on $\sim$parsec scales. For example, the repeating FRB\,20200120E was found to be associated with the star-forming, spiral galaxy M81 at 3.6~Mpc \citep{Bhardwaj21} commensurate with the global properties of more distant FRB hosts. However, upon its milliarcsecond localization through Very Long Baseline Interferometry (VLBI), the FRB was surprisingly associated to a $\sim$9.1~Gyr old globular cluster \citep{Kirsten21}. The origin in such an old stellar population demanded a delayed progenitor relative to star formation such as a magnetar formed via binary compact object mergers \citep{Lu+22} or the accretion induced collapse of a white dwarf \citep{Kremer2021}, and provided additional evidence that there may be multiple populations of FRB progenitors including those that do not directly trace star formation (e.g., \citealt{Li_Zhang_2020,Zhang_and_Zhang2022}). FRB\,20200120E aptly demonstrates that the environments on kiloparsec versus parsec scales can give conflicting constraints on the progenitor and the limited diagnostic power of global host environments. By providing observational support for a second progenitor formation channel besides CCSNe, it is still an open question as to what fraction of FRB progenitors are formed through alternative channels, though it is unlikely that every FRB originates via delayed channels \citep{Horowicz-Margalit2025}.

Precise positions of FRBs have also enabled studies which show that several coincide with the spiral arms of their host galaxies \citep{Chittidi21,Mannings+21,Woodland+24} where the formation of massive stars occurs at higher efficiency than the inner arm regions \citep{Kreckel+16}. This general finding is consistent with the locations of CCSNe \citep{Aramyan+16,Audcent-Ross+20} and their radial offset distributions \citep{Heintz+20,Mannings+21}, supporting a prompt formation channel relative to star formation. In a few cases, the combined power of high-resolution space-based imaging and milliarcsecond-scale VLBI localizations have pinpointed FRBs to be either coincident with star-forming regions \citep{Bassa+17,Kokubo+17,Marcote+17} or at small offsets from apparent star-formation \citep{Tendulkar+21,Dong+24}. However, such studies are typically limited to the most active repeaters (for which targeted VLBI follow-up observations can provide milliarcsecond-precision information), possibly leading to a biased understanding of the progenitor if repeating and non-repeating FRBs comprise distinct populations. Moreover, while incredibly valuable, space-based imaging is only available for a limited fraction of the FRB host population.

It is clear that FRB locations hold potential diagnostic power in quantifying the contributions of various formation channels to the FRB progenitor. Importantly, such studies can be performed on any FRB with a sub-arcsecond position and deep ground-based imaging. The Commensal Real-time ASKAP Fast Transient (CRAFT) survey \citep{mbb+10,Shannon+25} paved the way in sub-arcsecond FRB localizations enabled by interferometry, associating the first non-repeating FRB\,20180924B to its host galaxy \citep{Bannister+19}. Since then, CRAFT has continued to detect and localize both repeaters and non-repeaters to their hosts with sub-arcsecond precision which is sufficient to enable large-scale population studies of FRB host galaxy demographics \citep{Bhandari+20,Heintz+20,Gordon+23,Shannon+25}.

In this work, we leverage the large sample of FRBs detected with ASKAP/CRAFT (the majority of which are non-repeaters) and deep imaging with 4- to 10-meter class telescopes to determine the locations of FRBs within and about their host galaxies and compare to different host light distributions. In Section~\ref{sec:sample}, we introduce our sample, which comprises a total of 37~FRBs, 7 of which are newly presented here. In Section~\ref{sec:imaging}, we describe our optical and infrared imaging of their host galaxies, data reduction, and spectroscopy. In Section~\ref{sec:methods}, we describe the surface brightness profile fitting to derive smooth and residual light profiles from the data, the construction of synthetic spatial distributions of globular clusters, and our framework to statistically compare them. We also determine the projected FRB physical and host-normalized galactocentric offsets. In Section~\ref{sec:results} we present our results and the effects of offset and localization size. We then interpret our results in the context of the progenitor in Section~\ref{sec:discussion} and conclude in Section~\ref{sec:conclusions}. Throughout this work we assume WMAP9 cosmology \citep{WMAP9}.

\section{Sample Selection} \label{sec:sample}

To derive our sample, we begin with all FRBs detected by the CRAFT survey through the end of 2024, totaling 83 FRBs\footnote{Each FRB was discovered and localized by CRAFT (e.g., \citealt{Shannon+18,Shannon+25,Wang+24}) with the exception of the repeating FRB\,20201124A which was discovered by CHIME \citep{201124_discovery} and localized by CRAFT \citep{Day21,Fong+21}.}. We then narrow the sample using the following criteria:

\begin{enumerate}
    \item The host galaxy and its spectroscopic redshift are known. This removes 35 FRBs from the sample.
    \item The FRB localization (1$\sigma$) is $\leq 2\arcsec$ for robust association to the host galaxy \citep{Eftekhari2017}. This removes four additional FRBs from the sample.
    \item The semi-major axis of the FRB localization uncertainty is $\lesssim50\%$ of the approximate host galaxy size to enable association with features on sub-galactic scales (such as spiral arms, if present) and to robustly determine the FRB's galactocentric offset. This removes six additional FRBs from the sample.
    \item The FRB is not in close proximity to a bright star (following the methodology put forth in \citealp{Hjorth+12} for their study of GRB host galaxies). This removes two additional FRBs from the sample.
    \item The FRB sightline has low Galactic extinction, defined as E($B-V$) $<$ 0.1 and DM$_{\rm ISM}$ $<$ 100 pc~cm$^{-3}$ derived from NE2001 \citep{NE2001}, where DM$_{\rm ISM}$ is the Milky Way interstellar medium component of the FRB's dispersion measure (DM). This removes one additional FRB from the sample.
    \item Deep imaging of the field, reaching a 3$\sigma$ limit of $\gtrsim 25$~AB~mag, is available to resolve faint surface brightness features, if present. This removes one FRB from the sample.
\end{enumerate}

\noindent Applying these cuts results in a sample of 34 CRAFT FRBs, 27 of which have been previously published \citep{Bannister+19,Bhandari+20,Heintz+20,mpm+20,Fong+21,Bhandari+22,Bhandari+23,Gordon+23,Shannon+25}. Seven events, their host galaxies, and redshifts are newly presented here: FRBs 20231230D, 20240117B, 20240203C, 20240312D, 20240525A, 20240615B, and 20241027B. These FRBs were detected with one of two FRB detection systems: the incoherent sum detection system which has been used to detect and localize FRBs with ASKAP since 2018 \citep{Shannon+25} and the more recently commissioned (and more sensitive) array coherent detector, CRACO \citep{Wang+24}. Details of their detections and radio properties will be presented in future works.

We additionally include in our sample three FRBs that were not detected by CRAFT but are otherwise notable events$-$namely, the well-localized, repeating FRBs 20180916B \citep{Marcote_etal_2020}, 20200120E \citep{Bhardwaj21,Kirsten21}, and 20240209A \citep{Eftekhari+25,Shah+25} on account of being in spiral galaxies that meet our sample criteria or (candidate) globular cluster events. To our knowledge, no other FRBs in the literature have publicly available imaging that meets our sample requirements. While FRB\,20180916B is along a high-extinction sightline and does not formally meet criterion \#5, we perform all analysis on an infrared \textit{HST} image \citep{Tendulkar+21} to mitigate the effect of Galactic extinction (as the NIR is less affected by dust). Due to the large angular size of the host of FRB\,20200120E, we use a mosaicked $r$-band image from the Sloan Digital Sky Survey DR12 (SDSS; \citealt{SDSS}) for our analysis. This image does not reach our limiting magnitude cut of 25 AB mag, but the proximity of the host makes the existing image sufficient (and far deeper in terms of luminosity than any other imaging in our sample); thus we waive criteria \#6 for this burst. Lastly, we include FRB\,20240209A as a strong candidate for an FRB from a globular cluster given its localization to the outskirts of a massive, quiescent elliptical galaxy \citep{Shah+25,Eftekhari+25}. We thus proceed with a final sample of 37 FRB host galaxies which we detail in Table~\ref{tab:sample}.

\begin{figure}
    \centering
    \includegraphics[width=0.45\textwidth]{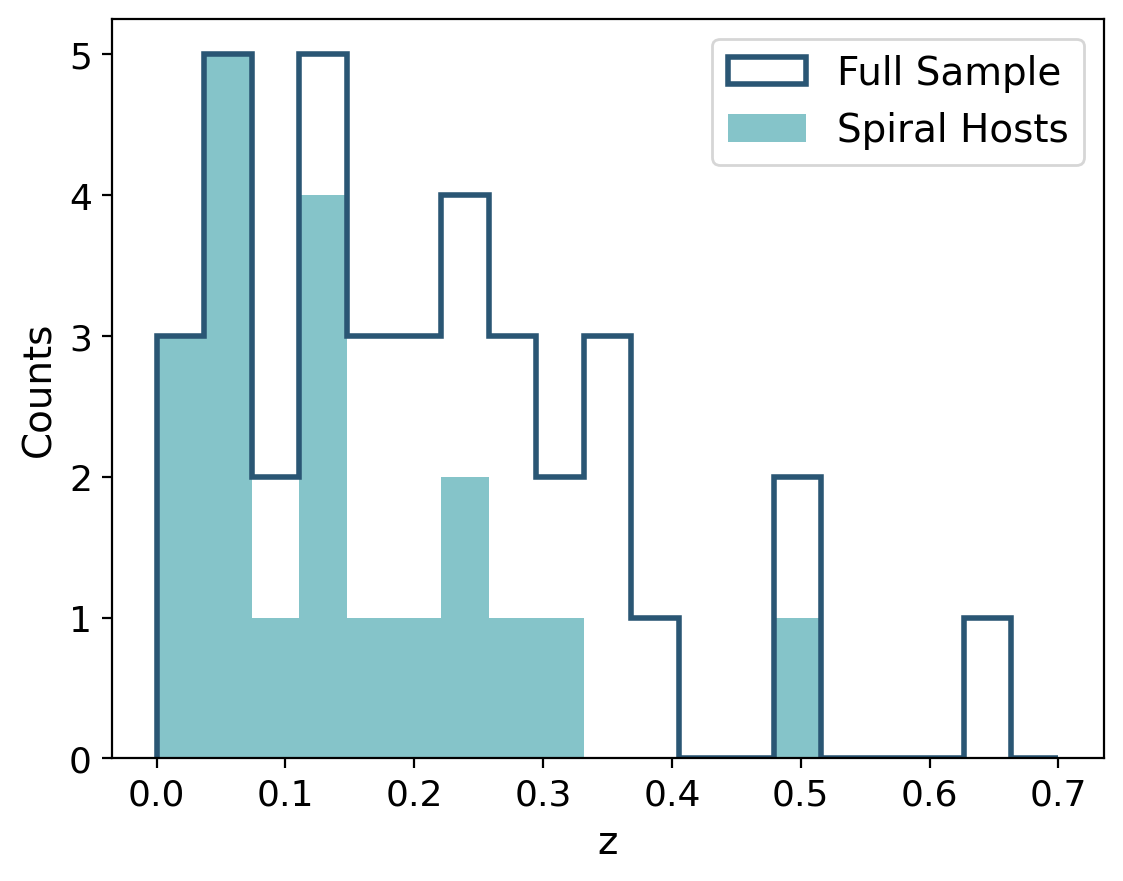}
    \caption{The redshift distribution of the FRB host galaxies in our sample which extends to $z \approx 0.64$. We denote the sub-sample of 20 hosts with clear spiral arms as solid light blue bars. There is a clear drop-off in the detection of spiral structure at $z\gtrsim 0.15$ which is likely an effect of the limit of our survey, as opposed to an intrinsic effect.}
    \label{fig:z_hist}
\end{figure}

The FRBs in our sample span a redshift range of $z=0.0008-0.643$ with a median redshift of $z\approx0.2$. The large majority are non-repeaters; to date, only FRBs 20180916B, 20200120E, 20201124A, and 20240209A are known to repeat. Overall, our sample comprises an assortment of spiral and disk galaxies (with only one known elliptical), but many are not easily classified from simple visual inspection. We comment further on the detailed morphological properties of the sample in Section~\ref{sec:galfit}. In Figure~\ref{fig:z_hist}, we show the redshift distribution of our sample, noting the 20 hosts which have clear spiral features. Notably, Figure~\ref{fig:z_hist} shows that spiral structure is near-ubiquitous in the host galaxy images at $z \lesssim 0.15$, but is less frequently visible at higher redshift. While we can see spiral structure out to $z \sim 0.5$ in some cases, the drop-off at higher redshift must be at least partly (if not wholly) explained by imaging depth and resolution limitations, although we consider the possible impact of redshift evolution of the FRB progenitor(s) in Section~\ref{sec:discussion_stars}.

\begin{deluxetable*}{l|cccccccc}[t]
\tabletypesize{\footnotesize}
\tablewidth{0pc}
\tablecaption{FRB Sample and Imaging Properties
\label{tab:sample}}
\tablehead{
\colhead{FRB} &
\colhead{R.A.$_{\rm FRB}$} & 
\colhead{Decl.$_{\rm FRB}$} & 
\colhead{R.A.$_{\rm Host}$} & 
\colhead{Decl.$_{\rm Host}$} & 
\colhead{$z$} &
\colhead{Imaging Facility/Instrument} &
\colhead{Filter(s)} & 
\colhead{References}
}
\startdata
20180916B & 01:58:00.75 & 65:43:00.3 & 01:58:00.27 & 65:42:53.2 & 0.034 & \textit{HST}/WFC3-IR & F110W & 1,2 \\
20180924B & 21:44:25.26 & $-$40:54:00.1 & 21:44:25.25 & $-$40:54:00.8 & 0.321 & \textit{HST}/WFC3-IR & F160W & 3,4,5 \\
20190102C & 21:29:39.76 & $-$79:28:32.5 & 21:29:39.58 & $-$79:28:32.5 & 0.291 & \textit{HST}/WFC3-IR & F160W & 4,5,6,7 \\
20190608B & 22:16:04.77 & $-$07:53:53.7 & 22:16:04.90 & $-$7:53:55.8 & 0.118 & \textit{HST}/WFC3-IR & F160W & 4,5,6 \\
20190714A & 12:15:55.13 & $-$13:01:15.6 & 12:15:55.00 & $-$13:01:16.0 & 0.236 & \textit{HST}/WFC3-IR & F160W & 4,5,8 \\
20191001A & 21:33:24.41 & $-$54:44:53.9 & 21:33:24.44 & $-$54:44:54.5 & 0.234 & \textit{HST}/WFC3-IR & F160W & 4,5,7 \\
20200120E & 09:57:54.68 & 68:49:08.0 & 09:55:33.06 & 69:03:54.1 & 0.001 & SDSS & $r$ & 9,10,11 \\
20200430A & 15:18:49.55 & 12:22:34.8 & 15:18:49.53 & 12:22:35.8 & 0.161 & VLT/FORS2 & $I$ & 5,8 \\
20200906A & 03:33:58.93 & $-$14:04:58.8 & 03:33:59.00 & $-$14:04:59.5 & 0.369 & VLT/FORS2 & $I$ & 5,12 \\
20201124A & 05:08:03.51 & 26:03:38.5 & 05:08:03.49 & 26:03:37.8 & 0.098 & \textit{HST}/WFC3-UVIS,WFC3-IR & F475X, F160W & 13,14,15 \\
20210117A & 22:39:55.01 & $-$16:09:05.2 & 22:39:55.07 & $-$16:09:05.4 & 0.214 & VLT/FORS2, HAWK-I & $I$, $K$ & 5,16 \\
20210320C & 13:37:50.10 & $-$16:07:21.6 & 13:37:50.13 & $-$16:07:21.2 & 0.280 & VLT/FORS2 & $I$ & 5,17 \\
20210807D & 19:56:53.11 & $-$00:45:44.1 & 19:56:52.81 & $-$00:45:44.5 & 0.129 & VLT/FORS2, Gemini GeMS+GSAOI & $I$, $K$ & 5,17,18,19 \\
20211127I & 13:19:14.11 & $-$18:50:16.5 & 13:19:13.97 & $-$18:50:16.7 & 0.047 & VLT/FORS2, Gemini GeMS+GSAOI & $I$, $K$ & 5,17,18,19 \\
20211203C & 13:38:15.00 & $-$31:22:49.0 & 13:38:15.03 & $-$31:22:48.5 & 0.344 & VLT/FORS2 & $R$ & 5,17 \\
20211212A & 10:29:24.19 & 01:21:37.5 & 10:29:24.23 & 01:21:38.9 & 0.071 & VLT/FORS2, Gemini GeMS+GSAOI & $I$, $K$ & 5,17,18,19 \\
20220105A & 13:55:12.81 & 22:27:58.4 & 13:55:12.92 & 22:27:59.4 & 0.278 & VLT/FORS2, HAWK-I & $R$, $K$ & 5,17 \\
20220725A & 23:33:15.65 & $-$35:59:24.9 & 23:33:15.69 & $-$35:59:24.9 & 0.192 & VLT/FORS2, HAWK-I & $R$, $K$ & 5 \\
20220918A & 01:10:22.11 & $-$70:48:41.0 & 01:10:22.01 & $-$70:48:41.0 & 0.491 & VLT/FORS2, HAWK-I & $R$, $K$ & 5 \\
20221106A & 03:46:49.15 & $-$25:34:11.3 & 03:46:49.10 & $-$25:34:10.9 & 0.204 & VLT/FORS2, HAWK-I & $R$, $K$ & 5 \\
20230526A & 01:28:55.83 & $-$52:43:02.4 & 01:28:55.85 & $-$52:43:02.6 & 0.157 & VLT/FORS2, HAWK-I & $R$, $K$ & 5 \\
20230708A & 20:12:27.73 & $-$55:21:22.6 & 20:12:27.74 & $-$55:21:22.7 & 0.105 & VLT/FORS2, HAWK-I & $R$, $K$ & 5 \\
20230902A & 03:28:33.55 & $-$47:20:00.6 & 03:28:33.60 & $-$47:20:00.4 & 0.362 & VLT/FORS2, HAWK-I & $R$, $K$ & 5 \\
20231226A & 10:21:27.30 & 06:06:36.9 & 10:21:27.33 & 06:06:35.0 & 0.157 & VLT/FORS2, HAWK-I & $R$, $K$ & 5 \\
20231230D & 07:55:02.63 & 08:30:51.8 & 07:55:02.59 & 08:30:50.8 & 0.506 & VLT/FORS2, HAWK-I & $R$, $K$ & 20, This Work \\
20240117B & 03:35:30.69 & $-$15:51:07.6 & 03:35:30.60 & $-$15:51:08.3 & 0.643 & VLT/FORS2, HAWK-I & $R$, $K$ & 20, This Work \\
20240201A & 09:59:37.34 & 14:05:16.89 & 09:59:37.48 & 14:05:19.4 & 0.043 & VLT/FORS2, HAWK-I & $R$, $K$ & 5 \\
20240203C & 00:35:35.41 & 15:50:27.1 & 00:35:35.43 & 15:50:26.6 & 0.243 & VLT/FORS2, MMT/MMIRS & $R$, $K$ & This Work \\
20240209A & 19:19:33.00 & 86:03:52.0 & 19:19:24.03 & 86:03:39.4 & 0.138 & Gemini/GMOS & $r$ & 21,22 \\
20240210A & 00:35:07.10 & $-$28:16:14.7 & 00:35:06.52 & $-$28:16:19.3 & 0.024 & VLT/FORS2, HAWK-I & $R$, $K$ & 5 \\
20240304A & 09:05:19.40 & $-$16:09:59.9 & 09:05:19.33 & $-$16:09:58.4 & 0.242 & VLT/FORS2, HAWK-I & $R$, $K$ & 5 \\
20240310A & 01:10:29.25 & $-$44:26:21.9 & 01:10:29.25 & $-$44:26:21.5 & 0.127 & VLT/FORS2, HAWK-I & $R$, $K$ & 5 \\
20240312D & 03:26:29.61 & $-$54:36:03.5 & 03:26:29.43 & $-$54:35:56.8 & 0.049 & SOAR/Goodman & $r$ & This Work \\
20240318A & 10:01:34.36 & 37:36:58.9 & 10:01:34.45 & 37:36:58.7 & 0.112 & Keck/LRIS, MMT/MMIRS & $R$, $K$ & 5, This Work \\
20240525A & 00:27:07.89 & $-$06:53:22.6 & 00:27:07.89 & $-$06:53:21.5 & 0.327 & VLT/FORS2 & $R$ & This Work \\
20240615B & 02:15:51.27 & $-$14:37:28.2 & 02:15:51.21 & $-$14:37:28.6 & 0.073 & VLT/FORS2, HAWK-I & $R$, $K$ & This Work \\
20241027B & 02:24:06.96 & $-$20:42:12.7 & 2:24:07.29 & $-$20:42:13.4 & 0.336 & SOAR/Goodman & $r$, $i$ & This Work \\
\enddata
\tablecomments{Details of the FRBs, host galaxies, and imaging data used in this work. All coordinates are given in the J2000 system. The host galaxy coordinates were derived via Source Extractor \citep{source_extractor} using the optical images, when available.}
References:
1. \citet{Marcote_etal_2020},
2. \citet{Tendulkar+21},
3. \citet{Bannister+19},
4. \citet{Mannings+21},
5. \citet{Shannon+25},
6. \citet{mpm+20},
7. \citet{Bhandari+20},
8. \citet{Heintz+20},
9. \citet{Bhardwaj21},
10. \citet{SDSS},
11. \citet{Kirsten21},
12. \citet{Bhandari+22},
13. \citet{Dong+24},
14. \citet{Fong+21},
15. \citet{Nimmo_201124},
16. \citet{Bhandari+23},
17. \citet{Gordon+23},
18. Deller et al. in prep.,
19. \citet{Woodland+24},
20. \citet{Muller+25},
21. \citet{Eftekhari+25},
22. \citet{Shah+25}
\end{deluxetable*}

\section{Observations and Data Analysis} \label{sec:imaging}
\subsection{Optical and Near-infrared Imaging}
For this work, we require deep optical and near-infrared (NIR) imaging to explore if there is a wavelength dependence on model preferences (as each filter probes different stellar populations). For every FRB host in our sample, we obtain or compile (from the literature) the deepest available imaging in both $R$- and $K$-bands. In cases where deep imaging was available but obtained with filters other than $R$ and $K$, we select images in immediately adjacent filters (e.g., $I$ or F160W); this occurs for 17 images in our sample. For 13 hosts, sufficiently deep imaging was not available in both wavelength regimes, so we proceed with only one image.

We newly present optical and NIR imaging, plus spectroscopy, of the hosts of ASKAP/CRAFT FRBs\,20231230D, 20240117B, 20240203C, 20240312D, 20240318A, 20240525A, 20240615B, and 20241027B through programs on the 8.2~m Very Large Telescope (PI Shannon, Program ID 108.21ZF) with FORS2 (\citealt{FORS2}), HAWK-I+GRAAL (\citealt{HAWKI,GRAAL}), and X-shooter (\citealt{X-shooter}); the 10~m Keck~I Telescope (PI Deller, Program ID W366) with LRIS (\citealt{LRIS}); the 6.5~m MMT (PI Fong, Program ID UAO-G213-24B) with MMIRS (\citealt{MMIRS}) and Binospec (\citealt{Binospec}); and the 4~m Southern Astrophysical Research Telescope (SOAR; PI Gordon, Program ID SOAR2024B-028) with the Goodman spectrograph (\citealt{SOAR}). We list all relevant details pertaining to these observations in Table~\ref{tab:new_obs}. 

We reduce all VLT data using \texttt{ESO Reflex} \citep{esoreflex} and the \texttt{CRAFT-OPTICAL-FOLLOWUP}\footnote{\url{https://github.com/Lachimax/craft-optical-followup/}} pipeline, as detailed in \citet{Marnoch+23}. This pipeline performs bias and flat-field corrections, image coaddition and mosaicing, astrometric calibration with respect to Gaia DR3 \citep{GaiaDR3}, and photometric calibration relative to the FORS2 Quality Control archive for the optical data and the 2MASS Point Source Catalog for the NIR data. For the data obtained with Keck and MMT, we use the \texttt{POTPyRI}\footnote{\url{https://github.com/CIERA-Transients/POTPyRI}} imaging reduction pipeline. \texttt{POTPyRI} performs flat-field, bias, and dark frame corrections and applies a World Coordinate System (WCS) calibrated to the Gaia DR3 catalog. The calibrated images are then stacked, and the pipeline calculates a zeropoint via automatic PSF photometry calibrated against Pan-STARRS DR1 in the optical or 2MASS in the infrared. For the SOAR imaging, we reduce our data with the \texttt{photpipe} pipeline \citep{Rest+05}. \texttt{photpipe} applies flat-field and bias corrections, then solves the WCS with reference to Gaia DR3. The frames are sky-subtracted and stacked using \texttt{swarp}, and PSF photometry is performed using a custom version of \texttt{DoPhot} \citep{Schechter93}. The pipeline then calculates the AB magnitude zeropoint of the final stacked image with reference to SkyMapper DR2. In total, we present nine new optical images and five new NIR images.

\begin{deluxetable*}{l|cccccc}[t]
\tablewidth{0pc}
\tablecaption{New Observational Data
\label{tab:new_obs}}
\tablehead{
\colhead{FRB} &
\colhead{Facility/Instrument} & 
\colhead{Observation Date} &
\colhead{Exp. Time [s]}  & 
\colhead{Grating/Grism or Filter} &
\colhead{Slit Width [\arcsec]} & 
\colhead{Program ID}
}
\startdata
20231230D & VLT/FORS2 & 2024 Jan 13 UTC & 4000 & $R$ &  & 108.21ZF \\
          & VLT/HAWK-I & 2024 Mar 2 UTC & 2400 & $K$ &  & 108.21ZF \\
\hline
20240117B & VLT/FORS2 & 2024 Feb 4 UTC & 4000 & $R$ &  & 108.21ZF \\
          & VLT/HAWK-I & 2024 Mar 1 UTC & 2400 & $K$ &  & 108.21ZF \\
\hline
20240203C & VLT/FORS2 & 2024 July 2 UTC & 4000 & $R$ &  & 108.21ZF \\
          & MMT/MMIRS & 2024 Sept 15 UTC & 2400 & $K$ &  & UAO-G213-24B \\
          & Keck/LRIS & 2024 Oct 5 UTC & 3690, 3600$^{a}$ & B400/3400, R400/8500$^{a}$ & 1.0 & O406 \\
          & VLT/X-shooter & 2024 July 13 UTC & 2468, 2516, 2400$^{b}$ &  & 1.3, 1.2, 1.2$^{b}$ & 108.21ZF \\
\hline
20240312D & SOAR/Goodman & 2024 Dec 24 UTC & 3600 & $r$ &  & SOAR2024B-028 \\
          & SOAR/Goodman & 2024 Mar 19 UTC & 1200 & R400, M1 & 1.0 & SOAR2024A-002 \\
\hline
20240318A & Keck/LRIS & 2025 Jan 2 UTC & 1800 & $R$ &  & W366 \\
          & MMT/MMIRS & 2025 Jan 13 UTC & 2400 & $K$ &  & UAO-G213-24B \\
          & MMT/Binospec & 2024 Nov 29 UTC & 7200 & 270 & 1.0 & UAO-G213-24B \\
\hline
20240525A & VLT/FORS2 & 2024 July 4 UTC & 4000 & $R$ &  & 108.21ZF \\
          & SOAR/Goodman & 2024 July 9 UTC & 3600 & R400, M2 & 1.0 & SOAR2024A-002 \\
          & VLT/X-shooter & 2024 July 13 UTC & 2468, 2516, 2400$^{b}$ &  & 1.3, 1.2, 1.2$^{b}$ & 108.21ZF \\
\hline
20240615B & VLT/FORS2 & 2024 Aug 7 UTC & 4000 & $R$ &  & 108.21ZF \\
          & VLT/HAWK-I & 2024 July 28 UTC & 2400 & $K$ &  & 108.21ZF \\
          & VLT/X-shooter & 2024 Aug 1 UTC & 2468, 2516, 2400$^{b}$ &  & 1.3, 1.2, 1.2$^{b}$ & 108.21ZF \\
\hline
20241027B & SOAR/Goodman & 2024 Nov 7 UTC & 1200 & $r$ & & SOAR-2024B-028 \\
          & SOAR/Goodman & 2024 Nov 7 UTC & 1200 & $i$ & & SOAR-2024B-028 \\
          & SOAR/Goodman & 2024 Nov 13 UTC & 3600 & R400, M2 & 1.0 & SOAR-2024B-028 \\
          & Keck/LRIS    & 2025 Jan 2 UTC & 2430, 2400$^{a}$ & B400/3400, R400/8500$^{a}$ & 1.0 & O406 \\
\enddata
\tablecomments{
$^{a}$ Values correspond to blue and red detectors, in order. \\
$^{b}$ Values correspond to UVB, VIS, and NIR arms, in order. \\
}
\end{deluxetable*}

\subsection{Literature and Survey Data}

The majority of the imaging data in this work comes from \citet{Shannon+25} who reported 43 FRBs detected by the CRAFT incoherent-sum survey; notably for our work, this included the first compilation of imaging from the \textbf{F}ast and \textbf{U}nbiased F\textbf{RB} Host Galax\textbf{Y} Survey (FURBY) Large Programme on the VLT (PI Shannon, Program ID 108.21ZF). Through this program, we obtained uniform and deep imaging in the $R$- and $K$-bands using FORS2, along with HAWK-I and its ground layer adaptive optics system GRAAL, respectively, as well as spectroscopy with X-shooter. Combined, these data create a homogeneous host galaxy dataset for select, well-localized CRAFT FRBs. Further details of the FURBY program, its goals and selection criteria, are detailed in \citet{Muller+25}. All 17 of the FRBs in our sample from 2022 onwards, with the exception of FRBs 20240209A, 20240312D, 20240318A, and 20241027B, were observed as part of the FURBY program.

We supplement the rich FURBY dataset with VLT/FORS2 $R$- and $I$-band imaging of CRAFT FRBs obtained prior to the FURBY program (PI Macquart, Program ID 105.204W, \citealt{Shannon+25}); \textit{Hubble Space Telescope (HST)} Wide Field Camera 3 (WFC3) NIR imaging in the F110W and F160W filters \citep{Mannings+21,Tendulkar+21,Dong+24}; \textit{HST}/WFC3-UVIS F475X imaging \citep{Dong+24}; $r$-band imaging from the Gemini Multi-Object Spectrograph (GMOS; \citealt{GMOS}) on Gemini North \citep{Eftekhari+25,Shah+25}; and $K$-band imaging taken with the Gemini South Adaptive Optics Imager (GSAOI, in combination with the Gemini Multi-conjugate adaptive optics System, GeMS; \citealt{GSAOI,GeMS}) from \citet{Woodland+24}. We use archival SDSS DR12 $r$-band imaging for the host of FRB\,20200120E, using \texttt{swarp} \citep{swarp} to create a stacked mosaic that encompasses the galaxy and FRB location. In total, we compile 23 optical and 23 NIR images from the literature (or archival surveys in the case of FRB\,20200120E), which we detail in Table~\ref{tab:sample}. 

\subsection{Astrometry and Positional Uncertainties} \label{sec:astrom}

To quantify the total uncertainty in the locations of the FRBs within their host galaxies, we compile or determine all relevant sources of uncertainty: FRB statistical and systematic uncertainties, the absolute astrometric uncertainty of the images, and the host positional uncertainty. In all cases, the uncertainty represented by the FRB localization is the dominant term in the total positional uncertainty. We report the individual uncertainties for each FRB (and image, when relevant) in Appendix~\ref{sec:astrometry_appendix} (Table~\ref{tab:astrometry}). References for the FRB statistical and systematic uncertainties can be found in Table~\ref{tab:sample}.

While the image reduction pipelines provide a robust astrometric solution in most cases, we find that three images needed further refinement after visual inspection relative to Gaia DR3 (e.g., common sources out of alignment by $\gtrsim 0.2\arcsec$). In these three cases (FRBs\,20221106A, 20230526A, and 20240201A, all $R$-band), we derive a new solution with respect to the Gaia DR3 catalog using a custom script that implements the \texttt{match\char`_coordinates\char`_sky} and \texttt{fit\char`_wcs\char`_from\char`_points} functions from \texttt{astropy} \citep{astropy}, ensuring that common sources are not saturated. Since there are optical and infrared images for most hosts, we first perform an astrometric tie between the optical image and Gaia (or use its existing astrometric solution which is also relative to Gaia). We then perform an astrometric tie between the optical and infrared images directly, as applicable.

We next estimate the positional uncertainty of the host galaxy coordinates by measuring their light centroids with \texttt{Source Extractor} \citep{source_extractor}. These terms are on order tens of milliarcseconds and are a minor contribution to the total positional uncertainty.

Finally, we combine each of these sources of uncertainty to derive the total uncertainty for each FRB on its respective image in quadrature, reporting the combined values in Table~\ref{tab:astrometry}. In all imaging figures in this work, we plot the total FRB uncertainties, represented by ellipses.

\subsection{Spectroscopy and Redshift Determination} \label{sec:spec}

We present long-slit optical spectroscopy of the hosts of FRBs\,20240203C, 20240312D, 20240318A, 20240525A, 20240615B, and 20241027B from which we derive their redshifts. These data were taken with the VLT (PI Shannon, Program ID 108.21ZF), Keck I (PI Liu, Program ID O406), MMT (PI Fong, Program ID UAO-G213-24B; \citealt{Binospec}), and SOAR (PI Gordon, Program ID SOAR2024A-002 and SOAR2024B-028). We list all relevant instruments and observational details in Table~\ref{tab:new_obs}.

To reduce our Keck and SOAR data, we use the Python Spectroscopic Data Reduction Pipeline (\texttt{PypeIt}; \citealt{pypeit:joss_pub,pypeit:zenodo}). \texttt{PypeIt} performs bias-subtraction, flat-fielding, cosmic ray masking, and wavelength calibrations on the individual frames. After extracting the 1D spectra corresponding to our host galaxies, we performed coaddition, flux calibration with a spectrophotometric standard star, and telluric correction. For the VLT data, we use \texttt{ESOReflex} for reduction \citep{esoreflex} which performs bias-subtraction and flat-fielding, followed by wavelength calibration and spatial distortion mapping along each \'{e}chelle order. After rectification, the orders are merged and a response function is determined from spectrophotometric standard star observations. Finally we extracted calibrated 1D spectra of the VLT data using \texttt{IRAF} \citep{IRAF}.

We determine the redshifts of the hosts via visual inspection of emission and absorption features which we detail in Appendix~\ref{sec:spec_appendix} (Table~\ref{tab:spec}). The uncertainty on the redshifts are on the order of $\pm0.001$.

Details of the spectroscopic observations, reductions, and redshift derivations for FRBs 20231230D and 20240117B are presented in \citet{Muller+25}. Redshifts for all other FRBs are taken from the literature (e.g., \citealt{Shannon+25}). 

\section{Methods} \label{sec:methods}

\subsection{Surface Brightness Profile Fitting} \label{sec:galfit}

To model the surface brightness profile of the FRB host galaxies and reveal any residual substructure, we use the light profile modeling software \texttt{Galfit} \citep{galfit_2002,galfit_2010}. \texttt{Galfit} inputs include the point spread function (PSF), science image, and a bad pixel mask (as applicable). It then produces a 2D surface brightness profile model of the FRB host galaxy and returns a residual image.

To determine the PSF for each science image, we use the \texttt{IRAFStarFinder} module of \texttt{photutils} \citep{photutils} to select the 50 brightest point sources within the image, with cuts to exclude saturated objects. After visually inspecting the selected sources to ensure there are no nearby contaminating objects, we generate the PSF using the \texttt{photutils} \texttt{EPSFBuilder} module.

We next trim the science images to only include $\approx$10--30\arcsec\ about the FRB host galaxies, adjusting the size to encompass sufficient sky background. If there are objects other than the host within the trimmed image, we create a bad pixel mask to exclude them from consideration during modeling. For the creation of the mask, we use the \texttt{SourceFinder} method of \texttt{photutils} to select all $>\!5\sigma$ sources that span a minimum of five connected pixels to generate a segmentation image. Then, we use the \texttt{make\char`_source\char`_mask} method to convert the segmentation image into a bad pixel mask.

We use a standard S\'ersic profile (\citealt{Sersic_1968}), given by

\begin{equation}
I(r) = I_e \times {\rm exp}(-b_n[(r/r_e)^{1/n} - 1]),
\end{equation}

\noindent where $I_e$ is the surface brightness at the effective radius $r_e$, $n$ is the S\'ersic index which describes the concentration of light, and $b_n$ is a coefficient related to $n$ that ensures half of the light is contained within $r_e$. We run \texttt{Galfit} with initial guesses for the following free parameters: the galaxy's central position, apparent magnitude, $r_e$, $n$, semi-minor/semi-major axis ratio ($b/a$), and position angle. Additionally, we include a sky background model with the default settings (i.e., sky background at the center of the fitting region is free, sky gradients in x and y are fixed to zero).  

For the majority of hosts in our sample, we fit a single S\'ersic profile allowing all parameters to be free during modeling, initialized with $n=1$. For six images (across four FRBs), \texttt{Galfit} returned non-physical values for the morphological properties or failed to converge within 100 iterations under these assumptions. For five of these images, we fix $n=1$, corresponding to an exponential disk profile\footnote{We additionally test fixing $n=4$, but the models provided much poorer fits in terms of $\chi^2_{\nu}$ than at $n=1$.}. For the remaining image, we find $n=4$ provides a better fit to the data than a fixed $n=1$ model; this is summarized in Table~\ref{tab:host_props}.

For hosts with visible disk and bulge components, we include a second S\'ersic profile; for 12 images, this returned a better model fit than using a single profile. For eight of these 12 images, we obtain the best model fit leaving all parameters free. However, the remaining four images within this category required fixing $n$ for one or both components to avoid non-physical values for the morphological properties or cases in which the residual image was visibly over-subtracted for large regions of the galaxy. We summarize our assumptions in Table~\ref{tab:host_props}.  

\begin{figure*}
\centering
\includegraphics[width=\textwidth]{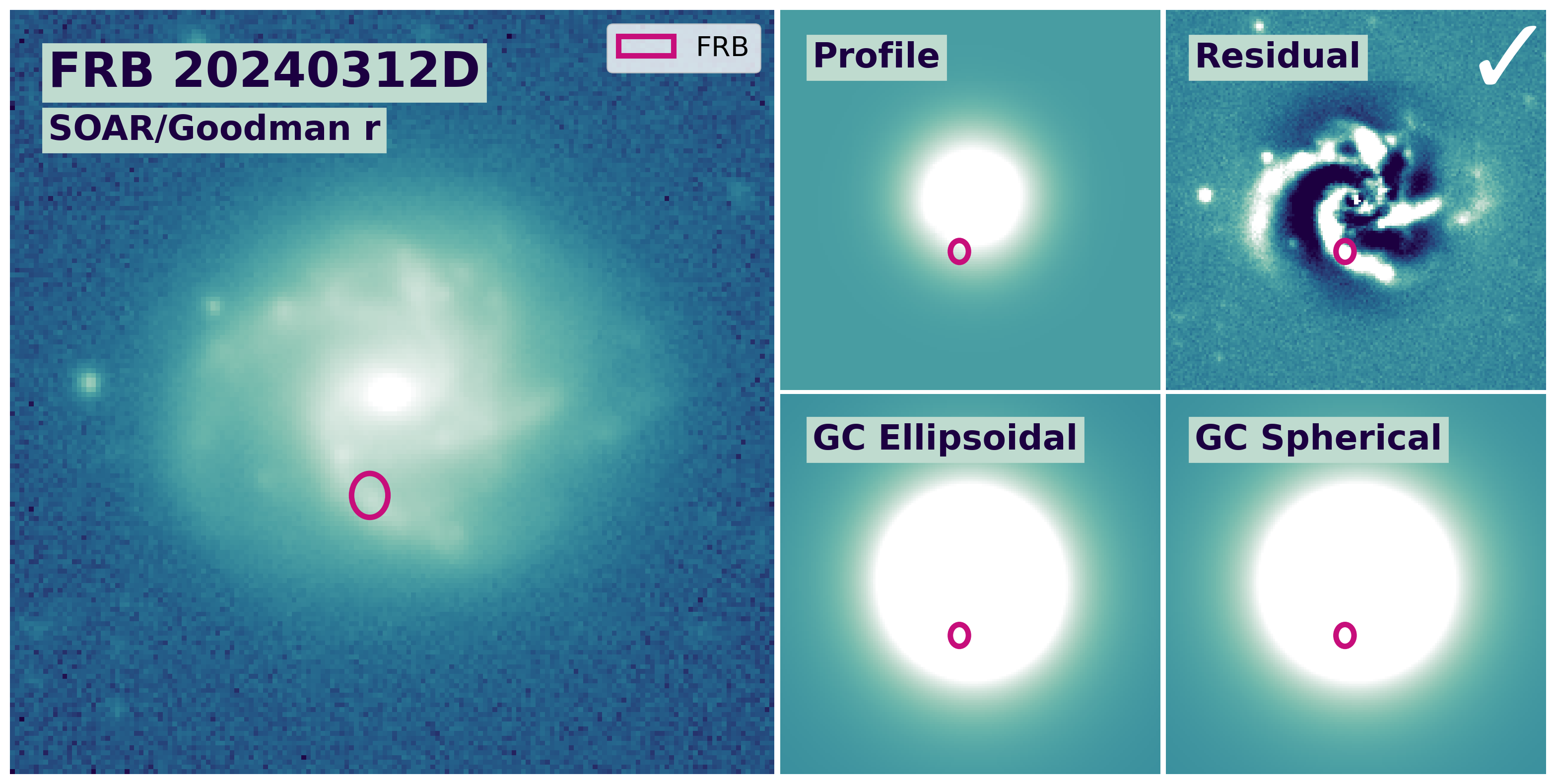}
\includegraphics[width=\textwidth]{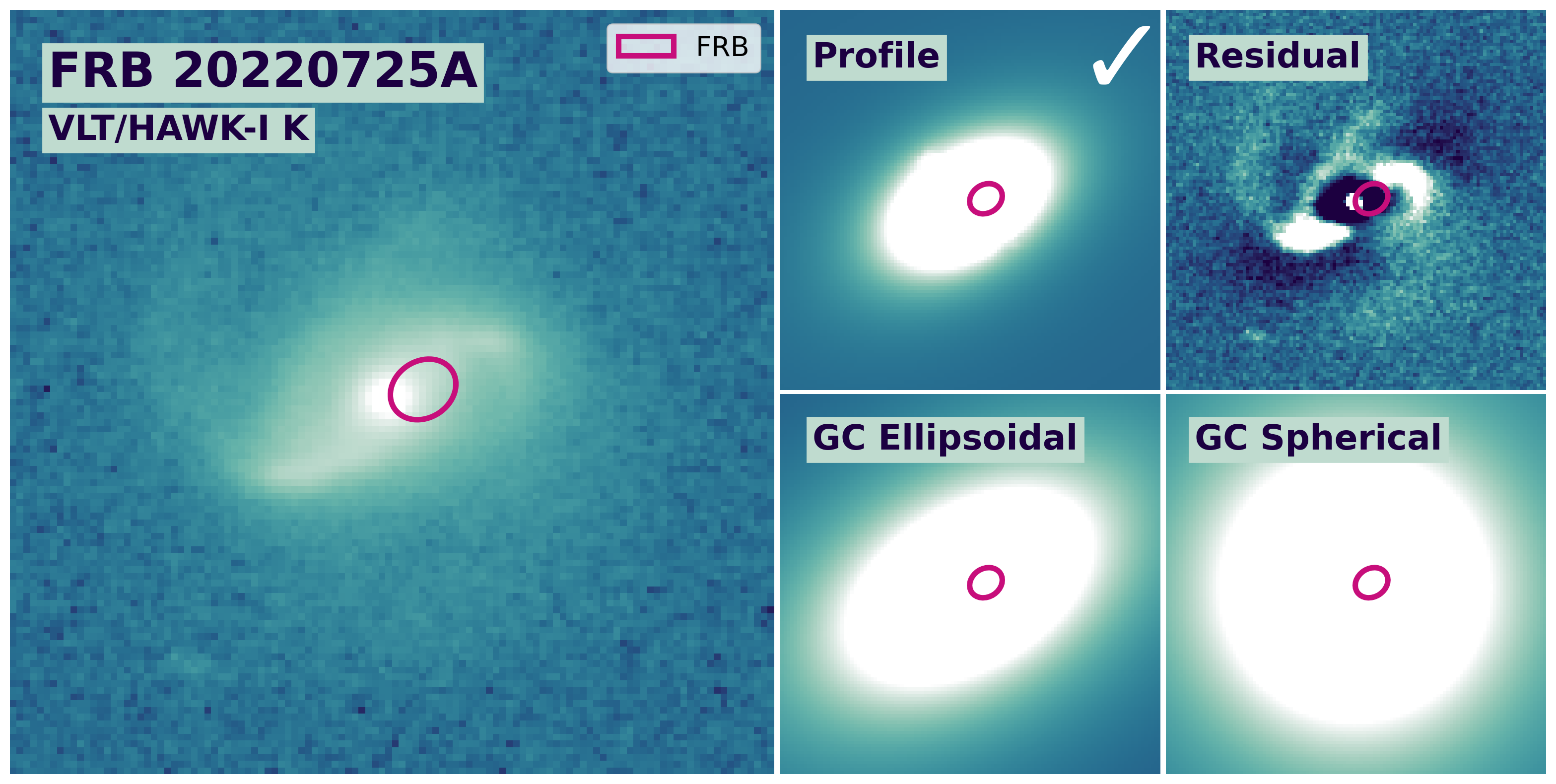}
\caption{Imaging and light profiles considered for two example events: FRBs\,20240312D (top panels) and 20220725A (lower panels), along with the FRB positions ($1\sigma$ total uncertainty; magenta ellipses). The smaller panels represent the \texttt{Galfit} surface brightness profile model (``Profile''), residual image (``Residual''), and 2D synthetic globular cluster distribution models (``GC Ellipsoidal''; ``GC Spherical''); the production of these images is described in Sections~\ref{sec:galfit}-\ref{sec:GC_maps}. We denote the preferred model for each FRB with a white check-mark. The images are oriented with North up and East to the left and span $\approx1$\arcmin\ on a side. It is clear that the GC models extend to larger radii, representative of their more extended spatial distribution compared to stars. The host of FRB\,20240312D has a face-on orientation, with no apparent difference in the two GC models, while the effects of the host inclination on the GC distributions can clearly be seen for the host of FRB\,20220725A.} 
    \label{fig:galfit_GC_example}
\end{figure*}

In all cases, we select the best-fit model by a combination of the $\chi^2_{\nu}$ statistic of the fit, visual inspection of the residual, and physicality of the derived morphological properties. For every host, we report the $r_e$ and $n$ values in Table~\ref{tab:host_props}, noting the corresponding disk and bulge parameters when applicable. In the top row (right) of Figure~\ref{fig:galfit_GC_example}, we show the best-fit \texttt{Galfit} model and residual image for the host of FRB\,20240312D. The FRB's placement on a spiral arm is more strongly apparent in the residual frame compared to the input science image (leftmost panel). We also show the VLT/HAWK-I image of FRB\,20220725A and its corresponding best-fit models for an example of a host with a smaller $b/a$. We present the science images, models, and residuals for the remainder of our sample in Figure~\ref{fig:all_models} (Appendix~\ref{sec:models_appendix}).

\begin{figure*}
    \centering
    \includegraphics[width=\textwidth]{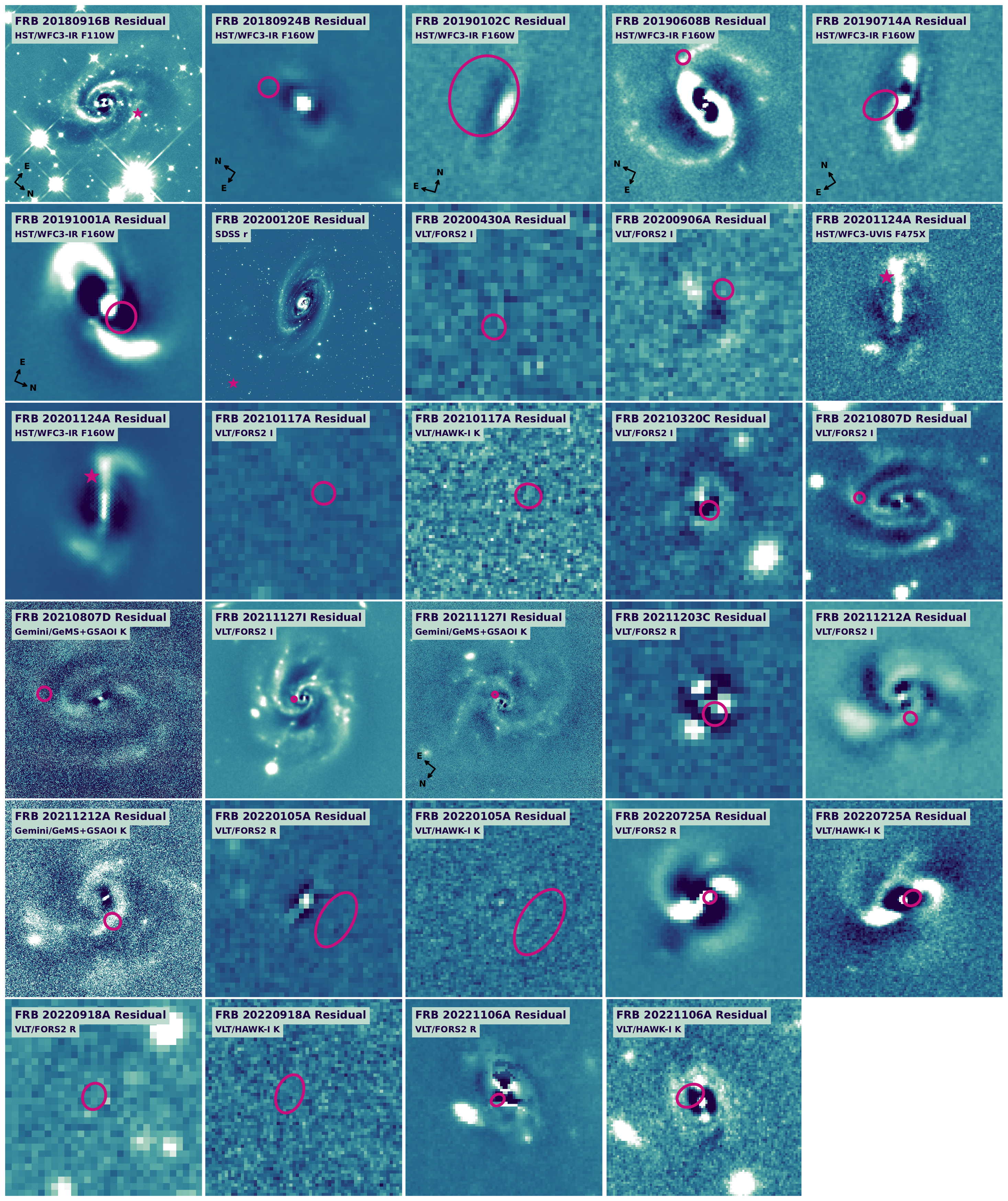}
    \caption{Compilation of residual images for our sample of FRB hosts, derived via \texttt{Galfit} surface brightness profile modeling (Section~\ref{sec:galfit}), along with the FRB localizations ($1\sigma$ total uncertainty; magenta ellipses). For clarity, FRBs with milliarcsecond-scale localizations are denoted by magenta stars as the localization is much smaller than the size of the symbol. The images have been scaled to highlight features of interest but span a range of sizes ($\approx$10\arcsec$-$0.75$^{\circ}$ across). The images are oriented with North up and East to left unless otherwise noted. There is broad diversity in the observed morphology of FRB hosts, with all at $z\lesssim 0.1$ showing clear spiral structure.}
    \label{fig:resid_comp1}
\end{figure*}

In Figure~\ref{fig:resid_comp1}, we show the residuals of each FRB host image in our sample. Twenty FRBs (30 images) show clear spiral structure, 8 FRBs (10 images) have ambiguous residual substructure that is not immediately apparent as spiral, and the remaining 9 FRBs (18 images) show no substructure, indicating that the smooth profile fit is an apt descriptor of the available data. Six FRBs return mixed results across the optical and NIR bands. Two hosts in our sample are purely described by a single elliptical profile and do not show evidence for needing a disk: FRB\,20240209A, as previously stated in \citet{Eftekhari+25}, and FRB\,20241027B, newly presented here. The remaining 35 hosts comprise spiral or disk galaxies with derived $n\approx1-2$. Among the spiral population, we find a diversity of morphologies upon visual inspection (Figure~\ref{fig:resid_comp1}), ranging from grand design (e.g., FRB\,20231226A) to multi-arm (e.g., FRB\,20240312D), flocculent (e.g., FRB\,20211127I) or barred (e.g., FRB\,20201124A) spirals. The FRB positions also vary widely in their relation to spiral substructure: some are directly on spiral arms, some are in close proximity to arms but not fully coincident, and others are clearly offset (e.g., FRB\,20191001A); we further quantify these trends in Section~\ref{sec:coflash_results}.

\renewcommand{\thefigure}{\arabic{figure} (Cont.)}
\addtocounter{figure}{-1}

\begin{figure*}
    \centering
    \includegraphics[width=\textwidth]{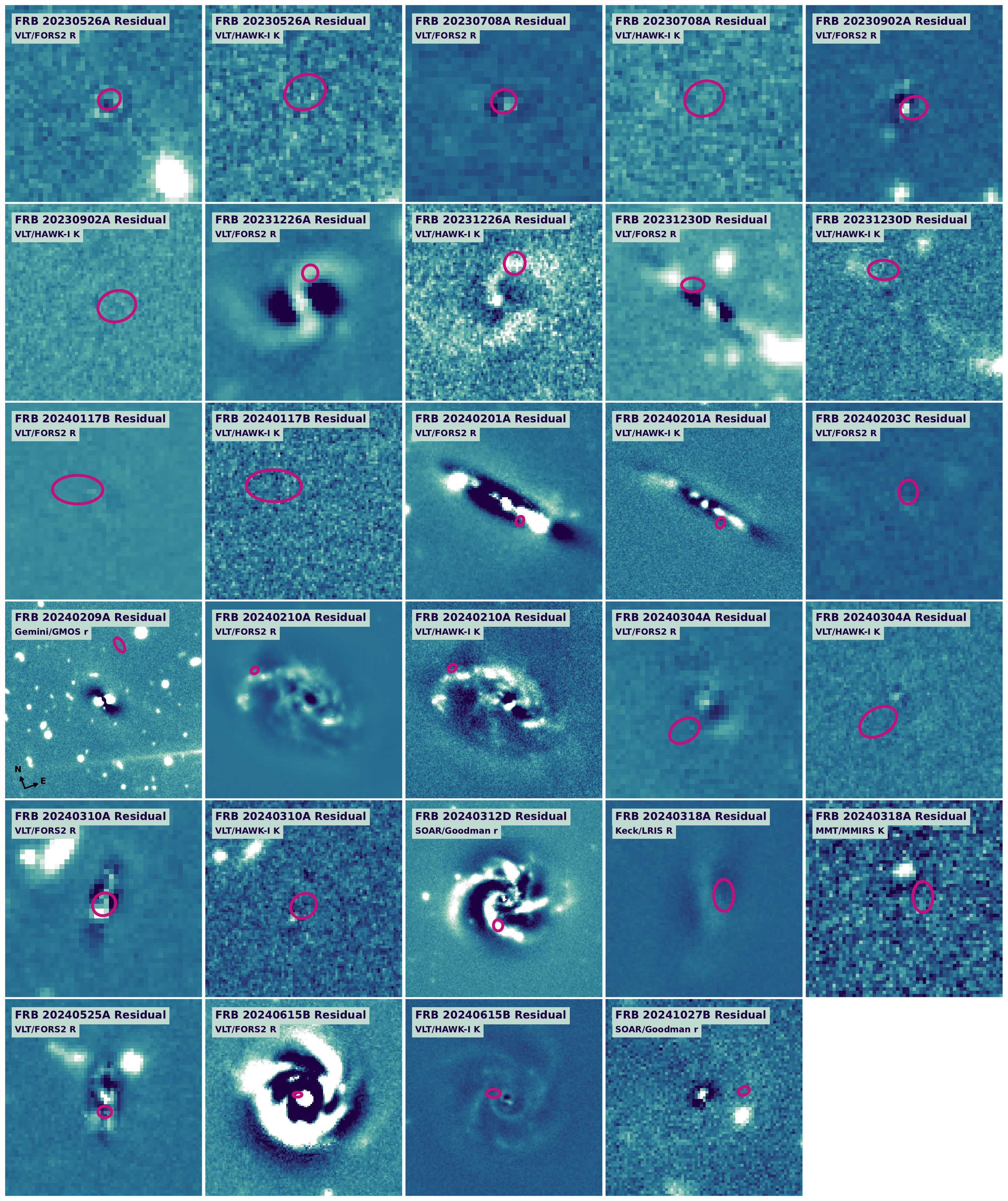}
    \caption{}
    \label{fig:resid_comp2}
\end{figure*}
\renewcommand{\thefigure}{\arabic{figure}}

\subsection{Synthetic Globular Cluster Spatial Distributions} \label{sec:GC_maps}

To test if the FRB positions are consistent with the spatial distribution of globular clusters (GCs) around their hosts, we next create synthetic spatial distributions of GCs given the properties of each host galaxy. As the vast majority of our sample is beyond the distance for which GCs can be directly observed (a few tens of Mpc; \citealt{vanDokkum18,Beasley2020}), we must infer where the GCs are most likely to exist. We adapt the work of \citet{Lim+24} who modeled the spatial distribution of directly-observed globular clusters around 118 nearby early-type galaxies\footnote{While our sample contains a majority of late-type galaxies, no similar large-scale work has been done to map the location of GCs about late-types due to the difficulty in robustly distinguishing GCs from star-forming regions and other inhomogeneities in the stellar disk. However, for the few galaxies where this work has been done, the M$_*$ and $r_{e,\rm GC}$ values appear to follow the same trends as the early-type galaxies. Thus, our adoption of the \citet{Lim+24} scaling relation is a reasonable extrapolation. We discuss the effects of this assumption further in Section~\ref{sec:BIC}.}. \citet{Lim+24} demonstrate that the spatial distribution of GCs correlates with host galaxy properties and follows a S\'ersic profile. In particular, we use their equation~8 which relates the effective radius of the GC spatial distribution, $r_{e,\rm GC}$, to the stellar mass of the host galaxy, M$_*$. 

Half of the galaxies in our sample have published stellar masses derived from spectral energy distribution (SED) modeling \citep{Bhardwaj21,Gordon+23,Eftekhari+25}. For the 17 hosts without published masses, we lack sufficient data to derive their stellar masses from SED modeling, and thus use a $K$-band mass-to-light ratio of 0.6 M$_{\odot}$/L$_{\odot}$ \citep{McGaugh+Schombert2014}; we find this method gives comparable results to the SED-modeled hosts for which both estimates are available\footnote{The published scatter in the mass-to-light relation is $\sim$0.1$-$0.3 dex \citep{McGaugh+Schombert2014,Conroy2013} which is also the difference we find between methods.}. We perform aperture photometry on $K$-band images of seven of these hosts using a custom script\footnote{https://github.com/charliekilpatrick/photometry} to derive their magnitudes. For the remaining ten, we use the $K$-band AB magnitudes reported in \citet{Shannon+25}. We derive the stellar mass for FRBs\,20241027B and 20230708A (see \citet{Muller+25} for the latter), using the SED modeling code \texttt{Prospector} \citep{Johnson+21} to jointly fit available photometry and spectroscopy. We report all M$_*$ and $r_{e,\rm GC}$ values in Table~\ref{tab:host_props}, noting which masses are newly derived here. 

With this information, we construct the 2D synthetic GC distributions (which is primarily set by $r_{e,\rm GC}$) using the \texttt{Sersic2D} class of \texttt{astropy} \citep{astropy} to generate the models. As the 3D distribution of GCs about their hosts is difficult to measure and thus not well understood\footnote{Further, we lack specific 3D information for our hosts as they are too distant to actually resolve clusters except for FRB\,20200120E.}, we create two GC models representing two possible 3D representations: ``GC Ellipsoidal'' and ``GC Spherical''. The GC Ellipsoidal model assumes that the GCs follow the orientation and ellipticity of the stellar disk; we use the position angle and $b/a$ from our surface brightness profile fitting (Section~\ref{sec:galfit}) to set these values. The GC Spherical model instead assumes the GCs are distributed spherically symmetrically about their host and is agnostic to the axis ratio and position angle. For both models, we assume $n=2$, adopting the median value of the GC distributions studied in \citet{Lim+24}, and $I_{e,\rm GC}=1$. In Figure~\ref{fig:galfit_GC_example}, we show the GC models associated with FRBs\,20240312D and 20220725A, representing two cases for different $b/a$ to demonstrate the changes induced in the GC models. We show the GC models for the remainder of our sample in Figure~\ref{fig:all_models} (Appendix~\ref{sec:models_appendix}).

\startlongtable
\begin{deluxetable*}{l|ccccccccc}
\tabletypesize{\footnotesize}
\tablewidth{0pc}
\tablecaption{Host Galaxy Properties
\label{tab:host_props}}
\tablehead{
\colhead{FRB} &
\colhead{Filter} & 
\colhead{$r_{e}$} &
\colhead{$r_{e}$} &
\colhead{$n$} & 
\colhead{log(M$_*$/M$_{\odot}$)} &
\colhead{$r_{e,\rm GC}$} &
\colhead{$r_{e,\rm GC}$} &
\colhead{Mass Reference} & \\
\colhead{} &
\colhead{} &
\colhead{[arcsec]} &
\colhead{[kpc]} &
\colhead{} & 
\colhead{} & 
\colhead{[arcsec]} &
\colhead{[kpc]} &
\colhead{}
}
\startdata
20180916B & F110W & 8.82, 2.80 & 5.98, 1.90 & 0.60, 2.13 & 9.91 & 5.40 & 3.66 & \citet{Gordon+23} \\
20180924B & F160W & 0.47 & 2.22 & 1.62 & 10.39 & 1.13 & 5.34 & \citet{Gordon+23} \\
20190102C & F160W & 0.97 & 4.27 & 0.56 & 9.69 & 0.70 & 3.08 & \citet{Gordon+23} \\
20190608B & F160W & 1.80, 0.57 & 3.87, 1.23 & 1$^{a}$, 3.79 & 10.56 & 2.84 & 6.12 & \citet{Gordon+23} \\
20190714A & F160W & 0.93 & 3.54 & 0.81 & 10.22 & 1.23 & 4.67 & \citet{Gordon+23} \\
20191001A & F160W & 1.07 & 4.03 & 1.24 & 10.73 & 1.93 & 7.28 & \citet{Gordon+23} \\
20200120E & $r$ & 246.73, 64.80 & 4.13, 1.09 & 1$^{a}$, 4$^{a}$ & 10.86 & 542.65 & 9.09 & \citet{Bhardwaj21} \\
20200430A & $I$ & 0.57 & 1.59 & 1.22 & 9.30 & 0.81 & 2.27 & \citet{Gordon+23} \\
20200906A & $I$ & 1.48 & 7.65 & 0.73 & 10.37 & 1.01 & 5.25 & \citet{Gordon+23} \\
20201124A & F475X & 0.97 & 1.77 & 0.42 & 10.22 & 2.55 & 4.67 & \citet{Gordon+23} \\
 & F160W & 0.95 & 1.75 & 0.96 & & & \\
20210117A & $I$ & 0.30 & 1.07 & 1$^{a}$ & 8.59 & 0.37 & 1.30 & \citet{Gordon+23} \\
 & $K$ & 0.56 & 1.96 & 1$^{a}$ & & & \\
20210320C & $I$ & 0.89 & 3.83 & 1.33 & 10.37 & 1.22 & 5.25 & \citet{Gordon+23} \\
20210807D & $I$ & 4.80, 1.42 & 11.18, 3.31 & 0.52, 2.92 & 10.97 & 5.19 & 12.10 & \citet{Gordon+23} \\
 & $K$ & 4.07, 1.64 & 9.49, 3.82 & 0.31, 2.05 & & & \\
20211127I & $I$ & 8.57, 0.93 & 7.97, 0.86 & 1$^{a}$, 1.04 & 9.48 & 2.81 & 2.62 & \citet{Gordon+23} \\
 & $K$ & 3.89 & 3.62 & 2.18 & & & \\
20211203C & $R$ & 0.46 & 2.28 & 0.60 & 9.76 & 0.66 & 3.26 & \citet{Gordon+23} \\
20211212A & $I$ & 3.86, 1.89 & 5.26, 2.58 & 0.21, 2.03 & 10.28 & 3.59 & 4.90 & \citet{Gordon+23} \\
 & $K$ & 3.85, 2.26 & 5.25, 3.08 & 0.23, 2.61 & & & \\
20220105A & $R$ & 0.79 & 3.37 & 0.75 & 10.01 & 0.93 & 3.96 & \citet{Gordon+23} \\
 & $K$ & 0.74 & 3.16 & 1.10 & & & & \\
20220725A & $R$ & 1.47 & 4.76 & 1.16 & 11.26 & 8.87 & 28.74 & This Work \\
 & $K$ & 1.95 & 6.31 & 2.60 & & & \\
20220918A & $R$ & 0.28 & 1.71 & 0.10 & 9.85 & 0.57 & 3.49 & This Work \\
 & $K$ & 0.39 & 2.39 & 2.04 & & & \\
20221106A & $R$ & 1.69 & 5.75 & 2.22 & 11.27 & 8.50 & 28.86 & This Work \\
 & $K$ & 1.62 & 5.51 & 2.65 & & & & \\
20230526A & $R$ & 0.75 & 2.05 & 0.32 & 10.73 & 2.66 & 7.30 & This Work \\
 & $K$ & 0.27 & 0.75 & 1.41 & & & \\
20230708A & $R$ & 0.31 & 0.59 & 1$^{a}$ & 7.97 & 0.41 & 0.80 & \citet{Muller+25} \\
 & $K$ & 0.40 & 0.78 & 2.69 & & & & \\
20230902A & $R$ & 0.46 & 2.37 & 0.99 & 10.25 & 0.94 & 4.79 & This Work \\
 & $K$ & 0.48 & 2.45 & 1.38 & & & \\
20231226A & $R$ & 1.83 & 5.02 & 0.79 & 10.34 & 1.87 & 5.13 & This Work \\
 & $K$ & 1.90 & 5.23 & 1.75 & & & \\
20231230D & $R$ & 1.92 & 11.98 & 1$^{a}$ & 10.96 & 1.89 & 11.77 & This Work \\
 & $K$ & 0.79 & 4.92 & 1$^{a}$ & & & \\
20240117B & $R$ & 0.81 & 5.68 & 0.28 & 10.85 & 1.27 & 8.95 & This Work \\
 & $K$ & 0.84 & 5.90 & 0.84 & & & \\
20240201A & $R$ & 4.38 & 3.73 & 0.68 & 10.22 & 5.48 & 4.66 & This Work \\
 & $K$ & 3.13 & 2.67 & 1.23 & & & \\
20240203C & $R$ & 1.69 & 6.53 & 0.94 & 9.50 & 0.69 & 2.66 & This Work \\
20240209A & $r$ & 3.06 & 7.56 & 4.06 & 11.35 & 15.01 & 37.08 & \citet{Eftekhari+25} \\
20240210A & $R$ & 7.64, 8.17 & 3.72, 3.98 & 0.15, 1.41 & 10.33 & 10.48 & 5.10 & This Work \\
 & $K$ & 7.65, 3.95 & 3.72, 1.92 & 0.14, 1.62 & & & \\
20240304A & $R$ & 0.95 & 3.68 & 1.16 & 9.99 & 1.01 & 3.91 & This Work \\
 & $K$ & 1.32 & 5.10 & 4.34 & & & \\
20240310A & $R$ & 0.92 & 2.11 & 0.93 & 9.97 & 1.67 & 3.83 & This Work \\
 & $K$ & 0.84 & 1.93 & 0.91 & & & \\
20240312D & $r$ & 6.78, 1.66 & 6.63, 1.62 & 0.51, 4$^{a}$ & 10.69 & 7.05 & 6.90 & This Work \\
20240318A & $R$ & 2.40 & 4.95 & 0.85 & 10.31 & 2.42 & 4.99 & This Work \\
 & $K$ & 2.13 & 4.38 & 0.94 & & & \\
20240525A & $R$ & 1.64 & 7.84 & 0.40 & 10.82 & 1.75 & 8.38 & This Work \\
20240615B & $R$ & 3.69 & 5.18 & 0.90 & 11.11 & 12.87 & 18.06 & This Work \\
 & $K$ & 4.05, 0.95 & 5.69, 1.33 & 0.53, 1.71 & & & \\
20241027B & $r$ & 2.63 & 12.79 & 4$^{a}$ & 11.55 & 13.85 & 67.47 & This Work \\
\enddata
\tablecomments{Properties of the host galaxies used in this work. All  values are derived here excepting some masses, as noted. $r_e$ is the host (stellar light) effective radius and $r_{e,\rm GC}$ is the effective radius of the synthetic globular cluster distribution. When two $r_e$ are given, the first refers to the galactic disk component and the second the bulge component (see Section~\ref{sec:galfit} for more information). \\
$^{a}$ Fixed value \\}
\end{deluxetable*}

\subsection{Model Likelihood Estimation} \label{sec:coflash_methods}

We now describe our framework to test the statistical associations of FRB locations with 2D underlying light or GC spatial distributions. The \texttt{coflash} (\textbf{Co}rrelating \textbf{F}RB \textbf{L}ocation \textbf{A}nd \textbf{S}ubstructure of \textbf{H}osts)\footnote{https://github.com/adamdeller/coflash} Python package for quantifying the statistical likelihood of observing an FRB at a given location within a host galaxy will be introduced in Deller et al. in prep. It computes the likelihood of association ($L$) against an assortment of models (Deller et al. in prep. will have a description of all supported models). In this work, we implement and test additional candidate progenitor model distributions and improve the ability to mask pixels in the observed images that are unrelated to the light from the host galaxy. We consider four models: the host's surface brightness profile and its residual substructure (``Profile'' and ``Residual'', described in Section~\ref{sec:galfit}) and two GC spatial distributions (GC Ellipsoidal and GC Spherical, described in Section~\ref{sec:GC_maps}). The addition of GC spatial distributions to \texttt{coflash} is newly presented here.

\texttt{coflash} takes as inputs the models, FRB coordinates, and one-sigma localization. It first generates a probability map for the FRB localization described by a 2D Gaussian. A bad pixel mask can also be passed; by default it is grown by three pixels in each direction and applied to the models. In cases where the default growth factor is not sufficient to remove all extraneous, non-galactic light, we increase the growth factor until only galaxy light remains. 

The Profile and Residual models require additional processing prior to calculating the $L$ values. For the Profile, we subtract the minimum value of the model from each pixel to ensure the probability goes to zero at the boundaries for a more direct comparison with the other models. In the Residual, we remove oversubtracted regions by setting all pixels with negative values to zero. As the Residual has a low dynamic range of pixel values, fluctuations in the sky background can strongly influence the likelihood of this model, resulting in lower $L$ values. To mitigate this effect, we apply a mask to exclude all pixels outside the 99.9\% light contour of the galaxy. We derive this contour by performing isophotal fitting on the Profile model for a given FRB image using the \texttt{fit\char`_image} method of \texttt{isophote.Ellipse} from \texttt{photutils}. We tested the impact of varying the exact value of the threshold applied to the residual image and the extent of galaxy outer limit mask, which balance the mitigation of image noise against the loss of signal. We found that our analysis was robust to variation of these parameters within reasonable range; tuning them could weakly impact the model preferences for some individual FRBs, but without any systematic change across the FRB ensemble. 

Upon correction, we normalize each model to unity to create probability density functions. To calculate the likelihoods, within \texttt{coflash}, we convolve the models with the FRB localization map and sum the resulting convolutions. While a conclusion cannot be derived from the absolute value of $L$, it is meaningful in a relative sense when compared against other $L$ values for the same host. The model that is most consistent with the location of the FRB will have the largest value of $L$.

\startlongtable
\begin{deluxetable*}{l|cccc}
\tabletypesize{\footnotesize}
\tablewidth{0pc}
\tablecaption{FRB Galactocentric Offsets
\label{tab:offsets}}
\tablehead{
\colhead{FRB} &
\colhead{Filter} & 
\colhead{Angular Offset} &
\colhead{Physical Offset} &
\colhead{Host-normalized Offset} \\
\colhead{} &
\colhead{} &
\colhead{[arcsec]} &
\colhead{[kpc]} & 
\colhead{[$r_e$]}
}
\startdata
20180916B & F110W & $7.73^{+ 0.02}_{- 0.02}$ & $5.25^{+0.01}_{-0.01}$ & $0.88^{+0.00}_{-0.00}$ \\
20180924B & F160W & $0.73^{+0.17}_{-0.18}$ & $3.47^{+0.81}_{-0.83}$ & $1.56^{+0.37}_{-0.38}$ \\
20190102C & F160W & $1.15^{+0.70}_{-0.57}$ & $5.06^{+3.08}_{-2.51}$ & $1.19^{+0.72}_{-0.59}$ \\
20190608B & F160W & $2.85^{+0.37}_{-0.37}$ & $6.14^{+0.79}_{-0.79}$ & $1.59^{+0.20}_{-0.20}$ \\
20190714A & F160W & $0.91^{+0.47}_{-0.43}$ & $3.46^{+1.79}_{-1.64}$ & $0.98^{+0.51}_{-0.46}$ \\
20191001A & F160W & $0.85^{+0.41}_{-0.38}$ & $3.22^{+1.56}_{-1.45}$ & $0.80^{+0.39}_{-0.36}$ \\
20200120E & $r$ & $1175.60^{+0.26}_{-0.26}$ & $19.70^{+0.00}_{-0.00}$ & $4.76^{+0.00}_{-0.00}$ \\
20200430A & $I$ & $1.49^{+0.42}_{-0.46}$ & $4.17^{+1.17}_{-1.28}$ & $2.63^{+0.74}_{-0.81}$ \\
20200906A & $I$ & $0.97^{+0.46}_{-0.45}$ & $5.01^{+2.41}_{-2.33}$ & $0.65^{+0.31}_{-0.30}$ \\
20201124A & F475X & $0.76^{+0.00}_{-0.00}$ & $1.40^{+0.01}_{-0.01}$ & $0.79^{+0.01}_{-0.01}$ \\
  & F160W & $0.76^{+0.01}_{-0.01}$ & $1.40^{+0.03}_{-0.03}$ & $0.80^{+0.02}_{-0.02}$ \\
20210117A & $I$ & $0.77^{+0.45}_{-0.29}$ & $2.72^{+1.60}_{-1.03}$ & $2.55^{+1.50}_{-0.97}$ \\
  & $K$ & $0.91^{+0.44}_{-0.39}$ & $3.20^{+1.54}_{-1.38}$ & $1.63^{+0.79}_{-0.70}$ \\
20210320C & $I$ & $0.83^{+0.41}_{-0.41}$ & $3.57^{+1.78}_{-1.76}$ & $0.93^{+0.46}_{-0.46}$ \\
20210807D & $I$ & $4.81^{+0.51}_{-0.50}$ & $11.21^{+1.20}_{-1.18}$ & $1.00^{+0.11}_{-0.11}$ \\
  & $K$ & $4.58^{+0.51}_{-0.51}$ & $10.69^{+1.19}_{-1.20}$ & $1.13^{+0.13}_{-0.13}$ \\
20211127I & $I$ & $2.38^{+0.50}_{-0.50}$ & $2.21^{+0.46}_{-0.47}$ & $0.28^{+0.06}_{-0.06}$ \\
  & $K$ & $2.13^{+0.51}_{-0.51}$ & $1.98^{+0.47}_{-0.47}$ & $0.55^{+0.13}_{-0.13}$ \\
20211203C & $R$ & $0.86^{+0.47}_{-0.36}$ & $4.26^{+2.34}_{-1.78}$ & $1.87^{+1.03}_{-0.78}$ \\
20211212A & $I$ & $1.67^{+0.46}_{-0.43}$ & $2.27^{+0.62}_{-0.59}$ & $0.43^{+0.12}_{-0.11}$ \\
  & $K$ & $1.53^{+0.43}_{-0.44}$ & $2.08^{+0.58}_{-0.60}$ & $0.40^{+0.11}_{-0.11}$ \\
20220105A & $R$ & $2.29^{+0.93}_{-0.86}$ & $9.80^{+4.00}_{-3.68}$ & $2.91^{+1.19}_{-1.09}$ \\
  & $K$ & $2.31^{+0.92}_{-0.85}$ & $9.90^{+3.92}_{-3.63}$ & $3.13^{+1.24}_{-1.15}$ \\
20220725A & $R$ & $0.63^{+0.40}_{-0.30}$ & $2.03^{+1.30}_{-0.97}$ & $0.43^{+0.27}_{-0.20}$ \\
  & $K$ & $0.70^{+0.41}_{-0.35}$ & $2.26^{+1.33}_{-1.14}$ & $0.36^{+0.21}_{-0.18}$ \\
20220918A & $R$ & $0.95^{+0.36}_{-0.45}$ & $5.83^{+2.19}_{-2.75}$ & $3.41^{+1.28}_{-1.61}$ \\
  & $K$ & $0.87^{+0.45}_{-0.41}$ & $5.31^{+2.76}_{-2.49}$ & $2.22^{+1.15}_{-1.04}$ \\
20221106A & $R$ & $1.33^{+0.67}_{-0.62}$ & $4.52^{+2.29}_{-2.10}$ & $0.79^{+0.40}_{-0.37}$ \\
  & $K$ & $1.16^{+0.67}_{-0.57}$ & $3.94^{+2.29}_{-1.94}$ & $0.72^{+0.42}_{-0.35}$ \\
20230526A & $R$ & $0.64^{+0.40}_{-0.34}$ & $1.76^{+1.09}_{-0.93}$ & $0.86^{+0.53}_{-0.45}$ \\
  & $K$ & $0.69^{+0.44}_{-0.35}$ & $1.90^{+1.20}_{-0.95}$ & $2.55^{+1.61}_{-1.27}$ \\
20230708A & $R$ & $0.57^{+0.37}_{-0.28}$ & $1.10^{+0.72}_{-0.54}$ & $1.86^{+1.22}_{-0.91}$ \\
  & $K$ & $0.60^{+0.38}_{-0.29}$ & $1.17^{+0.74}_{-0.56}$ & $1.50^{+0.94}_{-0.72}$ \\
20230902A & $R$ & $0.87^{+0.52}_{-0.44}$ & $4.44^{+2.66}_{-2.23}$ & $1.87^{+1.12}_{-0.94}$ \\
  & $K$ & $0.89^{+0.55}_{-0.44}$ & $4.53^{+2.83}_{-2.24}$ & $1.85^{+1.15}_{-0.92}$ \\
20231226A & $R$ & $1.76^{+0.51}_{-0.52}$ & $4.84^{+1.39}_{-1.44}$ & $0.97^{+0.28}_{-0.29}$ \\
  & $K$ & $1.95^{+0.55}_{-0.54}$ & $5.35^{+1.52}_{-1.49}$ & $1.02^{+0.29}_{-0.28}$ \\
20231230D & $R$ & $1.31^{+0.55}_{-0.53}$ & $8.15^{+3.44}_{-3.32}$ & $0.48^{+0.20}_{-0.20}$ \\
  & $K$ & $1.34^{+0.54}_{-0.51}$ & $8.33^{+3.39}_{-3.17}$ & $1.69^{+0.69}_{-0.64}$ \\
20240117B & $R$ & $1.88^{+1.15}_{-0.97}$ & $13.21^{+8.08}_{-6.77}$ & $2.33^{+1.42}_{-1.19}$ \\
  & $K$ & $1.85^{+1.12}_{-0.91}$ & $12.95^{+7.85}_{-6.36}$ & $2.19^{+1.33}_{-1.08}$ \\
20240201A & $R$ & $3.30^{+0.54}_{-0.46}$ & $2.81^{+0.46}_{-0.39}$ & $0.75^{+0.12}_{-0.10}$ \\
  & $K$ & $3.31^{+0.51}_{-0.49}$ & $2.82^{+0.44}_{-0.42}$ & $1.06^{+0.16}_{-0.16}$ \\
20240203C & $R$ & $0.66^{+0.45}_{-0.33}$ & $2.55^{+1.74}_{-1.29}$ & $0.39^{+0.27}_{-0.20}$ \\
20240209A & $r$ & $15.50^{+1.96}_{-1.94}$ & $ 38.31^{+4.85}_{-4.80}$ & $5.06^{+0.64}_{-0.6}$ \\
20240210A & $R$ & $9.00^{+0.54}_{-0.54}$ & $4.38^{+0.26}_{-0.26}$ & $1.18^{+0.07}_{-0.07}$ \\
  & $K$ & $8.94^{+0.52}_{-0.52}$ & $4.35^{+0.25}_{-0.25}$ & $1.17^{+0.07}_{-0.07}$ \\
20240304A & $R$ & $2.26^{+0.75}_{-0.78}$ & $8.73^{+2.91}_{-3.02}$ & $2.37^{+0.79}_{-0.82}$ \\
  & $K$ & $2.04^{+0.80}_{-0.76}$ & $7.88^{+3.08}_{-2.92}$ & $1.55^{+0.60}_{-0.57}$ \\
20240310A & $R$ & $0.95^{+0.54}_{-0.47}$ & $2.18^{+1.24}_{-1.07}$ & $1.04^{+0.59}_{-0.51}$ \\
  & $K$ & $0.84^{+0.50}_{-0.41}$ & $1.92^{+1.16}_{-0.94}$ & $1.00^{+0.60}_{-0.49}$ \\
20240312D & $r$ & $7.38^{+1.37}_{-1.33}$ & $7.22^{+1.34}_{-1.30}$ & $1.09^{+0.20}_{-0.20}$ \\
20240318A & $R$ & $1.27^{+0.58}_{-0.55}$ & $2.63^{+1.19}_{-1.14}$ & $0.53^{+0.24}_{-0.23}$ \\
  & $K$ & $1.24^{+0.62}_{-0.56}$ & $2.56^{+1.27}_{-1.14}$ & $0.59^{+0.29}_{-0.26}$ \\
20240525A & $R$ & $1.50^{+0.43}_{-0.45}$ & $7.17^{+2.07}_{-2.16}$ & $0.91^{+0.26}_{-0.28}$ \\
20240615B & $R$ & $1.29^{+0.65}_{-0.59}$ & $1.82^{+0.92}_{-0.83}$ & $0.35^{+0.18}_{-0.16}$ \\
  & $K$ & $1.23^{+0.66}_{-0.58}$ & $1.73^{+0.93}_{-0.82}$ & $0.30^{+0.16}_{-0.14}$ \\
20241027B & $r$ & $4.35^{+0.62}_{-0.60}$ & $21.22^{+3.04}_{-2.92}$ & $1.66^{+0.24}_{-0.23}$ \\
\enddata
\tablecomments{Measured angular, physical, and host-normalized offsets for the FRBs and hosts in this work.}
\end{deluxetable*}

\subsection{Galactocentric Offsets} \label{sec:offset}

We next determine the projected galactocentric offsets of the FRBs in our sample, calculating their angular (in units of arcsec), physical (kpc), and host-normalized ($r_e$) offsets. We use the total positional uncertainties derived in Section~\ref{sec:astrom} as they incorporate all relevant sources of uncertainty incurred when placing an FRB on an optical image. First, we create a probability map of the FRB localization described by a 2D Gaussian. We then calculate the separation between the host galaxy and every pixel within the 5$\sigma$ localization ellipse, constructing a distribution of angular offsets for each FRB. From this distribution, we derive the median and 68\% confidence interval, weighting by the FRB probability map. Using the host redshift, we then convert the angular offset and its weighted uncertainty to corresponding physical offset values. Finally, we use the disk $r_e$ derived in Section~\ref{sec:galfit} to calculate the host-normalized offset median and weighted uncertainty. 

\begin{figure*}[!t]
    \centering
\includegraphics[width=0.49\textwidth]{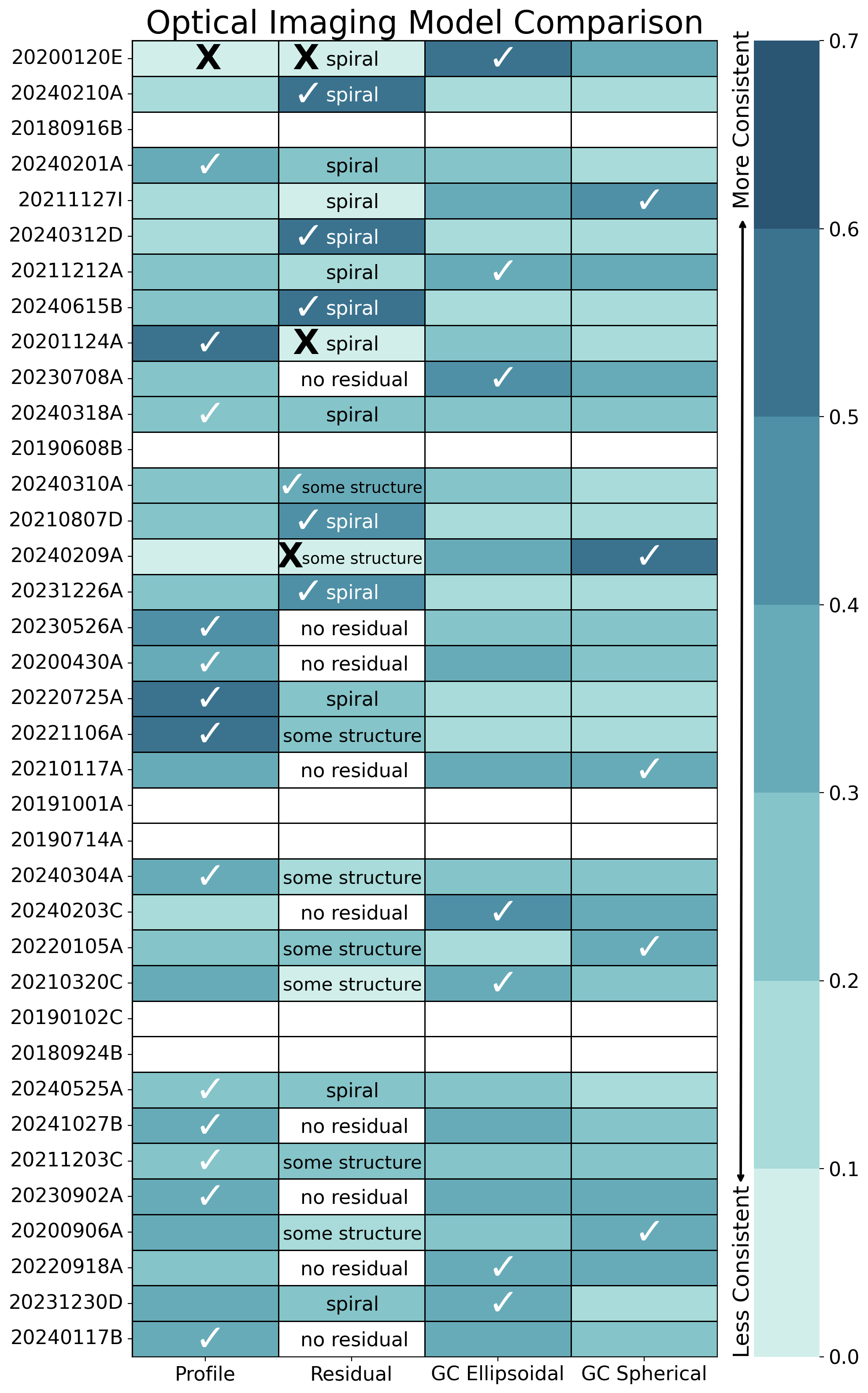}
\includegraphics[width=0.49\textwidth]{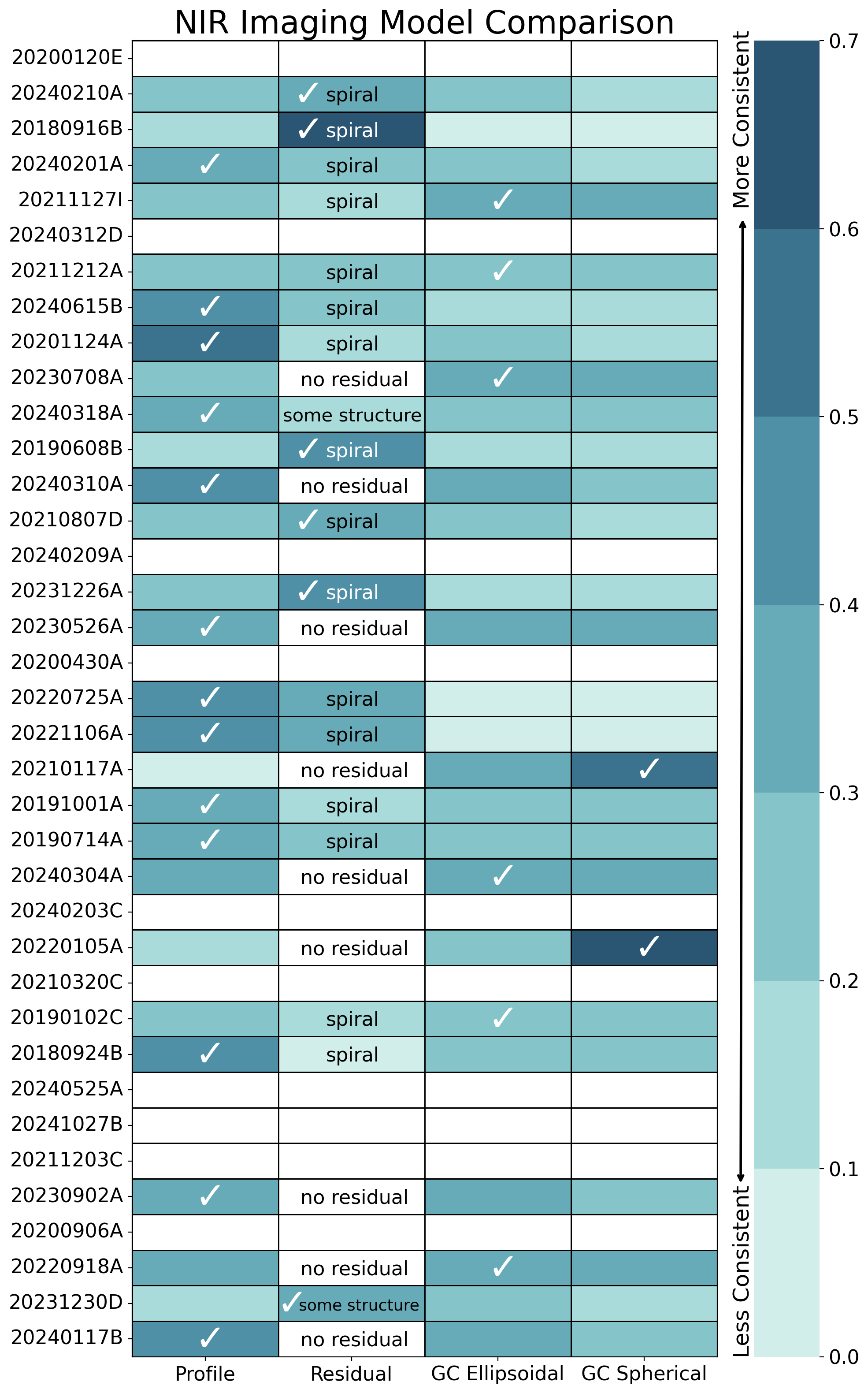}
\caption{Summary visualization of $L$ values from the four spatial distributions tested for each FRB image: Profile, Residual, GC Ellipsoidal, and GC Spherical. Each row corresponds to an FRB, which we sort in order of increasing redshift. Darker shades of blue represent stronger consistency with a given model. If no data are available for a given model, we leave the box blank. Preferred models are marked with white checkmarks. If a model is disfavored for a given FRB (see Section~\ref{sec:BIC}), we mark it with a black X. Lastly, we note in the Residual column if the residual substructure is clearly spiral (``spiral''), exists but is not easily classifiable as spiral (``some structure''), or if no residual remains (``no residual''). } 
    \label{fig:heatmap}
\end{figure*}

We report the median and 68\% confidence intervals on the angular, physical, and host-normalized offsets for each FRB image in Table~\ref{tab:offsets}. For FRBs\,20180916B and 20201124A, their milliarcsecond scale localizations are so precise that there were not enough values in their offset distributions to derive meaningful uncertainties. Thus, we calculate the median and 68\% confidence intervals without weighting. Fifteen FRBs in our sample have previously published offset measurements (c.f., \citealt{Bhardwaj21,Mannings+21,Bhandari+22,Bhandari+23,Dong+24,Woodland+24,Shah+25}). Upon comparison, we find our offset measurements are consistent within uncertainties to those in the literature with no systematic shift to smaller or larger values.

\section{Results} \label{sec:results}
\subsection{FRB Image Model Preferences} \label{sec:coflash_results}

We apply \texttt{coflash} to the optical and NIR images for each FRB in our sample using the total uncertainty ellipses described in Section~\ref{sec:astrom}. We consider the model with the largest $L$ to be the ``preferred'' model for a given FRB image, corresponding to the most consistent spatial distribution of its progenitor. Of the 31 FRBs with optical images, 13$\pm$3 prefer the Profile model (following the Wilson score interval corresponding to 68\% confidence), 6$\pm$2 prefer the Residual, 7$\pm$2 prefer GC Ellipsoidal, and 5$\pm$2 prefer GC Spherical. Similar trends appear in the NIR; from a sample of 27 FRBs, 13$\pm$3 prefer Profile, 6$\pm$2 Residual, 6$\pm$2 GC Ellipsoidal, and 2$\pm$1 GC Spherical. We visualize these results in Figure~\ref{fig:heatmap}, applying a flat normalization to the $L$ values to enable comparison of results across the sample; the data are split into optical (left panel) and NIR (right panel) regimes. For FRBs with both optical and NIR imaging, we find that the majority of FRBs have the same model preference across both bands. There are five exceptions: FRBs 20211127I, 20231230D, 20240304A, 20240310A, and 20240615B (see Figure~\ref{fig:heatmap}), but overall we find no systematic change between certain models. 

\begin{figure*}[t]
    \centering
\includegraphics[width=0.49\textwidth]{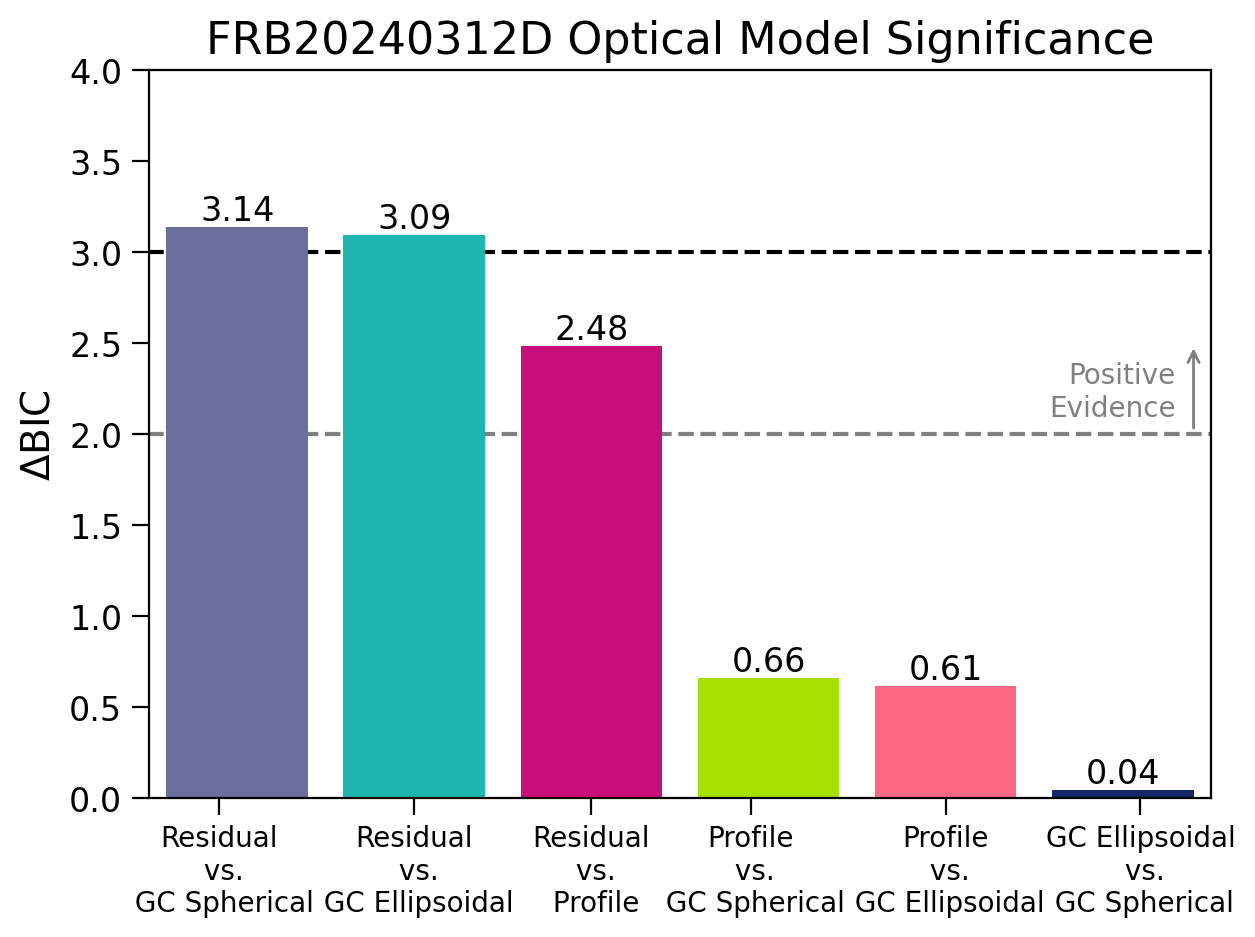}
\includegraphics[width=0.49\textwidth]{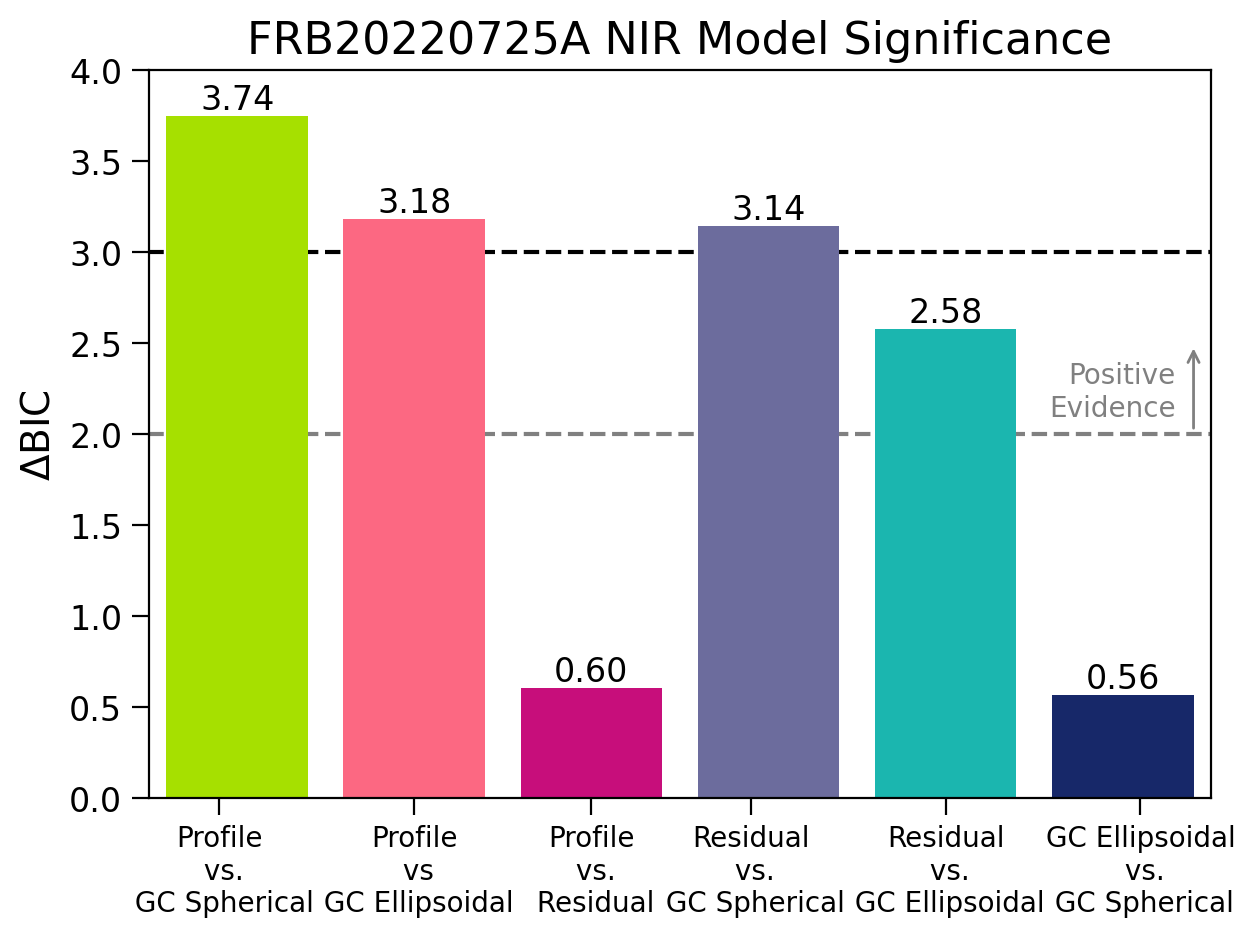}
    \caption{Comparison of the $\Delta$BIC between the four models for FRBs\,20240312D (left) and 20220725A (right). The three leftmost bars show how the preferred model compares against the other models. We denote the threshold for statistical interpretation, $\Delta$BIC\ $>3$, with a black dashed line. We show the minimum $\Delta$BIC for Positive Evidence (2) with a grey dashed line.}
    \label{fig:dBIC_example}
\end{figure*}

Our results show the majority of FRBs prefer the smooth galaxy light as represented by the Profile model in both $R$ and $K$ bands. However, there are a non-negligible fraction that prefer the GC models (combined, more often than the Residual preferences which trace galactic substructure), with 12/31 GC preferences in the optical and 8/27 in the NIR. We find only a few FRBs have apparently strong model preference. For instance, FRB\,20180916B strongly prefers the Residual model. This is consistent with its visually apparent coincidence with a spiral arm despite being 250 pc offset from the most nearby star-forming region within that arm \citep{Tendulkar+21}. Additionally, we find FRB\,20200120E strongly prefers the GC Ellipsoidal model which is concordant with its confirmed globular cluster origin \citep{Bhardwaj21,Kirsten21}. However, we also find cases of very weak model preference such as with FRB\,20230902A. This FRB prefers the Profile model, but its likelihood is only marginally greater than $L$ values of the other models. Indeed, other than a few FRBs with strong preference, Figure~\ref{fig:heatmap} shows that the vast majority of preferred models are only weakly selected compared to the other models. Our results additionally reveal a fairly low fraction of Residual preferences. However, since we are only reliably sensitive to spiral structure out to $z\sim0.15$ (e.g., Figure~\ref{fig:z_hist}), this fraction will be interpreted as a lower bound, hereafter. We further describe the importance of localization size in preference for spiral structure in Section~\ref{sec:loc_size}. 

\subsection{Statistical Interpretation of Model Preferences} \label{sec:BIC}

\begin{deluxetable*}{l|ccccc}
\linespread{1.4}
\tablewidth{0pc}
\tablecaption{FRBs with $\Delta$BIC\ $>3$ Between Models
\label{tab:stat_results}}
\tablehead{
\colhead{FRB}	 &
\colhead{Wavelength Regime} &
\colhead{Preferred Model} &
\colhead{Preferred Over} & 
\colhead{$\Delta$BIC} & 
\colhead{Disfavored Model}
 }
\startdata
20180924B & NIR & Profile & Residual  & 3.34 & \\
20201124A & Optical & Profile & Residual & 10.30 & Residual \\
          & NIR & Profile & Residual & 3.08 & \\
20220725A & Optical & Profile & GC Spherical & 3.24 & \\
          & NIR & Profile & GC Ellipsoidal, GC Spherical & 3.18, 3.74 & \\
20221106A & NIR & Profile & GC Spherical & 3.42 & \\
\hline
20180916B & NIR & Residual & Profile, GC Ellipsoidal, GC Spherical & 3.89, 4.70, 4.56 &  \\
20240210A & Optical & Residual & GC Spherical & 3.44 & \\
20240312D & Optical & Residual & GC Ellipsoidal, GC Spherical & 3.09, 3.14 &  \\
\hline
20200120E & Optical & GC Ellipsoidal & Profile, Residual & 6.12, $\infty^{a}$ & Profile, Residual \\
\hline
20210117A & NIR & GC Spherical & Profile & 3.31 &  \\
20220105A & NIR & GC Spherical & Profile, Residual & 3.57, 3.89 & \\
20240209A & Optical & GC Spherical & Profile, Residual & 3.81, 36.01 & Residual \\
\enddata
\tablecomments{Summary of the FRBs with preferred models that cross $\Delta$BIC$>$3, or with models that are disfavored. We do not include FRB\,20211127I in this summary (see Section~\ref{sec:BIC}). The $\Delta$BIC values correspond to the difference in BIC between the Preferred Model and the model(s) it is preferred over.\\
$^{a}$ The Residual likelihood is exactly 0.}
\end{deluxetable*}

In the previous section, we presented the model preferences based on the $L$ values alone, but this does not reveal if the preferences are demonstrably superior to the other models. To interpret the statistical significance of the model preferences, we use the difference in the Bayesian Information Criterion value ($\Delta$BIC; \citealp{Schwarz1978})\footnote{We note that one key assumption of the $\Delta$BIC test is that the true model exists within the set of candidate models.}. The BIC is defined as 

\begin{equation}
    \textnormal{BIC} = k\ln(n) - 2\ln(L).
\end{equation}

\noindent where $k$ is the number of parameters in the model (representing the pixel intensity values), $n$ is the number of data points (representing the number of pixels in the models), and $L$ is the likelihood. $\Delta$BIC is a measure of the statistical significance of one model compared to another, in which larger values correspond to higher degrees of supporting evidence for the superior model. As $n$ and $k$ are equivalent across our models, the $\Delta$BIC calculation is reduced to a comparison of the $L$ values. Following \citet{Kass_Raftery1995}, a 2$<$$\Delta$BIC$<$6 indicates Positive Evidence in support of the preferred model, 6$<$$\Delta$BIC$<$10 indicates Strong Evidence, and $\Delta$BIC$>$10 Very strong Evidence. Guided by these ranges - and mindful of the fact that the model light distributions often have similar support in the central regions of the hosts, making it impossible for one model to be strongly favored over another - we report on cases where $\Delta$BIC exceeds three (i.e., positive evidence in support of the preferred model, but with a higher threshold than the minimum of this range). We note when models reach higher degrees of statistical support.

To illustrate this method, in Figure~\ref{fig:dBIC_example}, we show the $\Delta$BIC comparisons between the four models for FRBs\,20240312D (left) and 20220725A (right); the images and models for these hosts are shown in Figure~\ref{fig:galfit_GC_example}. For FRB\,20240312D, the preferred Residual model crosses our $\Delta$BIC threshold when compared to the GC models, indicating positive evidence in support of the Residual. This can be interpreted as the Residual providing a better description of the likelihood of finding this FRB at its observed location than either of the GC models. While there is further positive evidence in support of the Residual model when compared to the Profile, this comparison does not exceed $\Delta{\rm BIC}>3$. We additionally show how the three non-preferred models compare against each other (the rightmost bars); all fall below the threshold for positive evidence, indicating no meaningful distinction between the models can be made. Similarly, for FRB\,20220725A in the NIR, the preferred Profile model has positive evidence when compared to the GC models, meaning the Profile model is a better description of the likelihood of finding the FRB at the observed location than either GC model. We also find support for the Residual as a better descriptor than the GC models even though it is not the preferred model. Thus, for neither of these FRBs can we state that a given light distribution is preferred to the exclusion of all others, but we {\em can} in both cases say that the GC models are disfavored. We show the $\Delta$BIC comparisons for the remainder of our sample in Figure~\ref{fig:all_models} (Appendix~\ref{sec:models_appendix}).

We now comment on the statistical interpretation of the model preferences across the sample. For the 12 FRBs (across 18 images) which have no visual host galaxy residuals, we do not calculate the $\Delta$BIC in relation to the Residual as support over a null model is not informative. We find only one FRB has a model preference that crosses our threshold when compared to all other models considered: the Residual of FRB\,20180916B. All others with model preferences with $\Delta$BIC\ $>3$ only do so with respect to one or two models. In general, this is not surprising: for a given model to have meaningful statistical support over all others, the model probability distributions must be substantially different within the region of the FRB localization. Many FRBs are found at offsets of $0.5-1.5r_e$, in which both the GC and Profile models generally have substantial light. In total, we find 11 FRBs (13 images) have model preferences with $\Delta$BIC\ $>3$ over at least one other model, roughly one third of our sample. For the optical (NIR) images, six (seven) preferences surpass this threshold, summarized in Table~\ref{tab:stat_results}. Lastly, we note that one FRB shows Strong Evidence for its preferred model (FRB\,20200120E), and two show Very Strong Evidence (20201124A in the optical and 20240209A); their $\Delta$BIC values are detailed in Table~\ref{tab:stat_results}.

We can also use this methodology to {\it disfavor} certain models (Table~\ref{tab:stat_results}); this requires each model to have $\Delta$BIC\ $>3$ over the disfavored model. Applying this criterion, we find three FRBs which have at least one disfavored model. For FRB\,20200120E, we find that both the Profile and Residual models are disfavored. However, as $\Delta\rm{BIC}\!<3$ between the two GC models, we cannot meaningfully distinguish between them for this FRB. For the optical image of FRB\,20201124A (which prefers the Profile), the Residual model is disfavored despite the presence of spiral structure in the host. Finally, the Residual model of FRB\,20240209A is disfavored. We denote the ruled out models for these FRBs with black X's in Figure~\ref{fig:heatmap}. 

Next, we are motivated to comment on the statistical interpretation of the FRBs that prefer the GC models. Of the 12 with GC preferences, only five reach $\Delta$BIC\ $>3$. These include FRB\,20200120E, confirmed to originate from a GC \citep{Kirsten21}; FRB\,20240209A, a highly-offset burst from a quiescent, elliptical host galaxy \citep{Shah+25,Eftekhari+25}; and FRBs\,20210117A and 20220105A (both in the NIR), which have large offsets and localizations (factors we explore further in Sections~\ref{sec:offset_results}$-$\ref{sec:loc_size}). The fifth, FRB\,20211127I, instead has a small offset from its host center. This host, along with five others that favor GC models (but not at the level of our threshold), have $r_{e,{\rm GC}}<r_e$. We find this is likely due to a limitation in the applicability of the \citet{Lim+24} $r_{e,\rm GC}-\rm M*$ relation to low-mass, late-type galaxies and marks an outlier in our sample. Their relation was calibrated on early-type galaxies which have smaller sizes for their mass compared to late-type galaxies (c.f., \citealt{van_der_Wel+14}). Thus, for cases of low-mass, late-type galaxies, we often derive $r_{e,\rm GC} \ll r_e$ which can drive a GC preference. Given the limitations of this method in such cases, we exclude FRB\,20211127I from our subsequent interpretation. We therefore consider only four GC model selections to have meaningful supporting evidence, corresponding to FRBs\,20200120E, 20210117A, 20220105A, and 20240209A, respectively.

Finally, we examine how the distribution of model preferences changes when considering the subset of FRB hosts at low redshift ($z\lesssim 0.15$). Of these 15 hosts, 46$\pm$12\% favor the Residual model, with 20$\pm$10\% at $\Delta$BIC\ $>3$. Meanwhile, only 20$\pm10$\% favor the Profile model, with a single FRB~host favoring the Profile at $\Delta$BIC\ $>3$; this is in contrast with the results of the entire sample where the Profile is most strongly preferred. The GC fractions remain similar. This highlights the sensitivity of this method to redshift, especially in determining the exact fractions of FRB locations which prefer smooth galaxy light versus spiral substructure.

However, based on the entire population, we find that the majority of FRBs are consistent with the stellar disks of their host galaxies, with only a few FRBs preferring the more extended spatial distribution of GCs. While three FRBs disfavor at least one model, we find no evidence for any particular model to be systematically disfavored across the entire FRB population. In other words, none of the four models considered here are truly poor descriptors of the spatial distribution of FRBs.

\subsection{Factors Influencing Model Preference}

\subsubsection{$\Delta$BIC Threshold}

Thus far, we have limited our interpretation to model preferences with positive evidence above a threshold of $\Delta$BIC\ $>3$. However, there exist several FRBs with $\Delta$BIC just under this threshold. We now explore how the above results would change if we relaxed the statistical threshold to preferred models with $\Delta$BIC $>2$, the minimum bound for positive evidence. We find a total of six additional FRBs with model preferences of $\Delta$BIC $>2$, half of which select the Residual. Including these FRBs with our previous results, there would be eight total Profile selections, seven Residual, two GC Ellipsoidal, and four GC Spherical encompassing both optical and NIR results. While the Profile is still the dominant model, the Residual is now a close second. This suggests that our small number of true associations (i.e., those with $\Delta$BIC $>3$) represents a lower limit on the potential number of statistical associations of FRBs to spiral arms. Upon visual inspection, we find that the FRBs with Residual preferences and $2<\Delta$BIC~$<3$ either have localizations larger than the underlying substructure (e.g., FRB\,20190608B) and/or localizations that only partially overlap with spiral structure (e.g., FRB\,20210807D). This implies that the lack of ability to claim supporting evidence in favor of association with sub-structure (residuals) may be limited by the depth of currently available data and/or the size of the FRB localization uncertainty. We further explore the effect of localization size on model selection in Section~\ref{sec:loc_size}. However, it is notable that this test of relaxing the significance threshold does not considerably increase the number of FRBs which select the GC models with positive statistical support. This further supports our findings that the locations of the majority of FRBs are inconsistent with the expected spatial distribution of GCs.

 \begin{figure*}[t]
    \centering
\includegraphics[width=\textwidth]{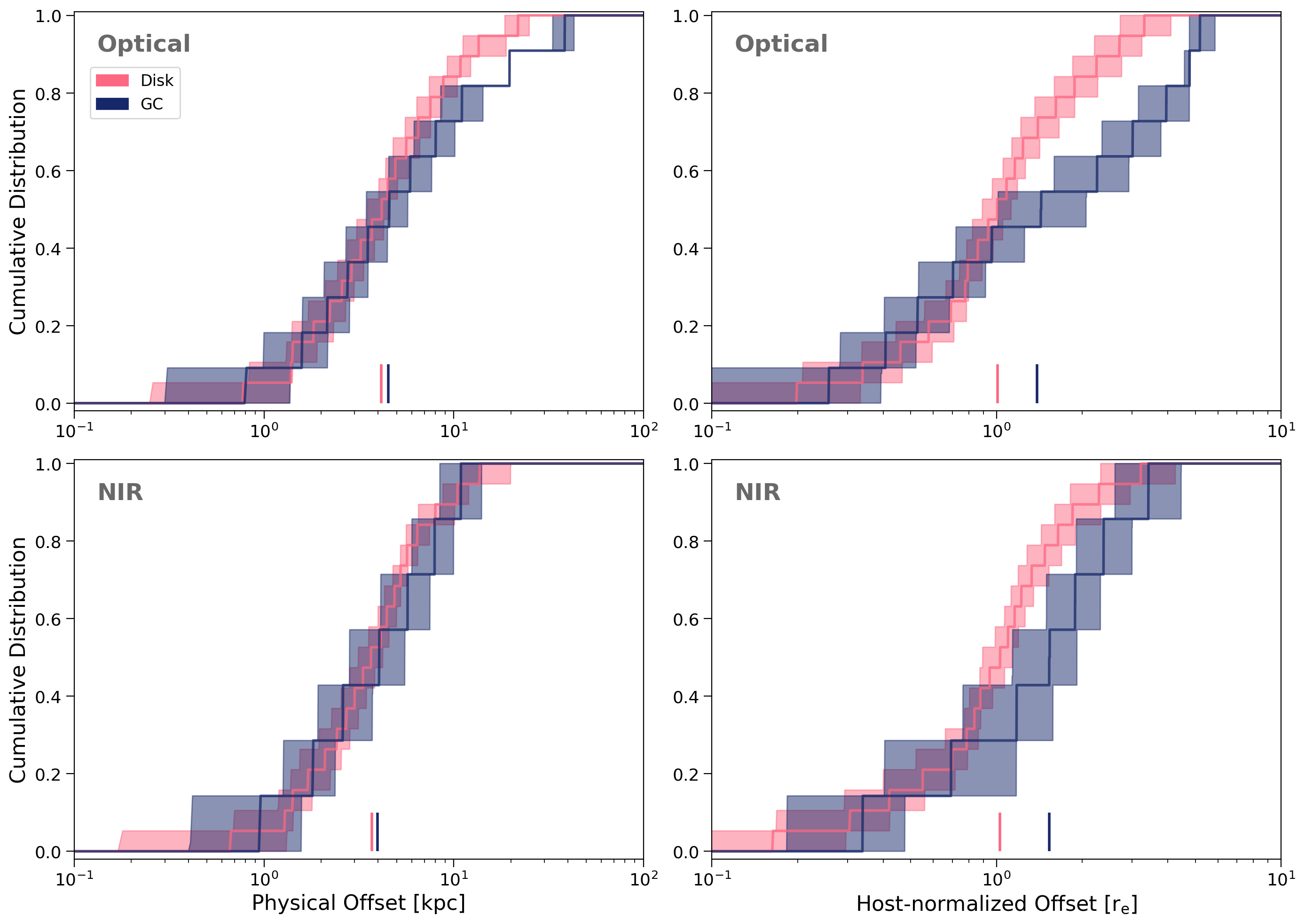}
    \caption{Cumulative distributions of physical (left column) and host-normalized (right column) offsets for our sample. For each wavelength regime (optical = top row, NIR = bottom row), we split the FRBs by their preferred model, grouping the FRBs that prefer the Profile and Residual models into Disk (pink) and GC Ellipsoidal and GC Spherical into GC (navy) populations. The shaded regions represent the 68\% confidence interval. We mark the medians of the distributions with short pink and navy vertical lines at the bottom of each panel.}
    \label{fig:offset_CDF}
\end{figure*}

\begin{figure}
    \centering
    \includegraphics[width=\linewidth]{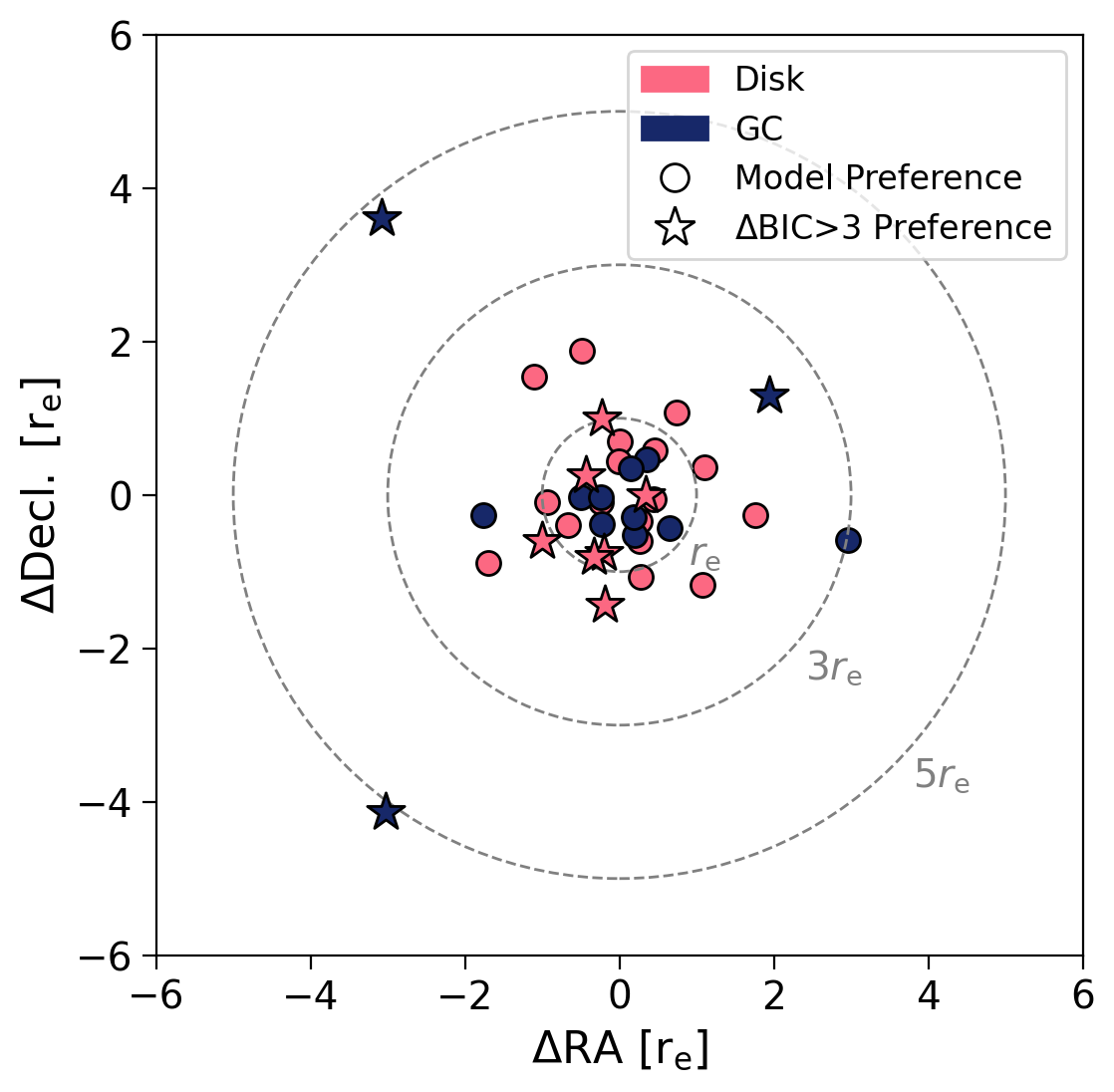}
    \caption{The host-normalized offsets of FRBs from their hosts in units of host effective radius. We group the FRBs by their model preferences into Disk (pink) and GC (blue) populations, and denote the model preferences with $\Delta$BIC\ $>3$ with stars. For size comparison, we show concentric circles of 1, 3, and 5$r_e$ in grey.}
    \label{fig:location}
\end{figure}

\begin{deluxetable*}{l|ccccc}
\linespread{1.4}
\tablewidth{0pc}
\tablecaption{Offsets by Model Preference
\label{tab:offset_stats}}
\tablehead{
\colhead{Offset Distribution}	 &
\colhead{Full Sample} &
\colhead{Disk Population, Optical} &
\colhead{Disk Population, NIR} &
\colhead{GC Population, Optical} &
\colhead{GC Population, NIR}
 }
\startdata
Angular [arcsec] & 1.37$^{+3.28}_{-0.73}$ & 1.51$^{+3.07}_{-0.79}$ & 1.26$^{+2.67}_{-0.63}$ & 1.16$^{+12.83}_{-0.62}$ & 1.21$^{+0.97}_{-0.63}$ \\
Physical [kpc] & 4.22$^{+5.67}_{-2.47}$ & 4.16$^{+5.68}_{-2.55}$ & 3.70$^{+3.28}_{-2.21}$ & 4.53$^{+15.17}_{-2.84}$ & 3.96$^{+5.02}_{-2.58}$ \\
Host-normalized [$r_e$] & 1.05$^{+1.48}_{-0.58}$ & 1.01$^{+1.00}_{-0.49}$ & 1.03$^{+0.71}_{-0.55}$ & 1.39$^{+3.37}_{-0.96}$ & 1.53$^{+1.17}_{-1.06}$ 
\enddata
\tablecomments{Median and 68\% CI of offsets split by category of preferred model and wavelength regime (see Section~\ref{sec:offset_results} for more information). For the full sample column, we use the optical measurements as applicable.}
\end{deluxetable*}

\subsubsection{Offset} \label{sec:offset_results}

We next explore how galactocentric offset correlates with the preferred model selection. As the GC models are generally more extended than the stellar disk, we naively expect FRBs with larger offsets to prefer the GC models. To robustly compare the behavior of the full offset distributions, we construct cumulative distribution functions (CDFs) of their physical and host-normalized offsets. For this analysis, we combine the FRBs into Disk (Profile and Residual) and GC (GC Ellipsoidal and Spherical) populations and split by their preferred model in each wavelength regime. We use the median and 68\% confidence intervals of the measured FRB offsets (see Section~\ref{sec:offset}) to construct asymmetric Gaussians representing the allowed distributions of offset measurements for each FRB. We then draw 1000 representative realizations from these distributions to construct the offset CDFs and estimate their uncertainties. We present the resulting CDFs in Figure~\ref{fig:offset_CDF}. In Table~\ref{tab:offset_stats}, we report the median angular, physical, and host-normalized offsets for each model category, split by wavelength regime. For both the physical and host-normalized distributions, we find the GC median offset is larger than the Disk median offset across all wavelengths.

In terms of physical offset, we find the Disk and GC distributions have large overlap, with only a few events at larger offset (optical only). However, for host-normalized offsets, the distributions more clearly diverge at large offsets ($r_e >1$). This effect is particularly strong in the optical band as the GC distribution includes FRBs\,20200120E and 20240209A which have large host-normalized offsets (4.76$r_e$ and 5.06$r_e$, respectively). As there is no NIR data for these FRBs, the distinction is less pronounced in the NIR panel. To quantify the degree to which the Disk and GC distributions are statistically distinct, we perform Anderson-Darling \citep{Anderson-Darling} tests on each of the 1000 iterations and record the number of times in which the $p$-value is $<0.05$. For the optical (NIR) physical offset CDFs, the two distributions have $p<0.05$ for 0.1\% (0.3\%) of the time. For host-normalized offsets, this occurs 9.1\% (15.6\%) of the time. Thus, we cannot reject the null hypothesis that the Disk and GC distributions are drawn from the same parent distribution for both the physical and host-normalized offsets and two wavelength regimes. This indicates that the FRBs in our sample are indeed consistent with being drawn from the same parent offset distribution, regardless of the model category that they prefer.

Given the slight deviations in the distributions, particularly at large offsets ($\gtrsim10$~kpc, $\gtrsim1 r_e$), we next explore how many more FRBs with large host-normalized offsets would be required to measure a statistical distinction in the Disk and GC offset CDFs. To test this, we randomly generate host-normalized offsets with values between $3-5r_e$ using a uniform distribution\footnote{We note that the most commonly used offset prior in the PATH (Probabilistic Association of Transients to their Hosts; \citealt{path}) algorithm for associating FRBs to their hosts assumes an exponential offset distribution truncated at 6$r_e$.} and assign 68\% confidence uncertainties ranging from $0.1-1r_e$ (symmetric errors). Assuming FRBs with these offsets would select a model in the GC Models category (as FRBs\,20200120E and 20240209A do), we add them to the observed GC Model optical distribution, regenerate the CDF, and perform Anderson-Darling testing. We find that the addition of one more FRB with large host-normalized offset returns $p<0.05$ ~25.1\% of the time, a considerable increase from 9.1\% for our real sample. Moreover, the addition of only two more FRBs at large offset would result in a majority of samples with $p<0.05$. We repeat this exercise with the NIR distributions and find similar trends. Thus, while FRBs with large host-normalized offsets are currently rare among the full FRB population, the addition of only a few more would lend statistical support that a distinction in offset is a main driver of model preference (and possibly FRB progenitor channel).

To visualize the effect of offset on model preference, in Figure~\ref{fig:location} we show the host-normalized offsets of the FRBs from their hosts colored by Disk and GC populations. For each FRB, we show one model preference, choosing the optical result when both wavelengths are available. We find approximately half (19/37) of FRBs are located within 1$r_e$ of their host. While the FRBs within this boundary select both Disk and GC models, none of the GC preferences for FRBs with offsets $\lesssim1r_e$ have $\Delta$BIC values surpassing 3. At $\sim2r_e$, only GC models are preferred, with those beyond $\sim3r_e$ consistently returning $\Delta$BIC\ $>3$ in support of the GC models. These results are complementary to the offset CDFs of Figure~\ref{fig:offset_CDF} which show a deviation in model selection at large $r_e$.

\begin{figure*}
    \centering
    \includegraphics[width=\textwidth]{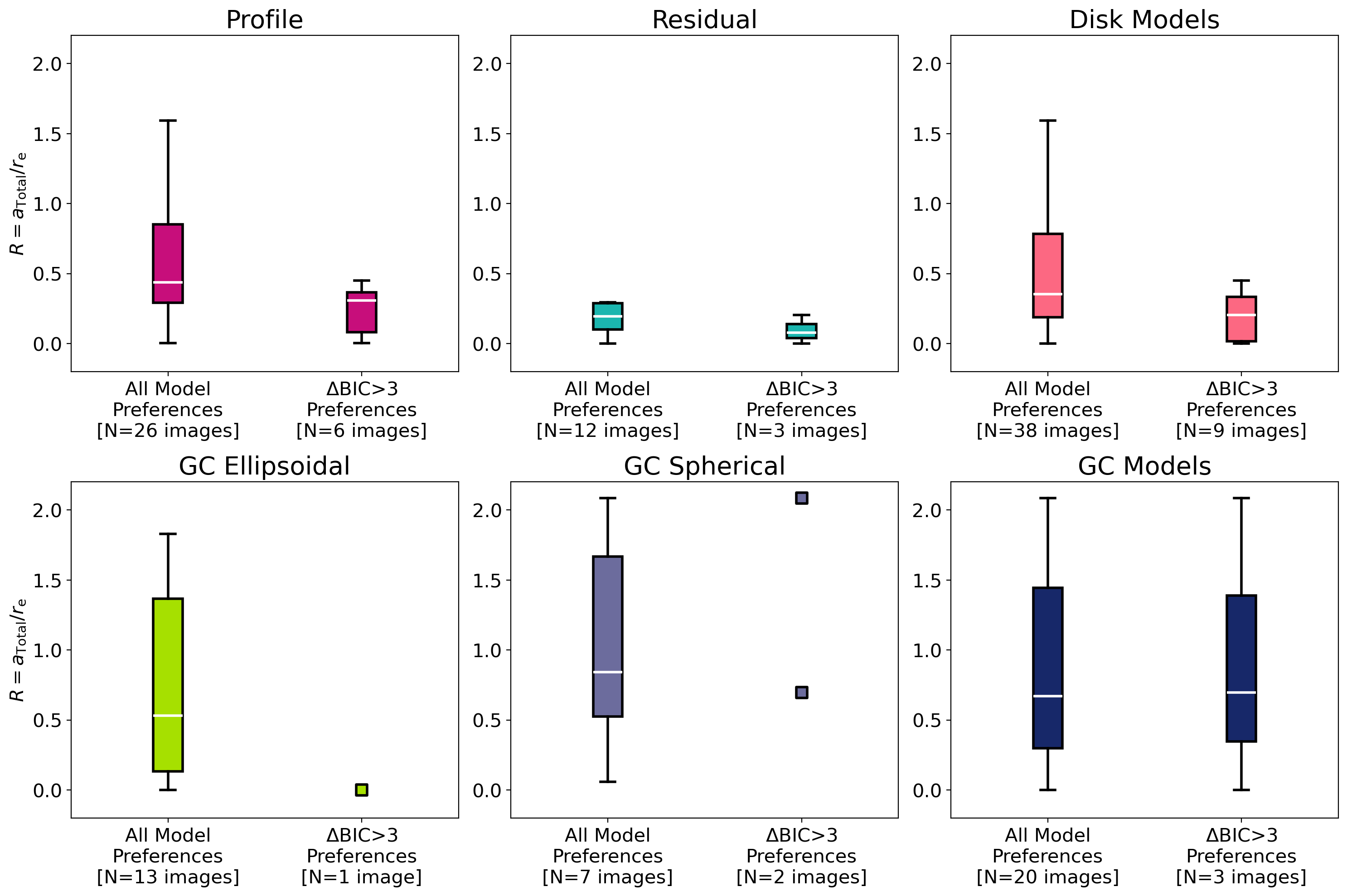}
    \caption{Boxplots of $R$, the ratio of total localization ellipse semi-major axis ($a_{\rm Total}$) to galaxy size ($r_e$), split by model preference and model category (rightmost column). We include all FRB images in the distributions, specifying the sample sizes in the x-axis labels. The white line in the boxes represents the distribution median. As the number of FRBs with GC Ellipsoidal and GC Spherical preferences that reach $\Delta$BIC\ $>3$ is small, we plot the $R$ values individually.}
    \label{fig:R_boxplot}
\end{figure*}

\subsubsection{Localization Size}
\label{sec:loc_size}

We next examine the role of localization size on model preference and degree of statistical support. To quantify the effective footprint of the localization on the host, we consider the ratio ($R$) of the total localization ellipse semi-major axis ($a_{\rm Total}$, see Table~\ref{tab:astrometry}) to the galaxy size ($r_e$; Table~\ref{tab:host_props}). In Figure~\ref{fig:R_boxplot}\footnote{We note that the criteria \#3 in Section~\ref{sec:sample} that nominally excludes events with $R>0.5$ was based on approximate values (FRB positional uncertainties alone and estimates of $r_e$). With updated uncertainties and accurate $r_e$ values, we have some events with $R \gtrsim 0.5$.}, we show the distribution of $R$ values for all FRBs that prefer a given model compared to the subset having preferences with $\Delta$BIC\ $>3$. Both optical and NIR selections for each FRB are included. We find that the FRBs that prefer the Residual have the lowest $R$, demonstrating that they require higher-precision localizations (relative to host size) than any other model tested. This can be naturally explained by the fact that substructure is of lower S/N and patchier than smooth light, so large localization uncertainties can hinder firm association. When combined into Disk and GC populations, we find the FRBs that trace disks have smaller $R$ than those that prefer GCs. 

When we compare the preferences with $\Delta$BIC\ $>3$ to the total model preferences, we find, unsurprisingly, that a higher localization precision relative to galaxy size (e.g., lower $R$) is required for positive statistical support. This is particularly true for the Residual, as no cases with $R>0.5$ return $\Delta$BIC\ $>3$ for the Residual model. The only exception to this trend is FRB\,20220105A whose large offset (3.1$r_e$) is likely driving its GC model preference despite having large $R$. Thus, $R$ is an important, but not the sole factor in driving the statistical support of model preferences.

\begin{figure*}
    \centering
\includegraphics[width=\textwidth]{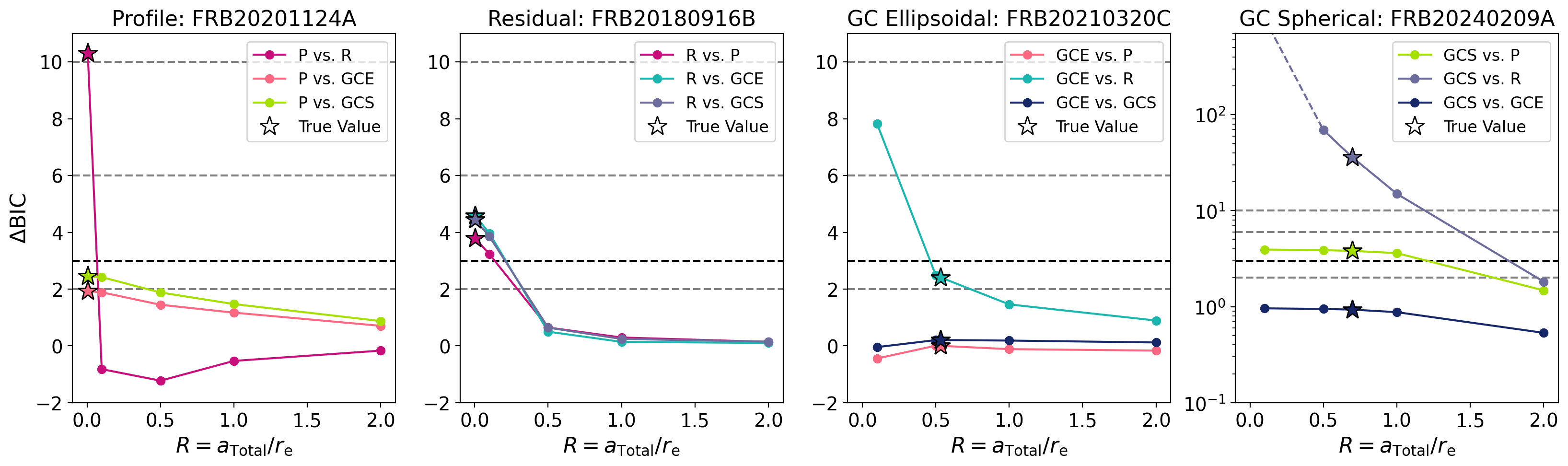}
\caption{Comparison of $\Delta$BIC for the preferred model against the other models as a function of $R$, the ratio of total localization ellipse semi-major axis ($a_{\rm Total}$) to galaxy size ($r_e$). Each panel corresponds to a representative model preference (from left to right: Profile, Residual, GC Ellipsoidal, GC Spherical). We denote the $\Delta$BIC at the actual $R$ with stars. The $\Delta$BIC\ $>3$ threshold for positive statistical evidence is shown by a dashed black line. The other thresholds for degrees of supporting evidence (see Section~\ref{sec:BIC}) are shown as grey dashed lines. We have abbreviated the model names in the legends for brevity (P=Profile, R=Residual, GCE = GC Ellipsoidal, GCS = GC Spherical). In the FRB\,20240209A panel, the first GC Spherical vs. Residual point is not shown as the Residual likelihood is exactly 0, hence returning a BIC of infinity; we represent the transition between that and the subsequent value as a dashed line.}
    \label{fig:R_testing}
\end{figure*}

As illustrative examples, in Figure~\ref{fig:R_testing} we show how the $\Delta$BIC of the preferred model changes with increasing $R$ for one representative FRB for each model preference: FRB\,20201124A (Profile), FRB\,20180916B (Residual), FRB\,20210320C (GC Ellipsoidal), and FRB\,20240209A (GC Spherical); we use the optical images when applicable. We find that the $\Delta$BIC values decrease with increasing $R$, demonstrating that we lose discriminating power for $R>1$. This is commensurate with the behavior shown in Figure~\ref{fig:R_boxplot}, in which the vast majority of FRBs with positive evidence in favor of their preferred model have $R<1$. This also affects the selection of various models differently: for the representative Profile and Residual FRBs in Figure~\ref{fig:R_testing}, $R<0.5$ is required to select a model with $\Delta$BIC\ $>3$, whereas this is not so stringent for the GC models.

Finally, while we find larger $R$ values for GC versus Disk model preferences, we emphasize that large localization sizes do not automatically translate to positive evidence in support of GC models. In fact, for FRBs\,20201124A and 20180916B (representative ``Disk'' FRBs), the $\Delta$BIC against the GC models never reaches a value of $-$3 (which would indicate positive support for the GC models instead) as $R$ increases, instead converging to 0. In addition, FRBs\,20210320C and 20240209A (representative FRBs with GC preferences) also lose supporting evidence at large $R$. Instead, other factors such as offset must also contribute (Section~\ref{sec:offset_results}). Lastly, this exercise shows that our original sample criterion of $R\lesssim0.5$ is a conservative requirement for attaining model preferences with positive statistical support, and our results are not strongly dependent on the exact cutoff.

\section{Discussion} \label{sec:discussion}

\subsection{Locations of FRBs relative to star formation and supernovae} \label{sec:discussion_stars}

Most FRB progenitor theories invoke magnetars (see \citealt{Platts+19} for a review). While there are many proposed ways to produce magnetars, the most commonly accepted channel is via the core collapse of massive stars. Motivated by the preliminary findings that FRBs align with regions of substructure (such as spiral arms) within their host galaxies \citep{Chittidi21,Mannings+21, Woodland+24}, we examine the interpretation that the majority of FRBs have progenitors associated with massive stars. 

Spiral arms are dense regions of gas, dust, and star formation within their galaxies (see \citealt{Sellwood-Masters2022} for a review). While spiral arms are visible across a wide range of wavelengths, there is a clear presence of young ($\lesssim$30 Myr), massive O- and B-type stars in arms, as traced by UV and H$\alpha$ imaging (e.g., \citealt{Sanchez-Gil+11,PHANGS-MUSE,PHANGS-HST}). Overall, star formation occurs in H$\textsc{\,ii}$ regions within giant molecular clouds (GMCs), which are spread throughout spiral galaxies and are thus not solely confined to arms (e.g., \citealt{Dobbs+06,Querejeta+21}). Indeed, studies of star formation tracers in nearby galaxies indicate that up to half of all star formation can occur in the interarm regions (i.e., any region outside of spiral arms; \citealt{Elmegreen-Elmegreen1986,Foyle+10,Kreckel+16}). Thus, even if 100\% of FRB progenitors depend on recent star formation, we should not expect 100\% of the hosts examined here to return a preference for the Residual model. With regard to young, massive stars, the GMCs in spiral arms are more massive than those in interarm regions, translating to larger gas reservoirs that in turn enable them to form a larger number of massive stars (\citealt{Kreckel+16};  although massive star formation can also occur outside of spiral arms). Thus, the fact that several FRBs are located visibly offset from the nearest spiral arms (Section~\ref{sec:BIC}) does not preclude a progenitor from a young massive star. However, as only $20-46\%$ of our sample (at $z\lesssim0.15$) favor association with spiral arms, this indicates that high star-formation efficiencies, particularly at the high-mass end, are not a strong driver in the production of FRB progenitors. 

Given that the most natural channel for magnetar formation is via core-collapse supernovae (CCSNe), we are motivated to compare how these transients trace spiral structure in their hosts. Indeed, studies on the distribution of CCSNe at $z \lesssim 0.15$ find the vast majority ($\gtrsim88\%$) occur on spiral arms \citep{Aramyan+16} in proximity to star-forming regions \citep{Galbany+14} and strongly trace H$\alpha$ within their host galaxies (a sensitive probe of massive O-type stars, \citealt{Anderson-James09}). If the FRB progenitor is also associated with massive stars, we would expect a similarly high fraction of FRBs to coincide with spiral arms. Instead, within the same redshift range ($z \lesssim 0.15$), we find that $\approx 20-46\%$ of FRBs favor spiral arms. From visual inspection alone, 7/15 FRBs ($\approx 46\%$) in this redshift range land on any residual substructure, still less than the CCSNe fraction.

We explore a few factors to see if they could reconcile these differences. First, if a massive O-type (B-type) star with a lifetime of 10 Myr (30$-$100 Myr) traveled 10 km~s$^{-1}$ from its birth site on a spiral arm, it could traverse a distance of $\sim0.1$~kpc (0.3$-$1 kpc) before its CCSN. These distances would correspond to an angular offset of $\sim0.05-0.5\arcsec$ at $z\approx0.1$ which could be observable at the upper bound given our localization sizes. However, the observed population of CCSNe should also be sensitive to such spatial drift. Given the large majority of CCSN occur on arms, it is instead more relevant to consider post core-collapse kicks of the magnetars themselves. Using the pulsar natal kick velocity distribution ($\sim 100-1000$~km~s$^{-1}$; \citealt{Hobbs+05,Deller+19}) and a fiducial $\sim10^4$~yr for the active lifetime of magnetars \citep{Kaspi-Beloborodov17}, we find a maximum travel distance of $\sim10$~pc from the nearest spiral arm. At $z \approx 0.1$, this translates to a $\sim\!5$~mas offset. This distance is much smaller than our FRB localization sizes and we would not be sensitive to such offsets. 

A second possibility is that in our framework, larger FRB localization sizes can preclude strong associations to spiral arms. Indeed, we have shown that precise localizations are required to return positive statistical support associating FRBs to spiral structure\footnote{As an example, FRB\,20190102C's large localization is not inconsistent with being on a spiral arm but a larger fraction of the FRB localization area overlaps with regions away from the arm than on it, and hence the Residual model is disfavored (Figure~\ref{fig:all_models}).} (Section~\ref{sec:loc_size}). However, at least three of the FRBs in our sample are clearly offset from spiral arms, even at the current localization precision. Given the difficulty of statistical association with low S/N light distributions, and the small numbers of FRBs known at $z\lesssim 0.15$ (where it is easiest to detect spiral arms), the current fraction of FRBs associated to visible spiral arms ($\approx 20-46\%$ at low redshift) can be interpreted as a lower bound. Thus, while the difference between FRBs and CCSNe locations with respect to spiral arms is notable, it is possible these fractions could be explained by a difference in methods and through a larger sample of FRBs at low-$z$.

If they cannot be reconciled, this could instead indicate that not all FRBs depend on massive star formation. Indeed, recently, \citet{Horowicz-Margalit2025} demonstrated that the host demographics of FRBs and Type Ia SNe (whose intermediate-age stellar progenitors are weak tracers of star formation) show considerable similarities and suggest they may even share similar progenitors. It is interesting to note that, like FRBs, SNe\,Ia occur both on and off spiral arms \citep{Aramyan+16,Audcent-Ross+20}, with 66\% (at $z\lesssim0.1$) occurring on spiral arms, modestly larger than our lower bound for FRBs. 

We also find that our results are consistent across both observed optical and NIR wavelengths (rest-frame range of $\sim$4000-6500\AA\ for $R$-band and $\sim$13000-21000\AA\ for $K$-band). NIR light traces the stellar mass density of older giant stars (e.g., \citealt{Elmegreen-Elmegreen84,Eskridge+02,Foyle+10}), whereas visible light is primarily dominated by A through F-type main sequence stars and G through K giant stars \citep{Kennicutt1998}. We do not find strong differences in the model preferences across wavelength regimes, nor any apparent differences in galaxy morphology. Thus, despite the differing types of stars traced by the two wavelengths, our sample is not sensitive to these differences. This is likely attributed to an imperfect sample to diagnose this; our sample, which extends to $z \approx 0.65$, smears the distribution of rest-frame light being probed. We also note that the model preferences and corresponding degrees of statistical support for repeating and apparently non-repeating FRBs do not show clear trends. While all repeaters do have model preferences with $\Delta$BIC\ $>3$, this is largely attributable to the enhanced localization precision afforded by repetition through VLBI.

A larger sample of precisely localized FRBs at low-$z$ (where the fraction of true spiral structure is essentially complete, and typical localization precision as a fraction of host galaxy size is the highest) would greatly aid in constraining the fraction of FRBs that are associated with spiral arms, its comparison to massive stars, and CCSN locations. We caution, however, that any straightforward extrapolation of the lowest redshift results to the broader FRB population rests on the assumption that there is no significant redshift dependence for the progenitor(s). If, for instance, multiple FRB progenitors existed with different luminosity functions, this assumption could easily be violated, and the higher redshift FRB population may have different properties on average. But taken at face value, the current results indicate that some (likely a majority, but not all) FRB progenitors are associated with massive star formation.

\subsection{Fraction of FRBs from Globular Clusters}
\label{sec:discussion_GCs}

To date, one FRB has been directly observed to originate in a globular cluster, FRB\,20200120E \citep{Bhardwaj21,Kirsten21}, only possible due to the proximity of its host galaxy M81 at 3.6~Mpc. Another FRB\,20240209A was strongly implied to have originated in a GC given its large offset and massive elliptical host \citep{Eftekhari+25,Shah+25}. However, the distance to this FRB implied by its redshift of $z=0.1384$ makes observational confirmation of a GC origin currently extremely challenging. As the majority of FRBs are beyond the horizon for which we can directly observe GCs (a few tens of Mpc, \citealt{vanDokkum18,Beasley2020}), indirect evidence is needed to infer their origins. The methodology we present in this work, which statistically compares the likelihood of FRB association with varying spatial distributions, is well-poised to address this problem. By leveraging precise localizations and host galaxy imaging, we can place meaningful, statistical constraints on GC origins. The strong statistical evidence in support of a GC origin for FRB 20200120E in our framework is reassuring and, taken together with meaningful support in favor of GC models in three other cases (FRBs\,20240209A, 20220105A, and 20210117A) indicates a minority of the observed cosmological FRB population does actually originate via this channel. This framework is also complementary to studies of host demographics which can provide further indirect evidence for GC origins (i.e., massive elliptical galaxies are known to have large populations of GCs, e.g., \citealt{Harris91}). Thus, our framework is a powerful tool to place constraints on GC origins for FRBs at higher redshifts, for which direct observational confirmation will require extreme depth and resolution. However, we note that while we do find some preferences for GCs at low ($\lesssim1r_e$) host-normalized offset, this method is more sensitive to preferring GCs for FRBs at large ($\gtrsim2r_e$) offset.

In addition to FRB\,20200120E, we find evidence for three more FRBs to favor GC models at $\Delta$BIC\ $>3$ (Section~\ref{sec:BIC}): FRBs\,20210117A, 20220105A, and 20240209A. While 13$\pm3$ FRBs (37\% of the sample) favor GC models, only four reach $\Delta$BIC\ $>3$ (11$\pm5$\%), indicating that the fraction of FRBs that are consistent with the spatial distribution of GCs is a minority of the overall FRB population. Nevertheless, this population of candidate and confirmed GC FRBs indicates that some fraction of FRB progenitors do not depend on massive star formation. As GCs are collections of old stellar populations and remnants (i.e., white dwarfs, neutron stars, black holes) and the active lifetime of magnetars is short, the magnetars that may have created these FRBs instead likely formed via dynamical channels such as close binary evolution or accretion-induced collapse of a white dwarf \citep{Moriya16,Margalit19,Kremer2021,Lu+22}.

The statistical support for four FRBs to originate in GCs motivates us to examine if only a minority originate from GCs, or if there is a population that could be missing from our sample due to selection effects. Since we have demonstrated that large offset ($\gtrsim 3r_e$) is a substantial factor to return positive statistical support in favor of GC models, in this discussion we use large projected galactocentric offset, which is relatively easy to determine, as a proxy for GC selection. We first note that there are no obvious reasons that should bias against detection of FRBs at large offsets. For instance, the outskirts of galaxy halos are less dense than central regions, implying that FRBs originating at larger offsets should be less scatter-broadened by the host galaxy ISM on average than those in more central regions.\footnote{However, if the luminosity of GC FRBs are considerably fainter than those not in GCs, this could introduce observational bias. \citet{Kirsten21} report that the luminosity of FRB\,20200120E is $\sim10^3$ times lower than other known FRBs, but the luminosity of FRB\,20240209A (which likely originates in a GC) is comparable to the known repeater population \citep{Shah+25}.}.
 
To examine whether selection effects play a role, we first consider a scenario in which we are missing highly offset FRBs due to a lack of deep optical followup. We focus on the CRAFT sample which comprises the majority of the FRBs in our work. Of the 83 FRBs that CRAFT has detected through the end of 2024, 57 have localizations that are sufficiently precise ($\lesssim2\arcsec$) to enable host association. Of these 57, only six have no apparent host galaxies: one due to close proximity to a bright star inhibiting followup, one due to lack of deep optical followup, and the remaining four remain without a clear host galaxy despite deep optical followup (due to a combination of high extinction and/or high DM, which generally corresponds to fainter, more distant galaxies). Thus, among the FRBs with localizations that enable host association, the CRAFT sample is largely complete with only a handful of indeterminate hosts.

Next, we consider the scenario in which our selection criteria bias us against large offset FRBs. There are 18 CRAFT FRBs that were excluded from our sample due to localizations either $\gtrsim2\arcsec$ or $\gtrsim50\%$ of their approximate host size (see Section~\ref{sec:sample} for further details of the sample criteria). Of these 18, the large majority (16) have localizations that either entirely overlap the disk of the host or are consistent with being both on and off the stellar disk, making detailed inference of their location impossible. The remaining two FRBs have positions that are clearly $\sim2-3r_e$ offset from their hosts. Thus, within the CRAFT sample, there are a total of six FRBs that could plausibly contribute to the GC population: four with indeterminate hosts and two with considerable offsets (but large localizations). In the scenario where these six FRBs all have large offsets and can be significantly associated with the distributions of GCs, the fraction of FRBs originating from GCs would rise from $\sim$11$\pm$5\% to a modest $\sim$23$\pm$6\%, still a minority of the full population. We note that this is a fairly crude exercise which equates large offsets to origins in GCs, and does not take into account FRBs which may originate from GCs at small offset. 

Either way, we find no support for a majority of observed FRBs arising from GCs\footnote{We note that our method assumes that FRBs trace a light distribution. Should an FRB originate in a globular cluster that is close to the center of the galaxy in projection, the higher central concentration of the smooth galaxy light would lead to the Profile model being favored over a GC origin for that individual FRB in our framework. More precisely estimating the fractional population contributions will require an approach that combines the probability origins of the entire FRB sample.}. Our results, derived from novel methods which solely examine the locations of FRBs within their hosts, are commensurate with inferences from the demographics of FRB host galaxies (e.g., \citealt{Gordon+23,Sharma+24,Horowicz-Margalit2025}) and rates for magnetar production in GCs (e.g., \citealt{Lu+22,Rao+25}), which suggest only a minority of FRBs are expected to occur in GCs. However, our analysis argues against FRB\,20200120E as the only detected FRB originating from a channel unrelated to recent massive star formation. Indeed, while our work strongly supports a GC origin for the repeating FRB\,20240209A, if FRB\,20210117A or 20220105A were confirmed to originate in a GC, it would be the first for a non-repeater. Taken together, our work supports multiple channels for magnetar formation to produce the observed FRB population, but with a minority formed through dynamical channels. 

\section{Conclusions and Future Work} \label{sec:conclusions}

Using a sample of 37 well-localized FRBs, 34 of which come from the CRAFT survey, we have presented a comprehensive analysis of the spatial distribution of FRBs within their host galaxies. We use the statistical framework \texttt{coflash} to test the consistency of the FRB locations with respect to four models: the smooth stellar light profile of their hosts, residual substructure (such as spiral arms), and two synthetic GC distributions (ellipsoidal and spherical). We perform this analysis on both optical and NIR imaging. We summarize our findings here:

\begin{itemize}
    \item We present seven new CRAFT FRBs: 20231230D, 20240117B, 20240203C, 20240312D, 20240525A, 20240615B, and 20241027B. The FRBs in our sample span a redshift range of $z=0.0008-0.643$ with a median $z=0.193$. The majority are apparent non-repeaters, with only 4/37 known to repeat. 
    \item We find that $70\%$ of our host galaxy sample exhibit clear disks, and $\approx 55\%$ show clear spiral arms. The spiral fraction rises to 86\% at $z \lesssim 0.15$, demonstrating the near-ubiquity of spiral structure in FRB hosts at low-$z$. Only two FRB hosts in our sample have elliptical morphologies. Despite the prevalence of spiral arms, FRB positions vary widely in relation to spiral substructure: some are coincident while others appear clearly offset.
    \item We calculate the projected angular, physical, and host-normalized galactocentric offsets for all FRBs in our sample, deriving population medians (with 68\% confidence intervals) of 1.4$^{+3.3}_{-0.7}$\arcsec, 4.2$^{+5.7}_{-2.5}$~kpc, and 1.1$^{+1.5}_{-0.6}~r_e$. Approximately half (19/37) of FRBs are located within 1$r_e$ of their host.
    \item Across the sample, the majority of FRBs favor the smooth galaxy light model, with only $\sim 8\%$ each that favor either the residual (spiral) or GC models. However, this changes when considering FRB hosts at low-redshifts ($z\lesssim 0.15$) only, for which $\approx 20-46\%$ favor an association with spiral arms, while only $\approx 7-20\%$ favor the smooth galaxy light.
    \item Overall, the majority of FRBs favor locations in the disk (smooth galaxy light or spiral structure), as opposed to the more extended distribution of GCs. While the galactocentric offset distributions of FRBs that favor disk versus GC models are not statistically distinct, large offsets of $\gtrsim 3r_e$ clearly and unsurprisingly drive a GC preference. The addition of only two more FRBs at such large offsets would lend statistical support for two distinct FRB offset distributions.
    \item We find no clear trends in how the locations of repeating and non-repeating FRBs trace the light distributions tested. The ultra-precise localizations of the four repeaters in our sample enable all to have positive statistical evidence for their preferred models.
    \item At comparable redshifts ($z\lesssim0.15$), a larger percentage of CCSNe occur on spiral arms ($\gtrsim88\%$) than FRBs which favor spiral structure in our sample ($\approx20-46\%$). We find no plausible physical reason to explain this discrepancy, beyond differences in methodology. Indeed, our methods are quite conservative and loosely depend on factors such as localization size and association to low S/N structures; thus our spiral fraction should be interpreted as a lower bound. 
    \item As star formation occurs both within and outside of spiral arms, with the youngest stars forming at highest efficiencies within arms, the modest fraction of spiral arm associations implies that high star formation efficiencies do not seem to be a strong factor in the production of the FRB progenitor. However, the fact that several FRBs favor smooth galaxy light and/or are visibly offset from spiral structure does not preclude association to recent star formation. 
    \item Beyond the sole confirmed FRB\,20200120E, we find support for three additional FRBs originating from GCs ($\approx$11$\pm$5\% across the population, or $\approx$13$\pm$9\% at low-redshift), although we find no support for a majority to originate from GCs based on their locations. Thus, our work supports multiple channels for magnetar formation to produce FRBs with only a minority coming from dynamical channels.
    \item We note that our results depend on the assumption that FRBs trace light distributions and that the true model is among the models that we test. However, as none of the tested models are systematically disfavored, none appear to be truly poor descriptors of the spatial distribution of FRBs.
\end{itemize}

The location of an FRB within its host galaxy is a powerful diagnostic for constraining the progenitor(s) and formation channels of FRBs. Higher precision localizations will help place stronger constraints on their locations, particularly in constraining the fraction associated with spiral arms. These constraints will be strongest for low-$z$ FRBs as spiral structure is more easily apparent in the local Universe. It is also possible to place more stringent constraints on the stellar populations that give rise to FRB progenitors by considering their position relative to substructure traced by different wavelengths of light, such as UV which traces young massive stars, or H$\alpha$ which maps the location of the most recent star formation. Additionally, with larger numbers of FRBs it would be possible to split these into smaller redshift bins, which could test if there is any redshift dependence on model preference, and potentially constrain progenitor ages by isolating specific stellar populations.

\texttt{coflash} is a flexible framework that is easily modified to test different spatial distributions. More complicated prescriptions for a progenitor that traces both star formation and stellar mass, for example, could be implemented. Further, this framework has already proved to be a strong diagnostic for testing GC origins complementary to global host demographics such as stellar mass, morphology, and degree of star formation. This makes it a powerful tool to constrain the fraction of FRBs associated with globular clusters, as direct observational confirmation will be feasible for only the most nearby events. By extending our framework from considering each source in isolation to jointly modeling the population of host galaxies, it will be possible to further increase the precision of this constraint.

\section{Acknowledgements}

A.C.G. thanks Chang Liu, Tim Miller, Anya Nugent, Noah Rogers, Vishwangi Shah, and Patrick Sheehan for helpful discussions and feedback on various aspects of this work.

A.C.G. and the Fong Group at Northwestern acknowledge support by the National Science Foundation under grant Nos. AST-1909358, AST-2206494, AST-2308182 and CAREER grant No. AST-2047919. W.F. gratefully acknowledges support by the David and Lucile Packard Foundation, the Alfred P. Sloan Foundation, and the Research Corporation for Science Advancement through Cottrell Scholar Award \#28284.

A.C.G., W.F., Y.D., T.E., C.D.K., J.X.P., S.S., and N.T. acknowledge support from NSF grants AST-1911140, AST-1910471 and AST-2206490 as members of the Fast and Fortunate for FRB Follow-up team.

S.S. is a joint NU-UC Brinson Postdoctoral Fellow and acknowledges the generous support of the Brinson Foundation. C.D.K. gratefully acknowledges support from the NSF through AST-2432037, the HST Guest Observer Program through HST-SNAP-17070 and HST-GO-17706, and from JWST Archival Research through JWST-AR-6241 and JWST-AR-5441. C.L. is supported by DoE award \#DE-SC0025599. R.M.S. acknowledges support through Australian Reserach Council Future Fellowship FT190100155 and Discovery Project DP220102305. M.G. is supported by the UK STFC Grant ST/Y001117/1. M.G. acknowledges support from the Inter-University Institute for Data Intensive Astronomy (IDIA). IDIA is a partnership of the University of Cape Town, the University of Pretoria and the University of the Western Cape. Y.D. is supported by the National Science Foundation Graduate Research Fellowship under grant No. DGE-2234667. N.T. acknowledges support by FONDECYT grant 1252229. J.J-S. acknowledges support through Australian Research Council Discovery Project DP220102305.

Based in part on observations obtained at the Southern Astrophysical Research (SOAR) telescope, which is a joint project of the Minist\'{e}rio da Ci\^{e}ncia, Tecnologia e Inova\c{c}\~{o}es (MCTI/LNA) do Brasil, the US National Science Foundation’s NOIRLab, the University of North Carolina at Chapel Hill (UNC), and Michigan State University (MSU).

W. M. Keck Observatory and MMT Observatory access was supported by Northwestern University and the Center for Interdisciplinary Exploration and Research in Astrophysics (CIERA).

This publication makes use of data obtained through Swinburne Keck program 2024B W366.

Some of the data presented herein were obtained at the W. M. Keck Observatory, which is operated as a scientific partnership among the California Institute of Technology, the University of California and the National Aeronautics and Space Administration. The Observatory was made possible by the generous financial support of the W. M. Keck Foundation.

The authors wish to recognize and acknowledge the very significant cultural role and reverence that the summit of Maunakea has always had within the indigenous Hawaiian community.  We are most fortunate to have the opportunity to conduct observations from this mountain.

Observations reported here were obtained at the MMT Observatory, a joint facility of the Smithsonian Institution and the University of Arizona.

Based on observations collected at the European Southern Observatory under ESO programmes 105.204W and 108.21ZF.

This scientific work uses data obtained from Inyarrimanha Ilgari Bundara / the Murchison Radio-astronomy Observatory. We acknowledge the Wajarri Yamaji People as the Traditional Owners and native title holders of the Observatory site. CSIRO’s ASKAP radio telescope is part of the Australia Telescope National Facility (https://ror.org/05qajvd42). Operation of ASKAP is funded by the Australian Government with support from the National Collaborative Research Infrastructure Strategy. ASKAP uses the resources of the Pawsey Supercomputing Research Centre. Establishment of ASKAP, Inyarrimanha Ilgari Bundara, the CSIRO Murchison Radio-astronomy Observatory and the Pawsey Supercomputing Research Centre are initiatives of the Australian Government, with support from the Government of Western Australia and the Science and Industry Endowment Fund.

This research has made use of the Keck Observatory Archive (KOA), which is operated by the W. M. Keck Observatory and the NASA Exoplanet Science Institute (NExScI), under contract with the National Aeronautics and Space Administration.

This research has made use of the NASA/IPAC Infrared Science Archive, which is funded by the National Aeronautics and Space Administration and operated by the California Institute of Technology.

Funding for SDSS-III has been provided by the Alfred P. Sloan Foundation, the Participating Institutions, the National Science Foundation, and the U.S. Department of Energy Office of Science. The SDSS-III web site is http://www.sdss3.org/.

SDSS-III is managed by the Astrophysical Research Consortium for the Participating Institutions of the SDSS-III Collaboration including the University of Arizona, the Brazilian Participation Group, Brookhaven National Laboratory, Carnegie Mellon University, University of Florida, the French Participation Group, the German Participation Group, Harvard University, the Instituto de Astrofisica de Canarias, the Michigan State/Notre Dame/JINA Participation Group, Johns Hopkins University, Lawrence Berkeley National Laboratory, Max Planck Institute for Astrophysics, Max Planck Institute for Extraterrestrial Physics, New Mexico State University, New York University, Ohio State University, Pennsylvania State University, University of Portsmouth, Princeton University, the Spanish Participation Group, University of Tokyo, University of Utah, Vanderbilt University, University of Virginia, University of Washington, and Yale University.

This work has made use of data from the European Space Agency (ESA) mission
{\it Gaia} (\url{https://www.cosmos.esa.int/gaia}), processed by the {\it Gaia}
Data Processing and Analysis Consortium (DPAC,
\url{https://www.cosmos.esa.int/web/gaia/dpac/consortium}). Funding for the DPAC
has been provided by national institutions, in particular the institutions
participating in the {\it Gaia} Multilateral Agreement.

The Pan-STARRS1 Surveys (PS1) and the PS1 public science archive have been made possible through contributions by the Institute for Astronomy, the University of Hawaii, the Pan-STARRS Project Office, the Max-Planck Society and its participating institutes, the Max Planck Institute for Astronomy, Heidelberg and the Max Planck Institute for Extraterrestrial Physics, Garching, The Johns Hopkins University, Durham University, the University of Edinburgh, the Queen's University Belfast, the Harvard-Smithsonian Center for Astrophysics, the Las Cumbres Observatory Global Telescope Network Incorporated, the National Central University of Taiwan, the Space Telescope Science Institute, the National Aeronautics and Space Administration under Grant No. NNX08AR22G issued through the Planetary Science Division of the NASA Science Mission Directorate, the National Science Foundation Grant No. AST-1238877, the University of Maryland, Eotvos Lorand University (ELTE), the Los Alamos National Laboratory, and the Gordon and Betty Moore Foundation.

This publication makes use of data products from the Wide-field Infrared Survey Explorer, which is a joint project of the University of California, Los Angeles, and the Jet Propulsion Laboratory/California Institute of Technology, and NEOWISE, which is a project of the Jet Propulsion Laboratory/California Institute of Technology. WISE and NEOWISE are funded by the National Aeronautics and Space Administration.


NOIRLab IRAF is distributed by the Community Science and Data Center at NSF NOIRLab, which is managed by the Association of Universities for Research in Astronomy (AURA) under a cooperative agreement with the U.S. National Science Foundation. 

This research made use of Regions, an Astropy package for region handling (Bradley et al. 2024).

%

\vspace{5mm}
\facilities{
\textit{HST} (WFC3),
Keck:1 (LRIS),
MMT (MMIRS, Binospec),
Sloan (SDSS),
SOAR (Goodman),
VLT:Antu (FORS2),
VLT:Kueyen (X-Shooter),
VLT:Yepun (HAWK-I),
}


\software{
\texttt{astropy} \citep{astropy},
\texttt{astroquery} \citep{astroquery},
\texttt{cmasher} \citep{cmasher},
\texttt{coflash}\footnote{https://github.com/adamdeller/coflash},
\texttt{CRAFT-OPTICAL-FOLLOWUP}\footnote{https://github.com/Lachimax/craft-optical-followup},
\texttt{DoPhot} \citep{Schechter93},
\texttt{dynesty} \citep{Speagle2020},
\texttt{ESOReflex} \citep{esoreflex},
\texttt{FRBs/FRB} \citep{F4_repo},
\texttt{galfit} \citep{galfit_2002,galfit_2010},
\texttt{IRAF} \citep{IRAF},
\texttt{linetools} \citep{linetools},
\texttt{matplotlib} \citep{matplotlib},
\texttt{numpy} \citep{numpy},
\texttt{pandas} \citep{pandas},
\texttt{photpipe} \citep{Rest+05},
\texttt{photutils} \citep{photutils},
\texttt{POTPyRI}\footnote{https://github.com/CIERA-Transients/POTPyRI},
\texttt{Prospector} \citep{Johnson+21},
\texttt{pypalettes}\footnote{https://github.com/JosephBARBIERDARNAL/pypalettes},
\texttt{PypeIt} \citep{pypeit:joss_pub,pypeit:zenodo},
\texttt{python-fsps} \citep{Conroy2009,Conroy2010},
\texttt{SAOImageDS9} \citep{DS9},
\texttt{regions} \citep{regions},
\texttt{scipy} \citep{scipy},
\texttt{sedpy} \citep{sedpy},
\texttt{sep} \citep{SEP},
\texttt{Source Extractor} \citep{source_extractor},
\texttt{swarp} \citep{swarp}
}
\clearpage


\clearpage
\appendix
\restartappendixnumbering

\section{Light Profile and Globular Cluster Distribution Models} \label{sec:models_appendix}

\begin{figure}[h]
    \centering
    \includegraphics[width=0.6\textwidth]{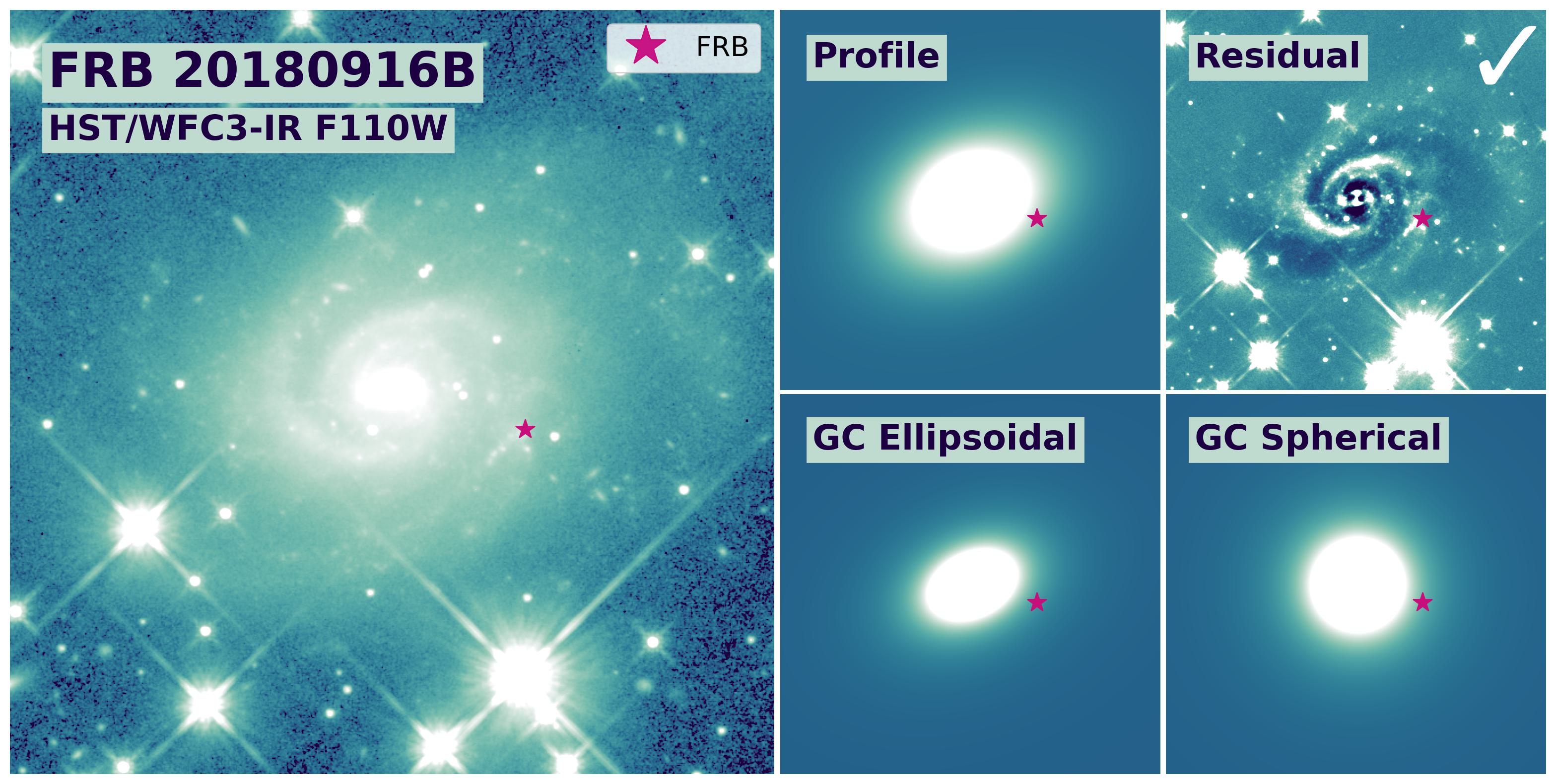}
    \includegraphics[width=0.39\textwidth]{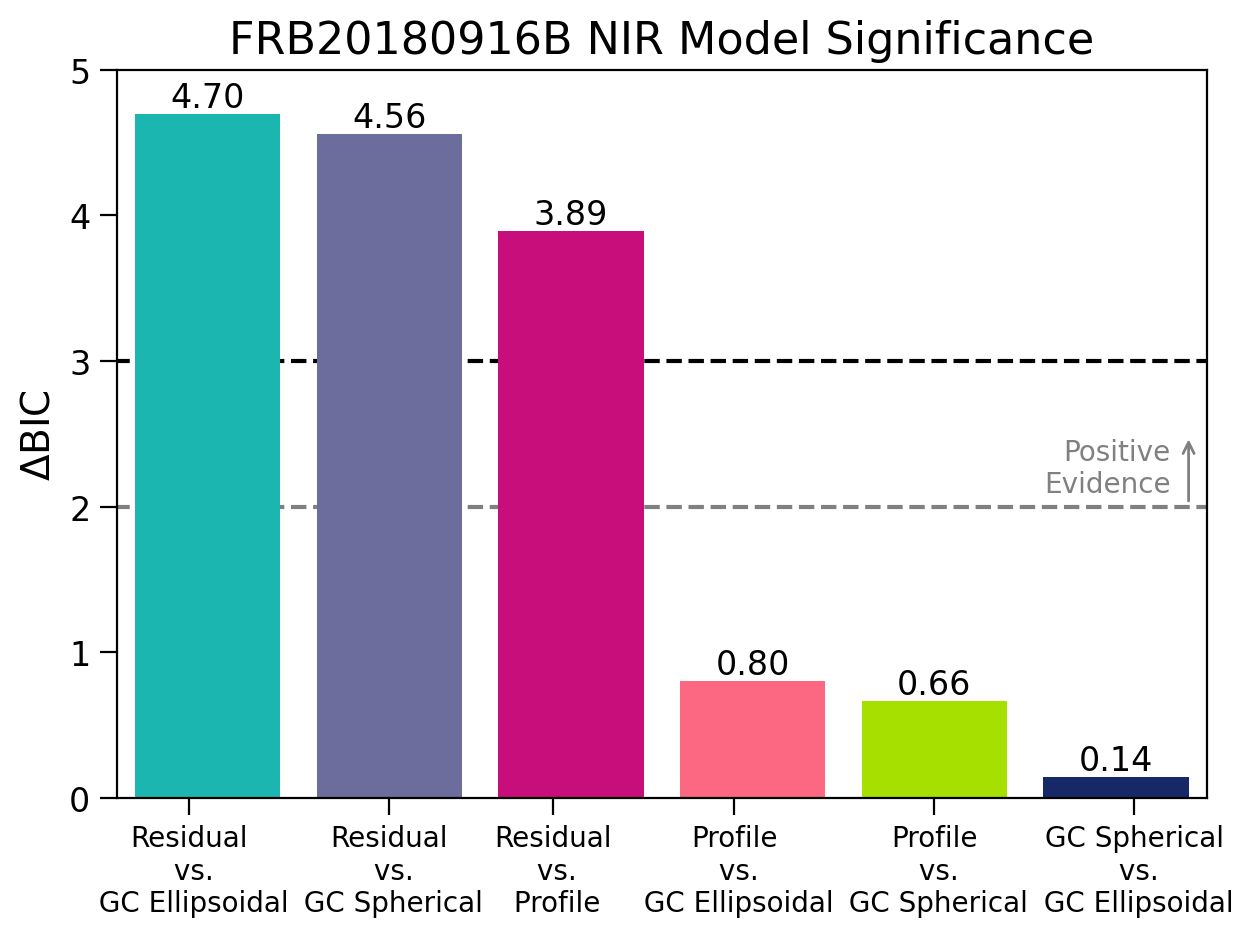}
    \caption{\textbf{(Left:)} Imaging and light profiles for FRB\,20180916B in the NIR. \textbf{(Right:)} Comparison of the $\Delta$BIC between the four models for FRB\,20180916B in the NIR. The complete figure set (57 images) is available in the online journal.}
    \label{fig:all_models}
\end{figure}

We present the models and statistical significance comparisons for each FRB host in this appendix in Figure~\ref{fig:all_models}.

In this preprint version, we show all figures contained in Figure Set~\ref{fig:all_models} below, spanning Figures~\ref{fig:appendix_180924}$-$\ref{fig:appendix_241027_opt}.

\begin{figure}
    \centering
    \includegraphics[width=0.6\textwidth]{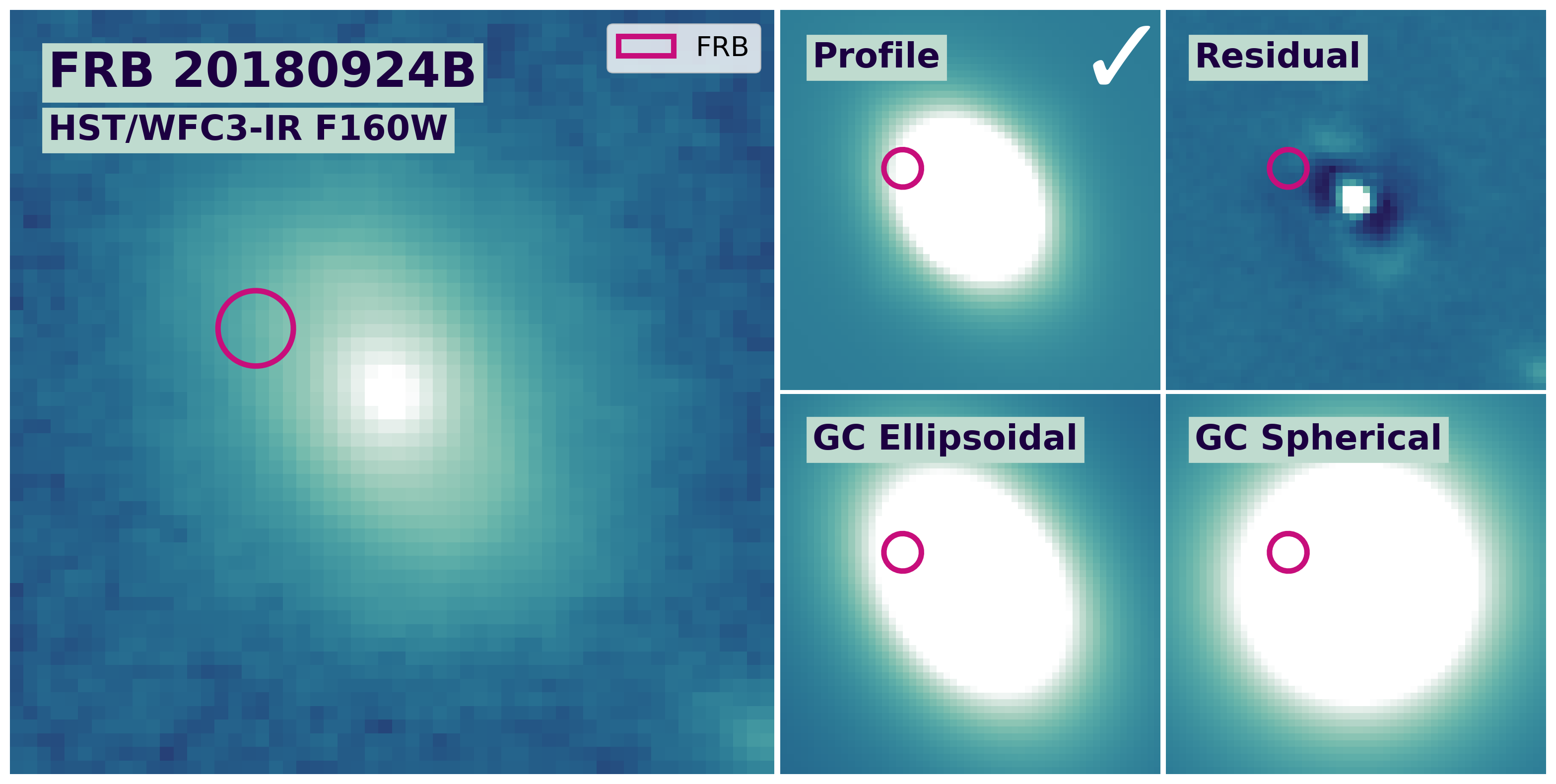}
    \includegraphics[width=0.39\textwidth]{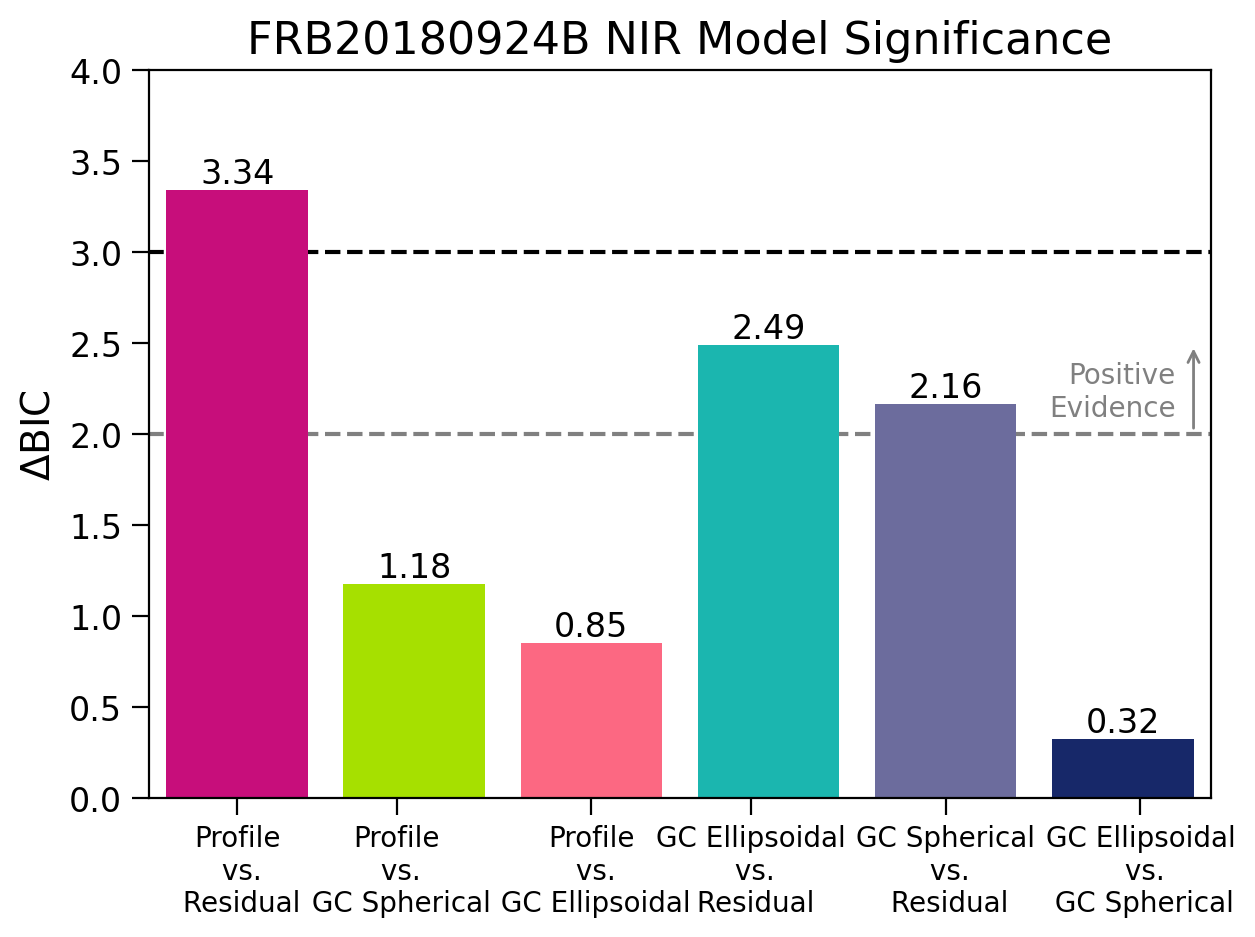}
    \caption{\textbf{(Left:)} Imaging and light profiles for FRB\,20180924B in the NIR. \textbf{(Right:)} Comparison of the $\Delta$BIC between the four models for FRB\,20180924B in the NIR.}
    \label{fig:appendix_180924}
\end{figure}

\begin{figure}
    \centering
    \includegraphics[width=0.6\textwidth]{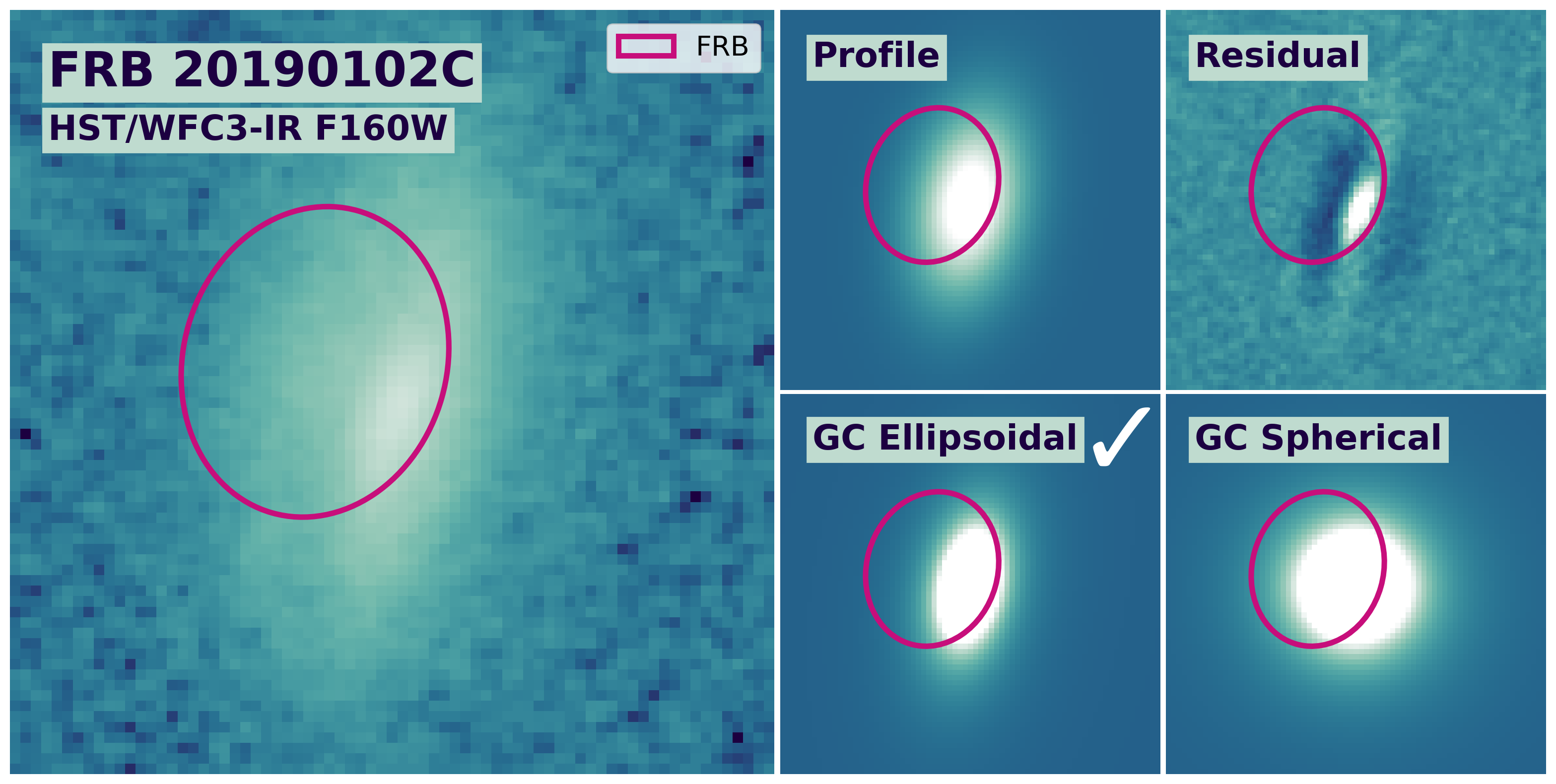}
    \includegraphics[width=0.39\textwidth]{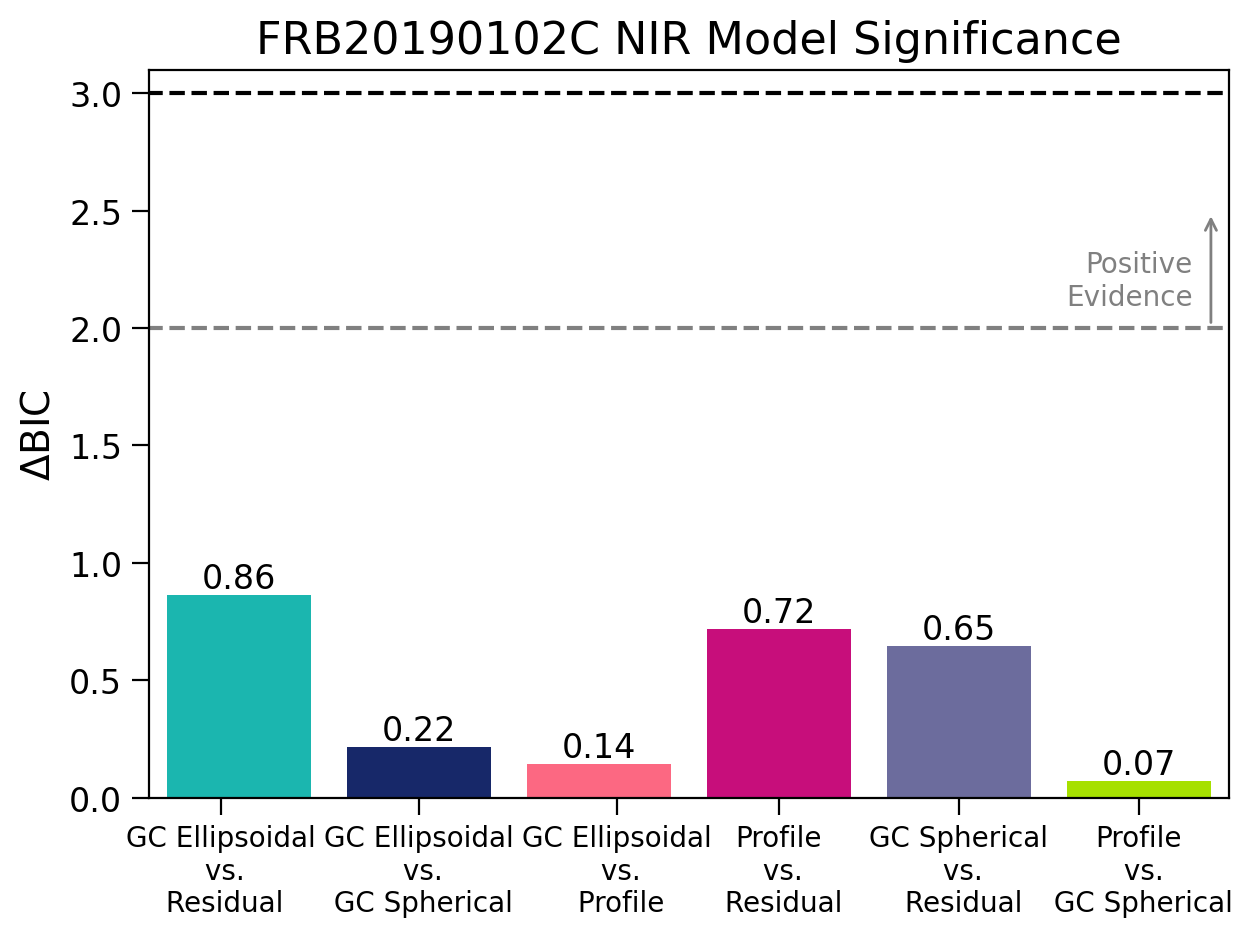}
    \caption{\textbf{(Left:)} Imaging and light profiles for FRB\,20190102C in the NIR. \textbf{(Right:)} Comparison of the $\Delta$BIC between the four models for FRB\,20190102C in the NIR.}
    \label{fig:appendix_190102}
\end{figure}

\begin{figure}
    \centering
    \includegraphics[width=0.6\textwidth]{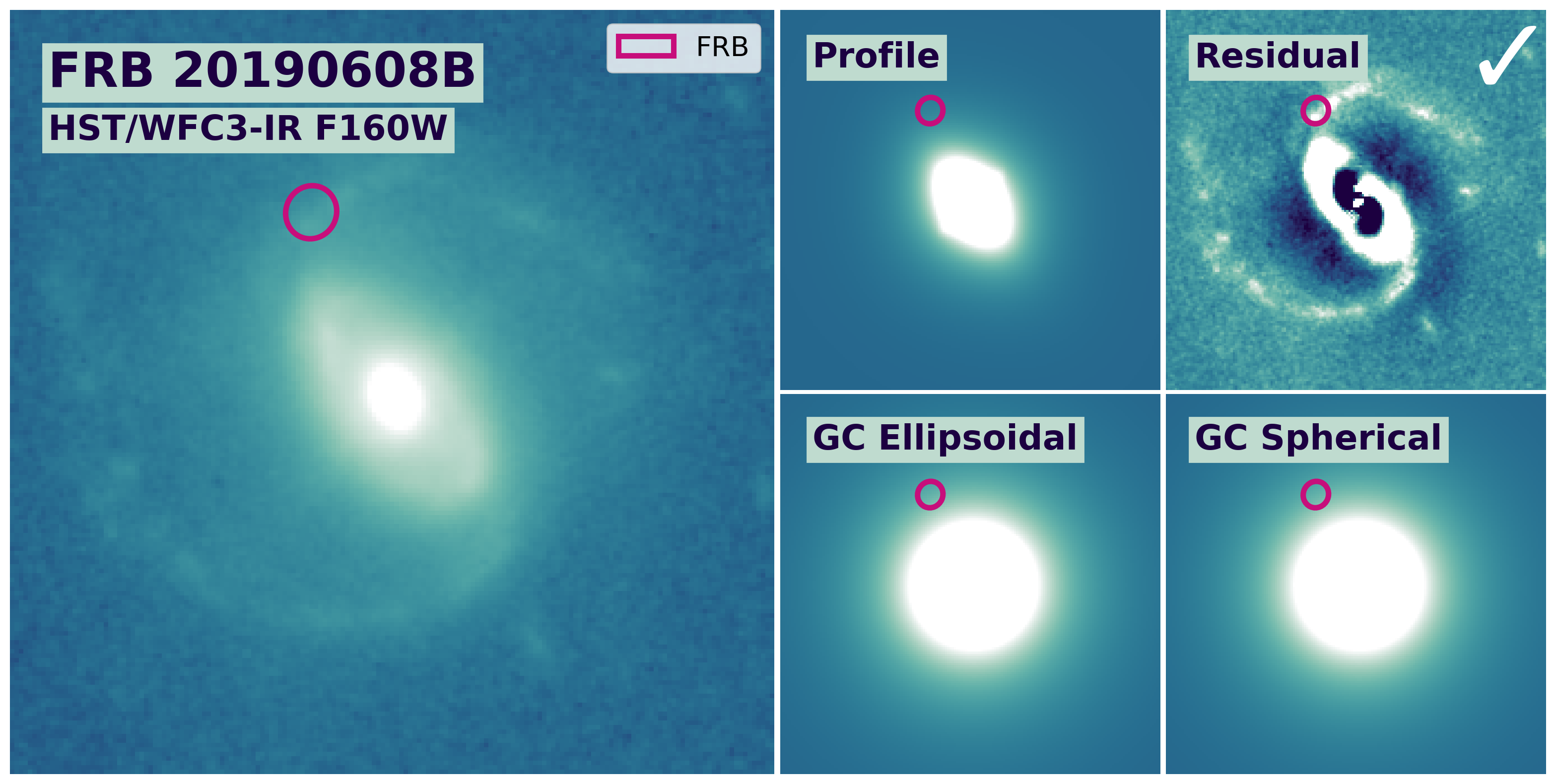}
    \includegraphics[width=0.39\textwidth]{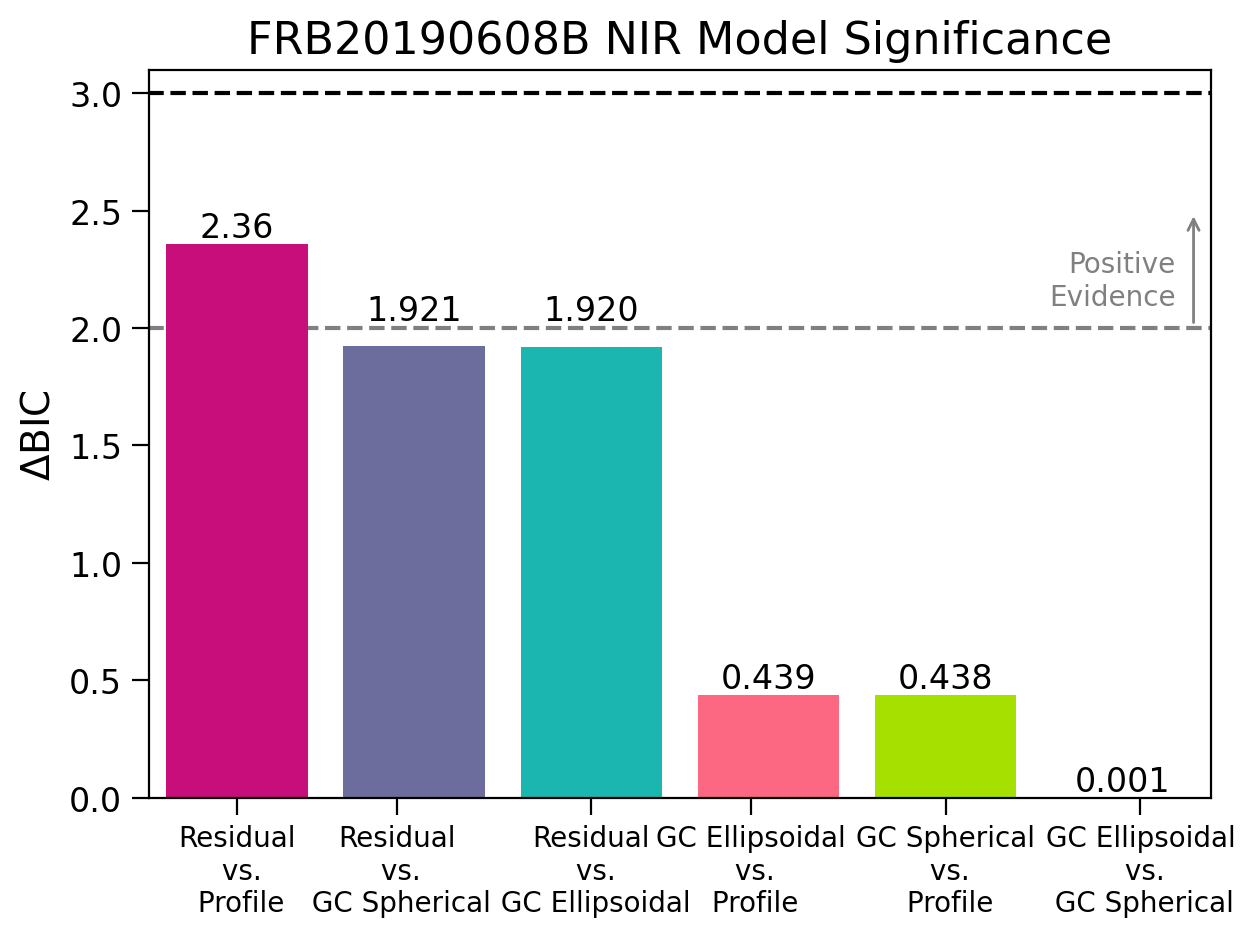}
    \caption{\textbf{(Left:)} Imaging and light profiles for FRB\,20190608B in the NIR. \textbf{(Right:)} Comparison of the $\Delta$BIC between the four models for FRB\,20190608B in the NIR.}
    \label{fig:appendix_190608}
\end{figure}

\begin{figure}
    \centering
    \includegraphics[width=0.6\textwidth]{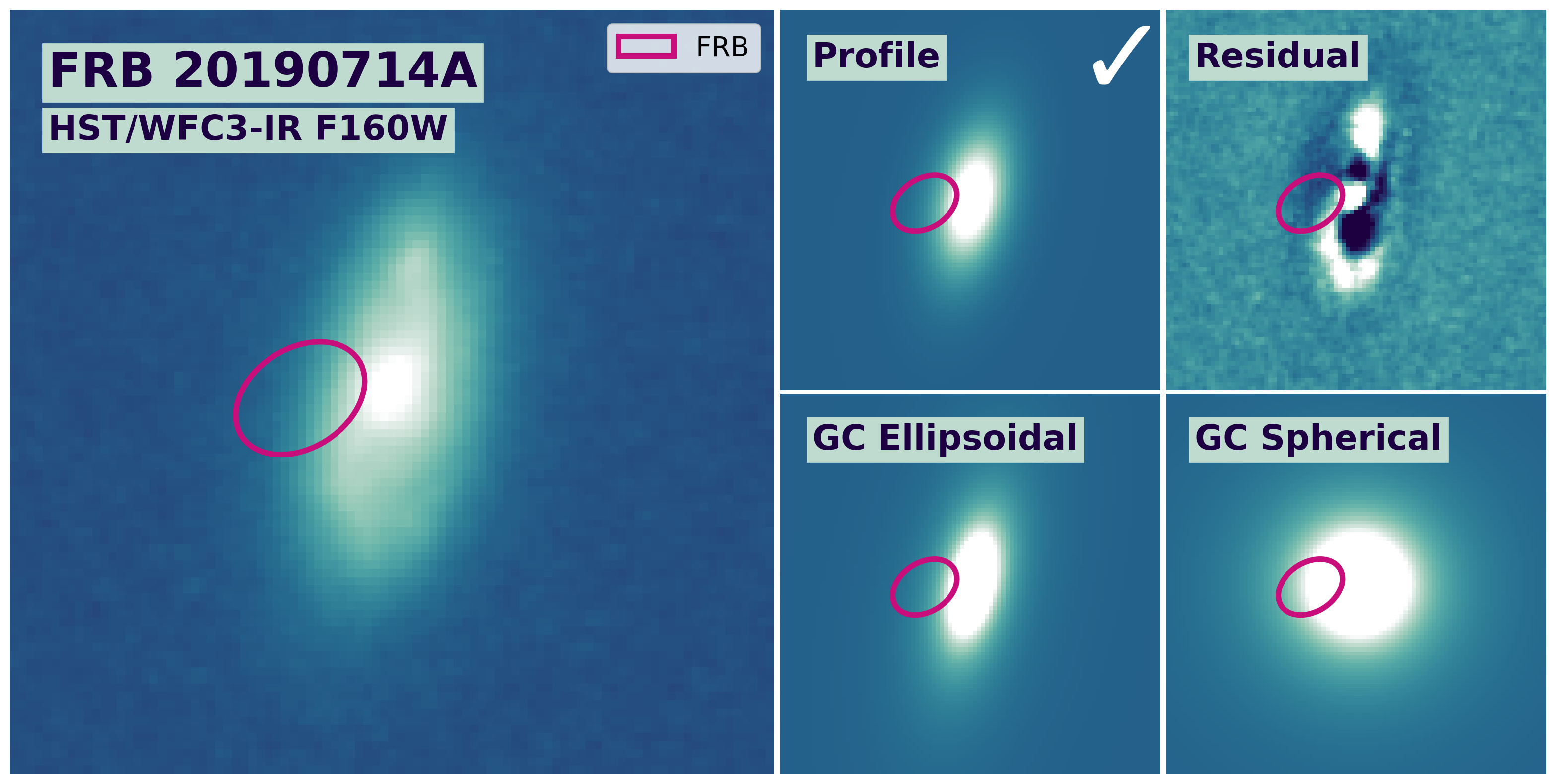}
    \includegraphics[width=0.39\textwidth]{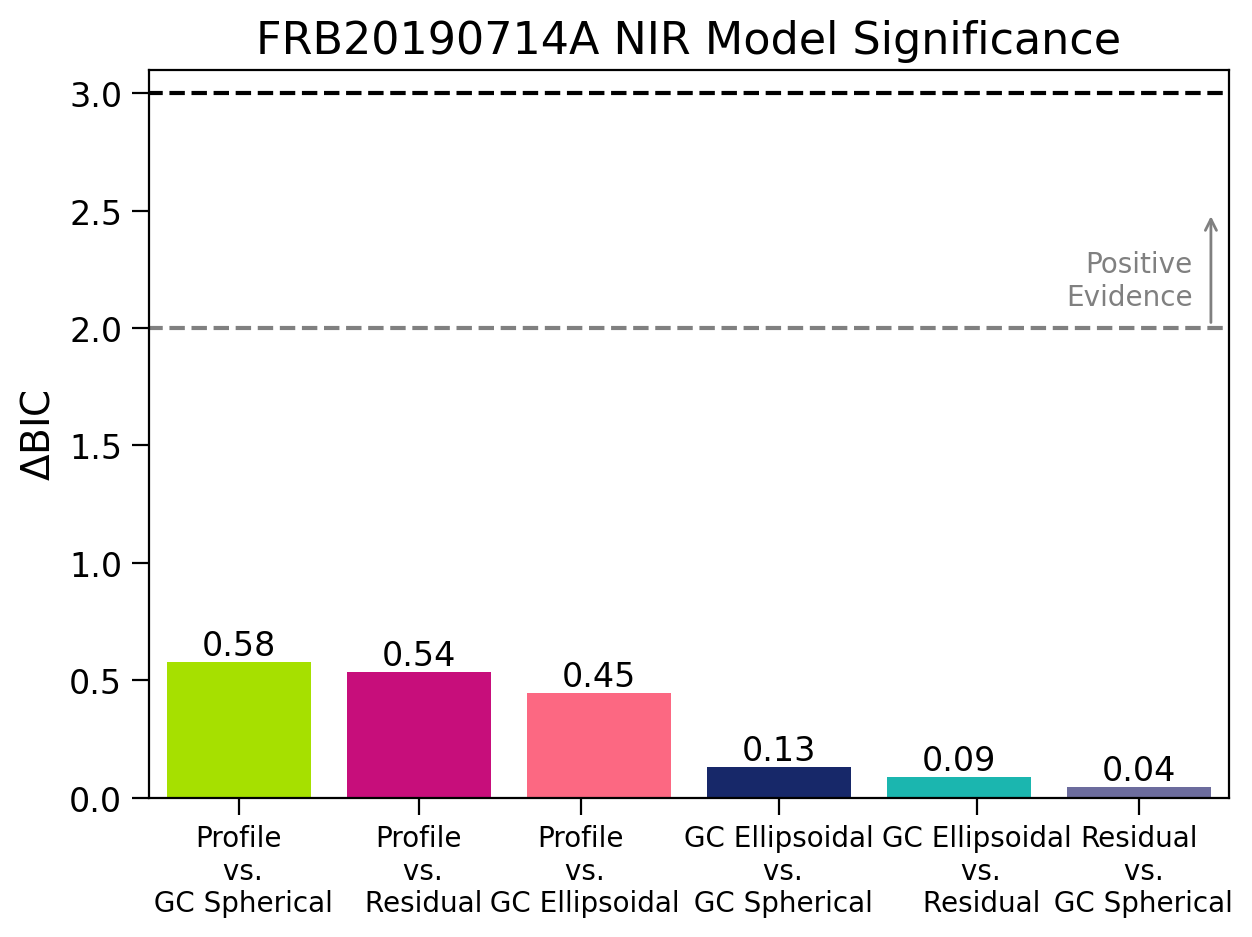}
    \caption{\textbf{(Left:)} Imaging and light profiles for FRB\,20190714A in the NIR. \textbf{(Right:)} Comparison of the $\Delta$BIC between the four models for FRB\,20190714A in the NIR.}
    \label{fig:appendix_190714}
\end{figure}

\begin{figure}
    \centering
    \includegraphics[width=0.6\textwidth]{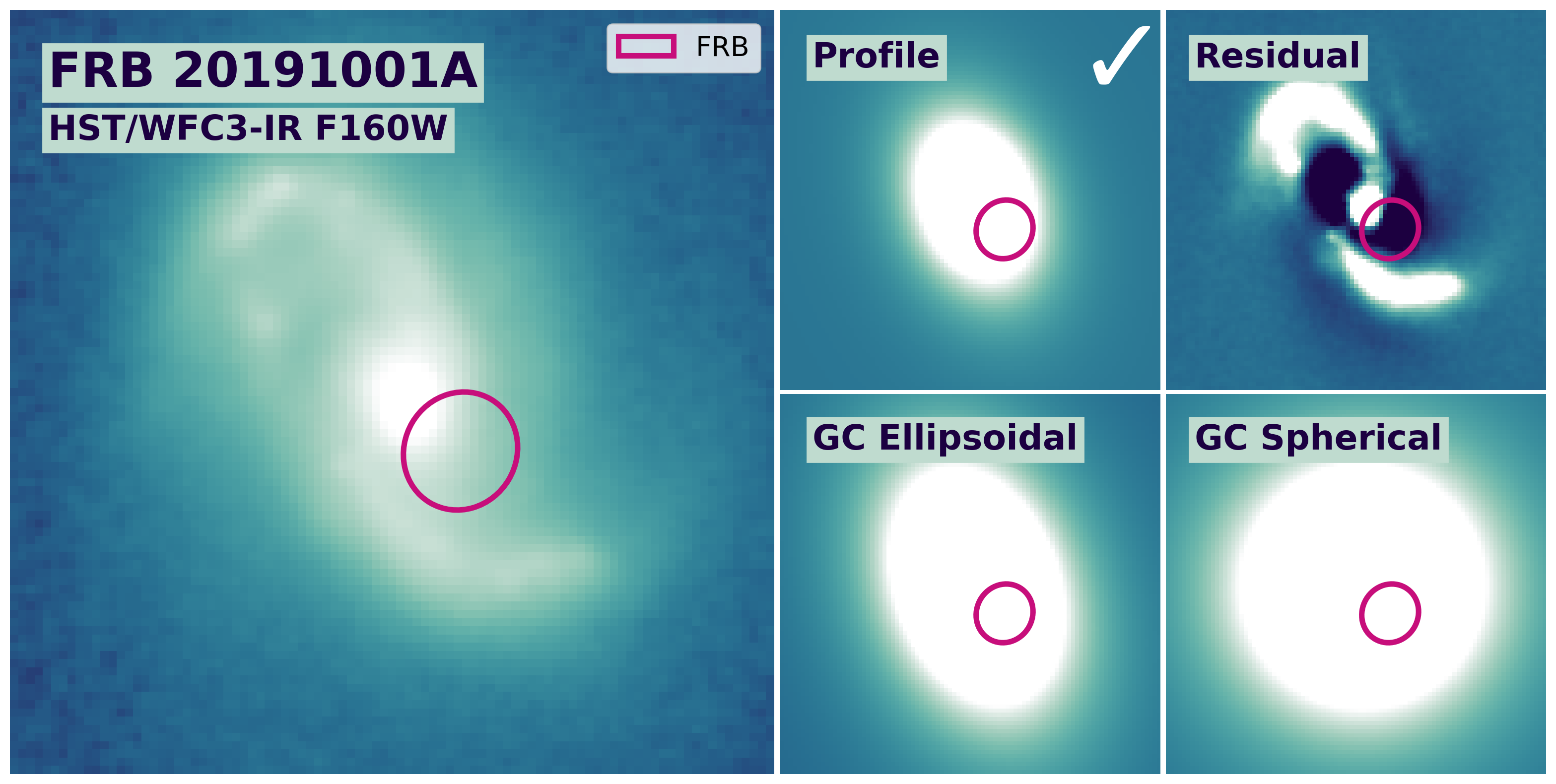}
    \includegraphics[width=0.39\textwidth]{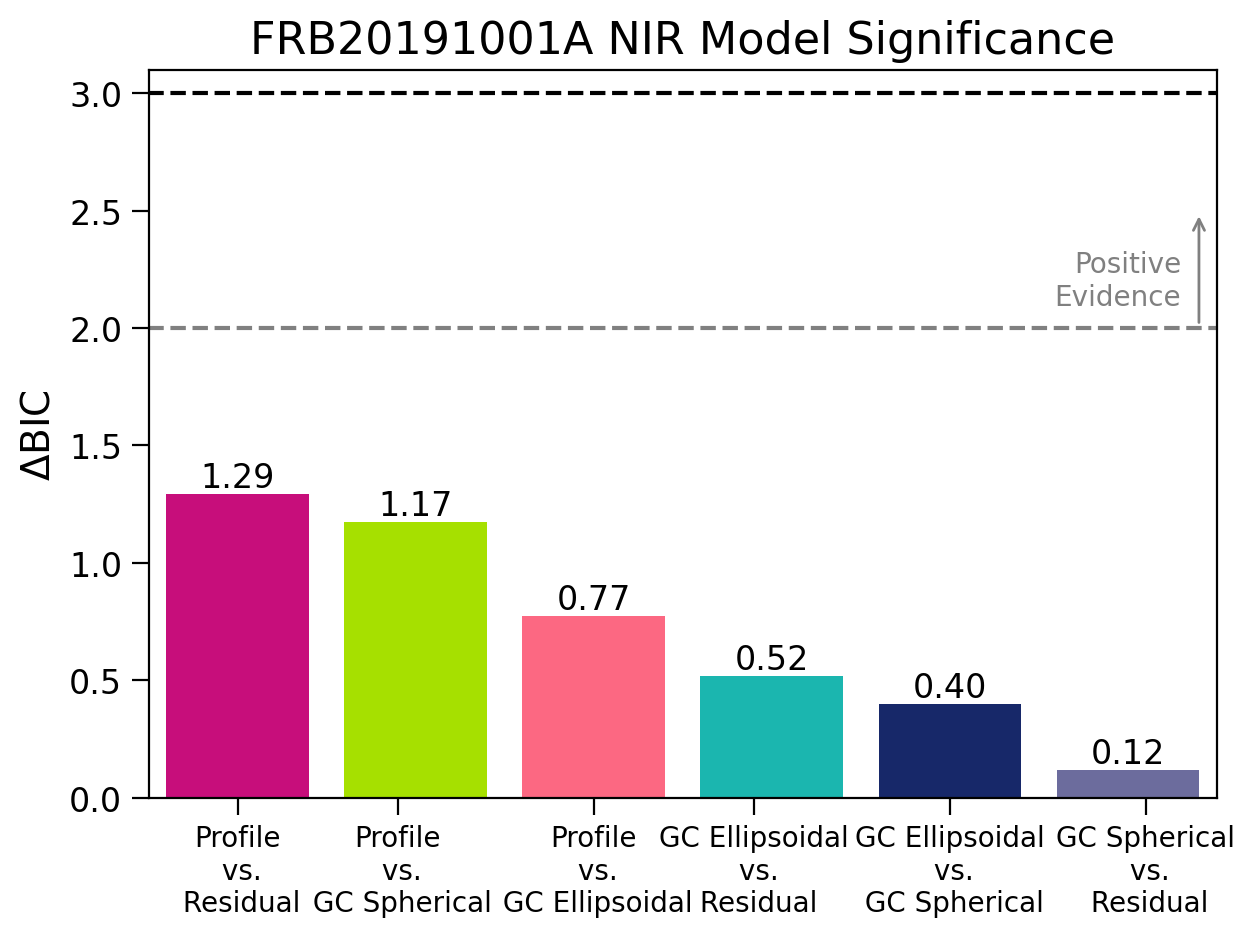}
    \caption{\textbf{(Left:)} Imaging and light profiles for FRB\,20191001A in the NIR. \textbf{(Right:)} Comparison of the $\Delta$BIC between the four models for FRB\,20191001A in the NIR.}
    \label{fig:appendix_191001}
\end{figure}

\begin{figure}
    \centering
    \includegraphics[width=0.6\textwidth]{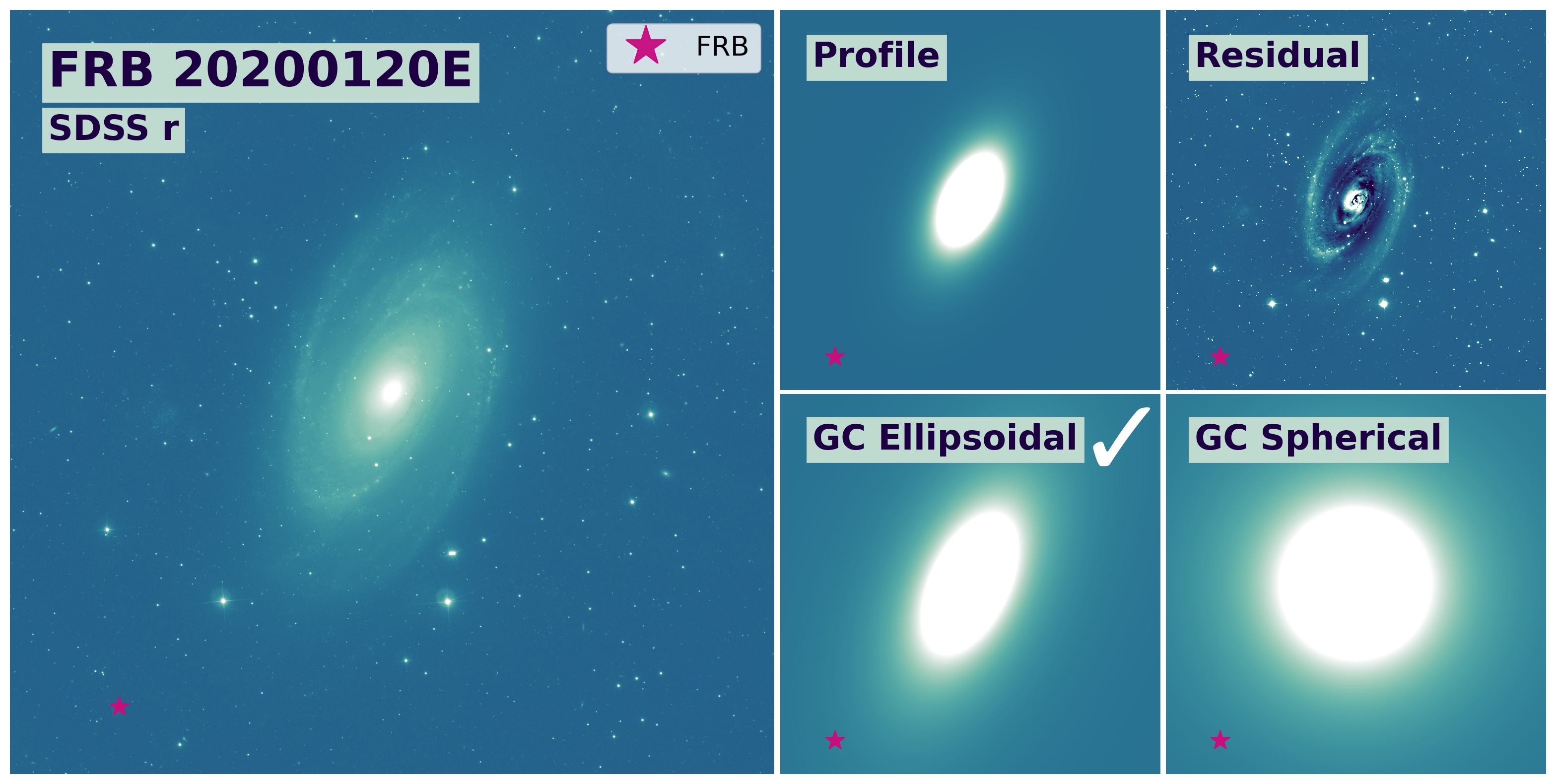}
    \includegraphics[width=0.39\textwidth]{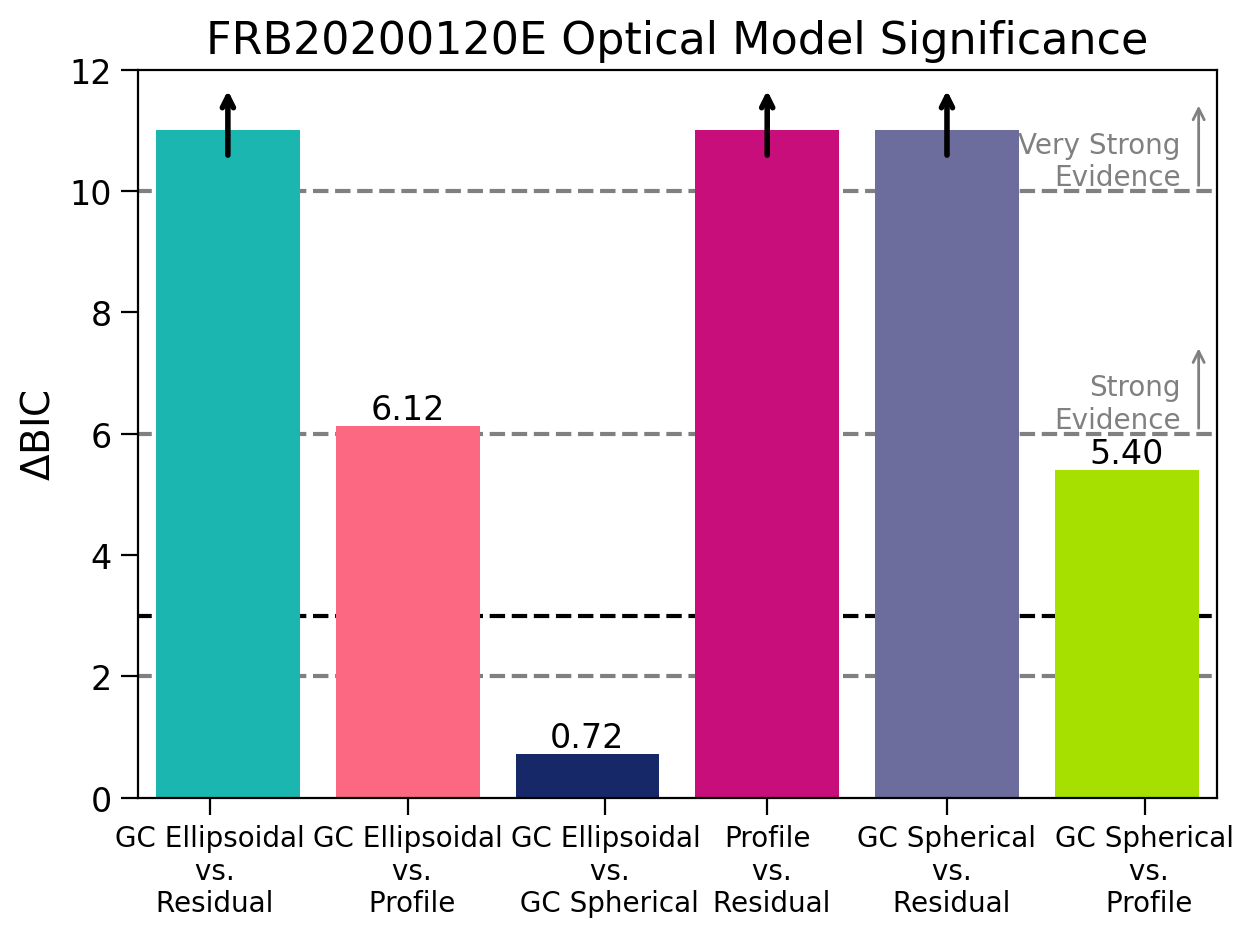}
    \caption{\textbf{(Left:)} Imaging and light profiles for FRB\,20200120E in the optical. \textbf{(Right:)} Comparison of the $\Delta$BIC between the four models for FRB\,20200120E in the optical.}
    \label{fig:appendix_200120}
\end{figure}

\begin{figure}
    \centering
    \includegraphics[width=0.6\textwidth]{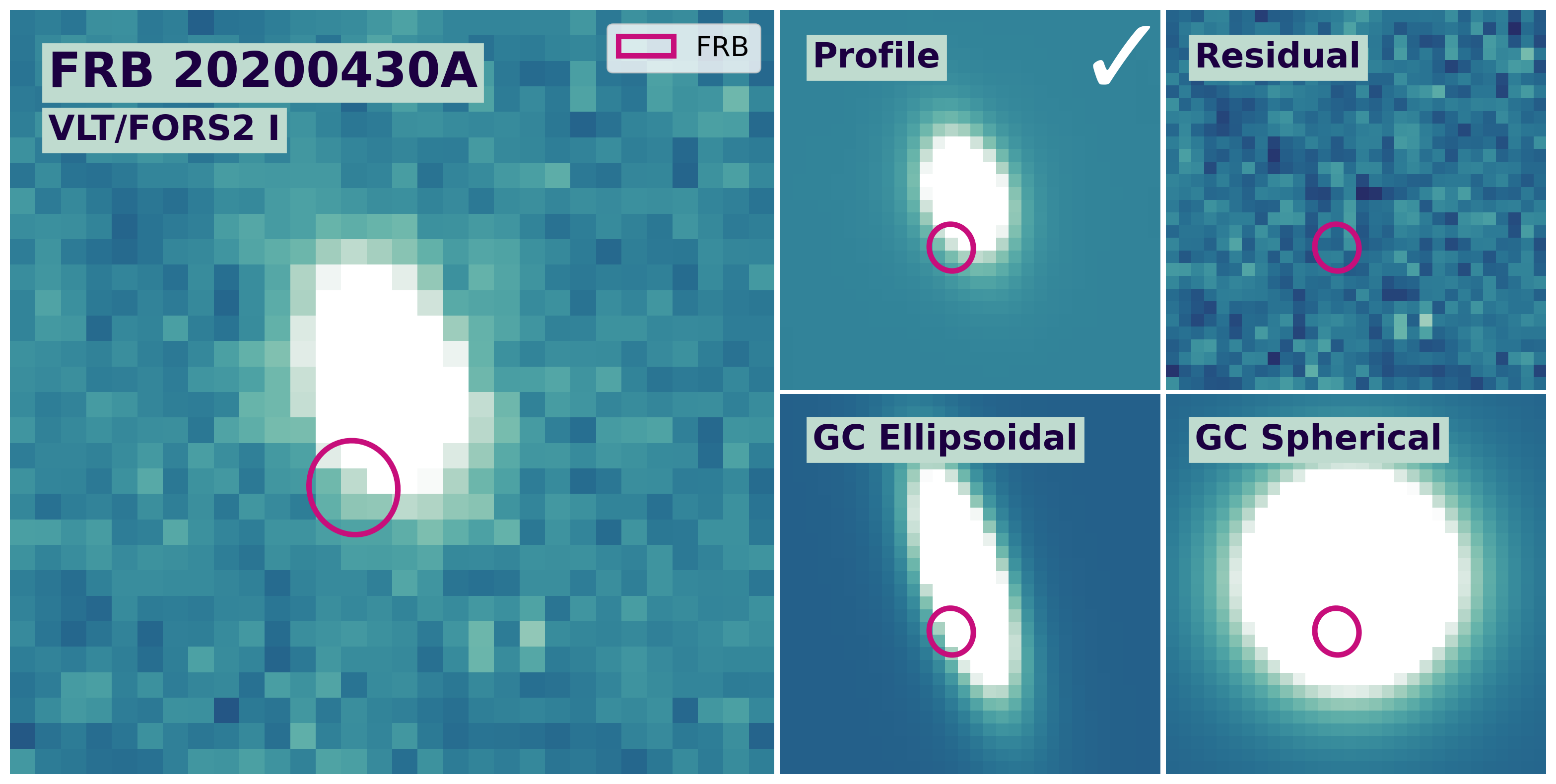}
    \includegraphics[width=0.39\textwidth]{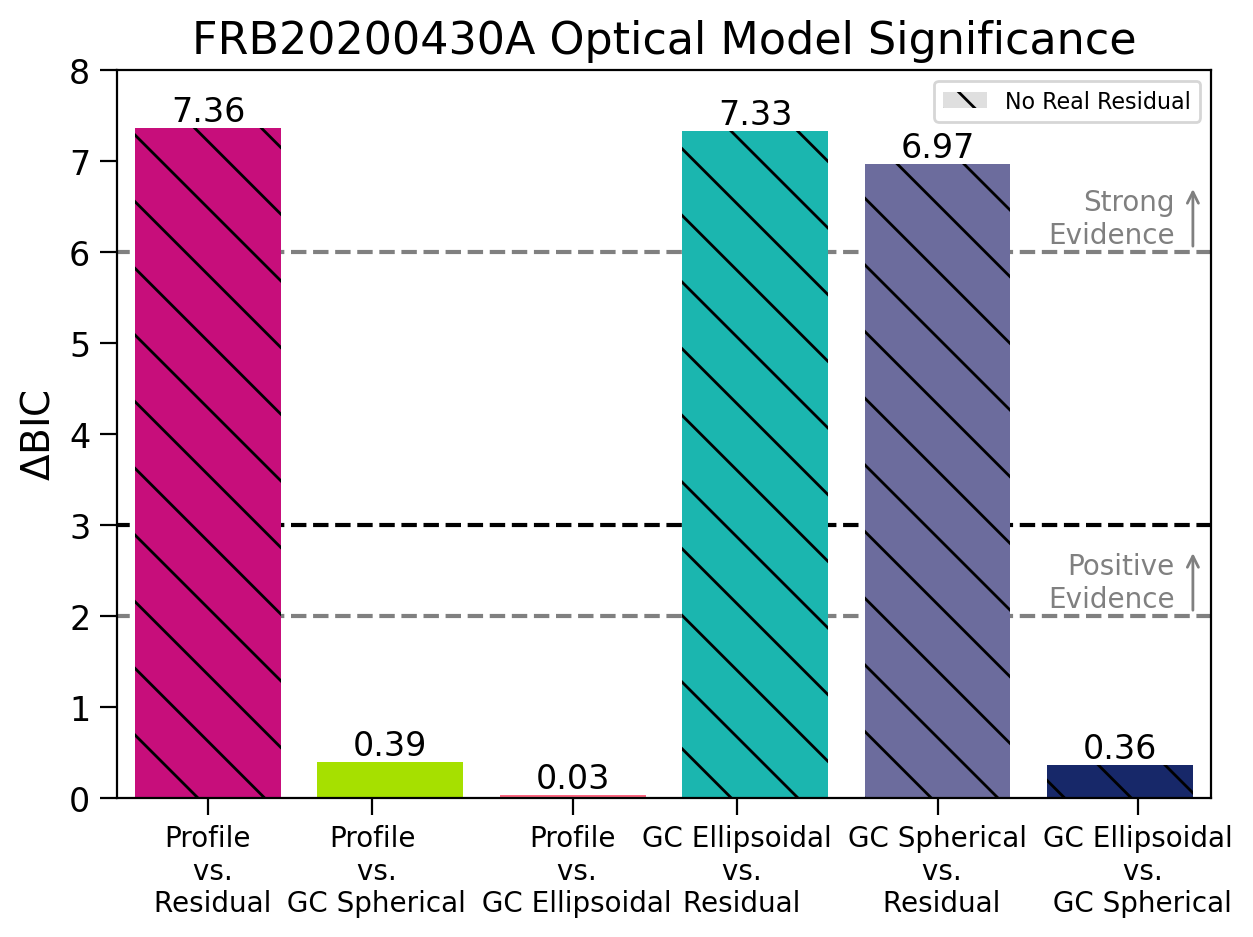}
    \caption{\textbf{(Left:)} Imaging and light profiles for FRB\,20200430A in the optical. \textbf{(Right:)} Comparison of the $\Delta$BIC between the four models for FRB\,20200430A in the optical.}
    \label{fig:appendix_200430}
\end{figure}

\begin{figure}
    \centering
    \includegraphics[width=0.6\textwidth]{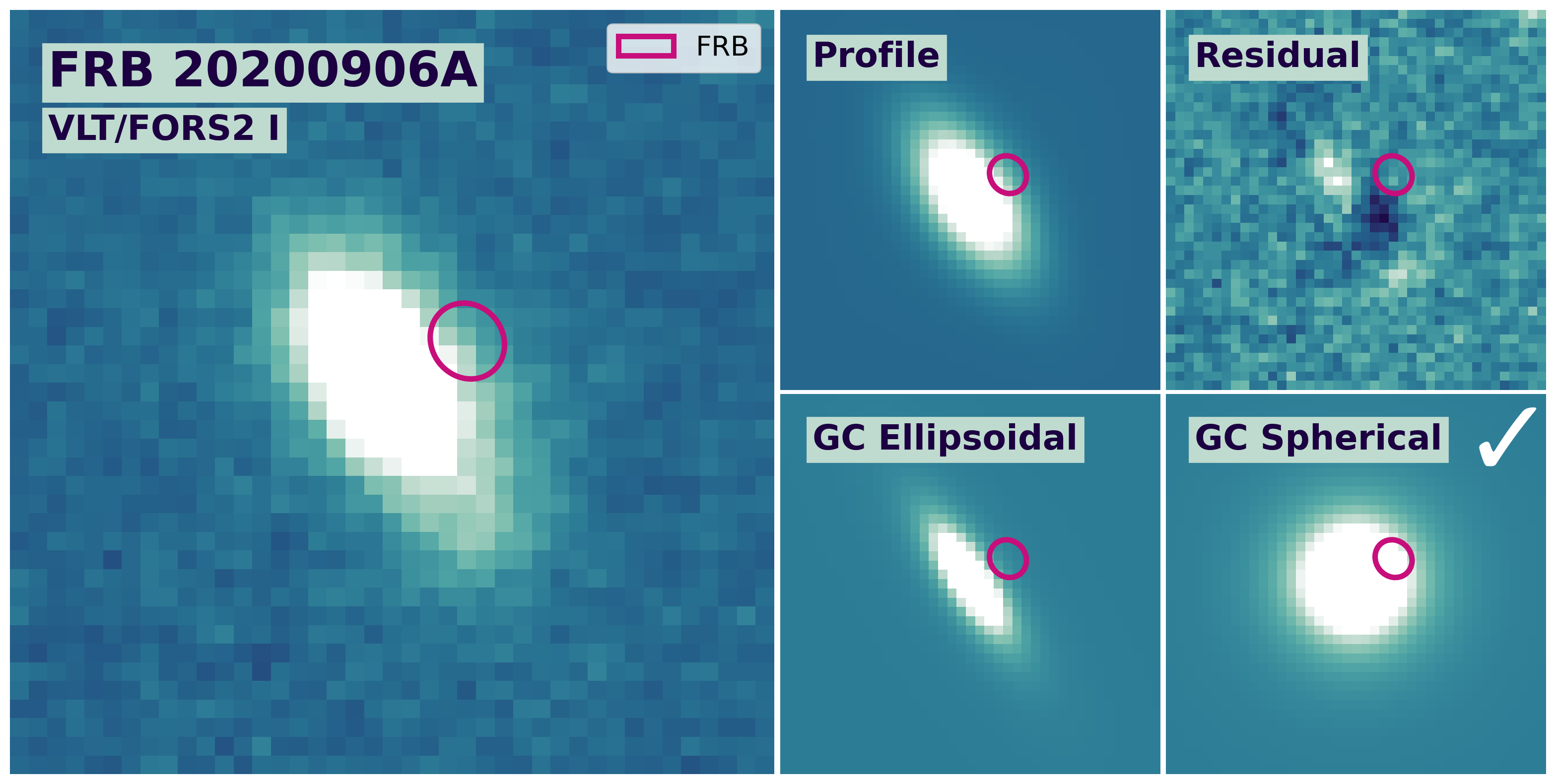}
    \includegraphics[width=0.39\textwidth]{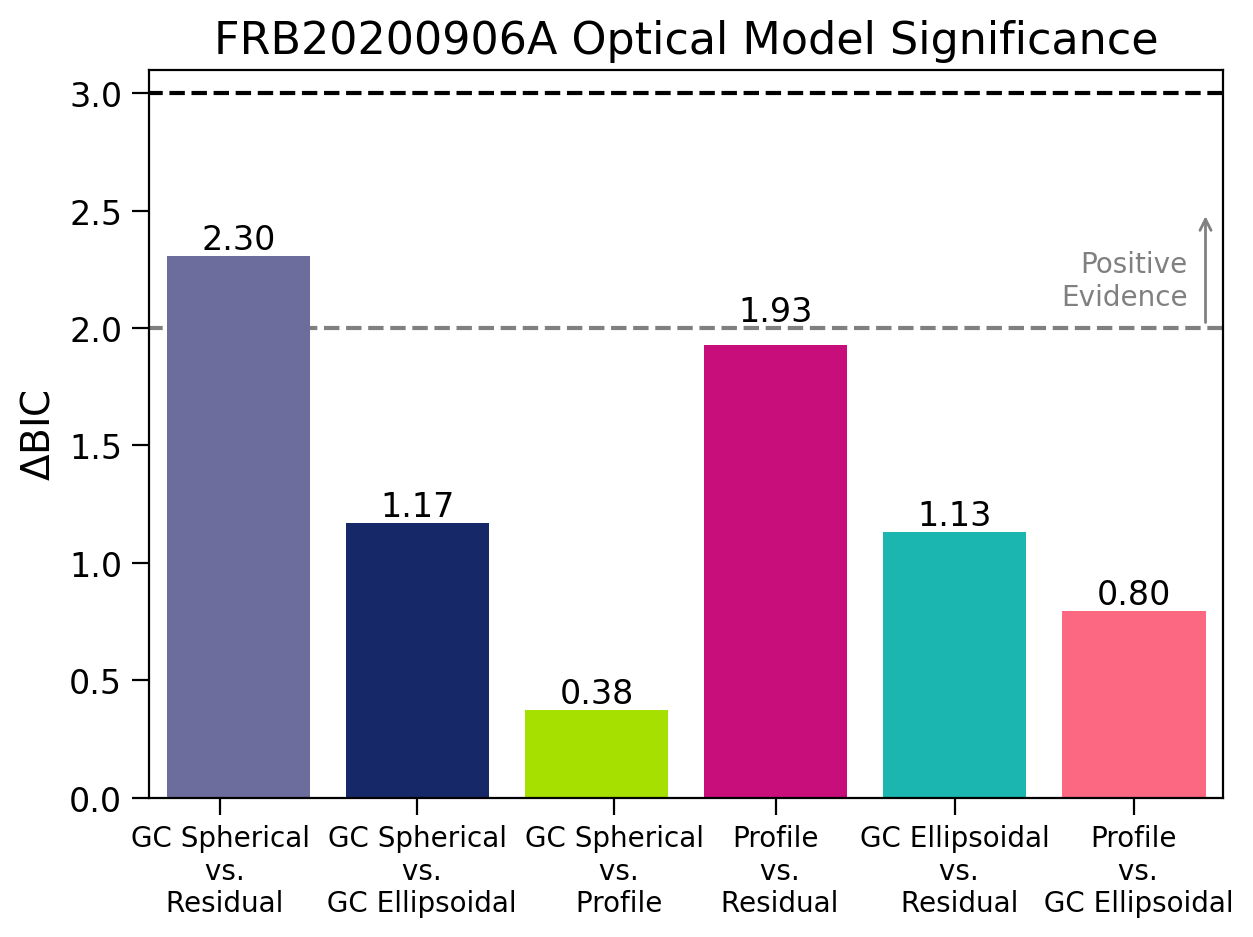}
    \caption{\textbf{(Left:)} Imaging and light profiles for FRB\,20200906A in the optical. \textbf{(Right:)} Comparison of the $\Delta$BIC between the four models for FRB\,20200906A in the optical.}
    \label{fig:appendix_200906}
\end{figure}

\begin{figure}
    \centering
    \includegraphics[width=0.6\textwidth]{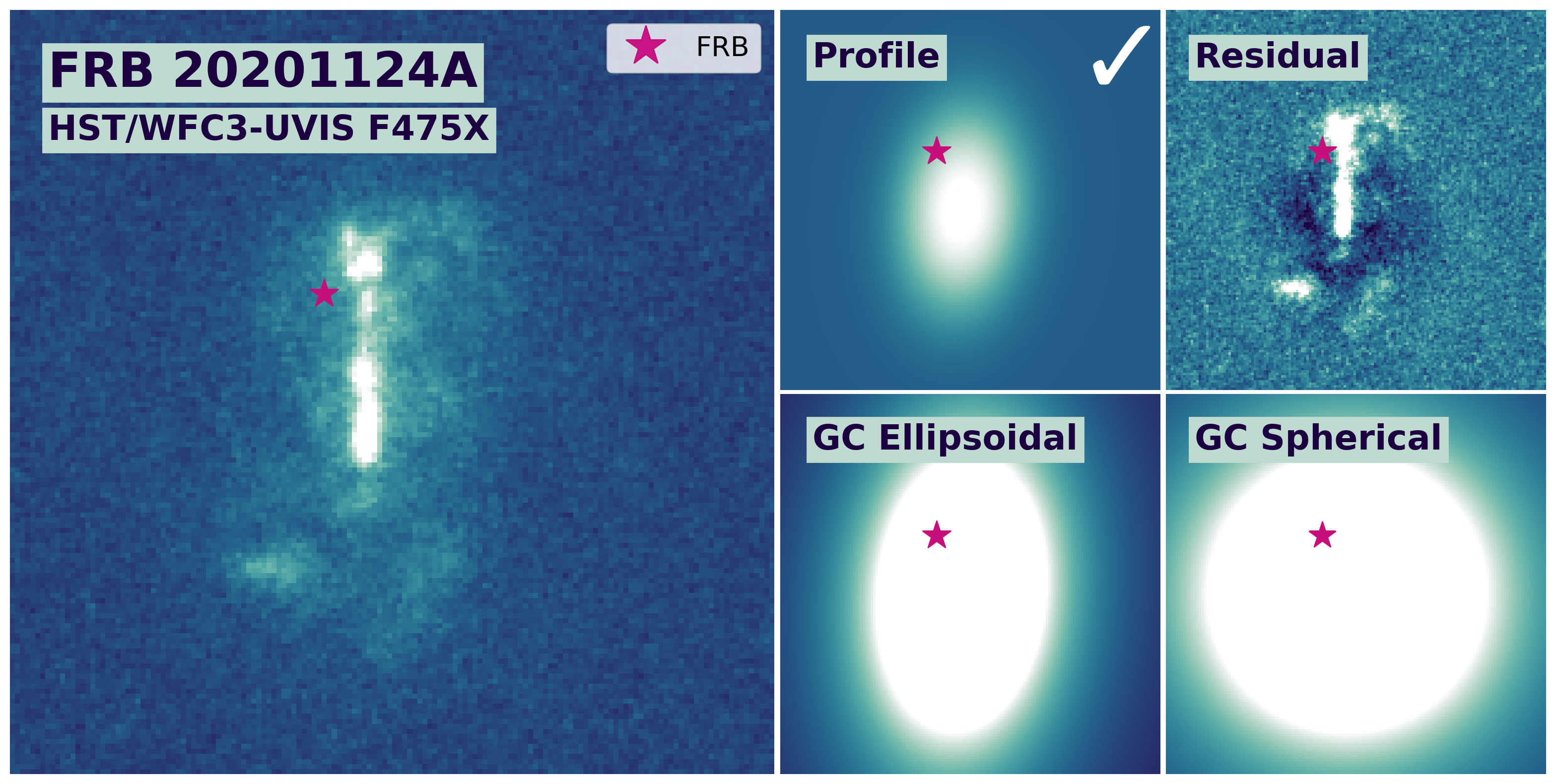}
    \includegraphics[width=0.39\textwidth]{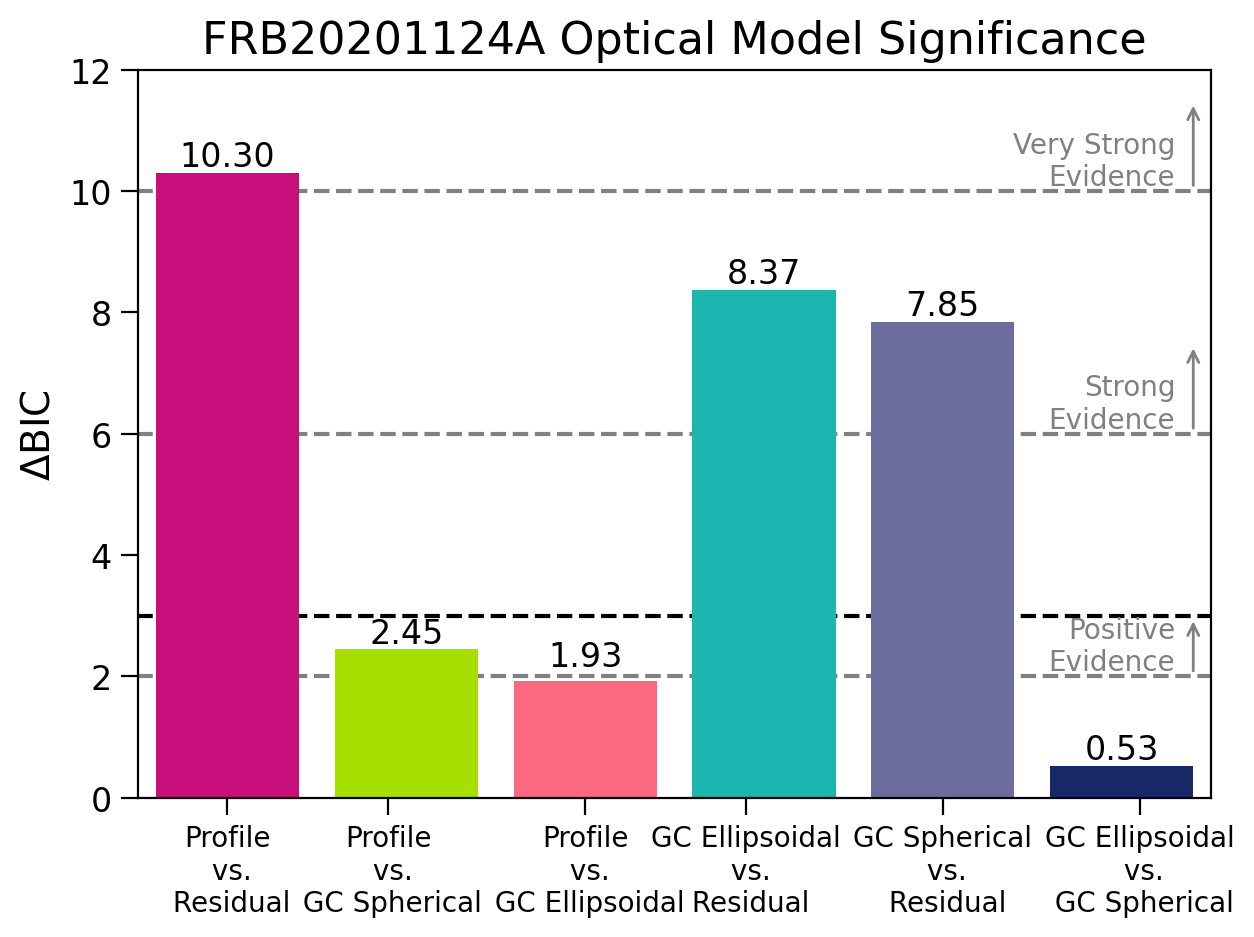}
    \caption{\textbf{(Left:)} Imaging and light profiles for FRB\,20201124A in the optical. \textbf{(Right:)} Comparison of the $\Delta$BIC between the four models for FRB\,20201124A in the optical.}
    \label{fig:appendix_201124_opt}
\end{figure}

\begin{figure}
    \centering
    \includegraphics[width=0.6\textwidth]{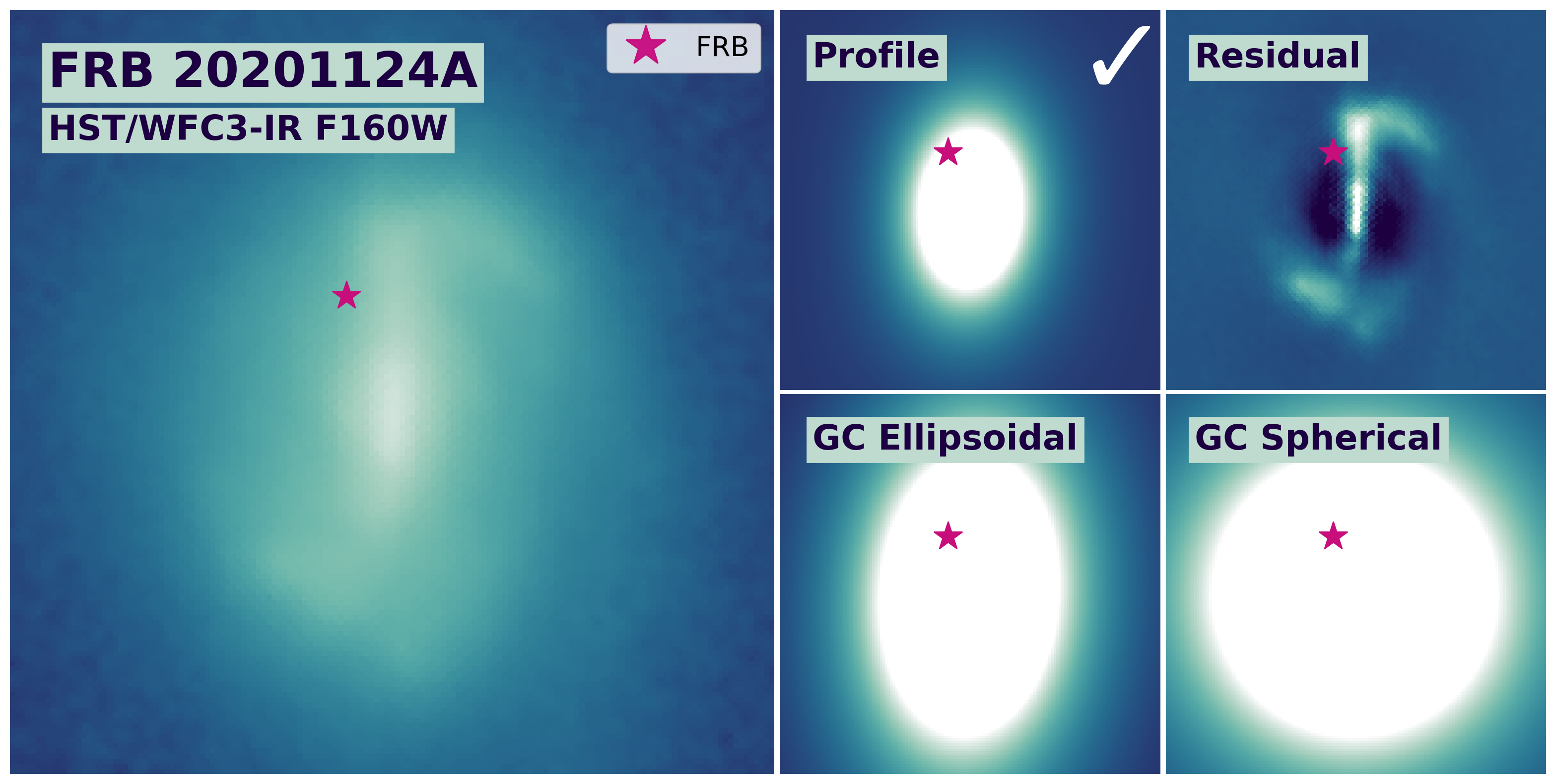}
    \includegraphics[width=0.39\textwidth]{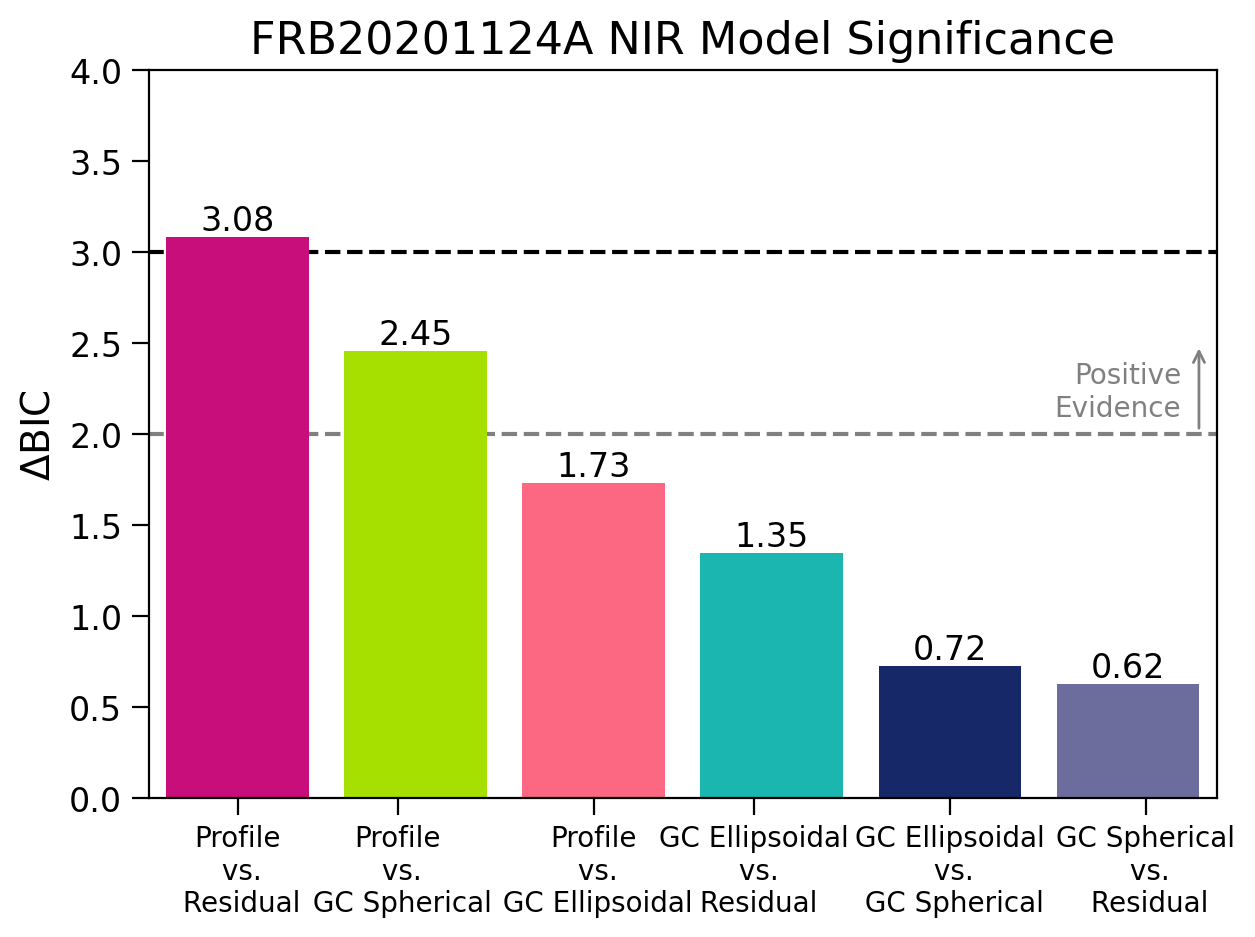}
    \caption{\textbf{(Left:)} Imaging and light profiles for FRB\,20201124A in the NIR. \textbf{(Right:)} Comparison of the $\Delta$BIC between the four models for FRB\,20201124A in the NIR.}
    \label{fig:appendix_201124_NIR}
\end{figure}

\begin{figure}
    \centering
    \includegraphics[width=0.6\textwidth]{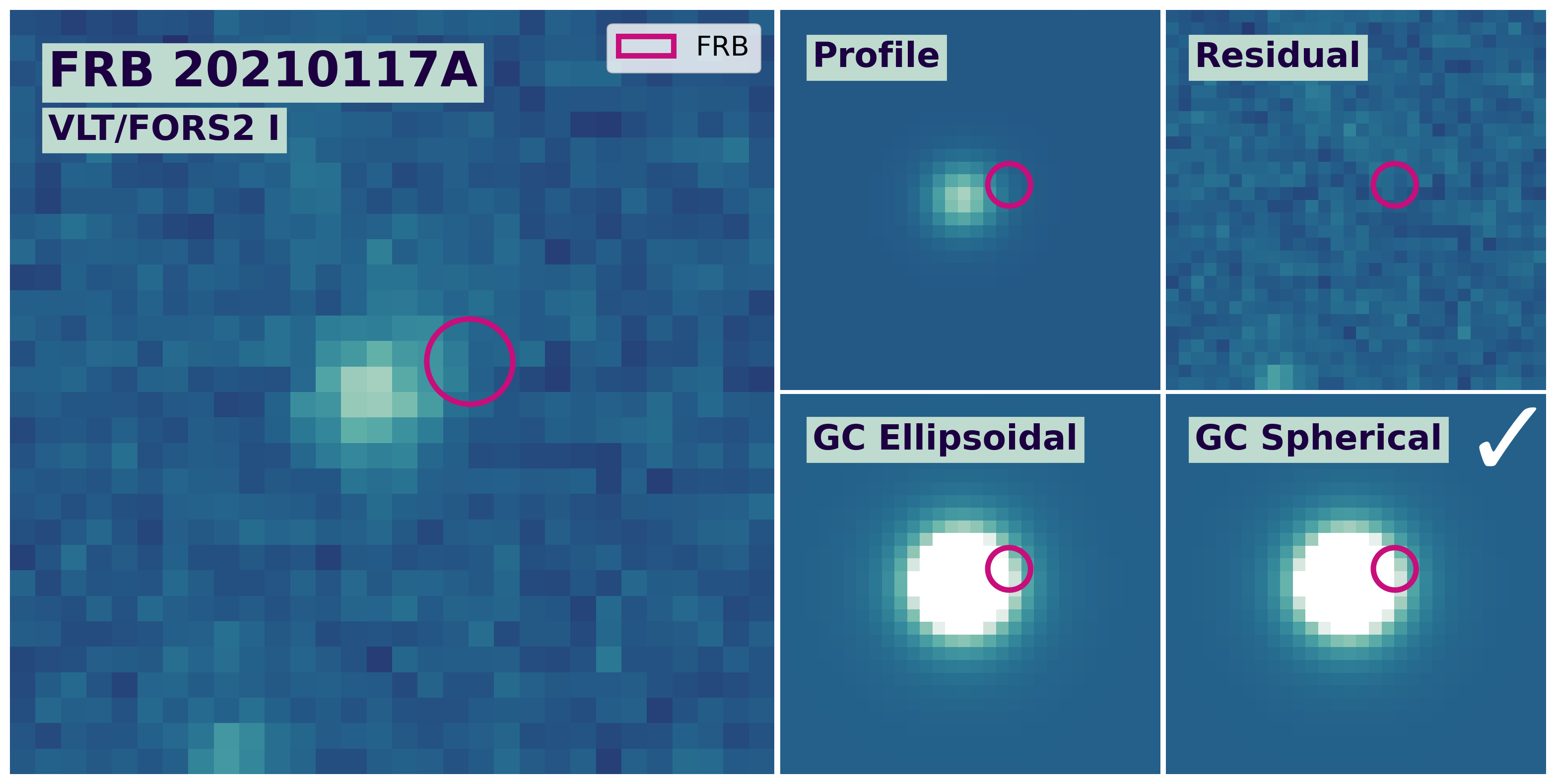}
    \includegraphics[width=0.39\textwidth]{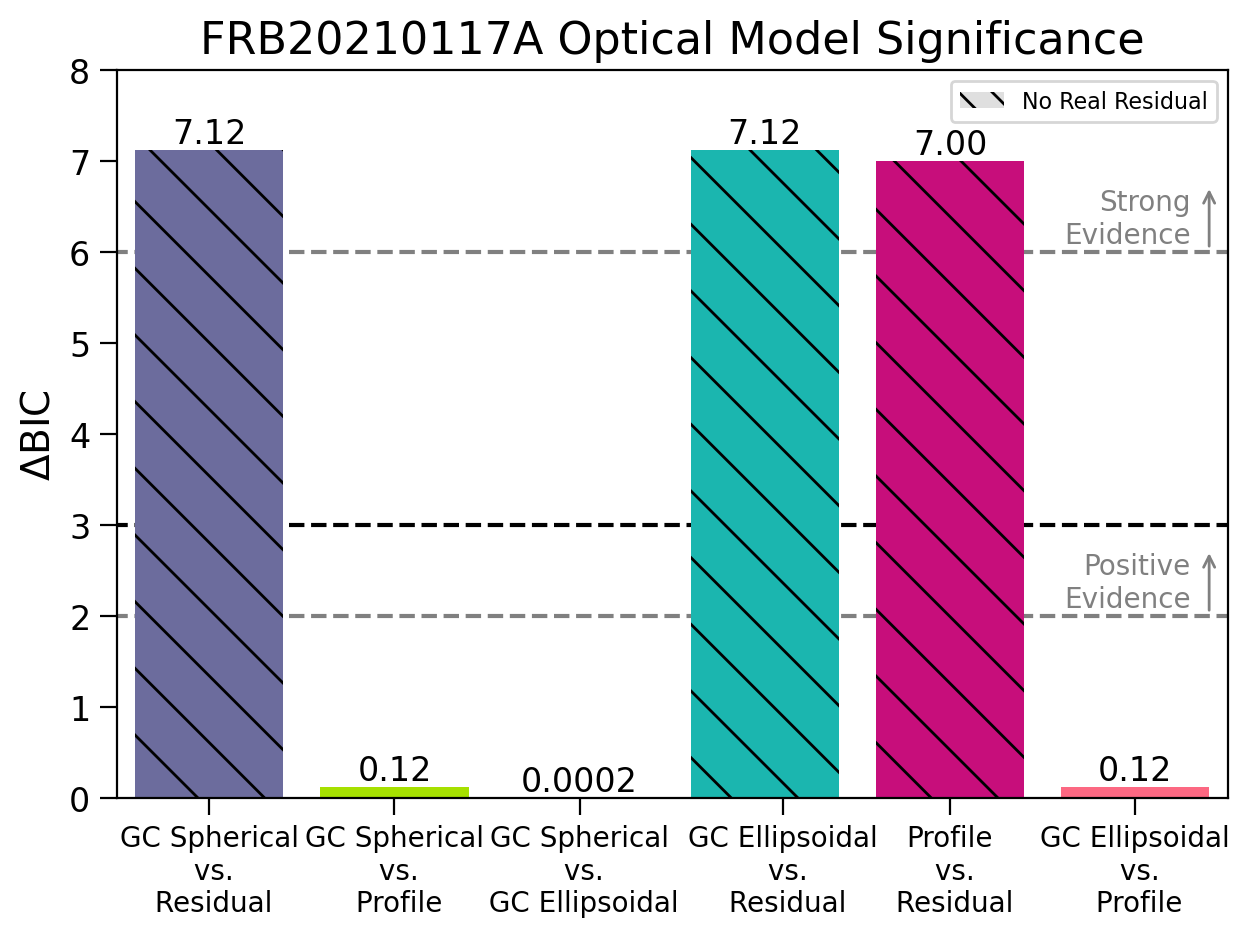}
    \caption{\textbf{(Left:)} Imaging and light profiles for FRB\,20210117A in the optical. \textbf{(Right:)} Comparison of the $\Delta$BIC between the four models for FRB\,20210117A in the optical.}
    \label{fig:appendix_210117_opt}
\end{figure}

\begin{figure}
    \centering
    \includegraphics[width=0.6\textwidth]{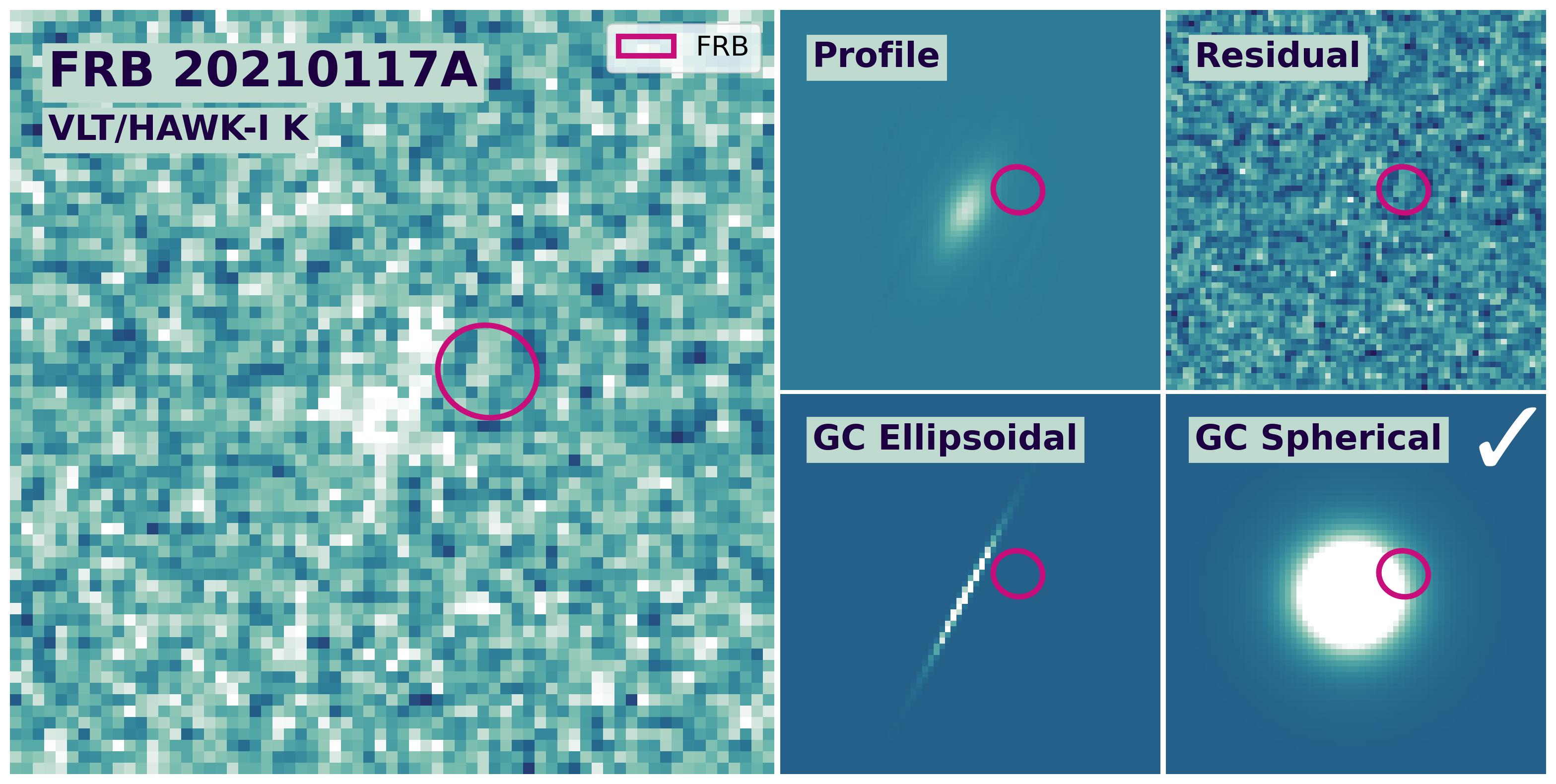}
    \includegraphics[width=0.39\textwidth]{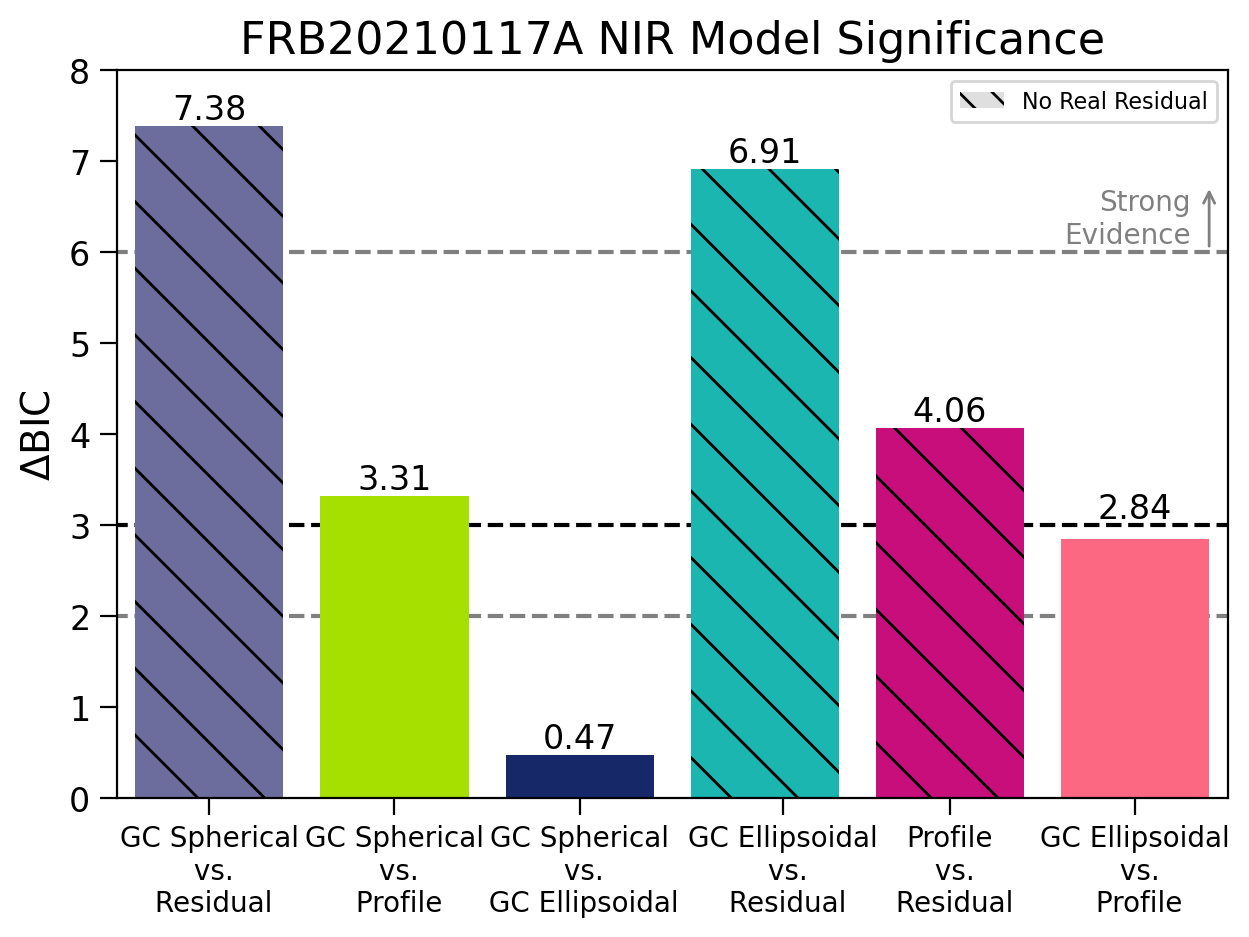}
    \caption{\textbf{(Left:)} Imaging and light profiles for FRB\,20210117A in the NIR. \textbf{(Right:)} Comparison of the $\Delta$BIC between the four models for FRB\,20210117A in the NIR.}
    \label{fig:appendix_210117_NIR}
\end{figure}

\begin{figure}
    \centering
    \includegraphics[width=0.6\textwidth]{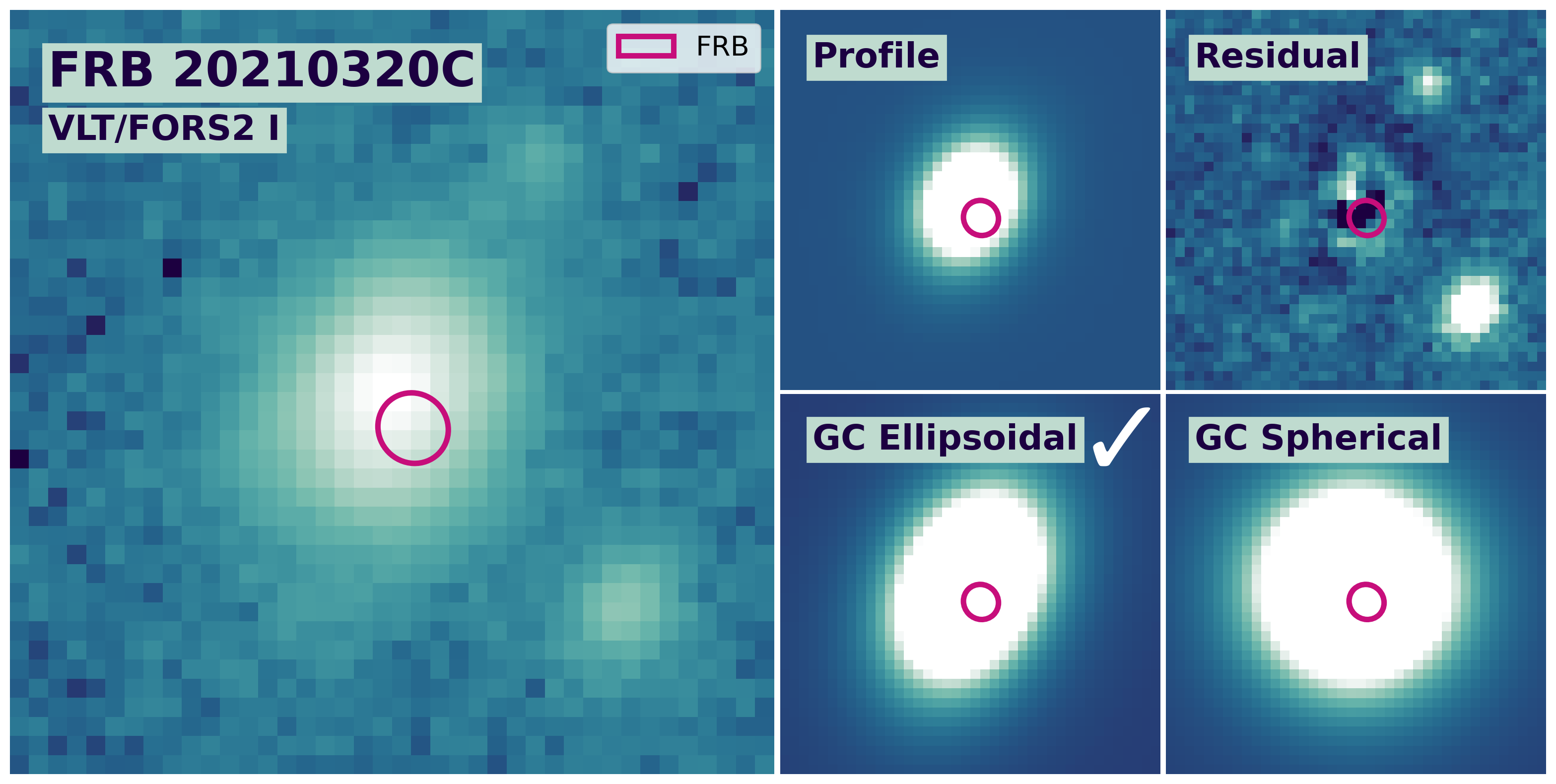}
    \includegraphics[width=0.39\textwidth]{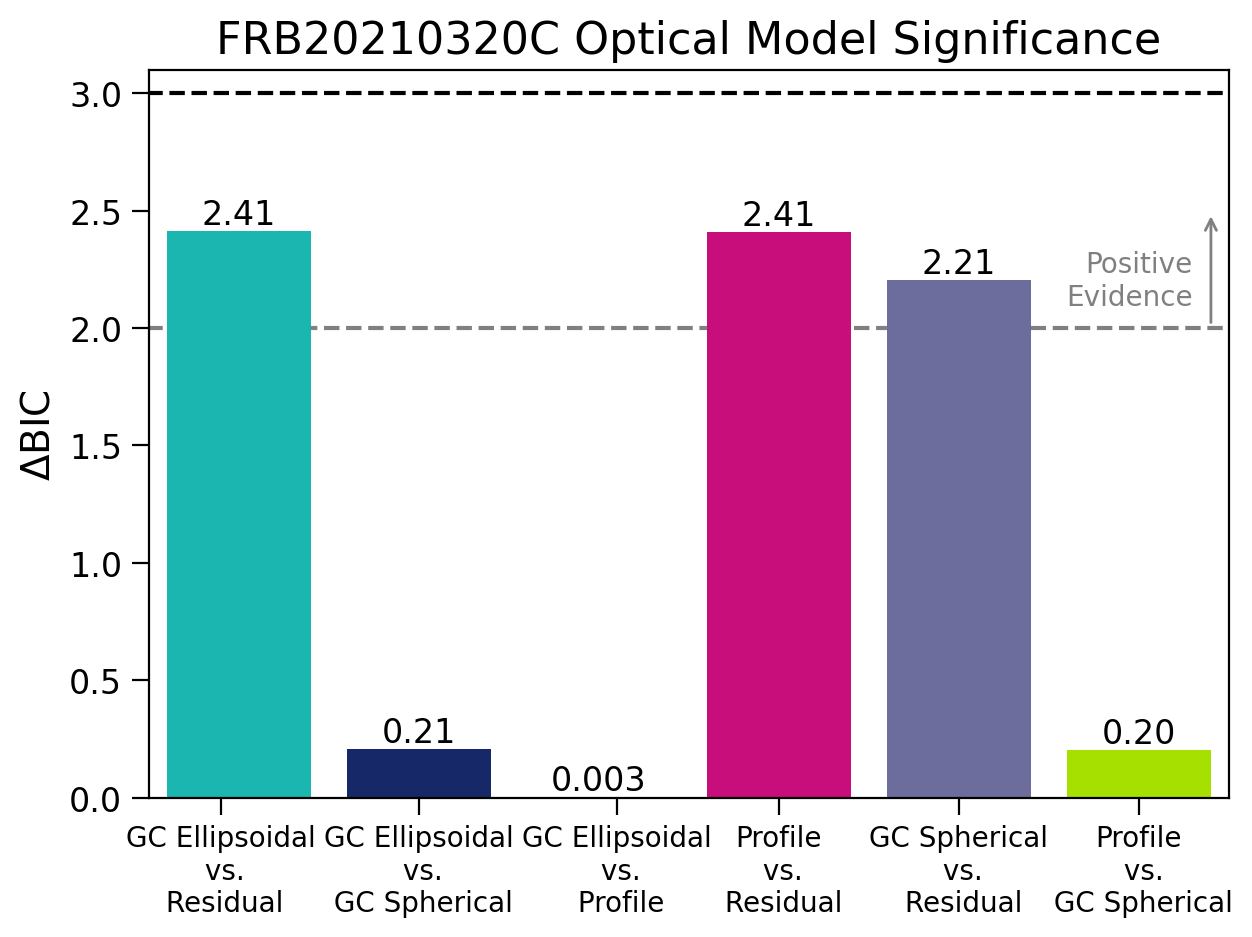}
    \caption{\textbf{(Left:)} Imaging and light profiles for FRB\,20210320C in the optical. \textbf{(Right:)} Comparison of the $\Delta$BIC between the four models for FRB\,20210320C in the optical.}
    \label{fig:appendix_210320_opt}
\end{figure}

\begin{figure}
    \centering
    \includegraphics[width=0.6\textwidth]{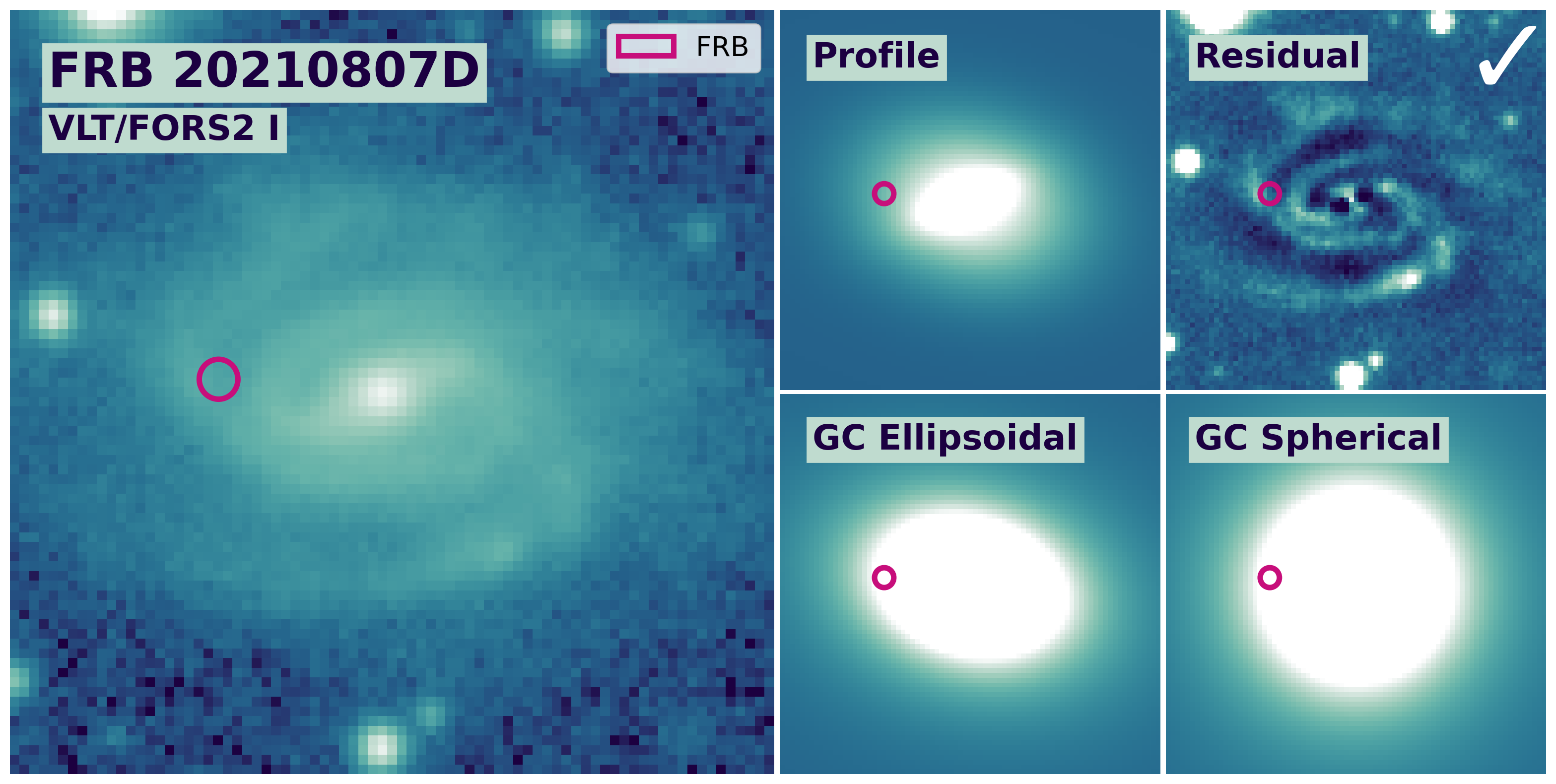}
    \includegraphics[width=0.39\textwidth]{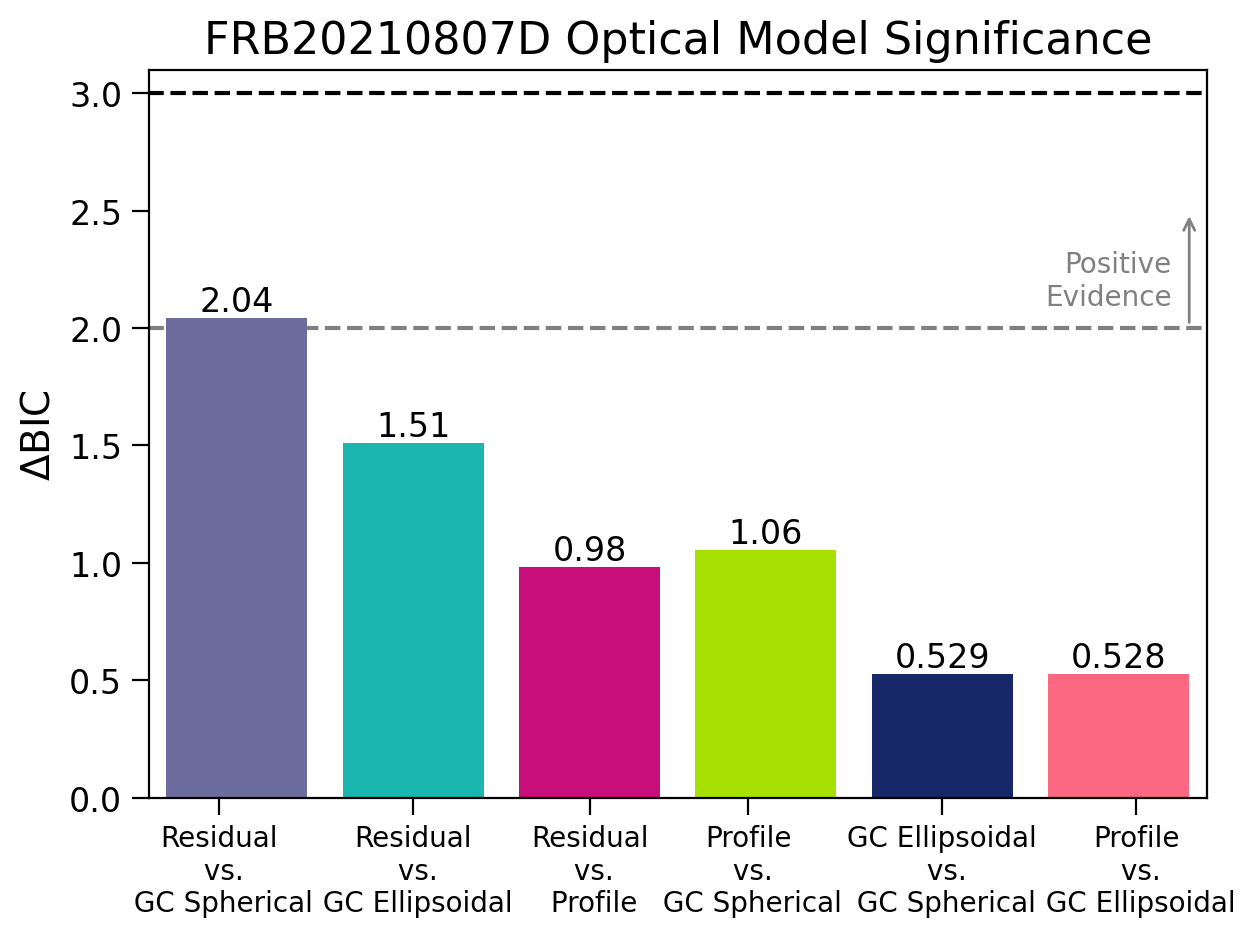}
    \caption{\textbf{(Left:)} Imaging and light profiles for FRB\,20210807D in the optical. \textbf{(Right:)} Comparison of the $\Delta$BIC between the four models for FRB\,20210807D in the optical.}
    \label{fig:appendix_210807_opt}
\end{figure}

\begin{figure}
    \centering
    \includegraphics[width=0.6\textwidth]{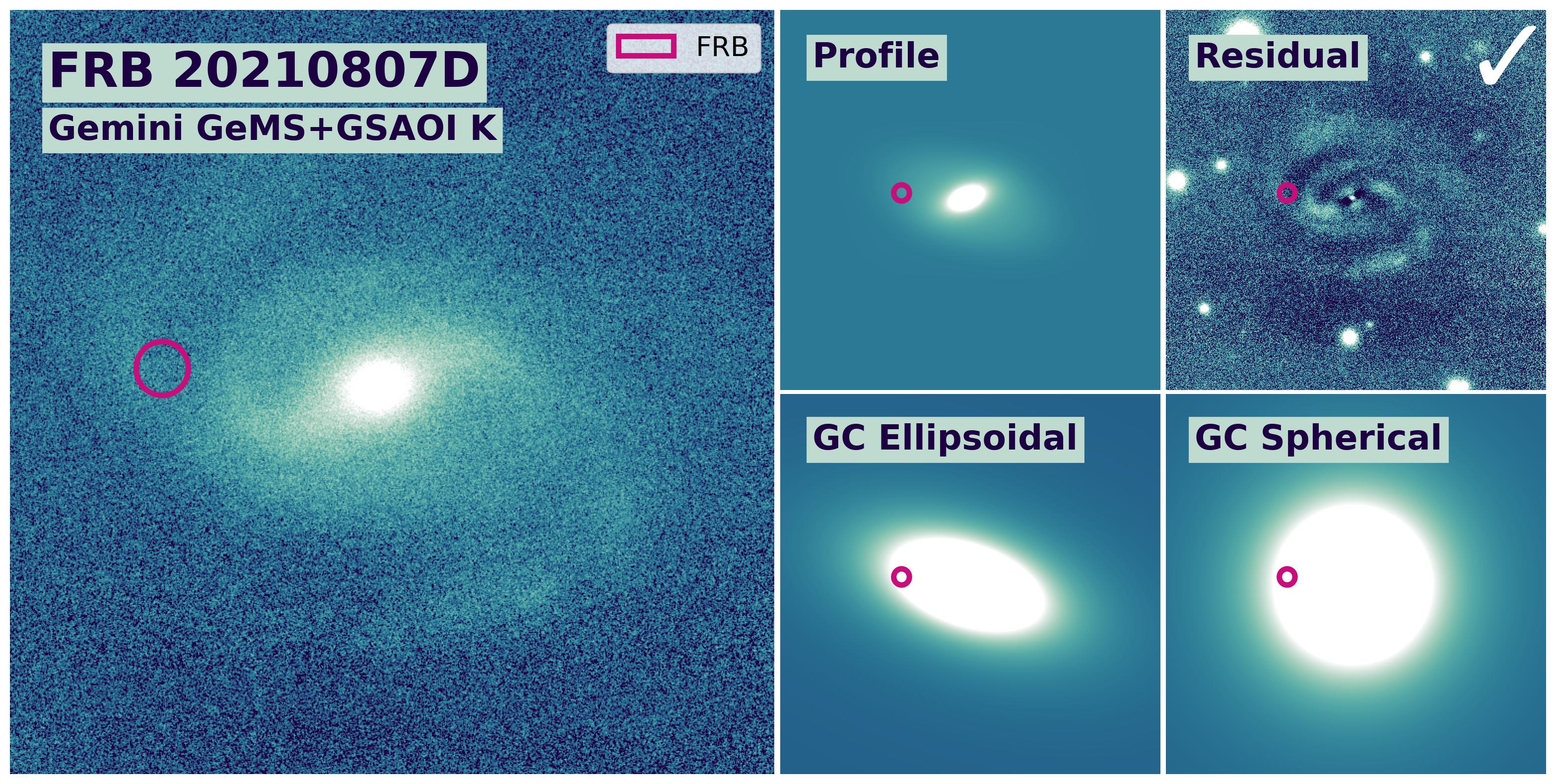}
    \includegraphics[width=0.39\textwidth]{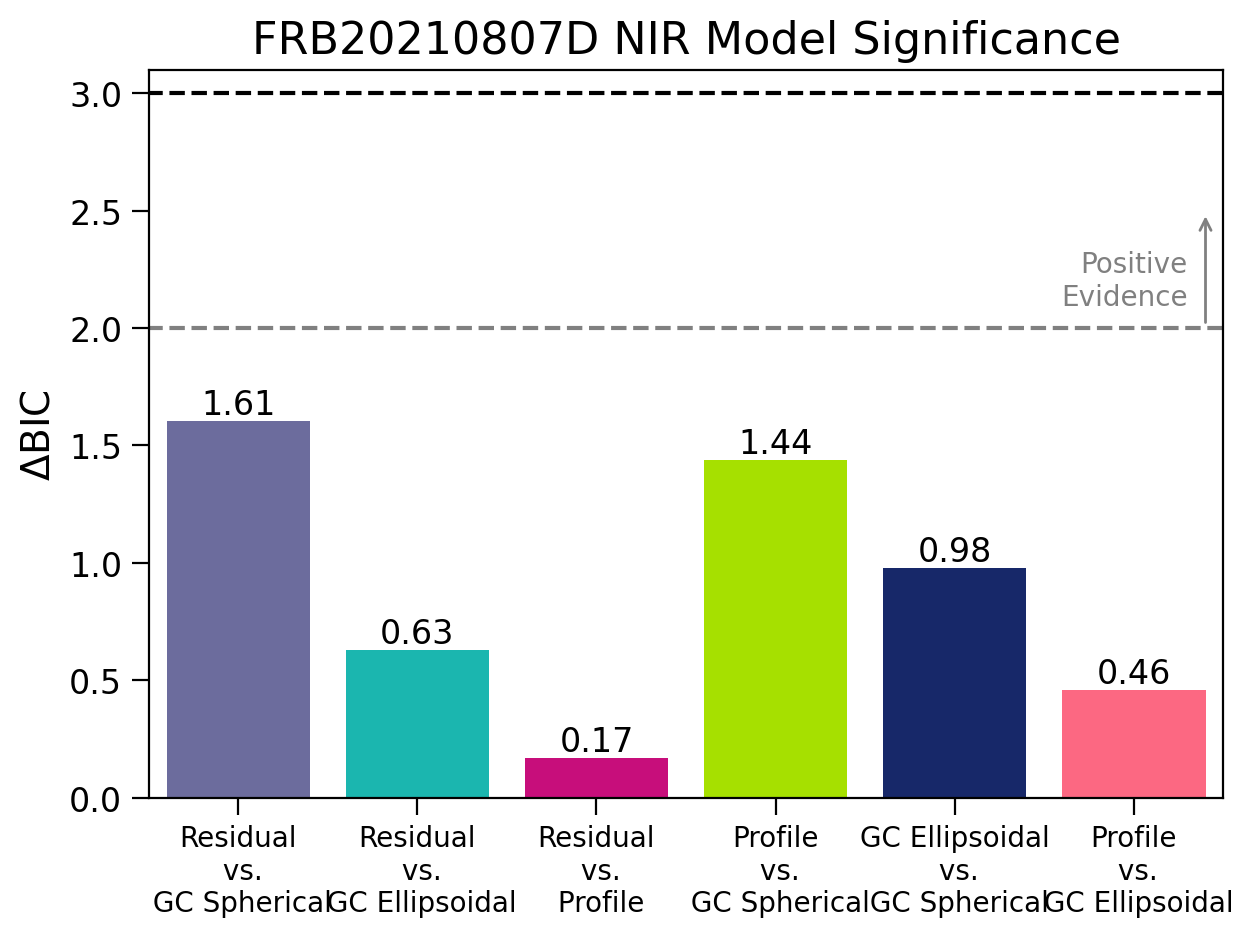}
    \caption{\textbf{(Left:)} Imaging and light profiles for FRB\,20210807D in the NIR. \textbf{(Right:)} Comparison of the $\Delta$BIC between the four models for FRB\,20210807D in the NIR.}
    \label{fig:appendix_210807_NIR}
\end{figure}

\begin{figure}
    \centering
    \includegraphics[width=0.6\textwidth]{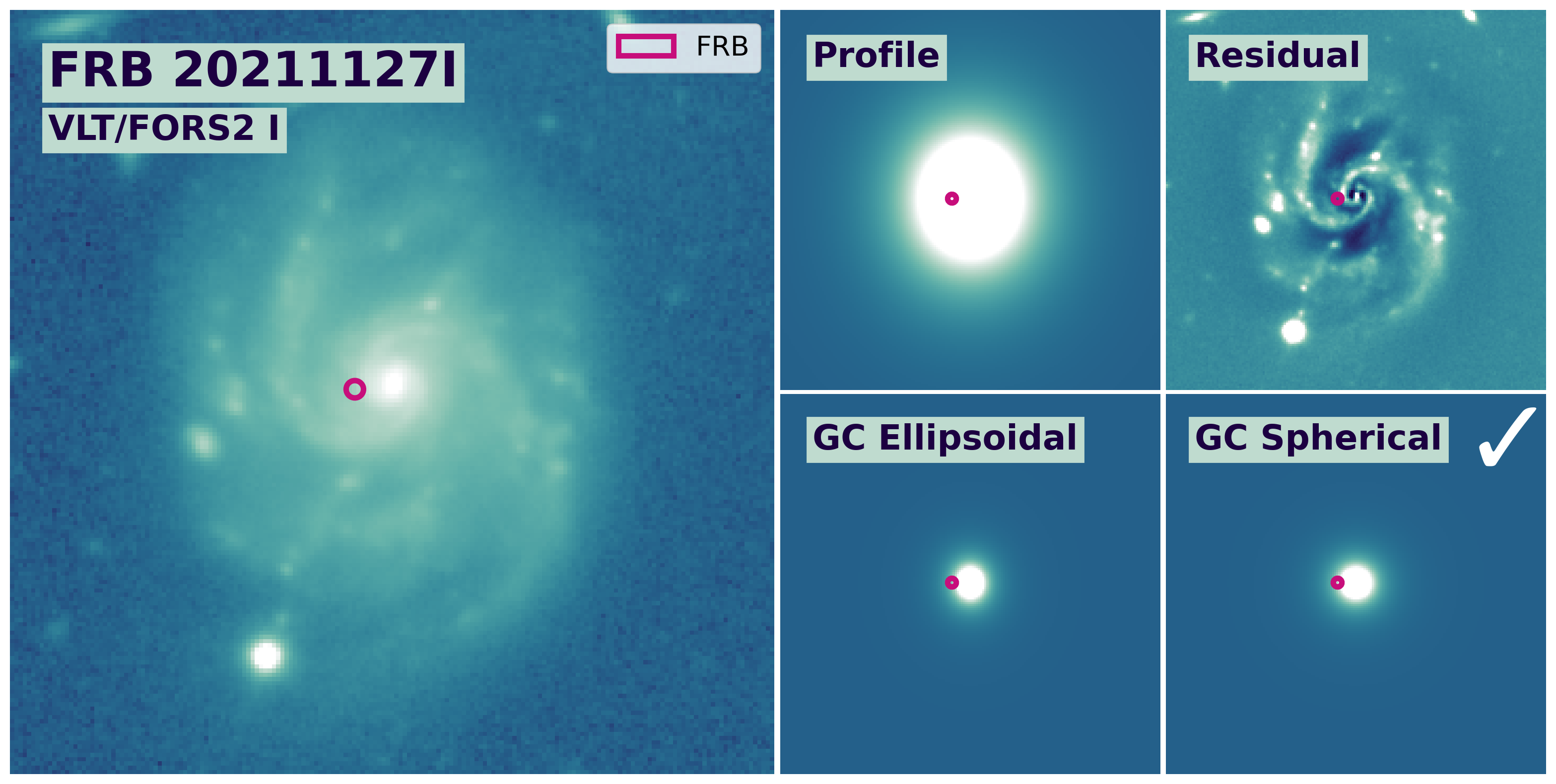}
    \includegraphics[width=0.39\textwidth]{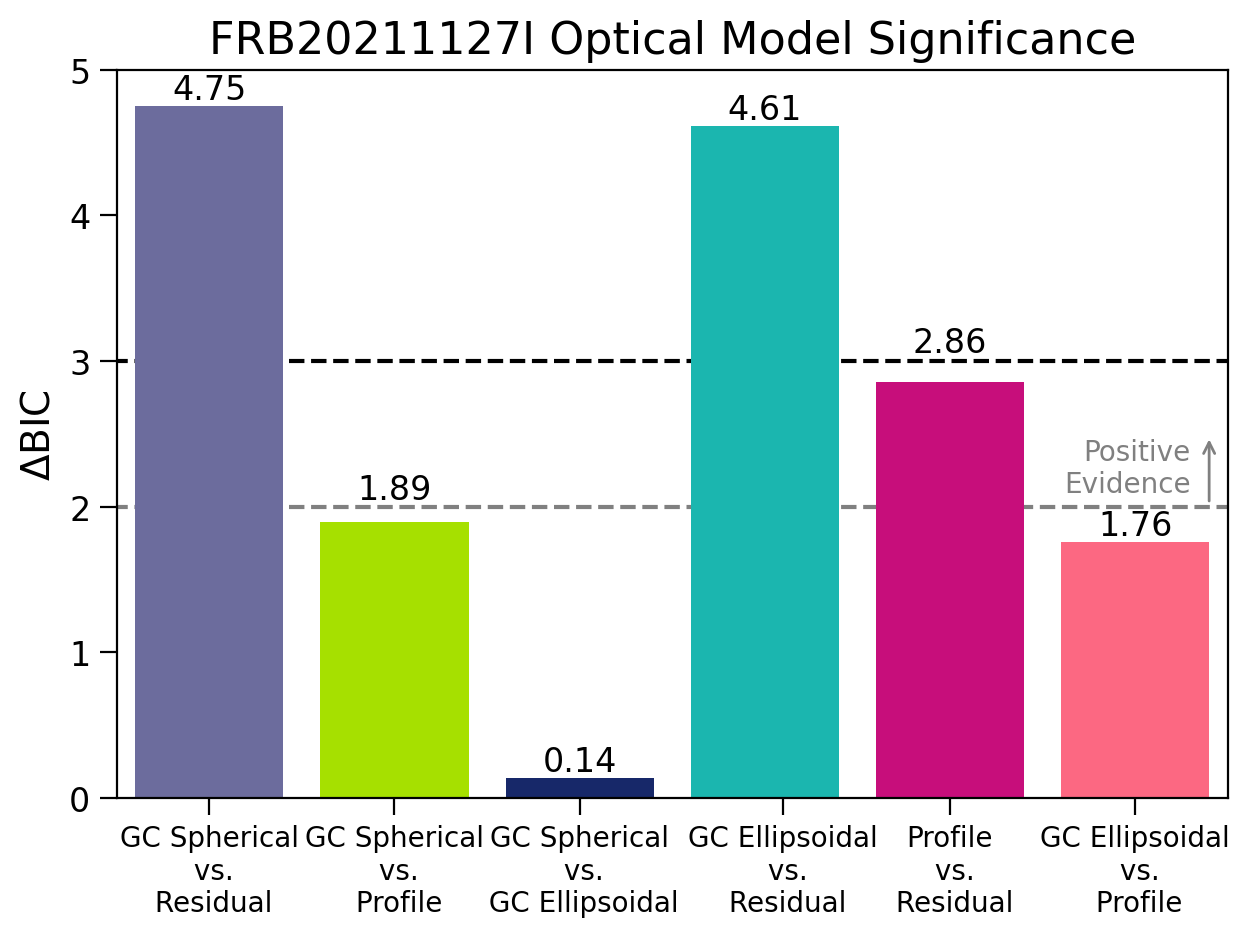}
    \caption{\textbf{(Left:)} Imaging and light profiles for FRB\,20211127I in the optical. \textbf{(Right:)} Comparison of the $\Delta$BIC between the four models for FRB\,20211127I in the optical.}
    \label{fig:appendix_211127_opt}
\end{figure}

\begin{figure}
    \centering
    \includegraphics[width=0.6\textwidth]{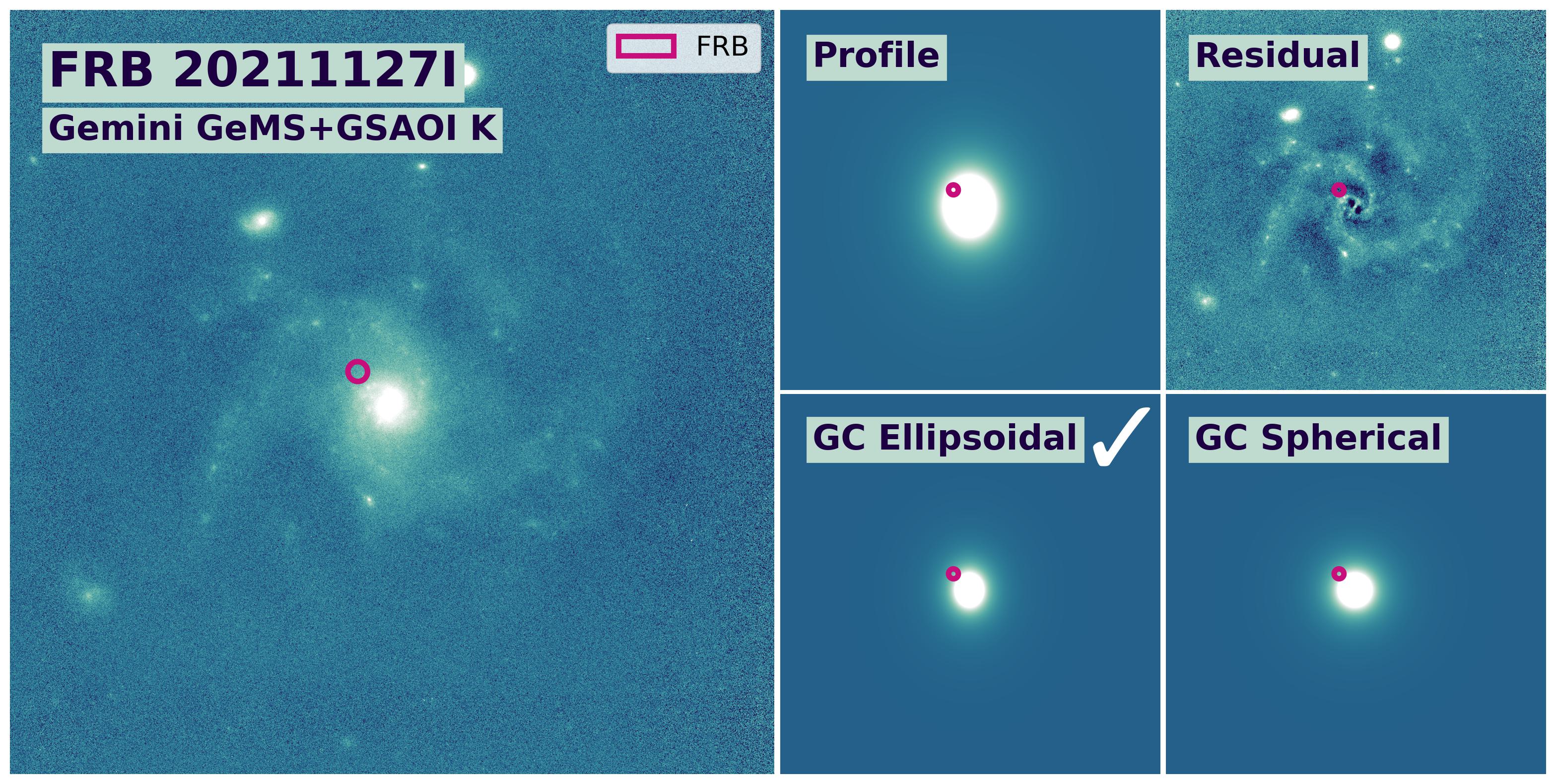}
    \includegraphics[width=0.39\textwidth]{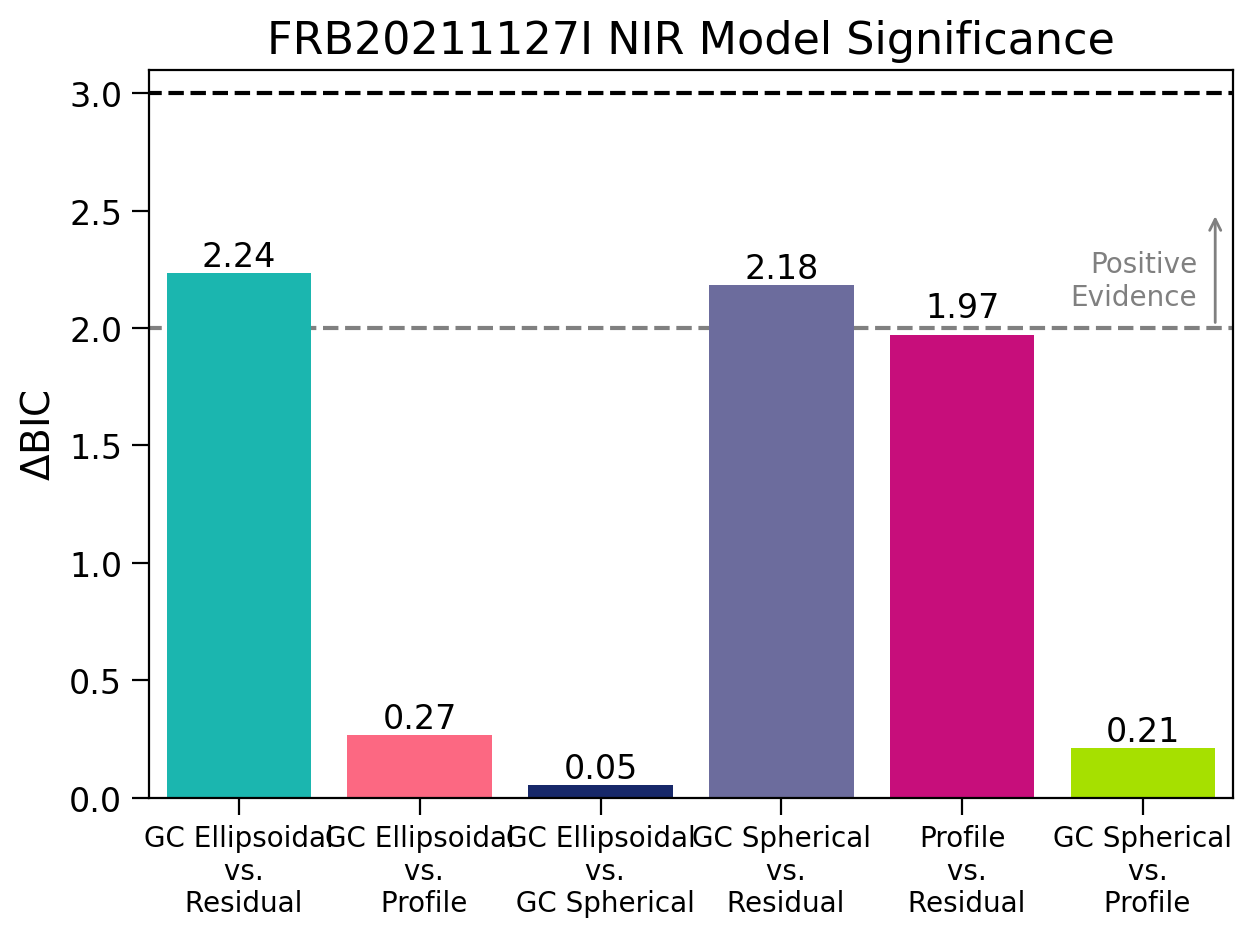}
    \caption{\textbf{(Left:)} Imaging and light profiles for FRB\,20211127I in the NIR. \textbf{(Right:)} Comparison of the $\Delta$BIC between the four models for FRB\,20211127I in the NIR.}
    \label{fig:appendix_211127_NIR}
\end{figure}

\begin{figure}
    \centering
    \includegraphics[width=0.6\textwidth]{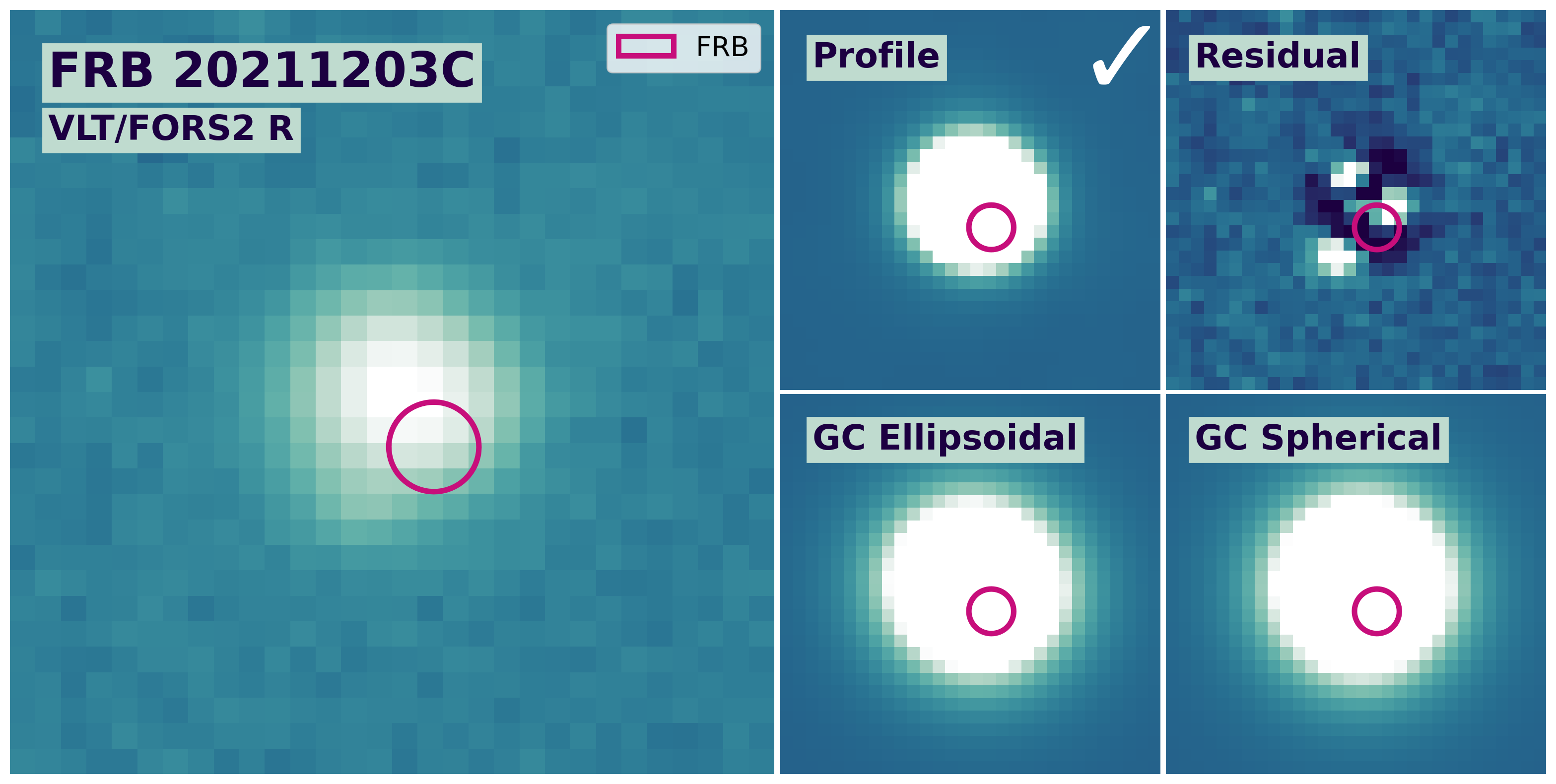}
    \includegraphics[width=0.39\textwidth]{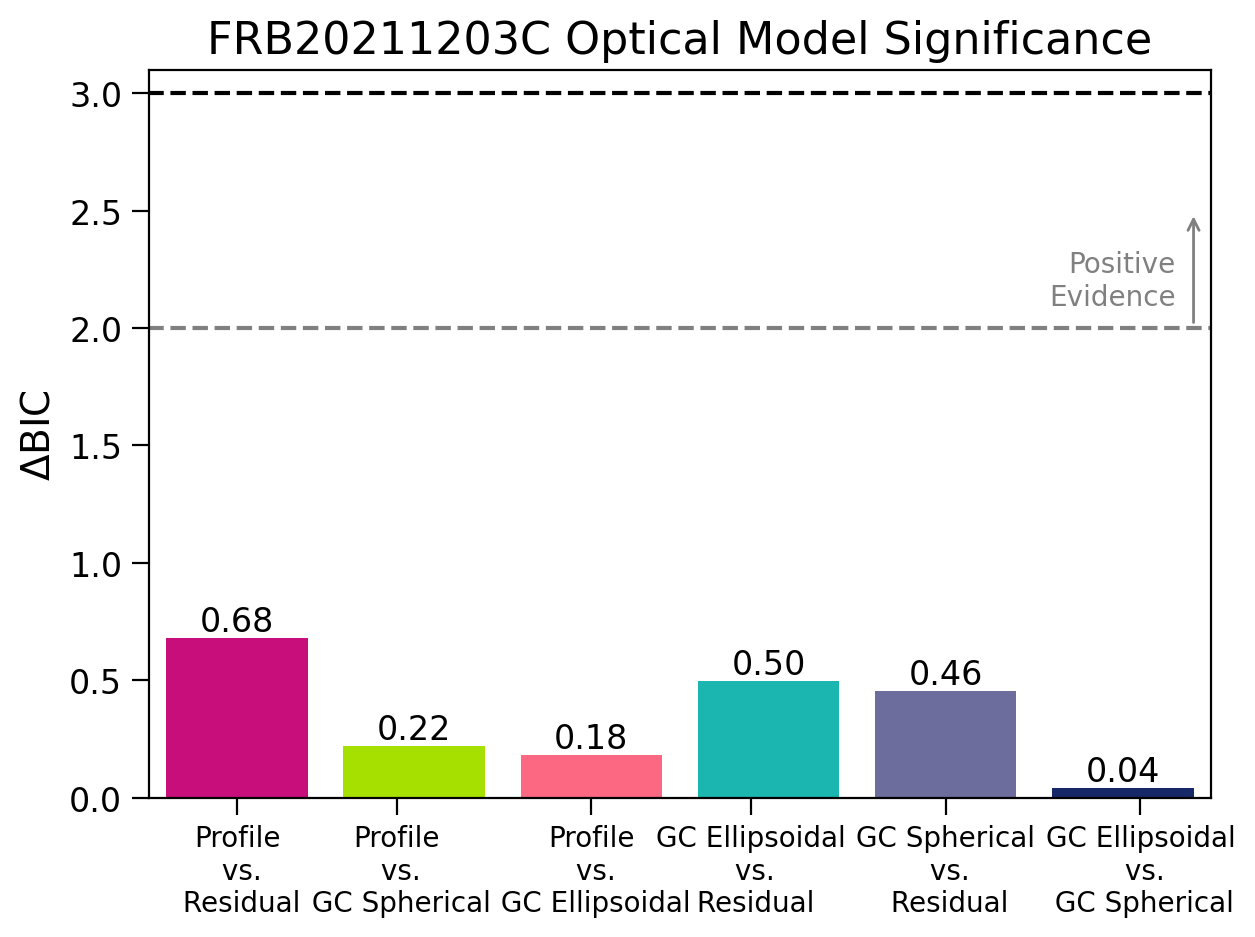}
    \caption{\textbf{(Left:)} Imaging and light profiles for FRB\,20211203C in the optical. \textbf{(Right:)} Comparison of the $\Delta$BIC between the four models for FRB\,20211203C in the optical.}
    \label{fig:appendix_211203_opt}
\end{figure}

\begin{figure}
    \centering
    \includegraphics[width=0.6\textwidth]{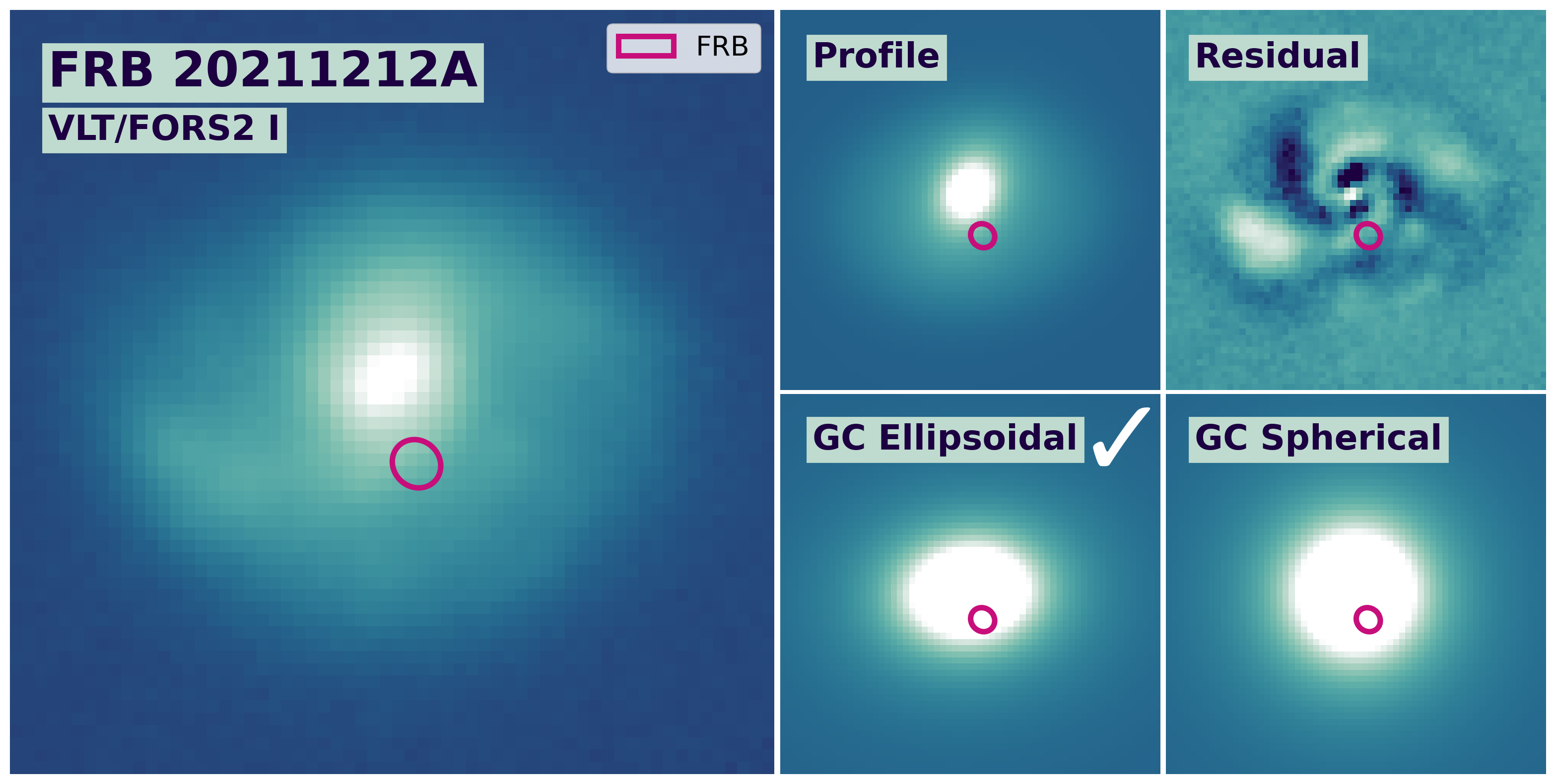}
    \includegraphics[width=0.39\textwidth]{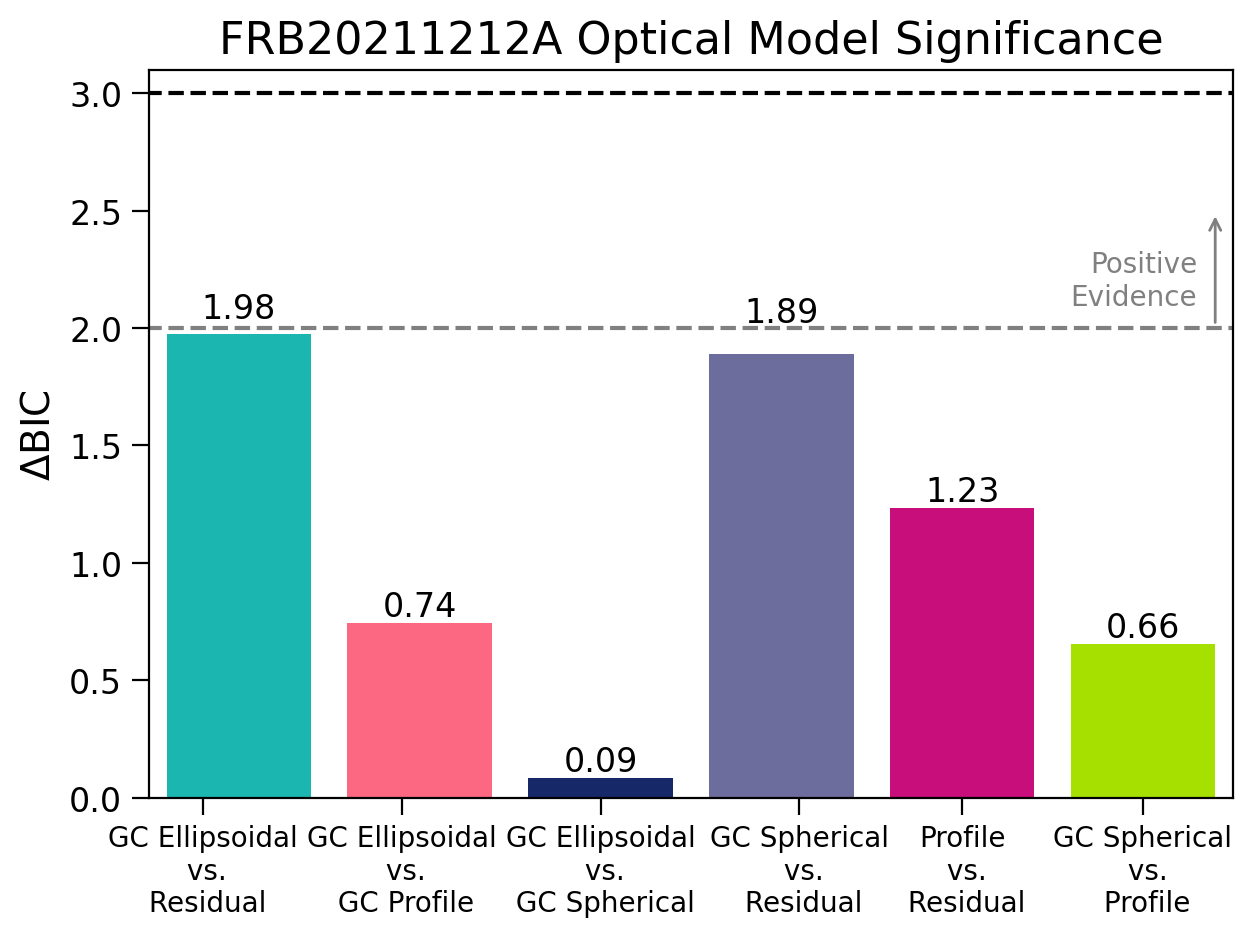}
    \caption{\textbf{(Left:)} Imaging and light profiles for FRB\,20211212A in the optical. \textbf{(Right:)} Comparison of the $\Delta$BIC between the four models for FRB\,20211212A in the optical.}
    \label{fig:appendix_211212_opt}
\end{figure}

\begin{figure}
    \centering
    \includegraphics[width=0.6\textwidth]{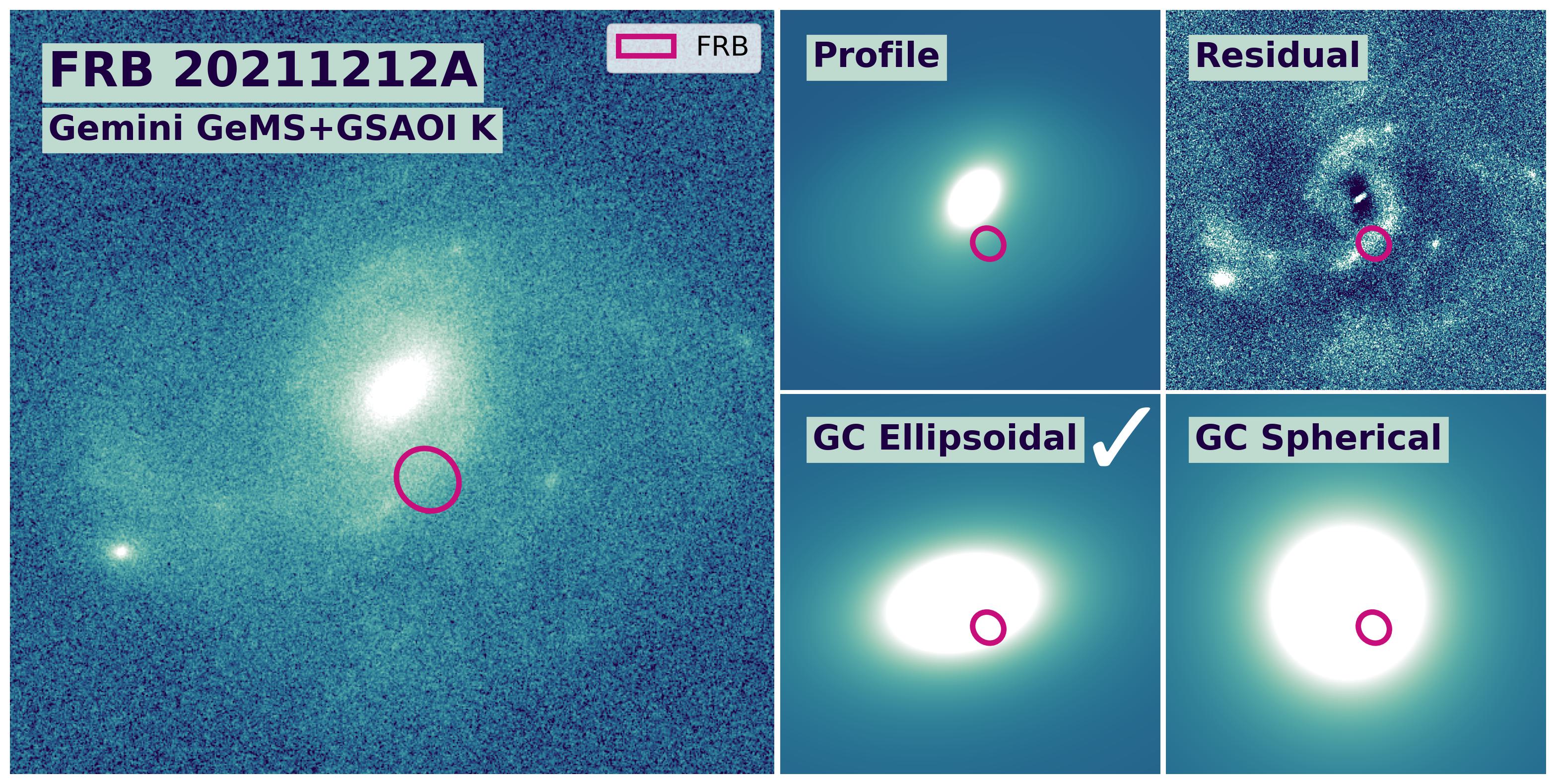}
    \includegraphics[width=0.39\textwidth]{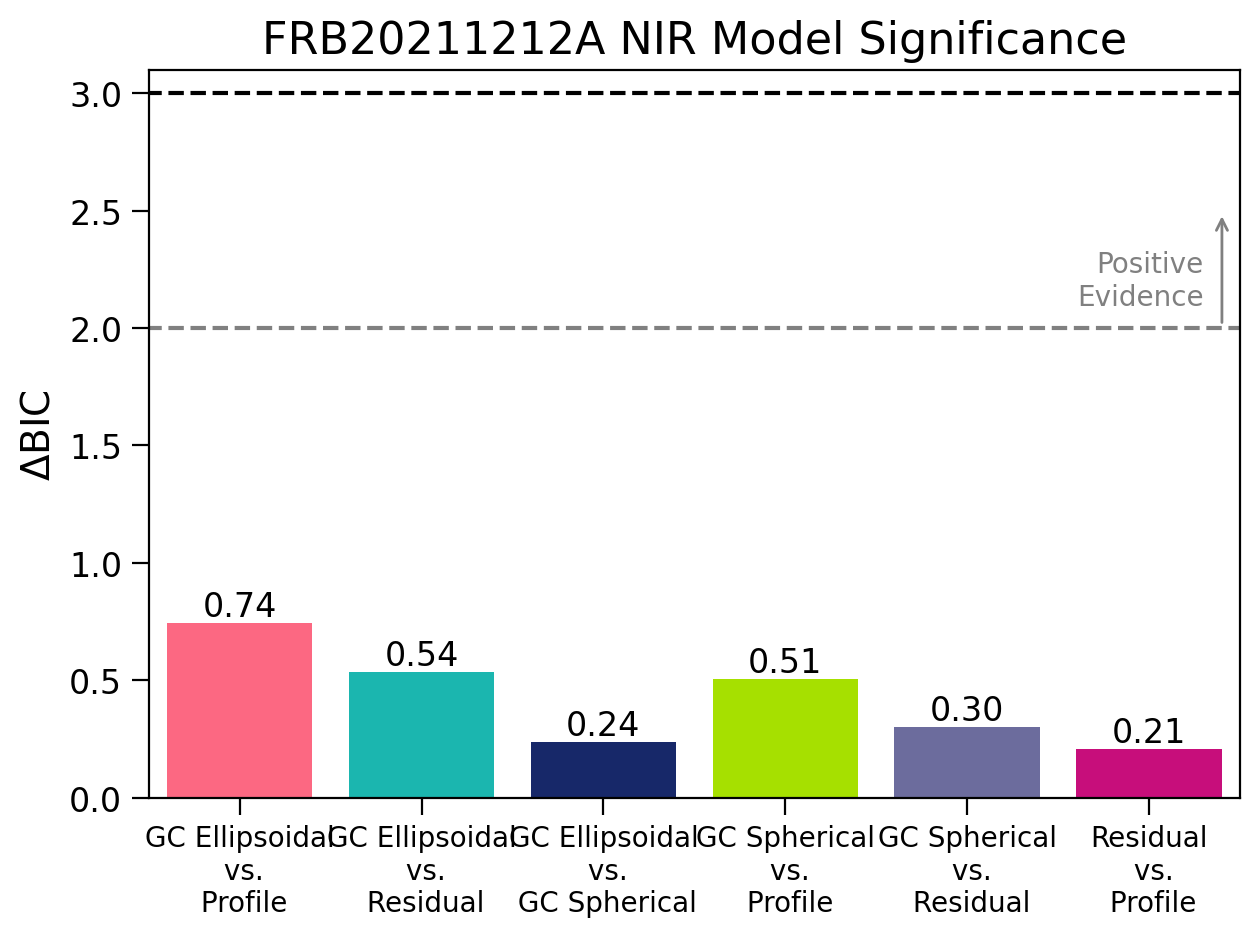}
    \caption{\textbf{(Left:)} Imaging and light profiles for FRB\,20211212A in the NIR. \textbf{(Right:)} Comparison of the $\Delta$BIC between the four models for FRB\,20211212A in the NIR.}
    \label{fig:appendix_211212_NIR}
\end{figure}

\begin{figure}
    \centering
    \includegraphics[width=0.6\textwidth]{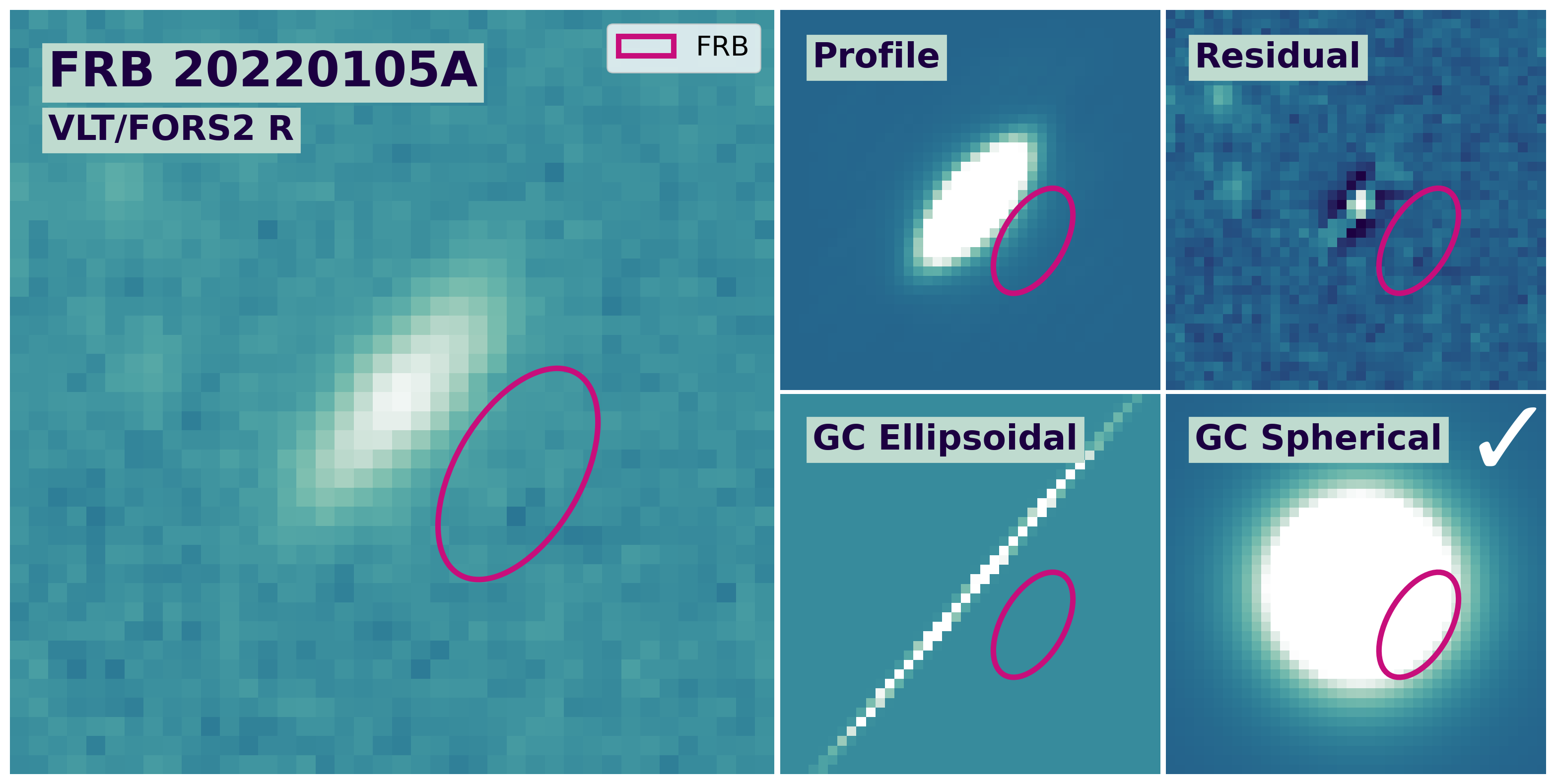}
    \includegraphics[width=0.39\textwidth]{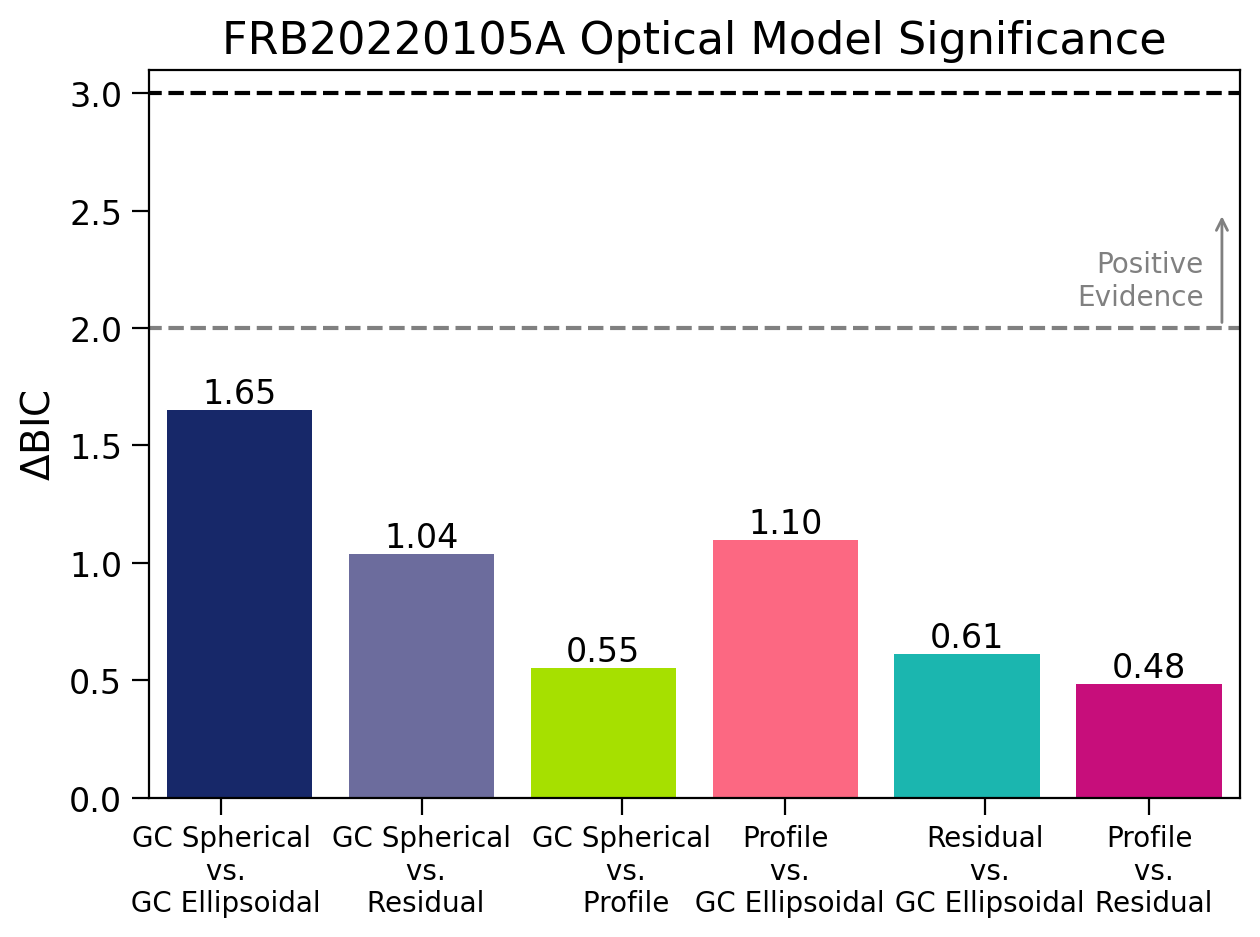}
    \caption{\textbf{(Left:)} Imaging and light profiles for FRB\,20220105A in the optical. \textbf{(Right:)} Comparison of the $\Delta$BIC between the four models for FRB\,20220105A in the optical.}
    \label{fig:appendix_220105_opt}
\end{figure}

\begin{figure}
    \centering
    \includegraphics[width=0.6\textwidth]{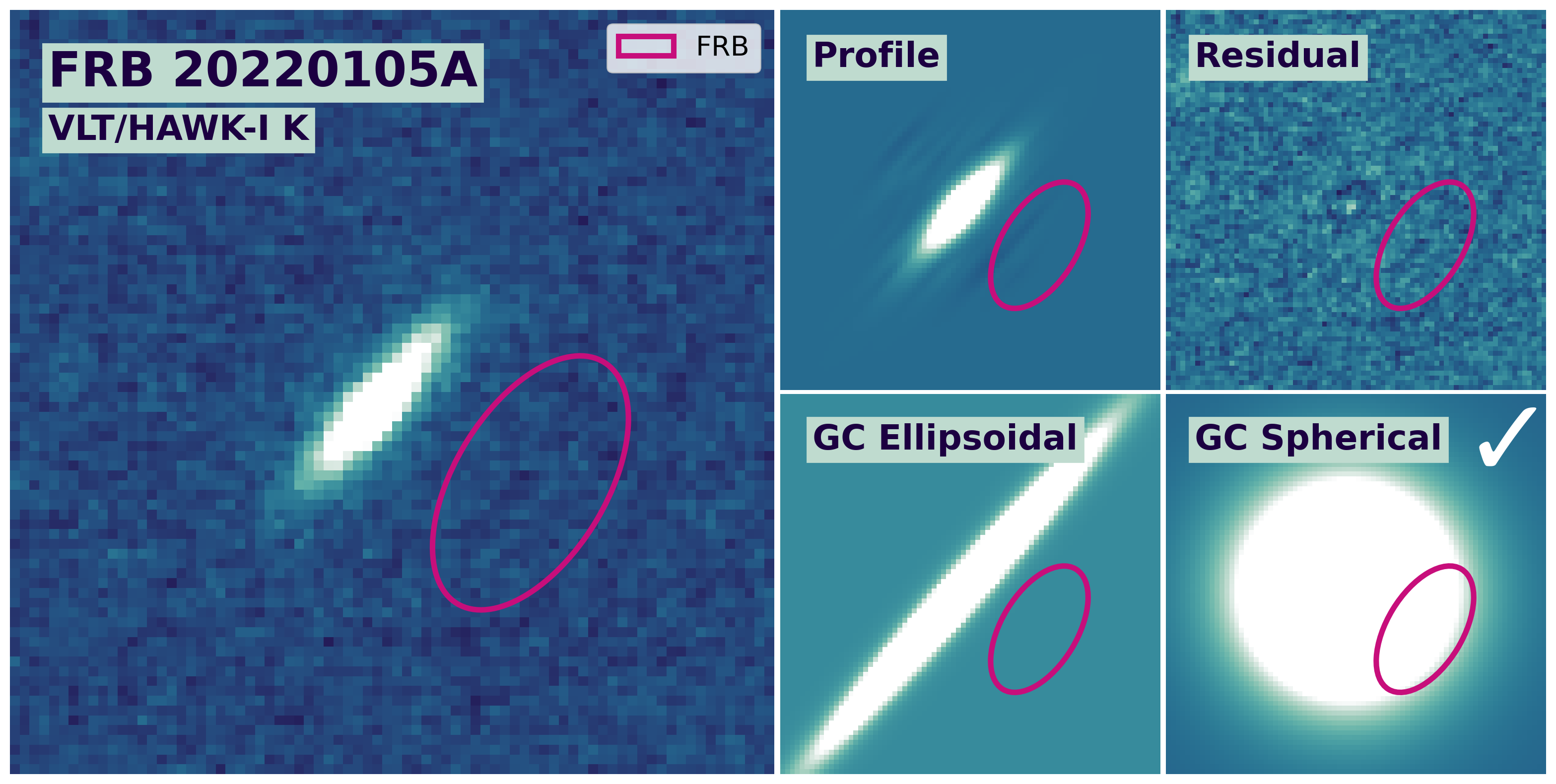}
    \includegraphics[width=0.39\textwidth]{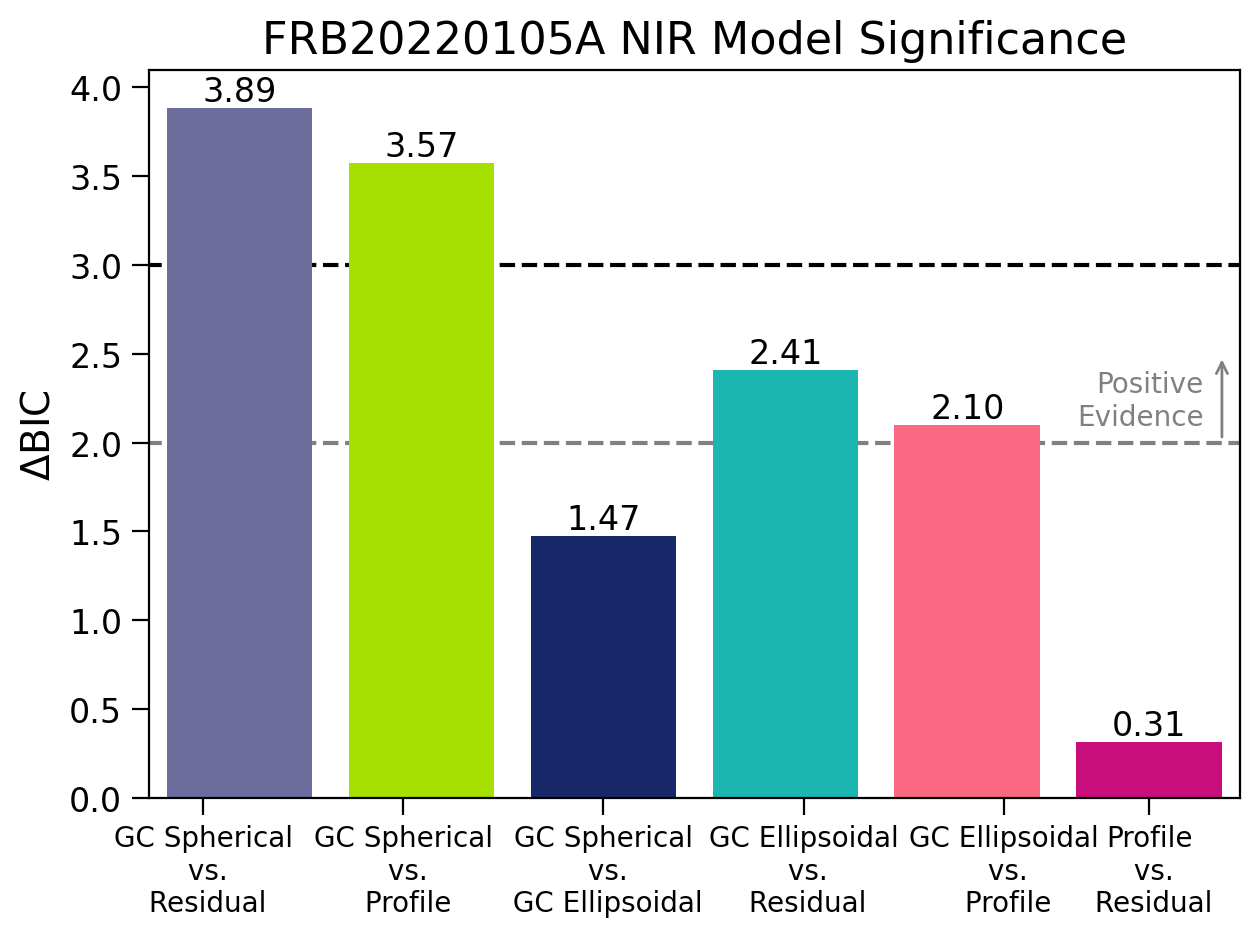}
    \caption{\textbf{(Left:)} Imaging and light profiles for FRB\,20220105A in the NIR. \textbf{(Right:)} Comparison of the $\Delta$BIC between the four models for FRB\,20220105A in the NIR.}
    \label{fig:appendix_220105_NIR}
\end{figure}

\begin{figure}
    \centering
    \includegraphics[width=0.6\textwidth]{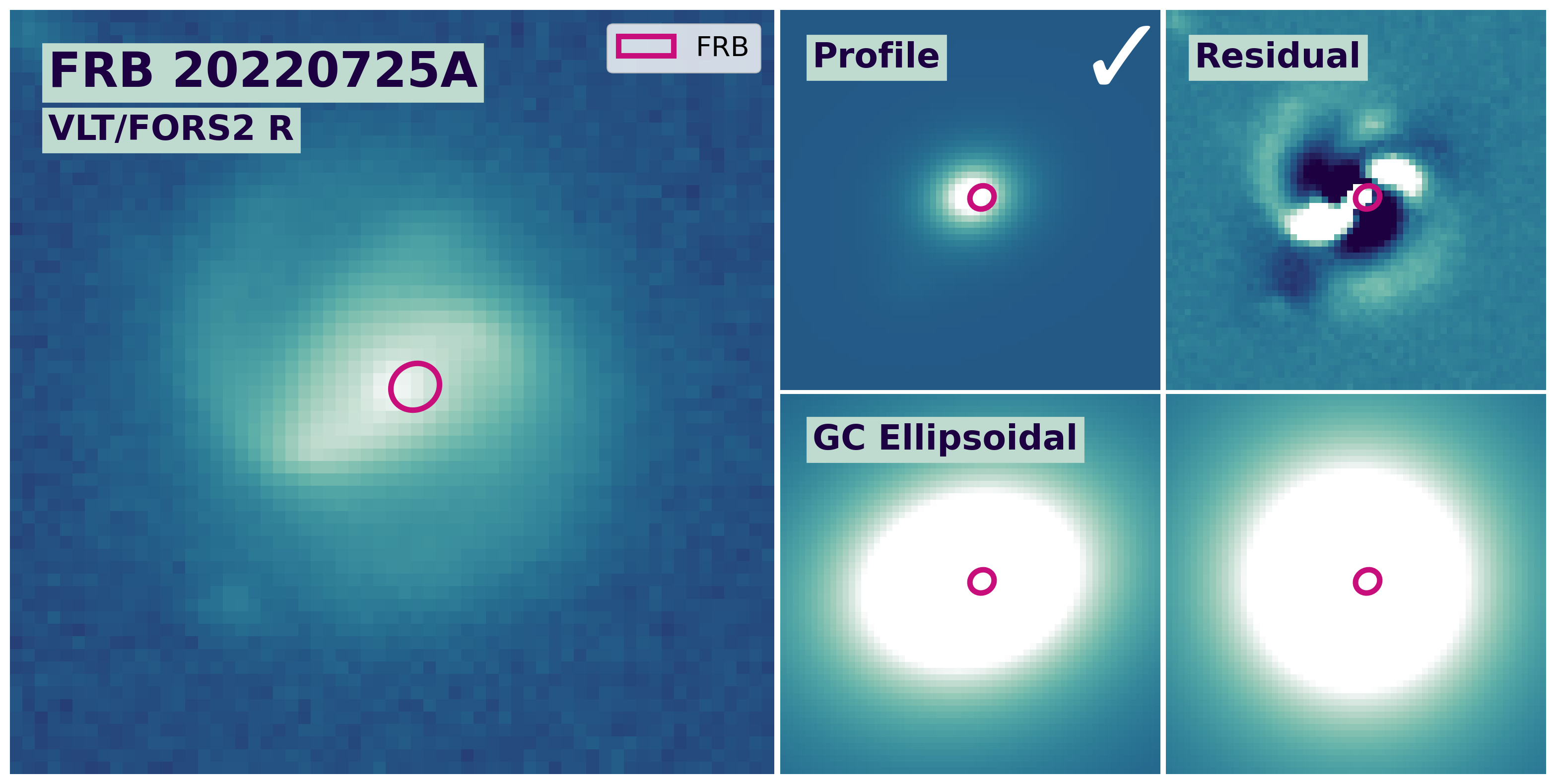}
    \includegraphics[width=0.39\textwidth]{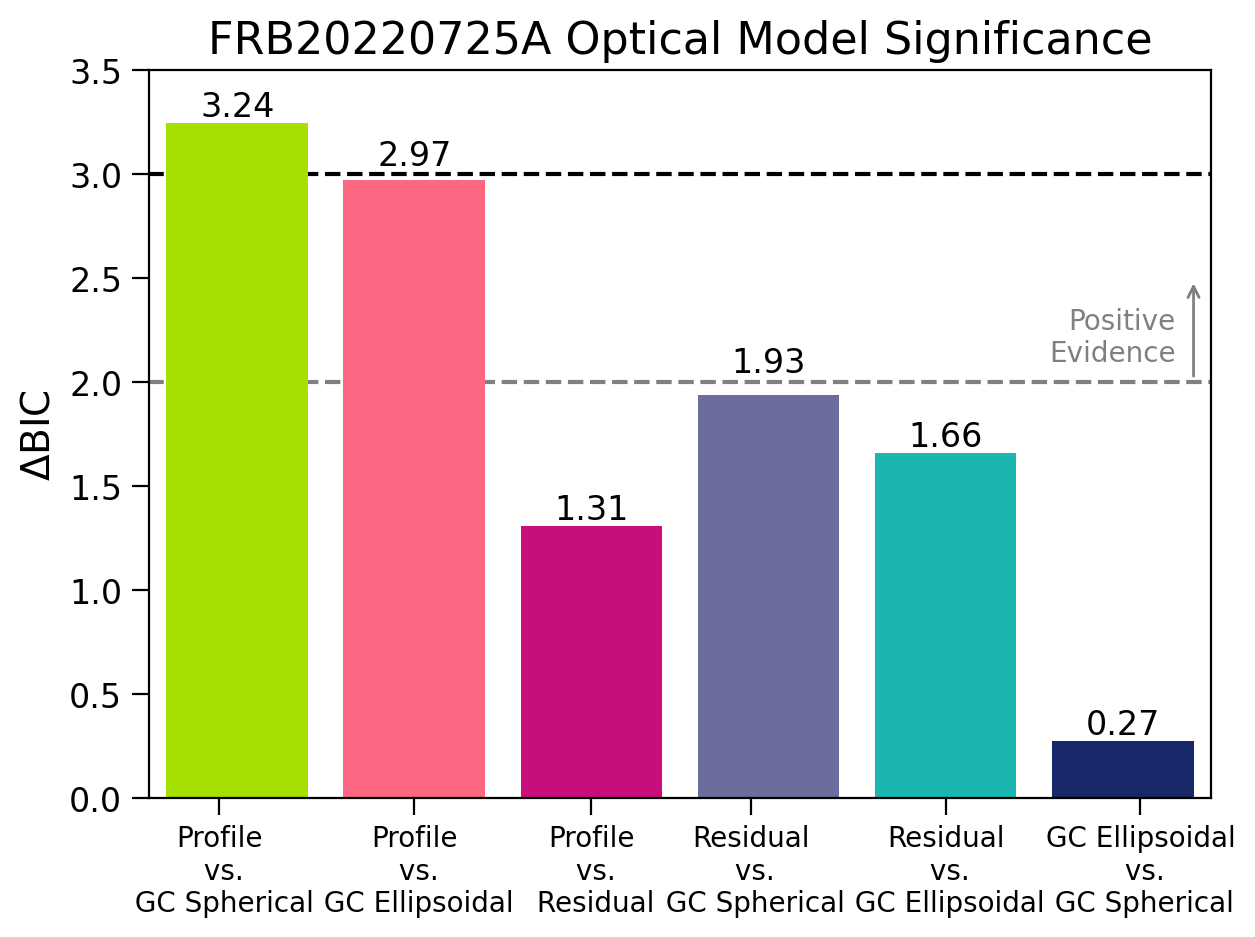}
    \caption{\textbf{(Left:)} Imaging and light profiles for FRB\,20220725A in the optical. \textbf{(Right:)} Comparison of the $\Delta$BIC between the four models for FRB\,20220725A in the optical.}
    \label{fig:appendix_220725_opt}
\end{figure}

\begin{figure}
    \centering
    \includegraphics[width=0.6\textwidth]{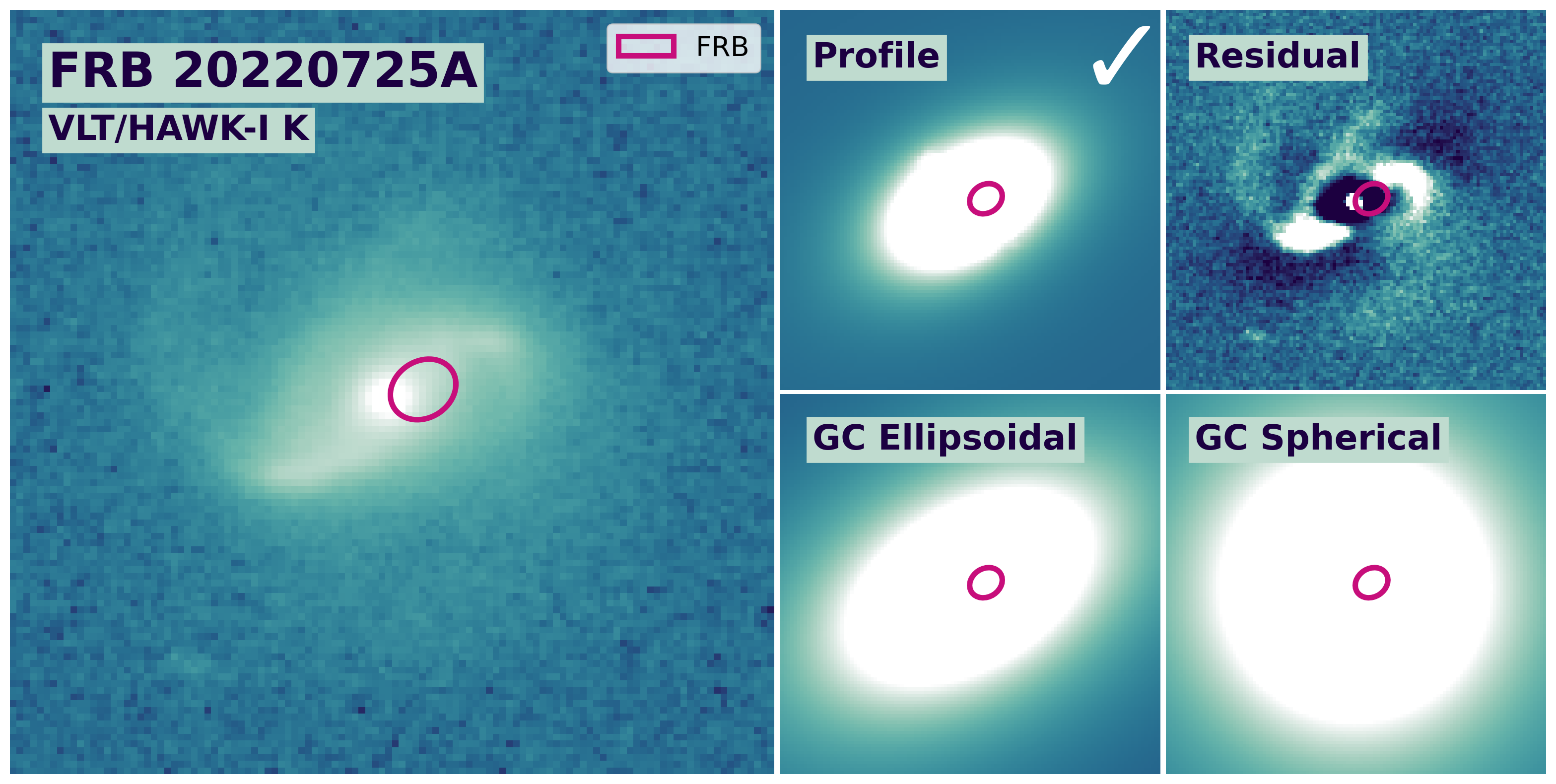}
    \includegraphics[width=0.39\textwidth]{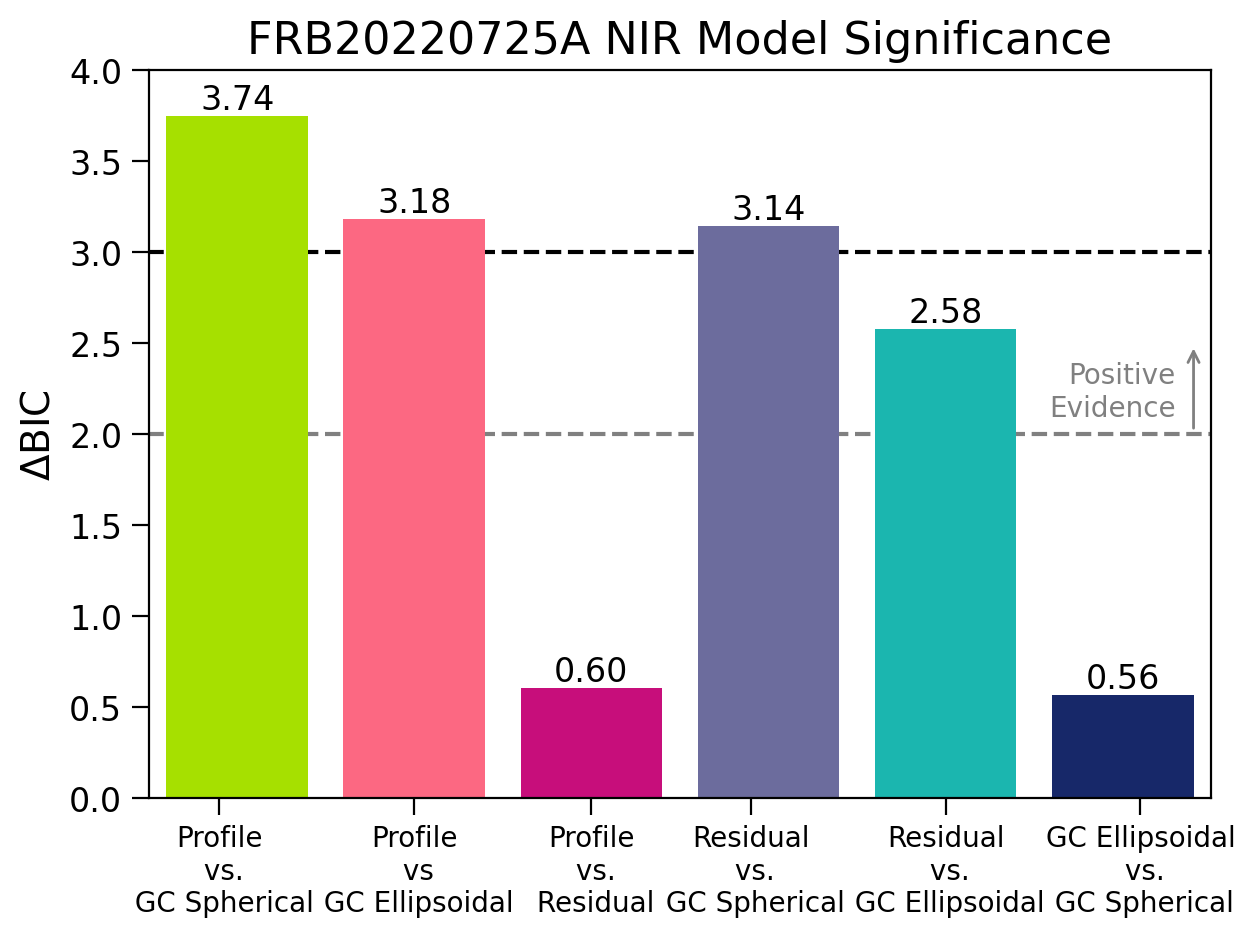}
    \caption{\textbf{(Left:)} Imaging and light profiles for FRB\,20220725A in the NIR. \textbf{(Right:)} Comparison of the $\Delta$BIC between the four models for FRB\,20220725A in the NIR.}
    \label{fig:appendix_220725_NIR}
\end{figure}

\begin{figure}
    \centering
    \includegraphics[width=0.6\textwidth]{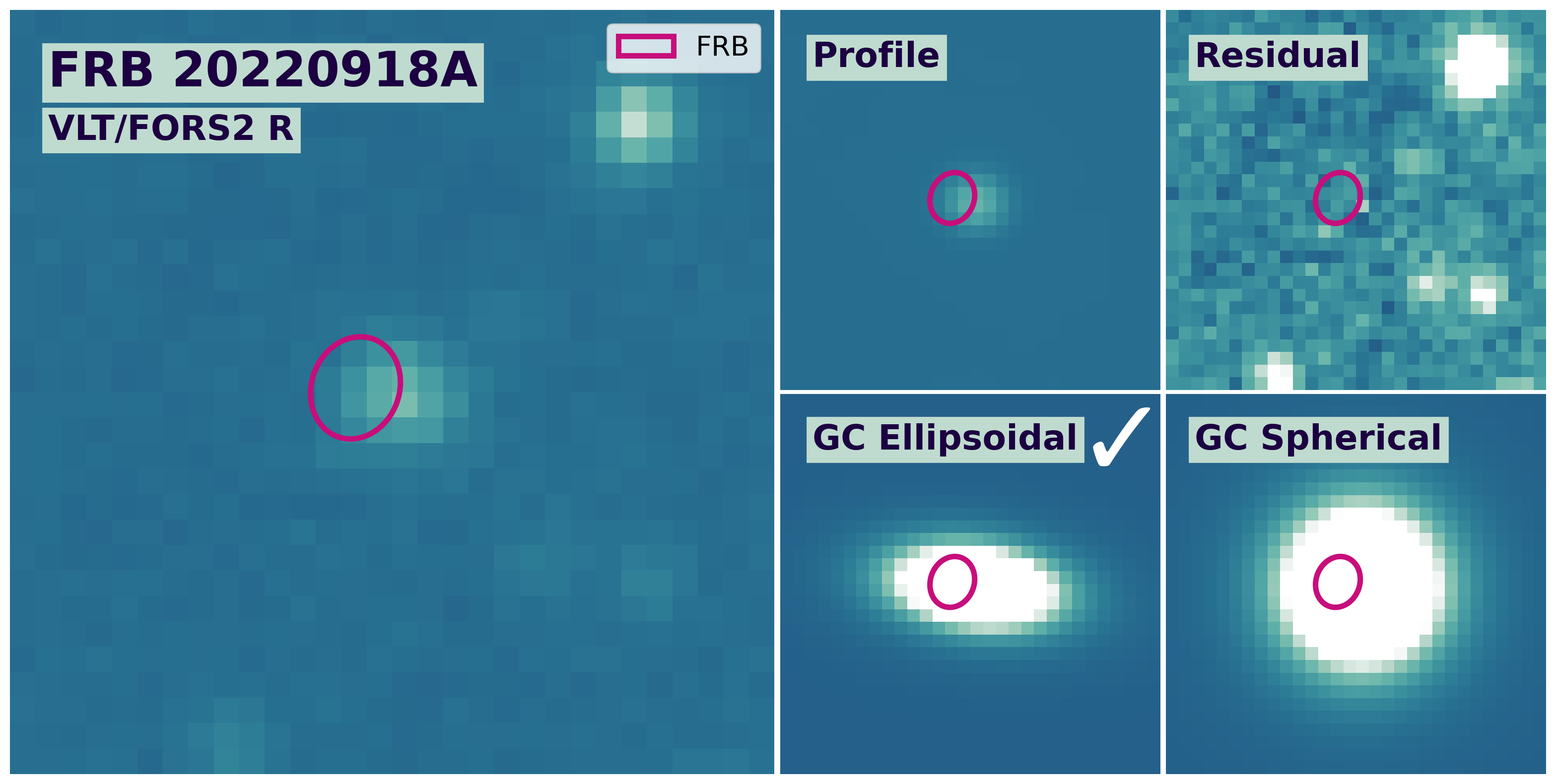}
    \includegraphics[width=0.39\textwidth]{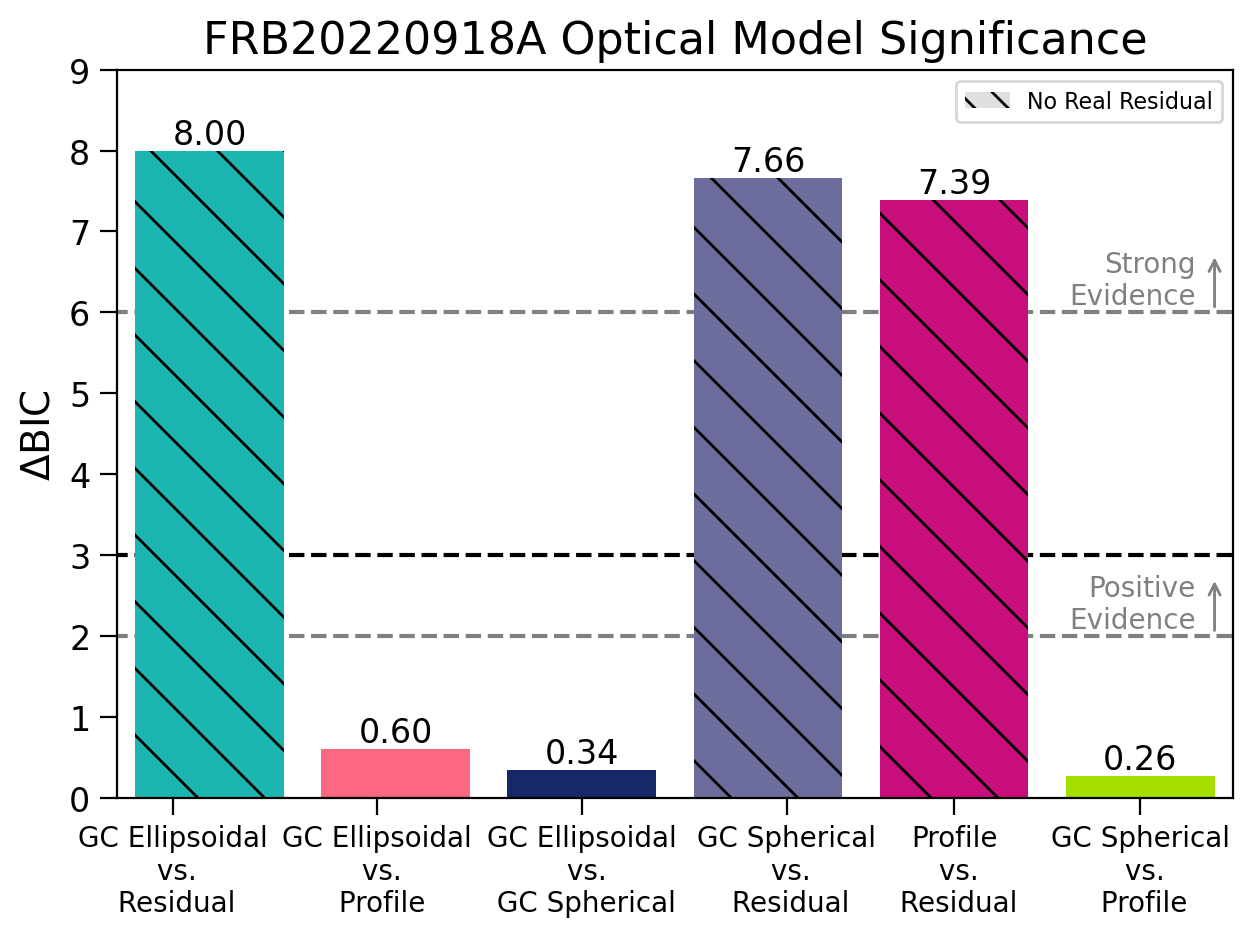}
    \caption{\textbf{(Left:)} Imaging and light profiles for FRB\,20220918A in the optical. \textbf{(Right:)} Comparison of the $\Delta$BIC between the four models for FRB\,20220918A in the optical.}
    \label{fig:appendix_220918_opt}
\end{figure}

\begin{figure}
    \centering
    \includegraphics[width=0.6\textwidth]{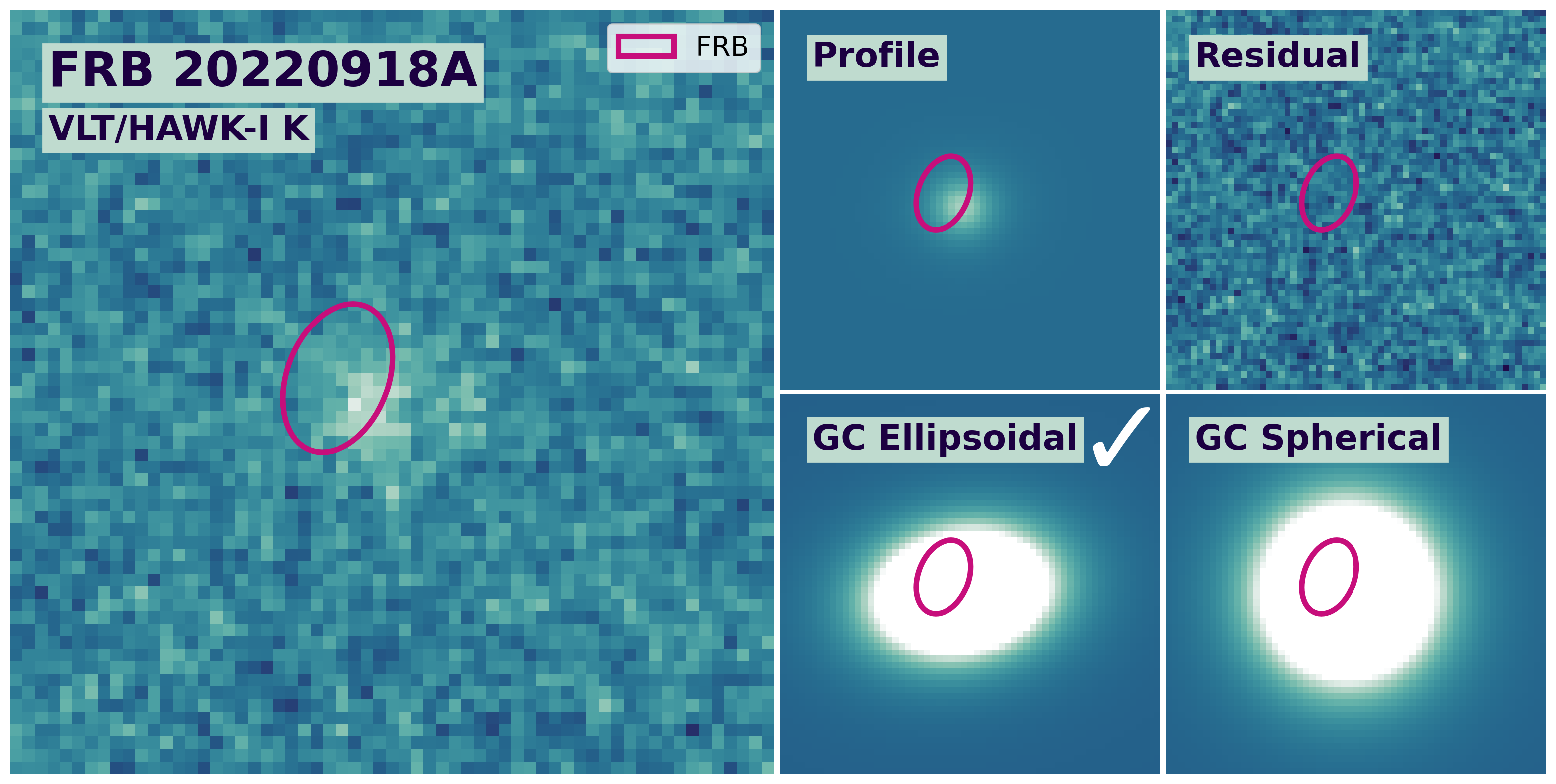}
    \includegraphics[width=0.39\textwidth]{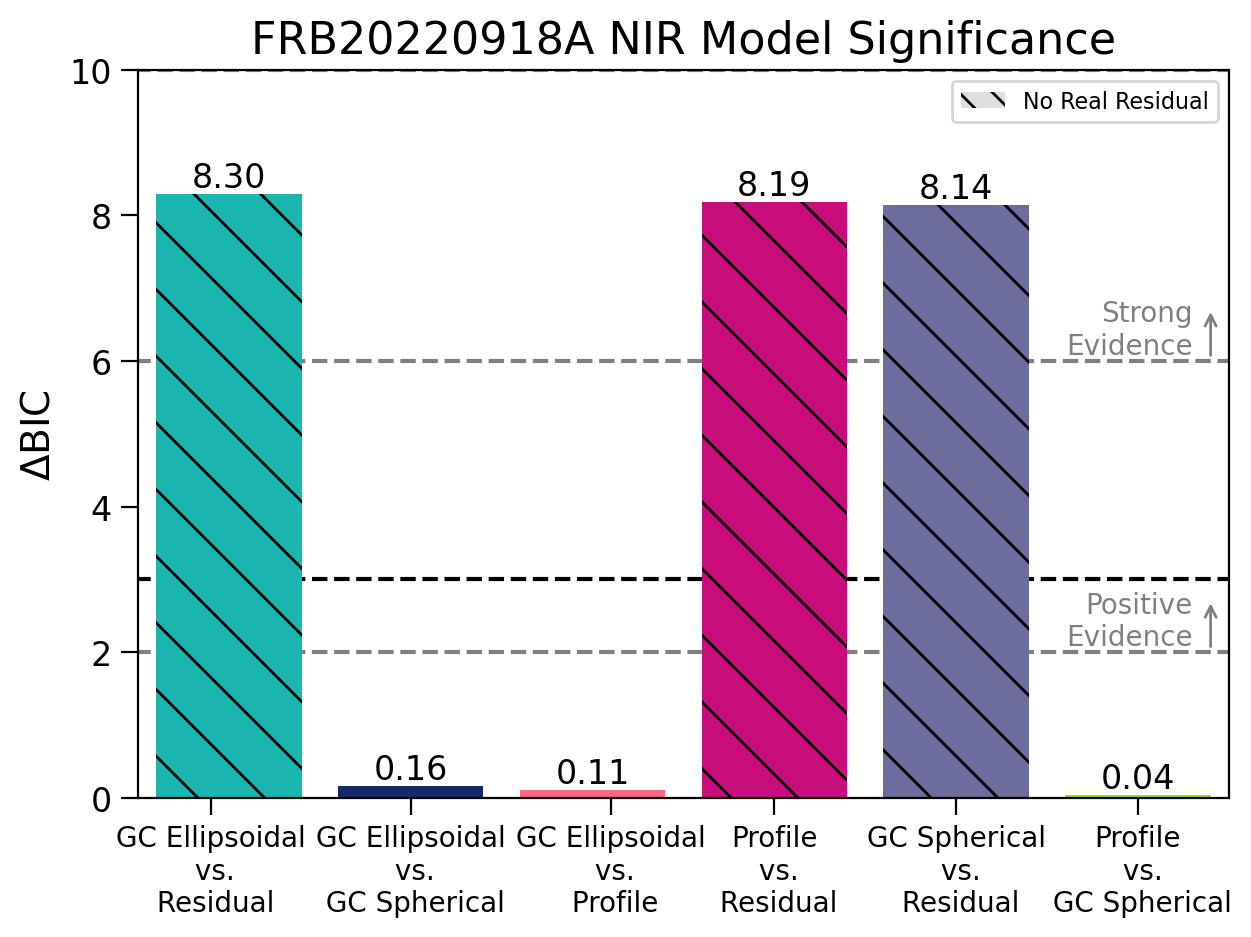}
    \caption{\textbf{(Left:)} Imaging and light profiles for FRB\,20220918A in the NIR. \textbf{(Right:)} Comparison of the $\Delta$BIC between the four models for FRB\,20220918A in the NIR.}
    \label{fig:appendix_220918_NIR}
\end{figure}

\begin{figure}
    \centering
    \includegraphics[width=0.6\textwidth]{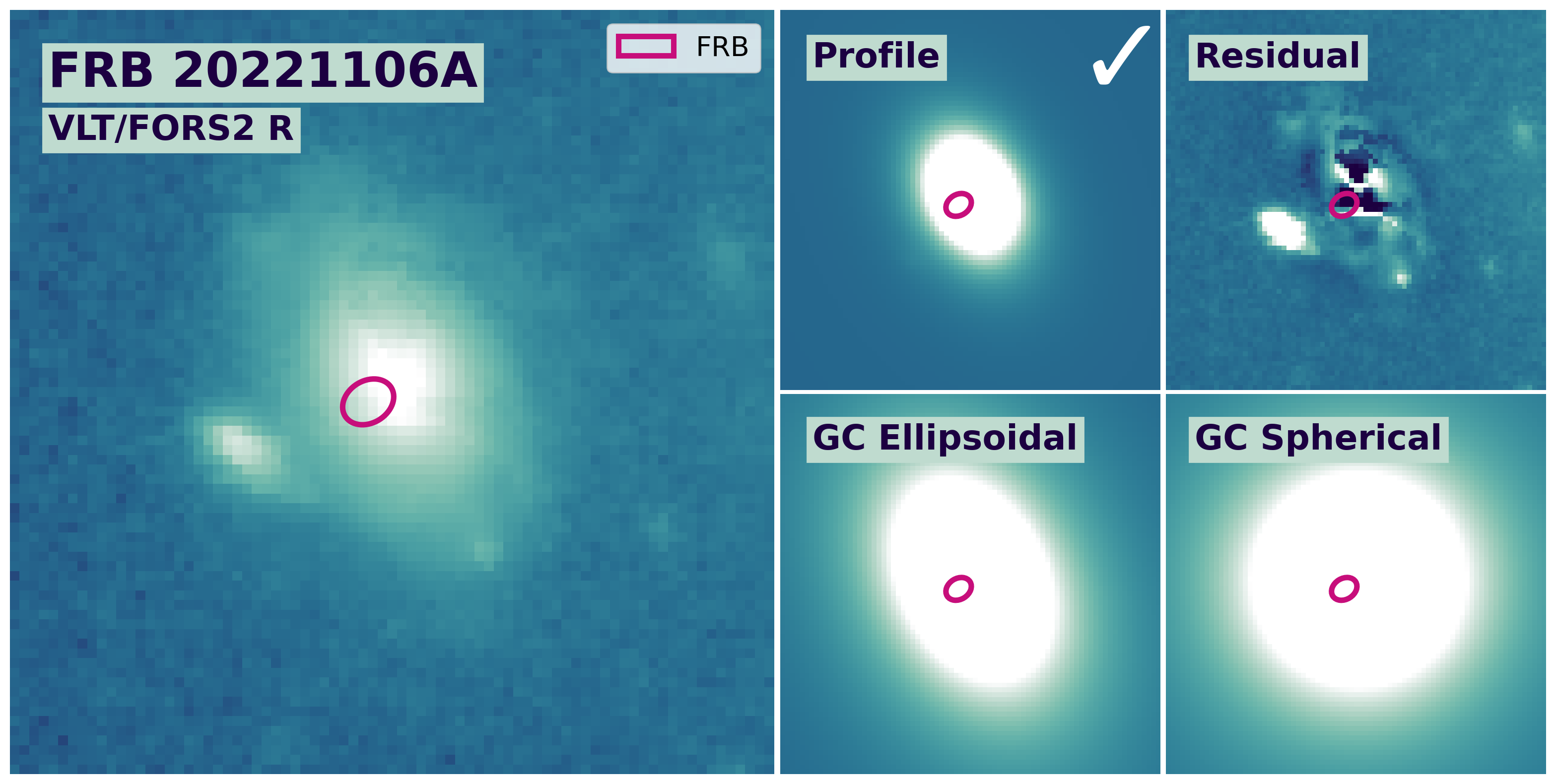}
    \includegraphics[width=0.39\textwidth]{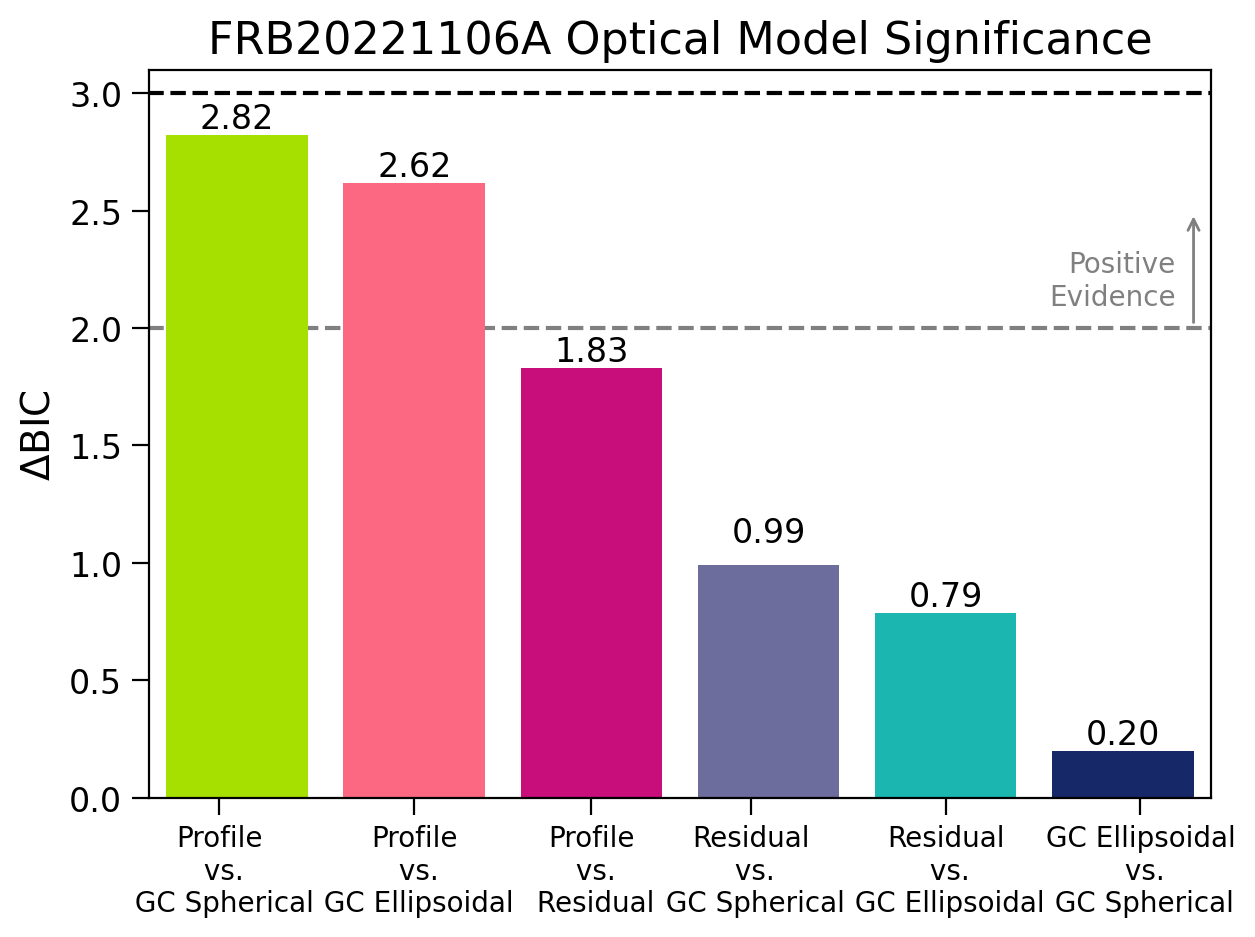}
    \caption{\textbf{(Left:)} Imaging and light profiles for FRB\,20221106A in the optical. \textbf{(Right:)} Comparison of the $\Delta$BIC between the four models for FRB\,20221106A in the optical.}
    \label{fig:appendix_221106_opt}
\end{figure}

\begin{figure}
    \centering
    \includegraphics[width=0.6\textwidth]{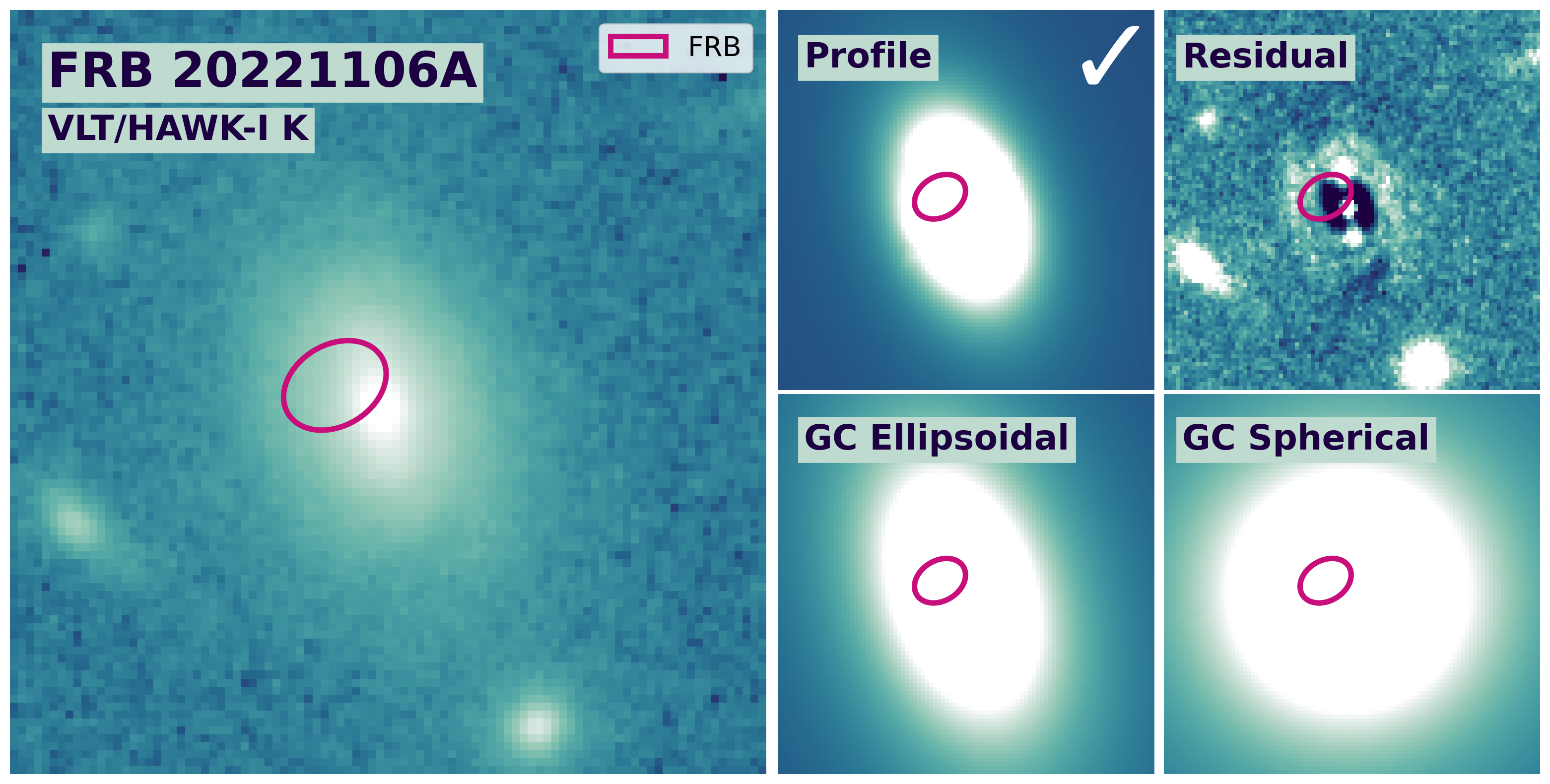}
    \includegraphics[width=0.39\textwidth]{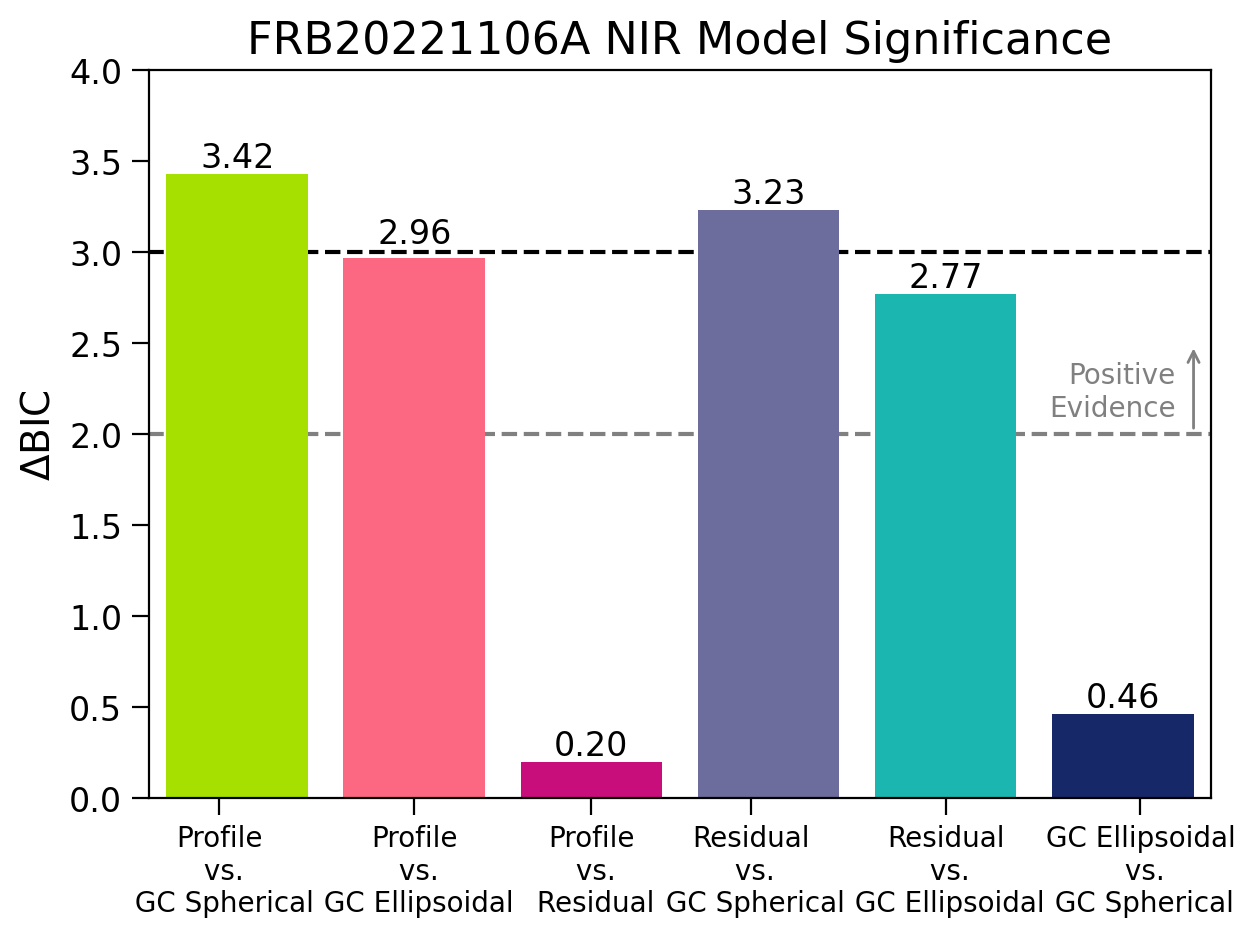}
    \caption{\textbf{(Left:)} Imaging and light profiles for FRB\,20221106A in the NIR. \textbf{(Right:)} Comparison of the $\Delta$BIC between the four models for FRB\,20221106A in the NIR.}
    \label{fig:appendix_221106_NIR}
\end{figure}

\begin{figure}
    \centering
    \includegraphics[width=0.6\textwidth]{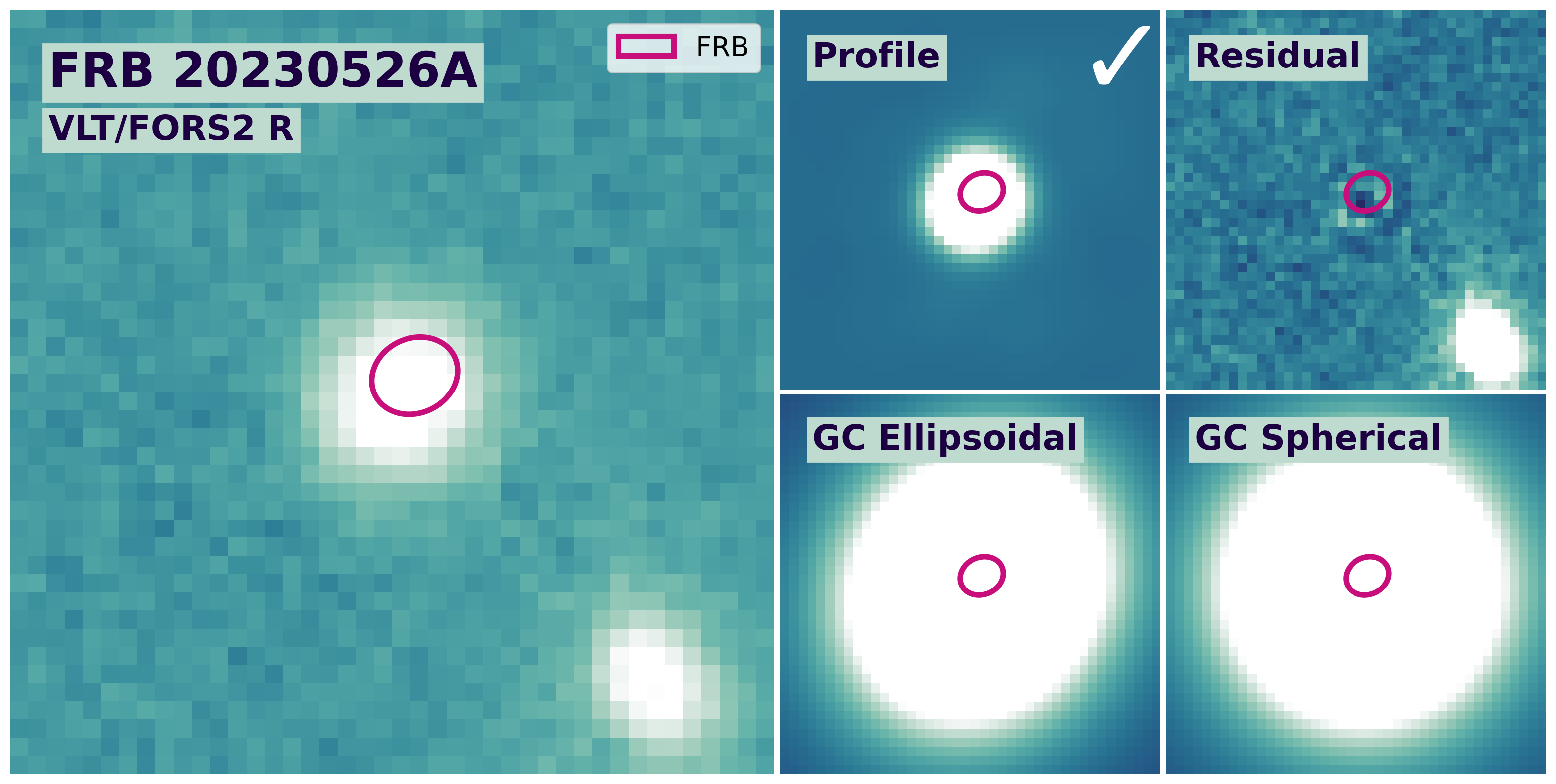}
    \includegraphics[width=0.39\textwidth]{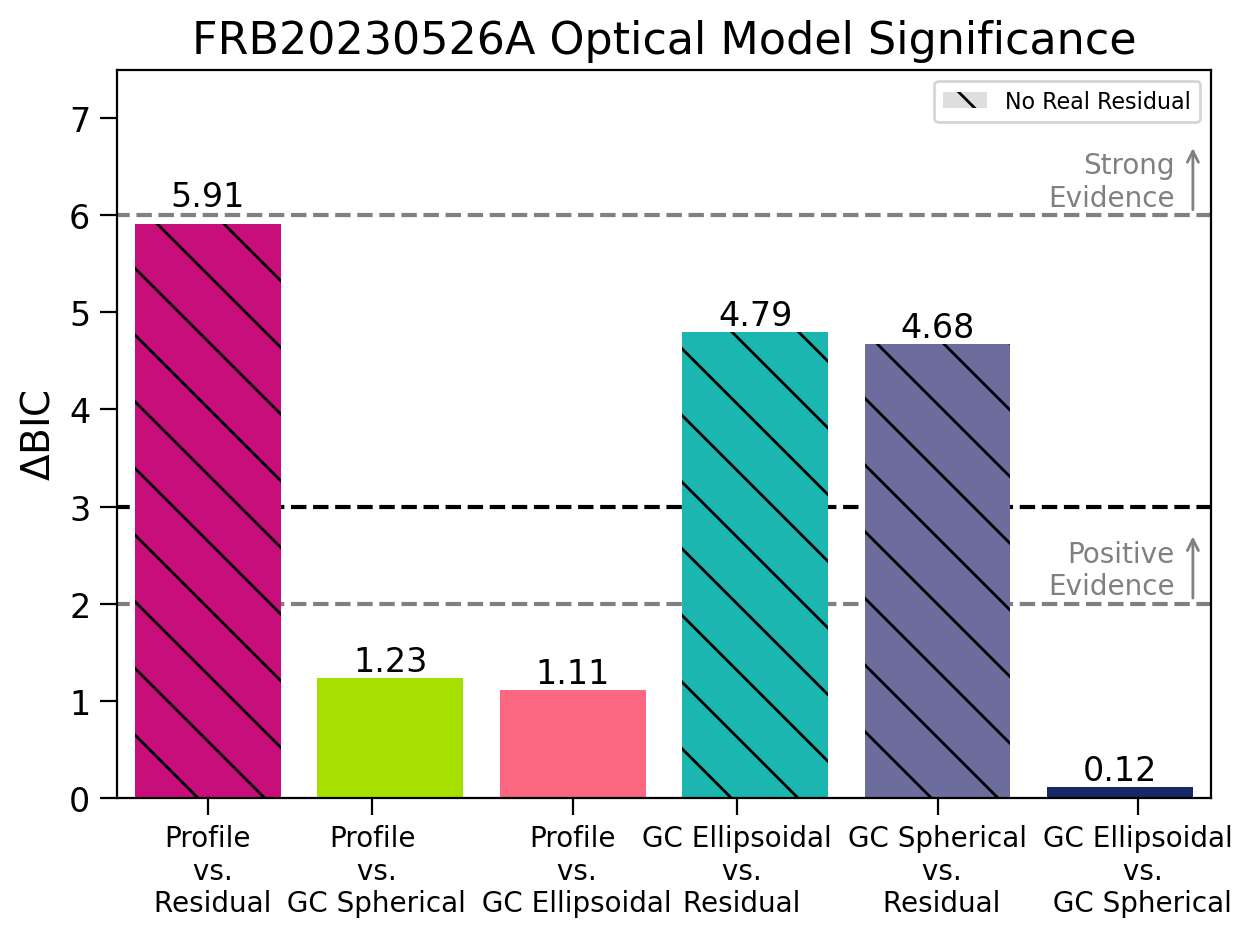}
    \caption{\textbf{(Left:)} Imaging and light profiles for FRB\,20230526A in the optical. \textbf{(Right:)} Comparison of the $\Delta$BIC between the four models for FRB\,20230526A in the optical.}
    \label{fig:appendix_230526_opt}
\end{figure}

\begin{figure}
    \centering
    \includegraphics[width=0.6\textwidth]{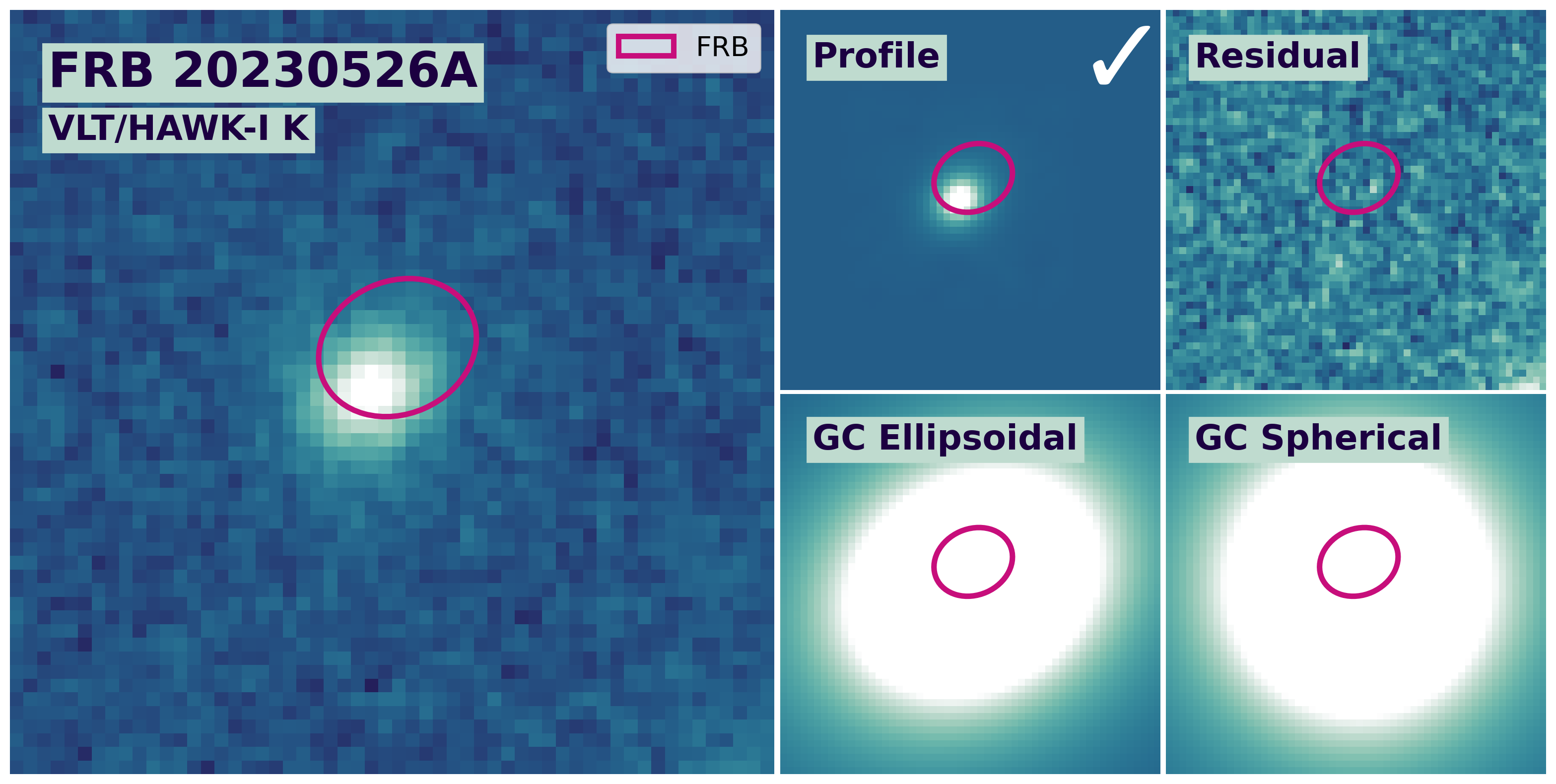}
    \includegraphics[width=0.39\textwidth]{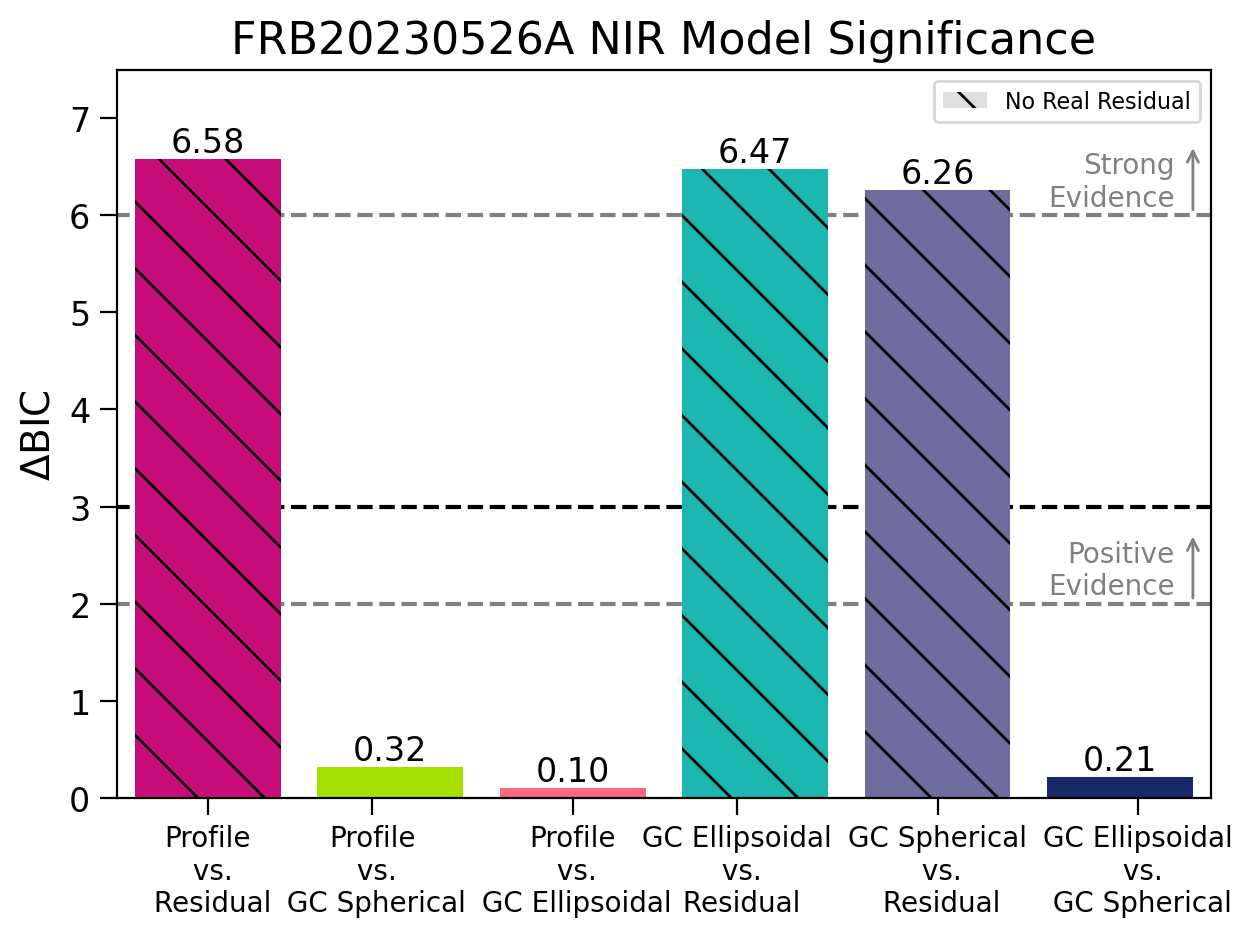}
    \caption{\textbf{(Left:)} Imaging and light profiles for FRB\,20230526A in the NIR. \textbf{(Right:)} Comparison of the $\Delta$BIC between the four models for FRB\,20230526A in the NIR.}
    \label{fig:appendix_230526_NIR}
\end{figure}

\begin{figure}
    \centering
    \includegraphics[width=0.6\textwidth]{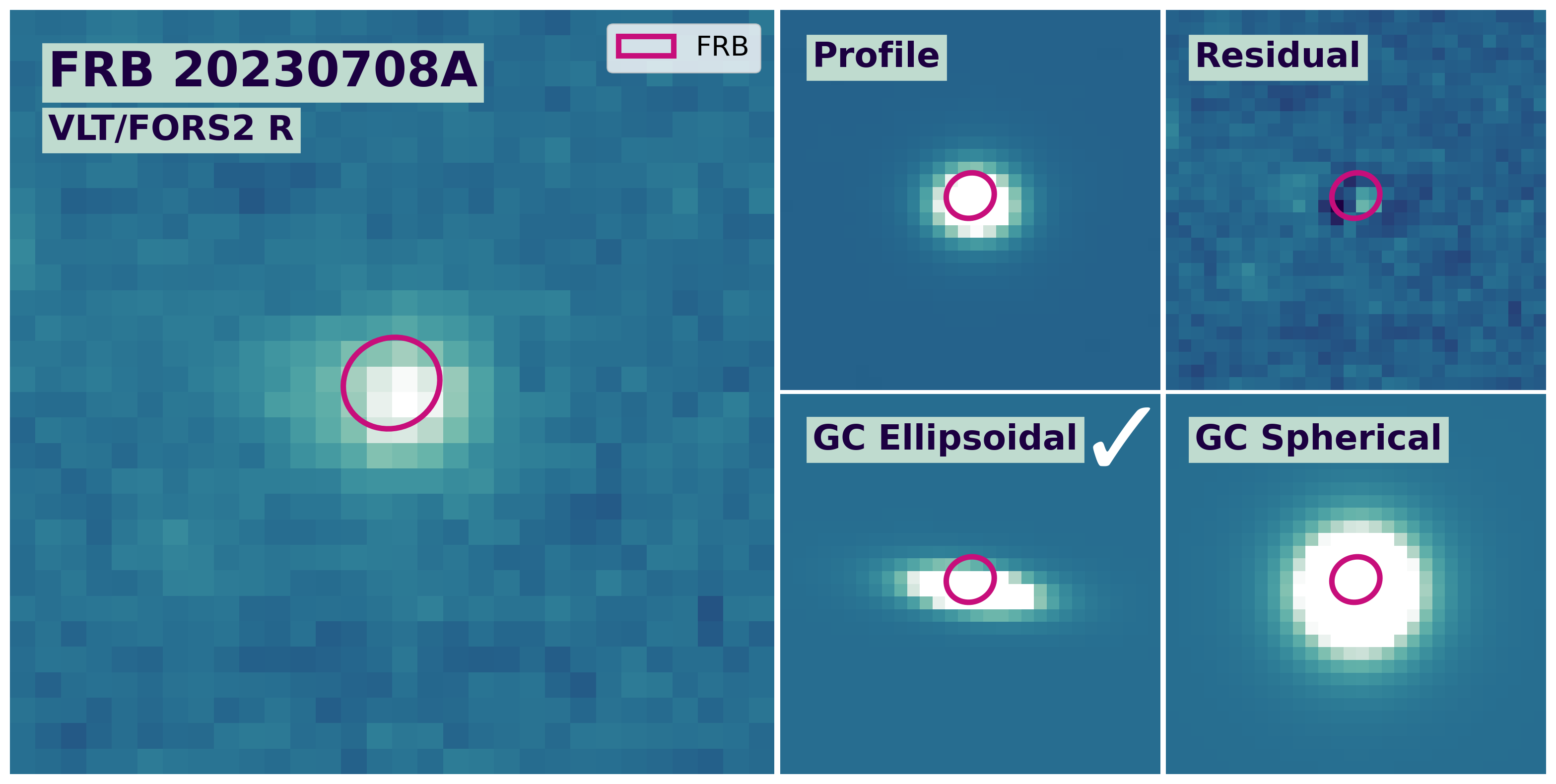}
    \includegraphics[width=0.39\textwidth]{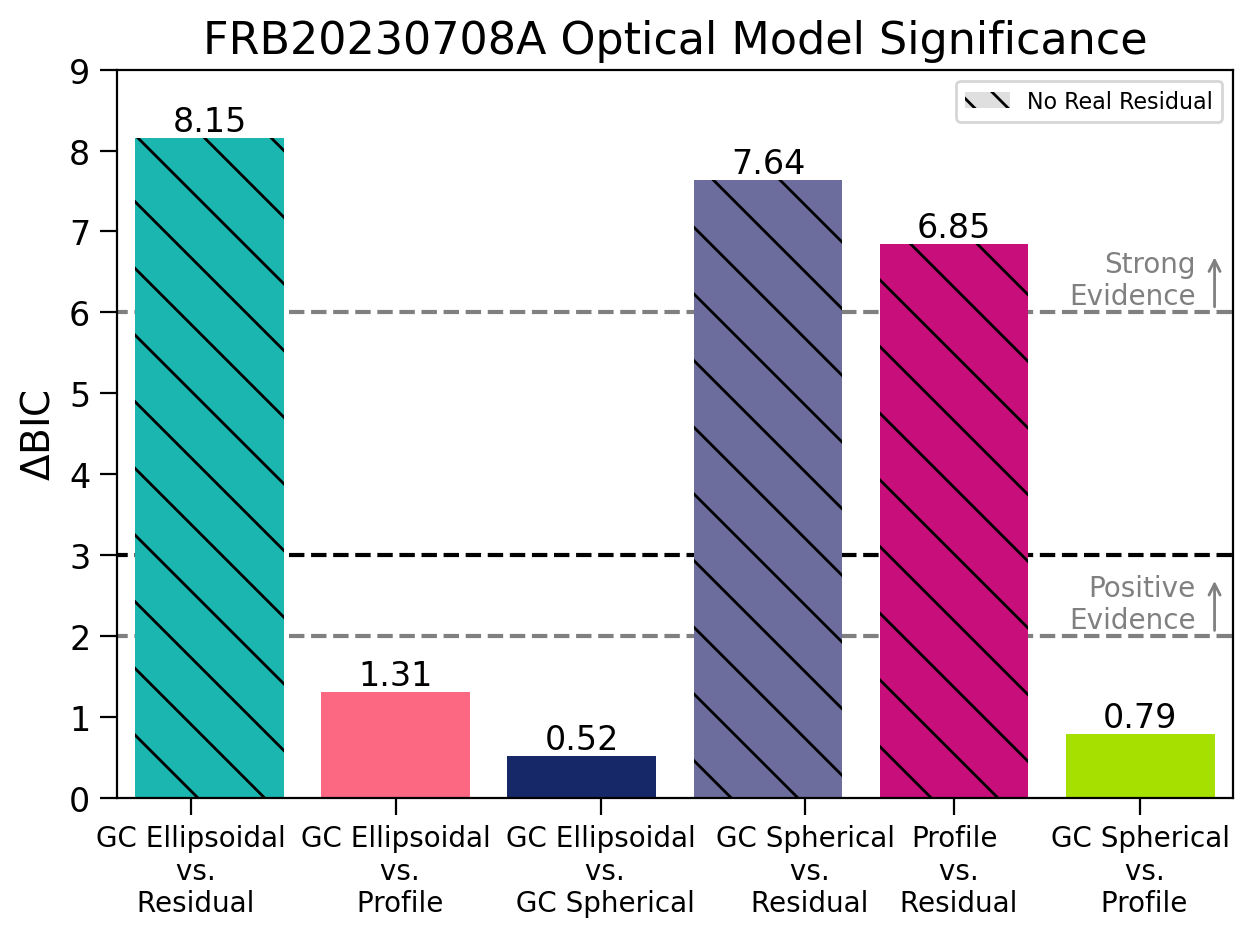}
    \caption{\textbf{(Left:)} Imaging and light profiles for FRB\,20230708A in the optical. \textbf{(Right:)} Comparison of the $\Delta$BIC between the four models for FRB\,20230708A in the optical.}
    \label{fig:appendix_230708_opt}
\end{figure}

\begin{figure}
    \centering
    \includegraphics[width=0.6\textwidth]{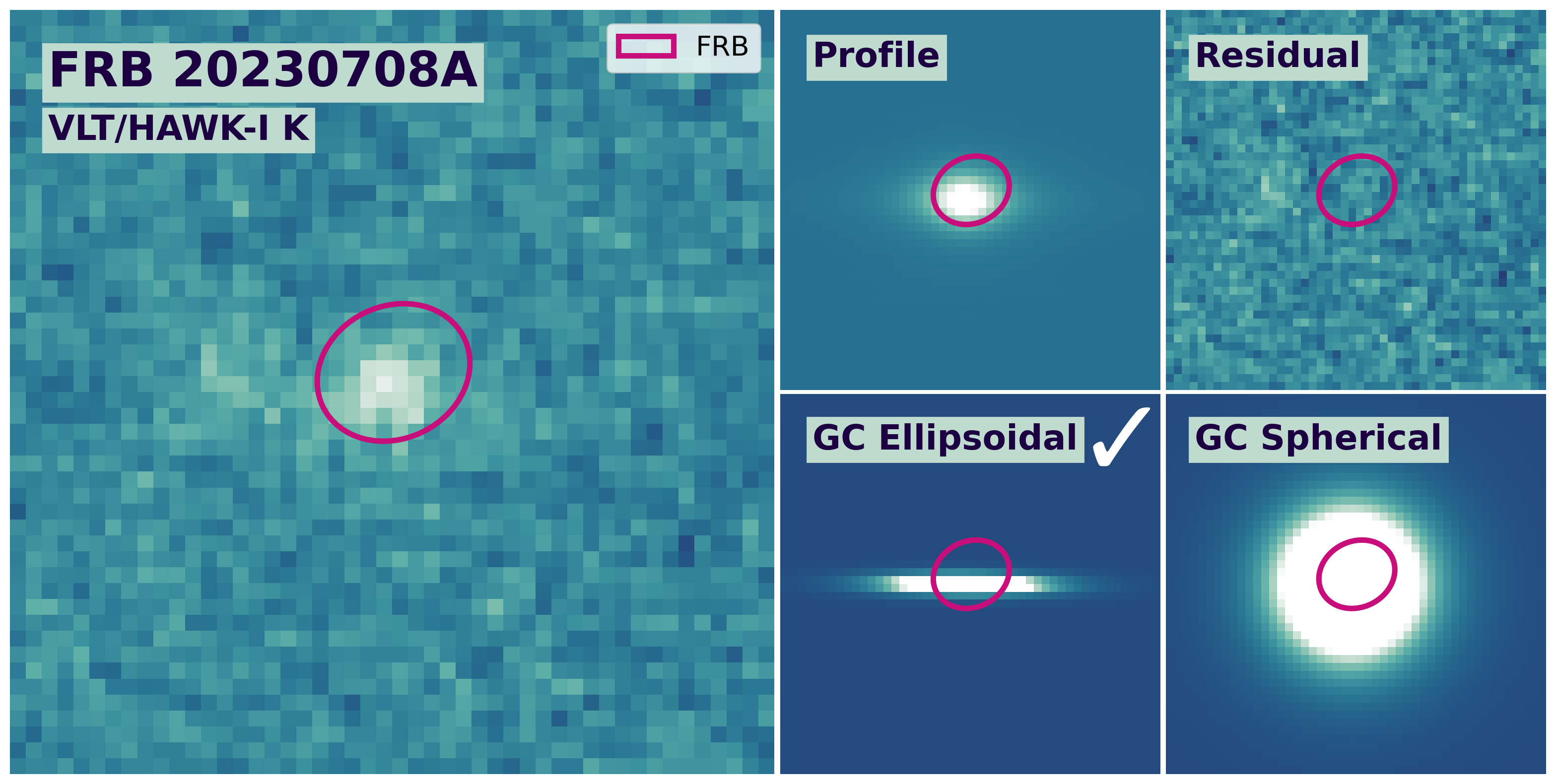}
    \includegraphics[width=0.39\textwidth]{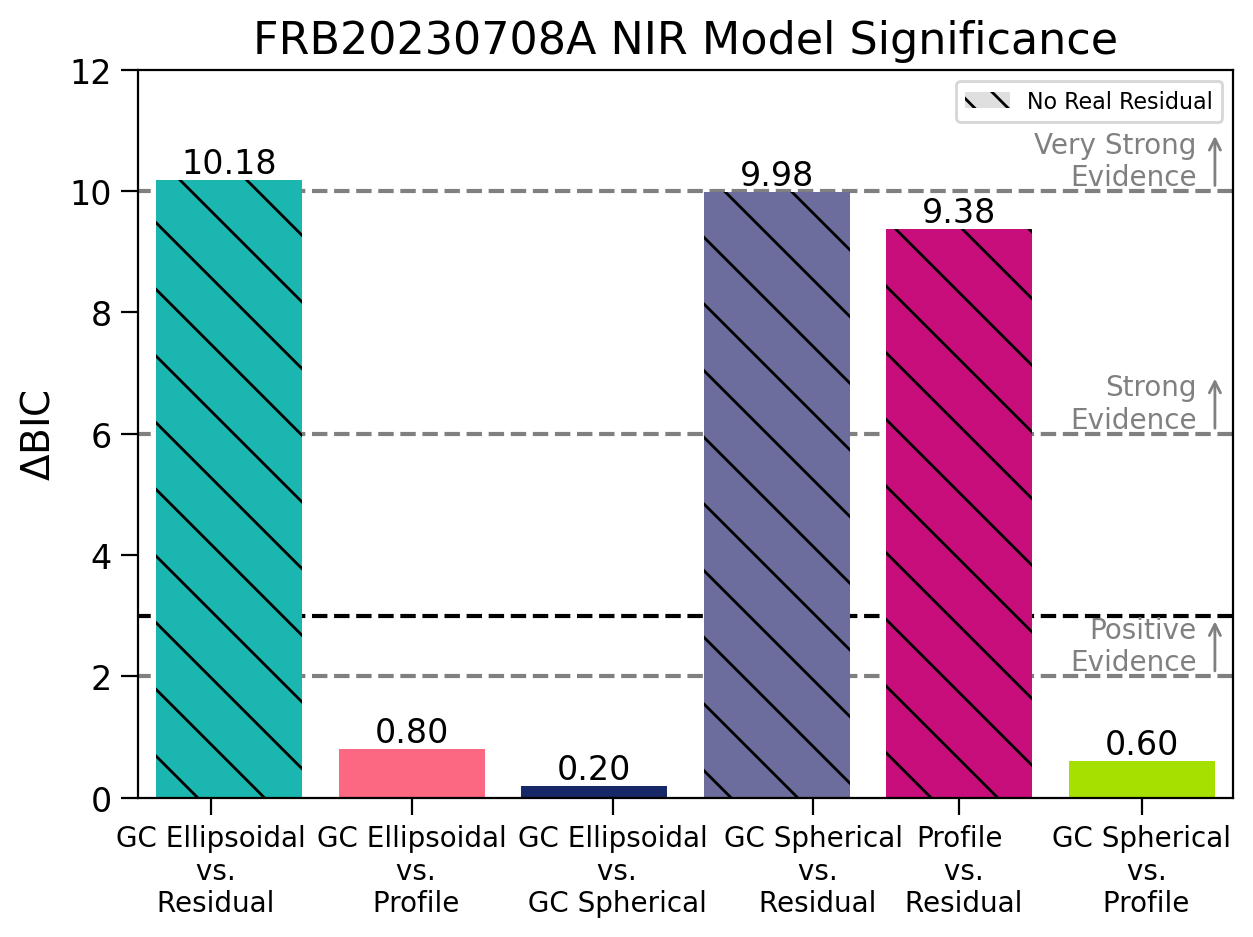}
    \caption{\textbf{(Left:)} Imaging and light profiles for FRB\,20230708A in the NIR. \textbf{(Right:)} Comparison of the $\Delta$BIC between the four models for FRB\,20230708A in the NIR.}
    \label{fig:appendix_230708_NIR}
\end{figure}

\begin{figure}
    \centering
    \includegraphics[width=0.6\textwidth]{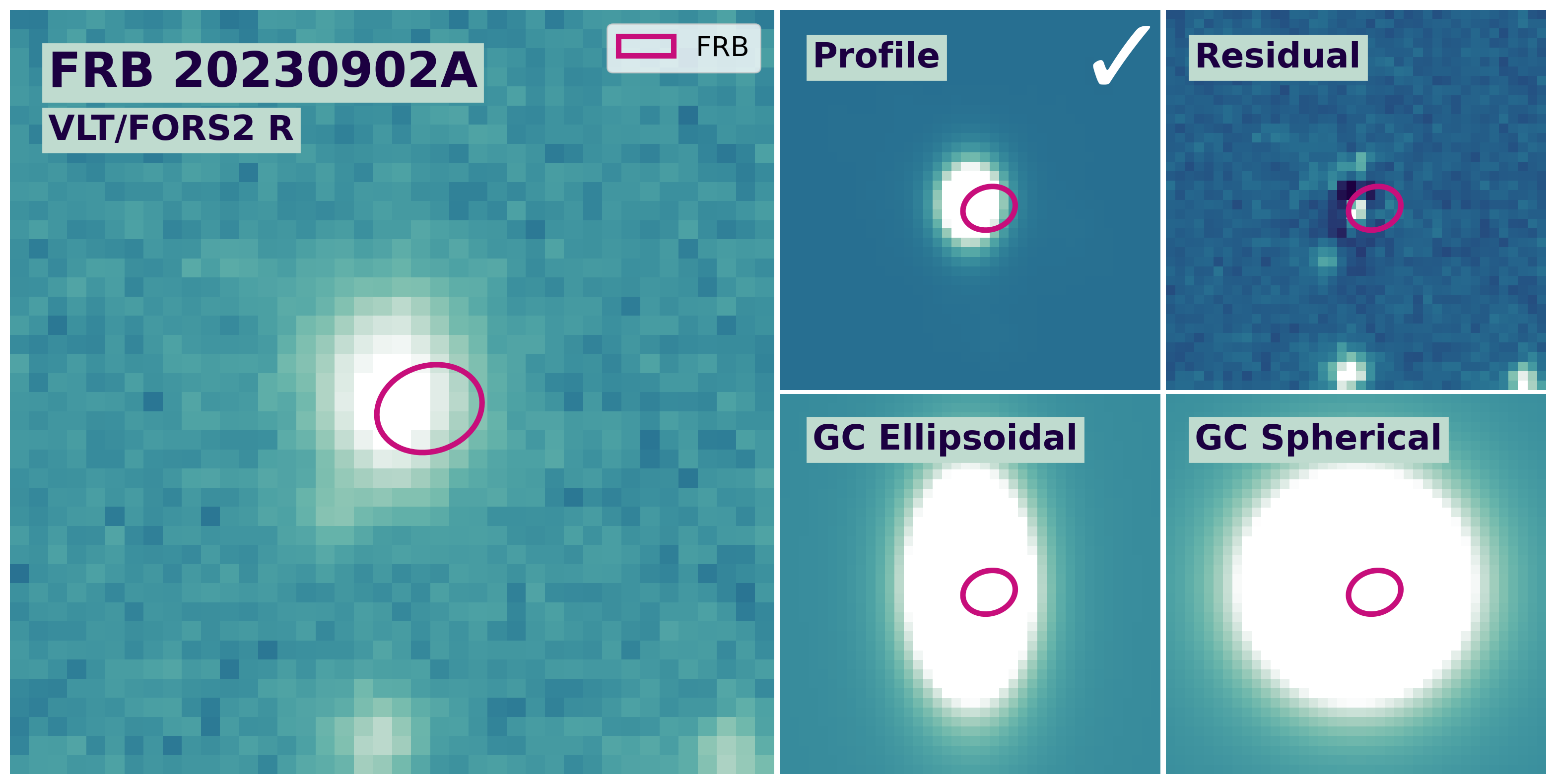}
    \includegraphics[width=0.39\textwidth]{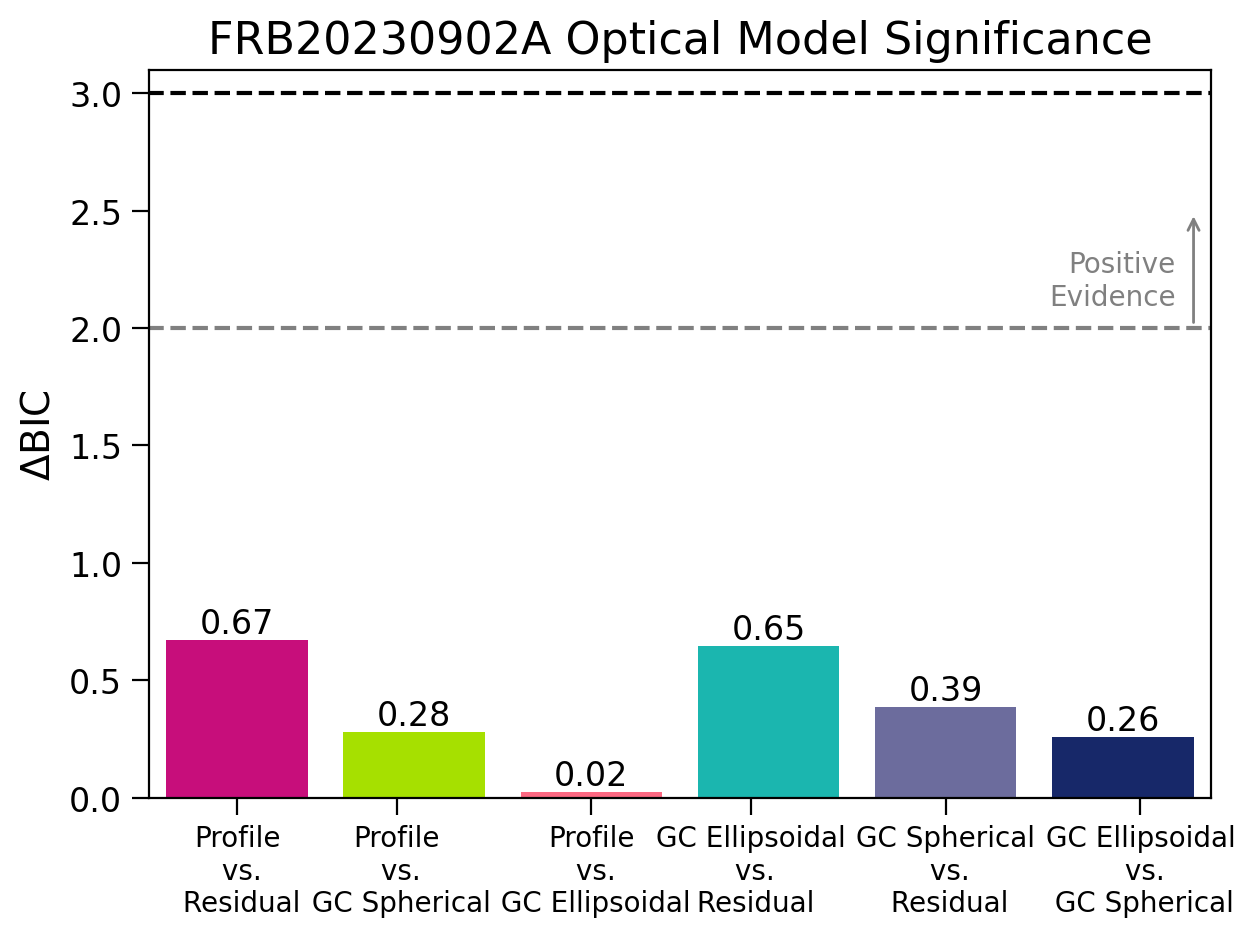}
    \caption{\textbf{(Left:)} Imaging and light profiles for FRB\,20230902A in the optical. \textbf{(Right:)} Comparison of the $\Delta$BIC between the four models for FRB\,20230902A in the optical.}
    \label{fig:appendix_230902_opt}
\end{figure}

\begin{figure}
    \centering
    \includegraphics[width=0.6\textwidth]{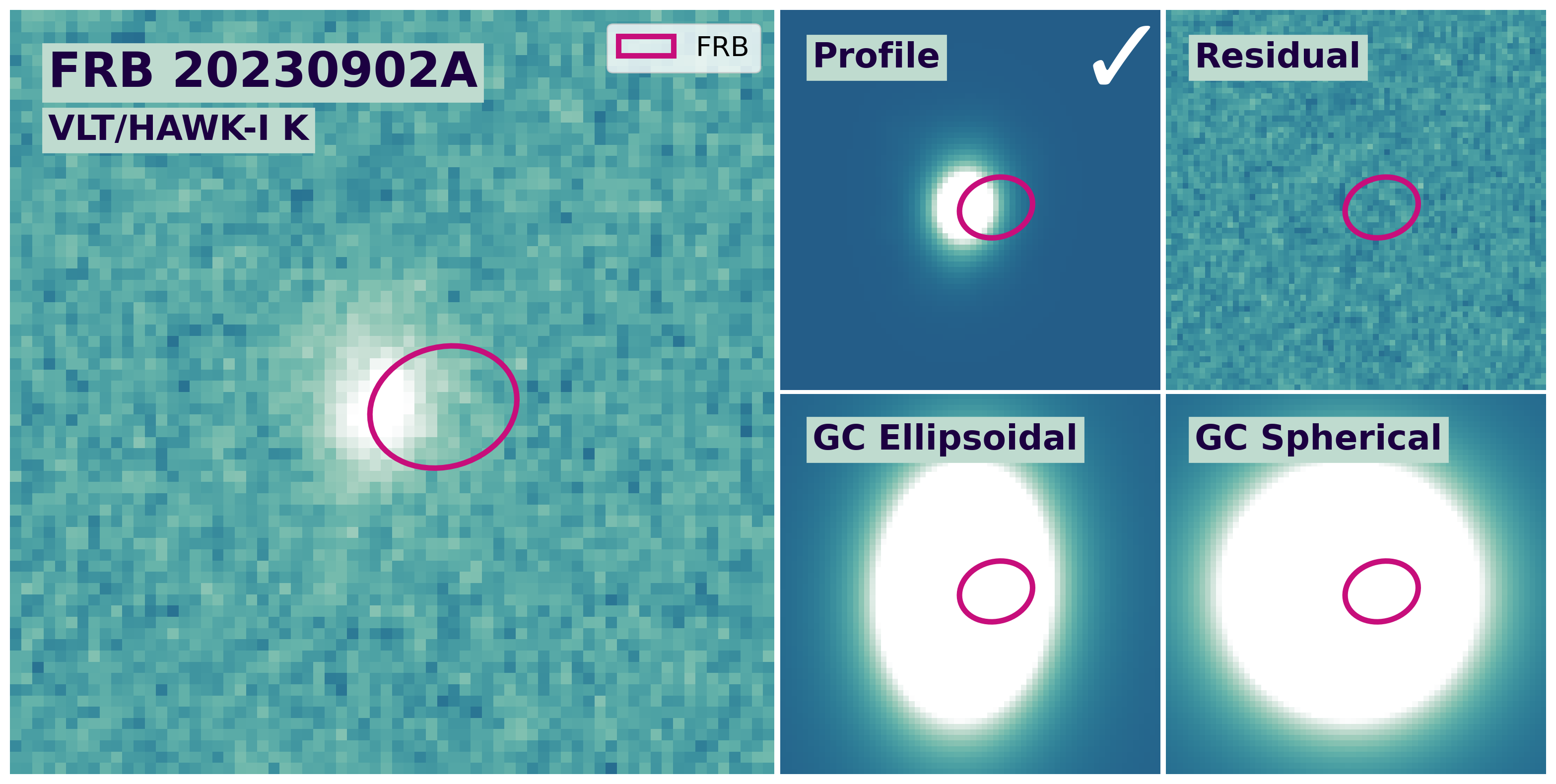}
    \includegraphics[width=0.39\textwidth]{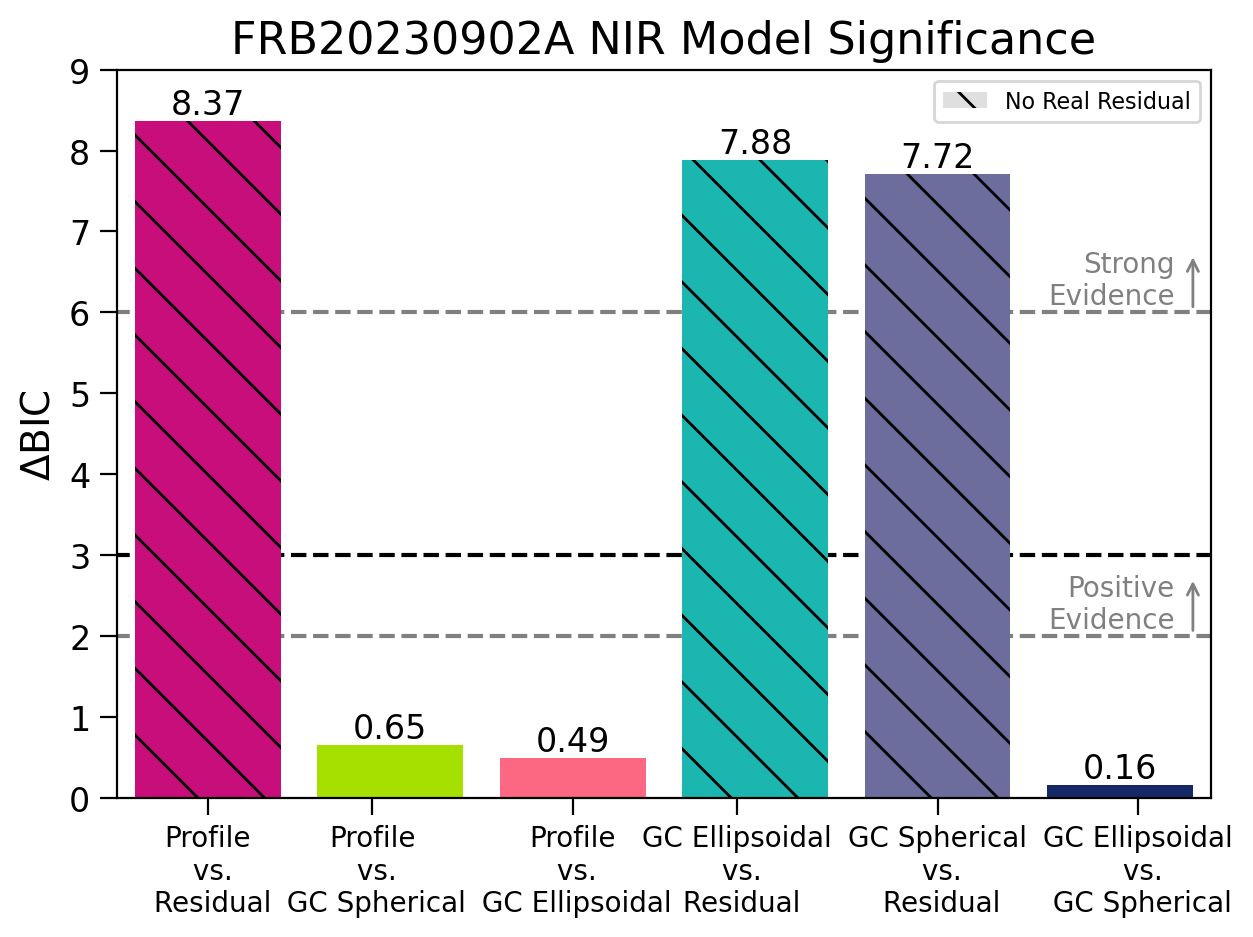}
    \caption{\textbf{(Left:)} Imaging and light profiles for FRB\,20230902A in the NIR. \textbf{(Right:)} Comparison of the $\Delta$BIC between the four models for FRB\,20230902A in the NIR.}
    \label{fig:appendix_230902_NIR}
\end{figure}

\begin{figure}
    \centering
    \includegraphics[width=0.6\textwidth]{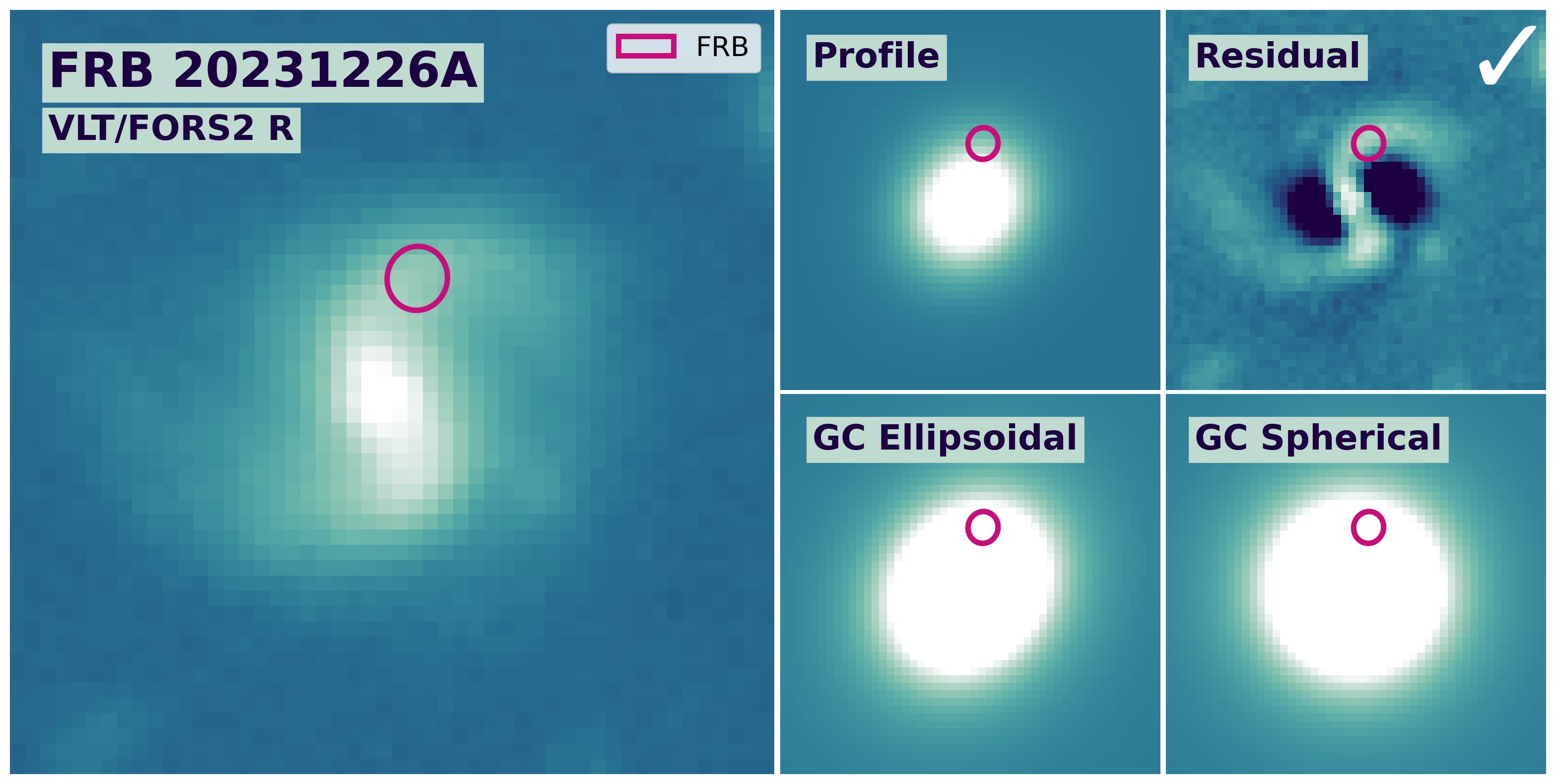}
    \includegraphics[width=0.39\textwidth]{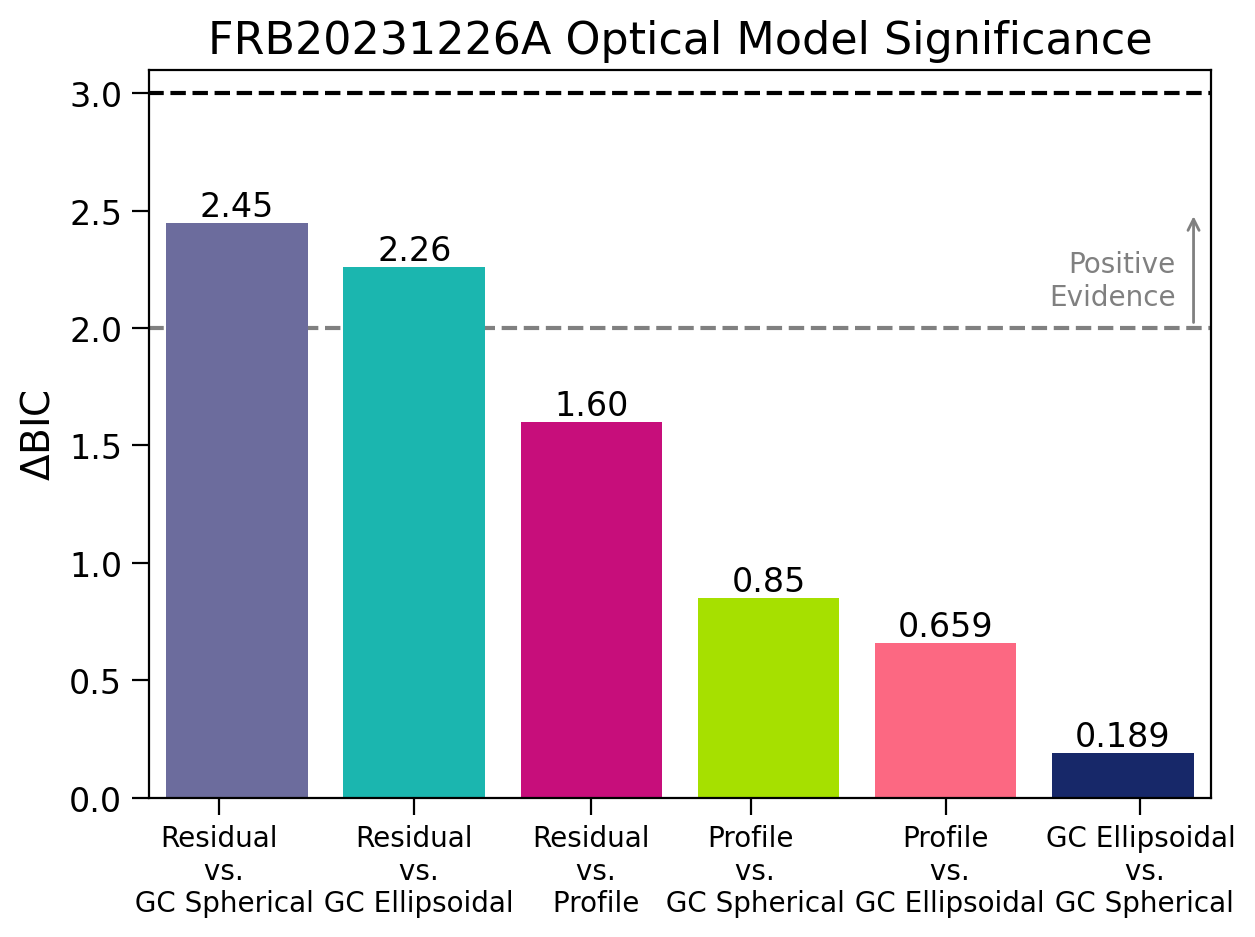}
    \caption{\textbf{(Left:)} Imaging and light profiles for FRB\,20231226A in the optical. \textbf{(Right:)} Comparison of the $\Delta$BIC between the four models for FRB\,20231226A in the optical.}
    \label{fig:appendix_231226_opt}
\end{figure}

\begin{figure}
    \centering
    \includegraphics[width=0.6\textwidth]{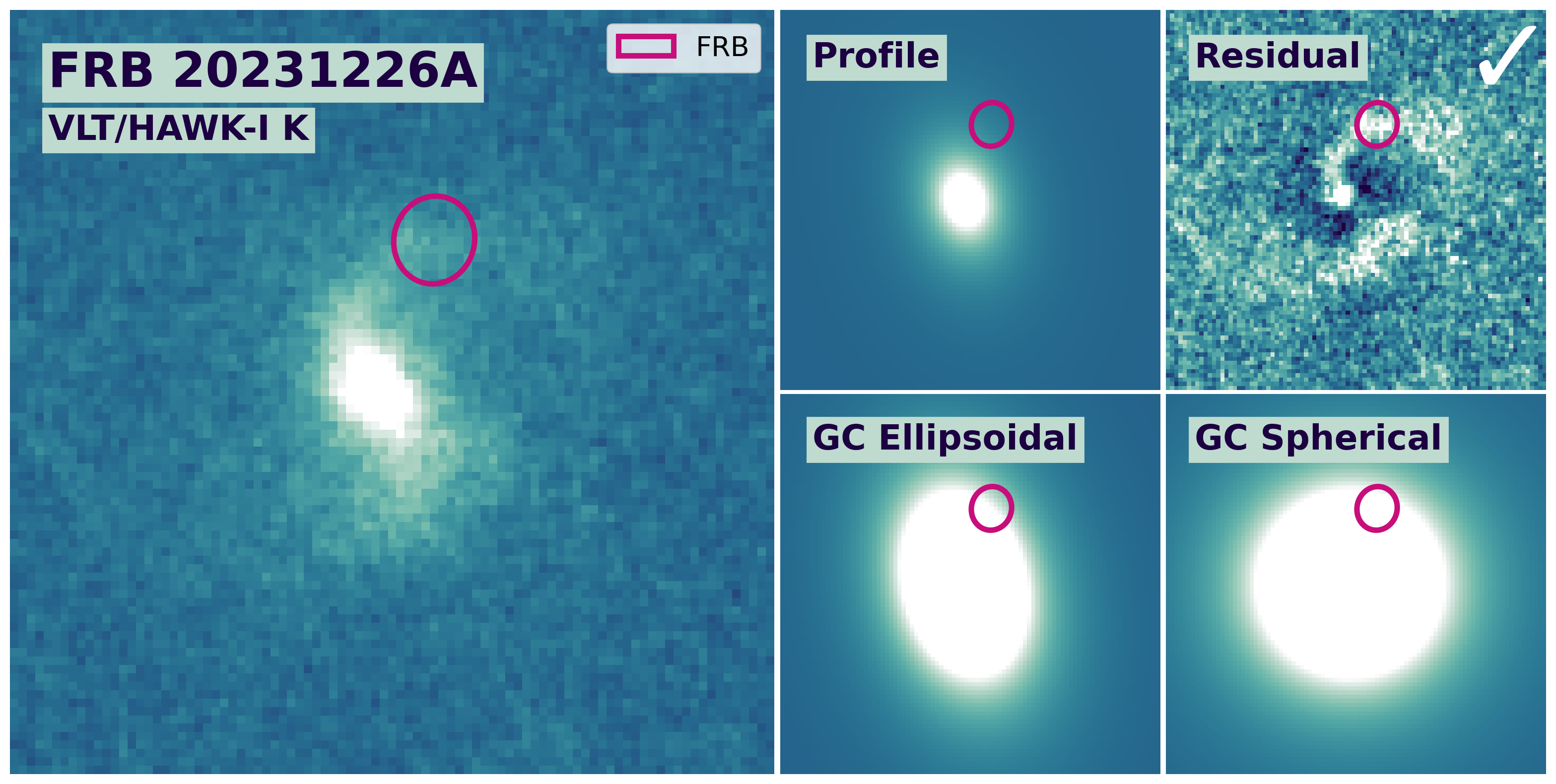}
    \includegraphics[width=0.39\textwidth]{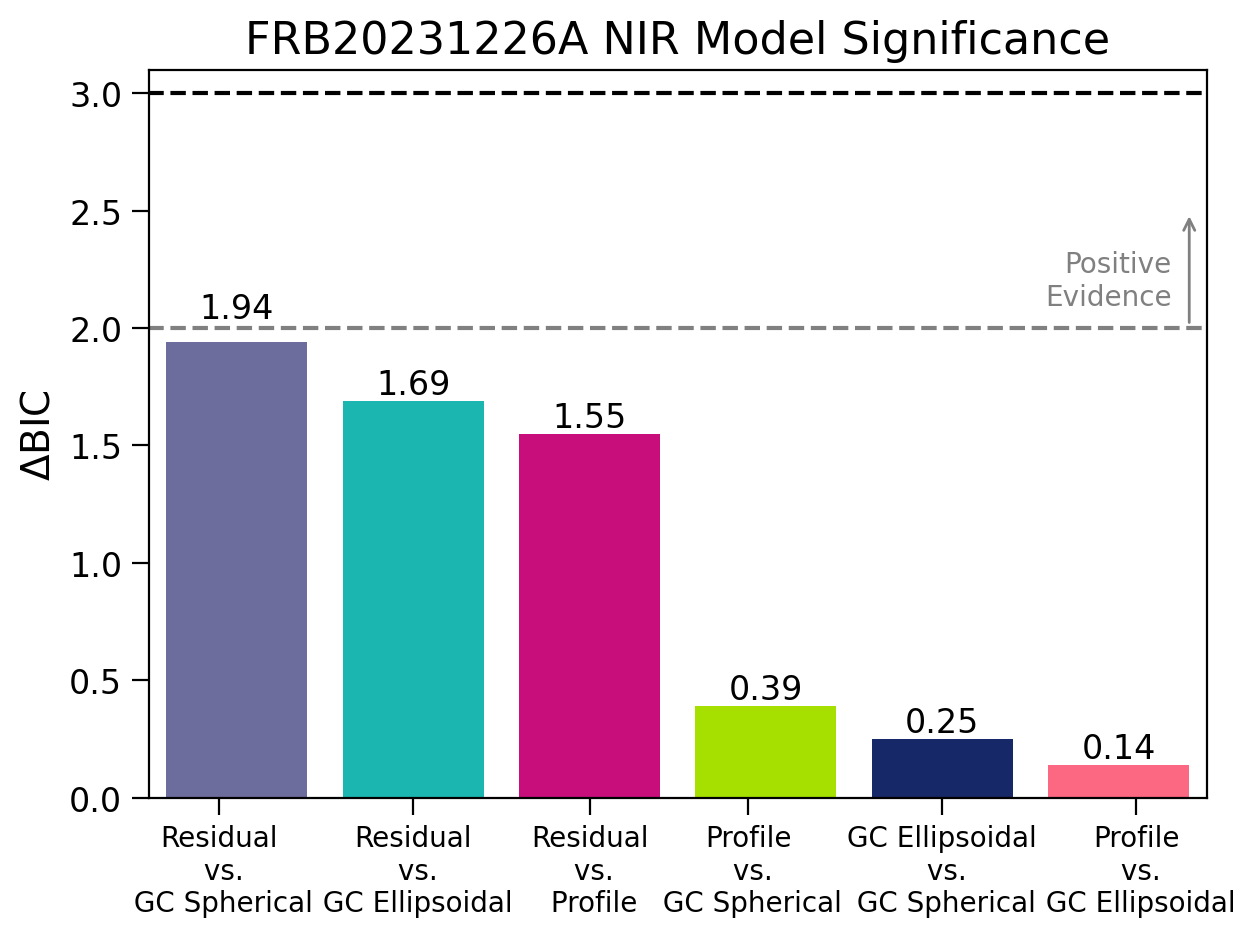}
    \caption{\textbf{(Left:)} Imaging and light profiles for FRB\,20231226A in the NIR. \textbf{(Right:)} Comparison of the $\Delta$BIC between the four models for FRB\,20231226A in the NIR.}
    \label{fig:appendix_231226_NIR}
\end{figure}

\begin{figure}
    \centering
    \includegraphics[width=0.6\textwidth]{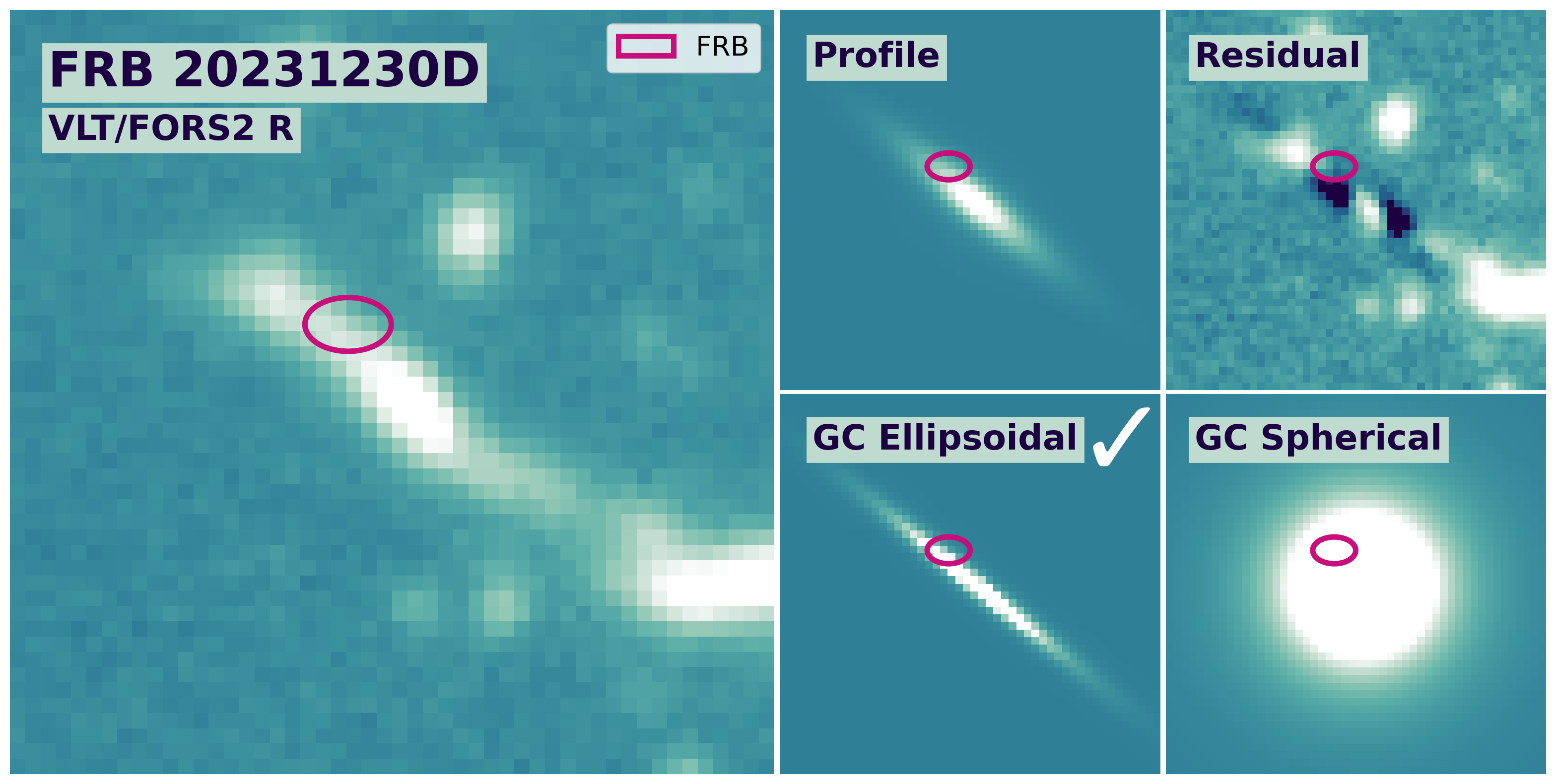}
    \includegraphics[width=0.39\textwidth]{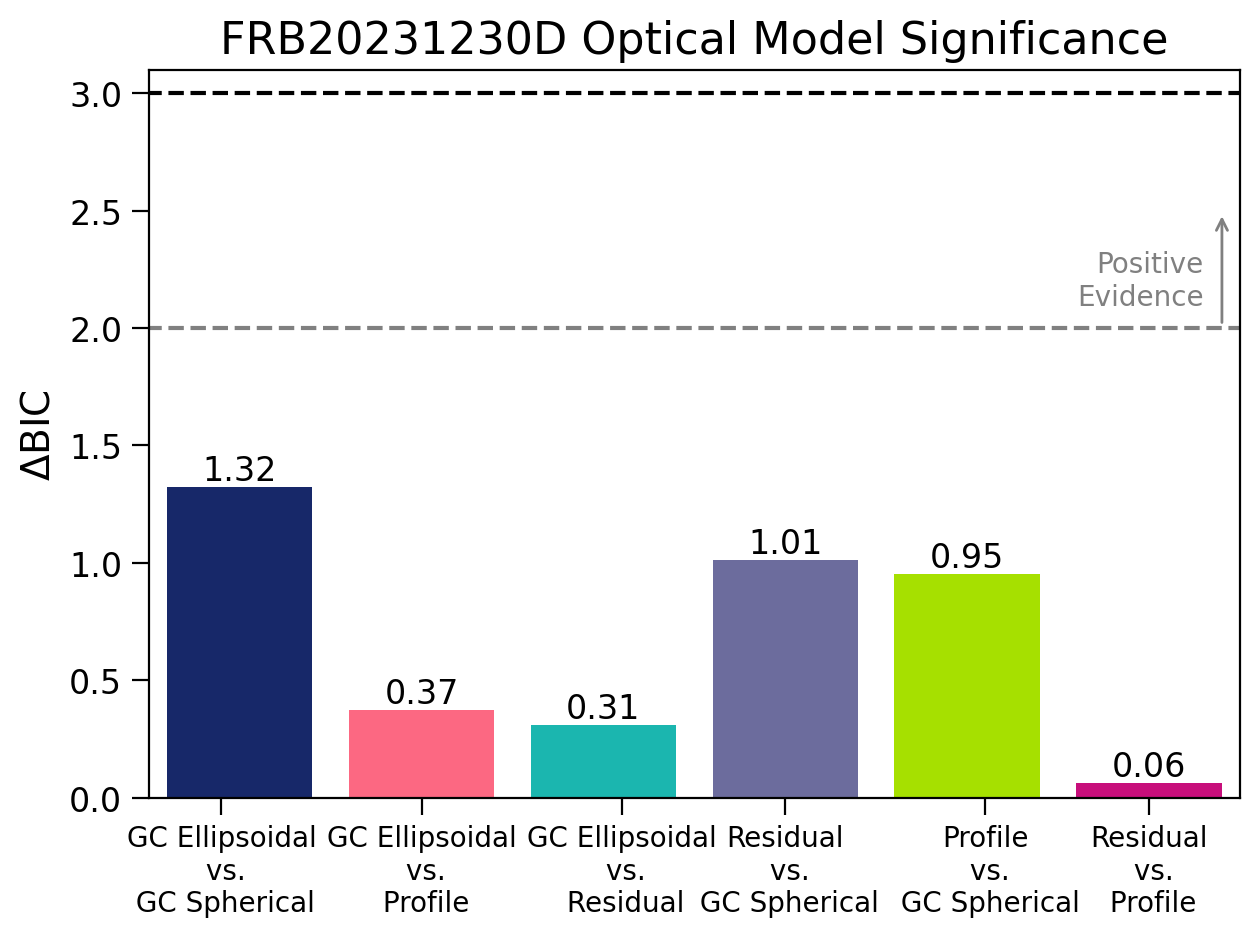}
    \caption{\textbf{(Left:)} Imaging and light profiles for FRB\,20231230D in the optical. \textbf{(Right:)} Comparison of the $\Delta$BIC between the four models for FRB\,20231230D in the optical.}
    \label{fig:appendix_231230_opt}
\end{figure}

\begin{figure}
    \centering
    \includegraphics[width=0.6\textwidth]{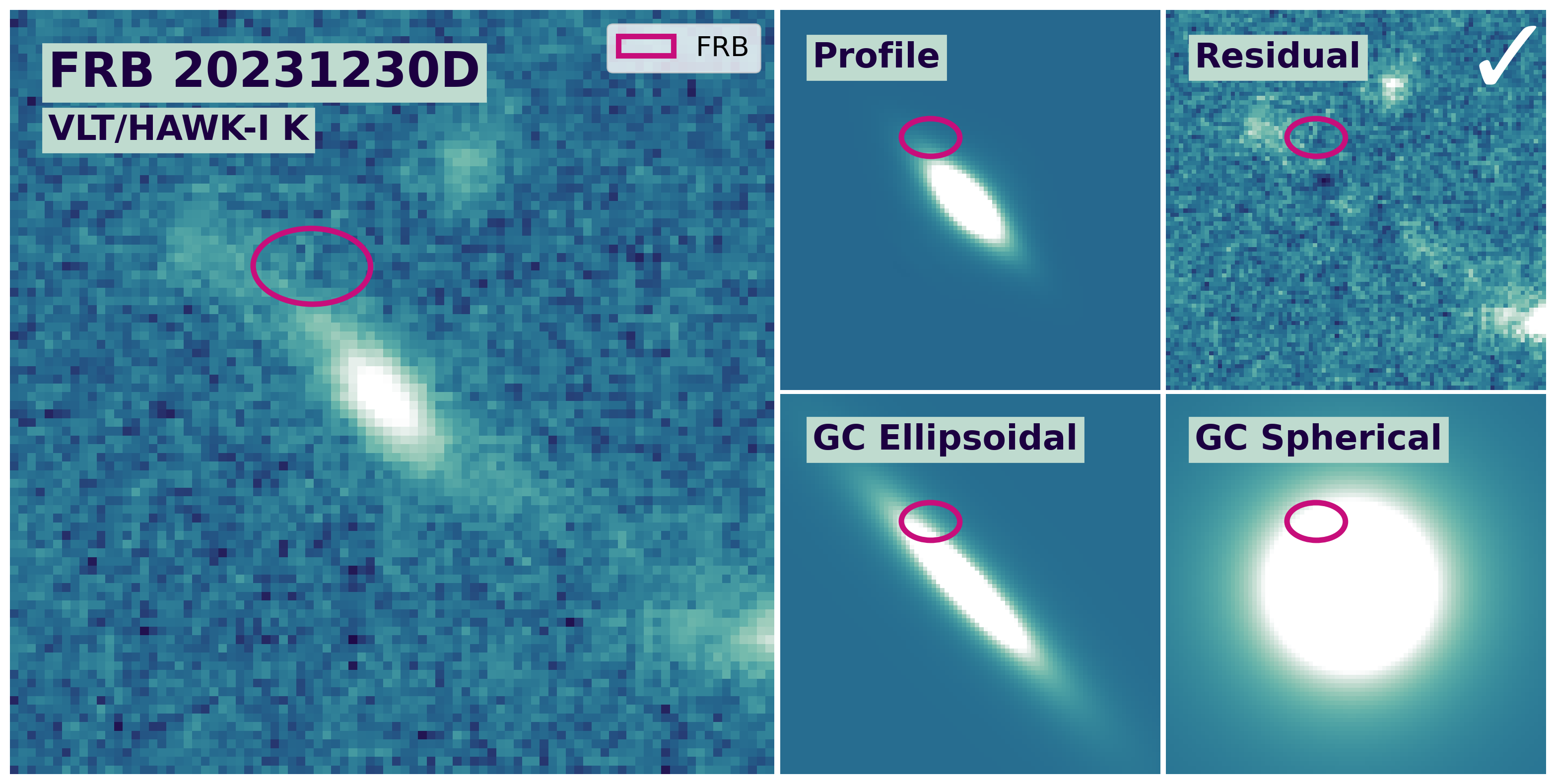}
    \includegraphics[width=0.39\textwidth]{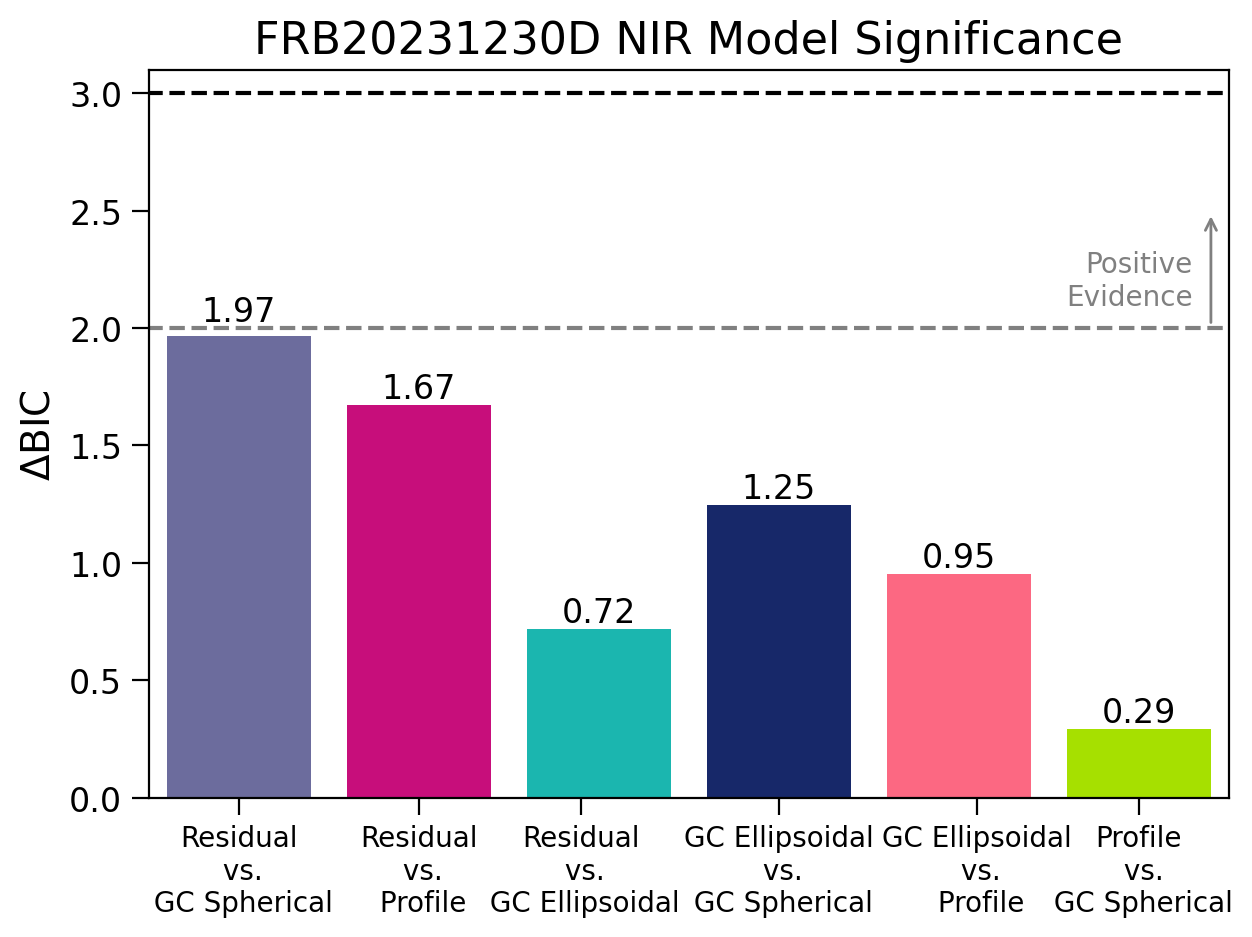}
    \caption{\textbf{(Left:)} Imaging and light profiles for FRB\,20231230D in the NIR. \textbf{(Right:)} Comparison of the $\Delta$BIC between the four models for FRB\,20231230D in the NIR.}
    \label{fig:appendix_231230_NIR}
\end{figure}

\begin{figure}
    \centering
    \includegraphics[width=0.6\textwidth]{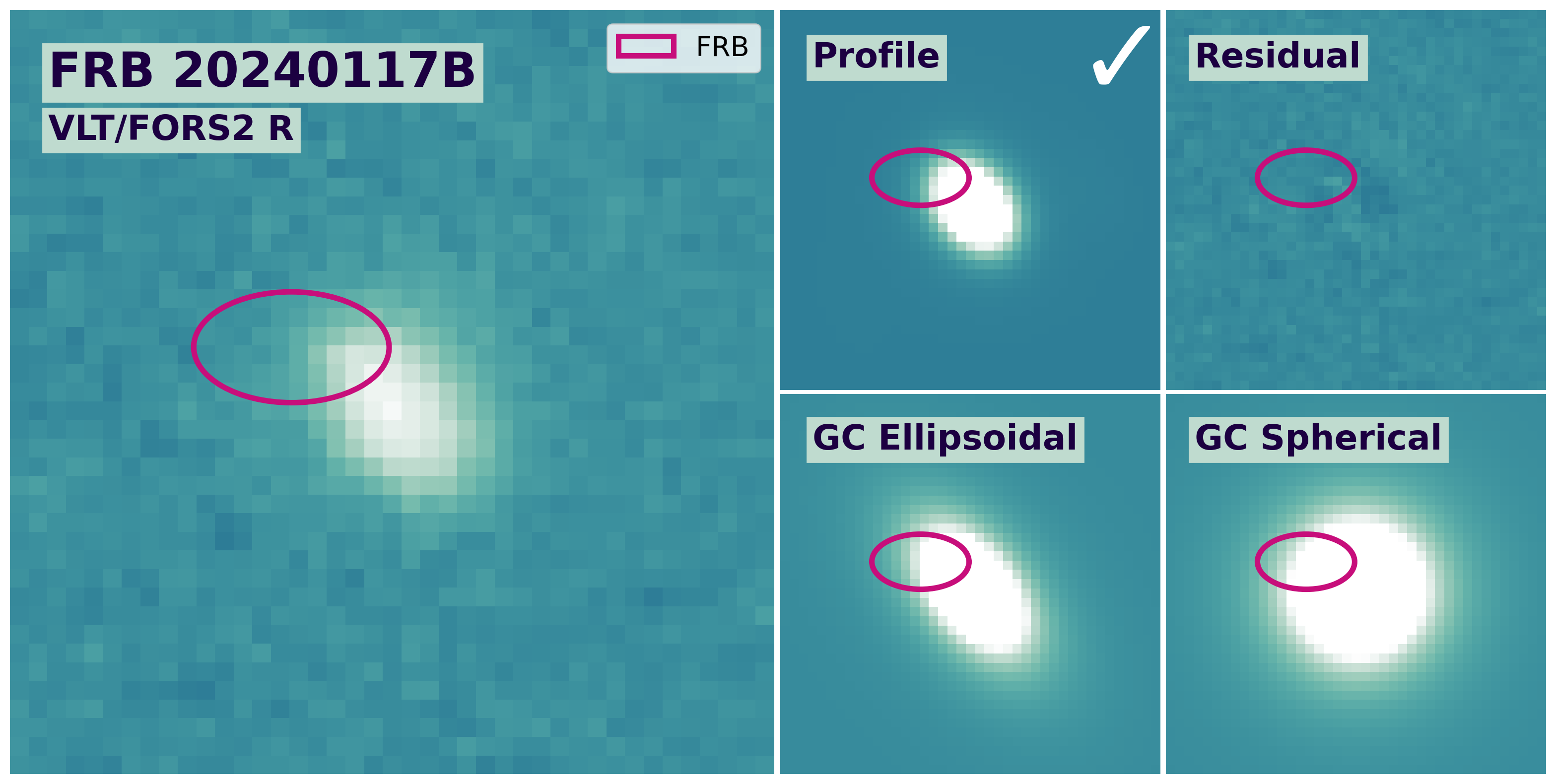}
    \includegraphics[width=0.39\textwidth]{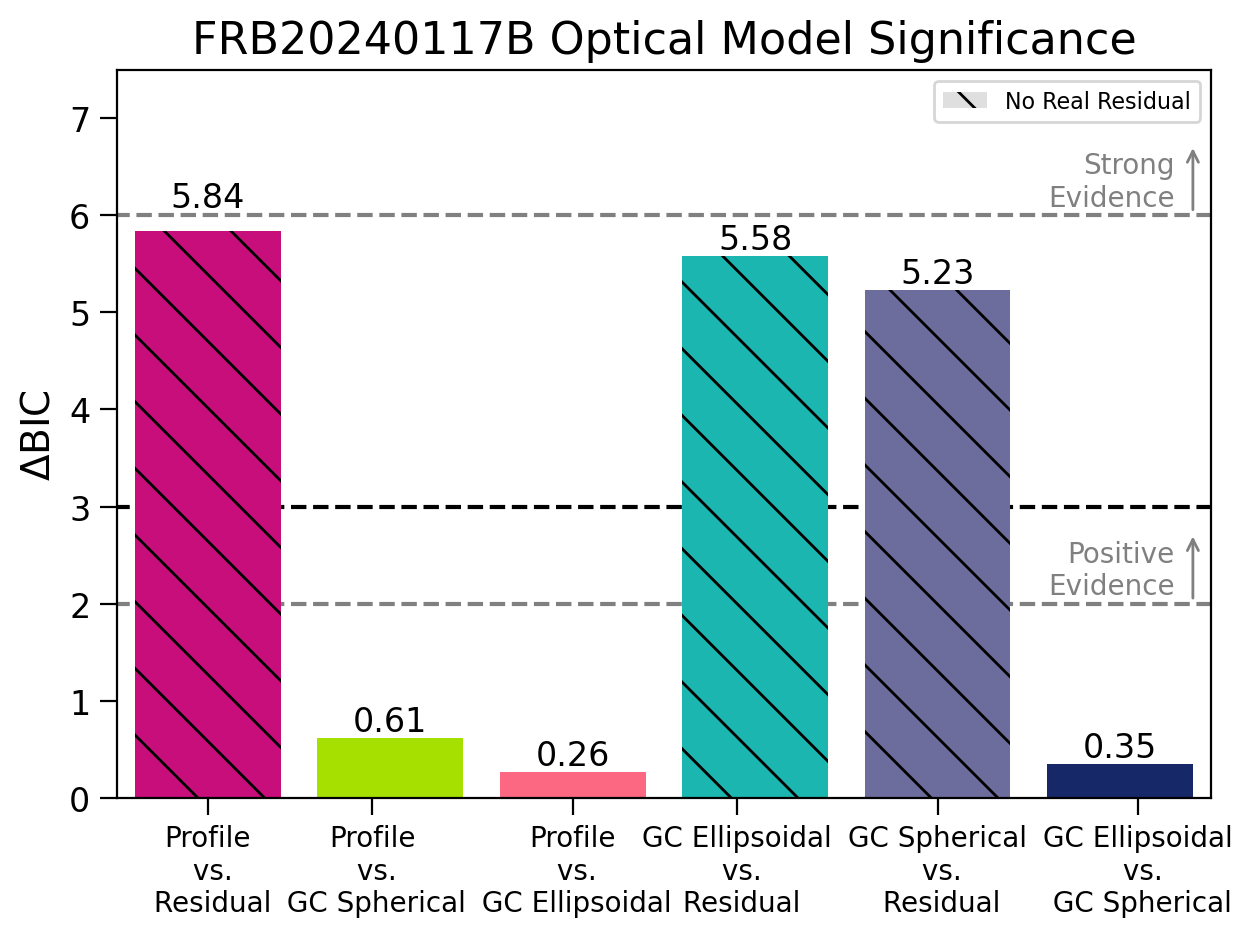}
    \caption{\textbf{(Left:)} Imaging and light profiles for FRB\,20240117B in the optical. \textbf{(Right:)} Comparison of the $\Delta$BIC between the four models for FRB\,20240117B in the optical.}
    \label{fig:appendix_240117_opt}
\end{figure}

\begin{figure}
    \centering
    \includegraphics[width=0.6\textwidth]{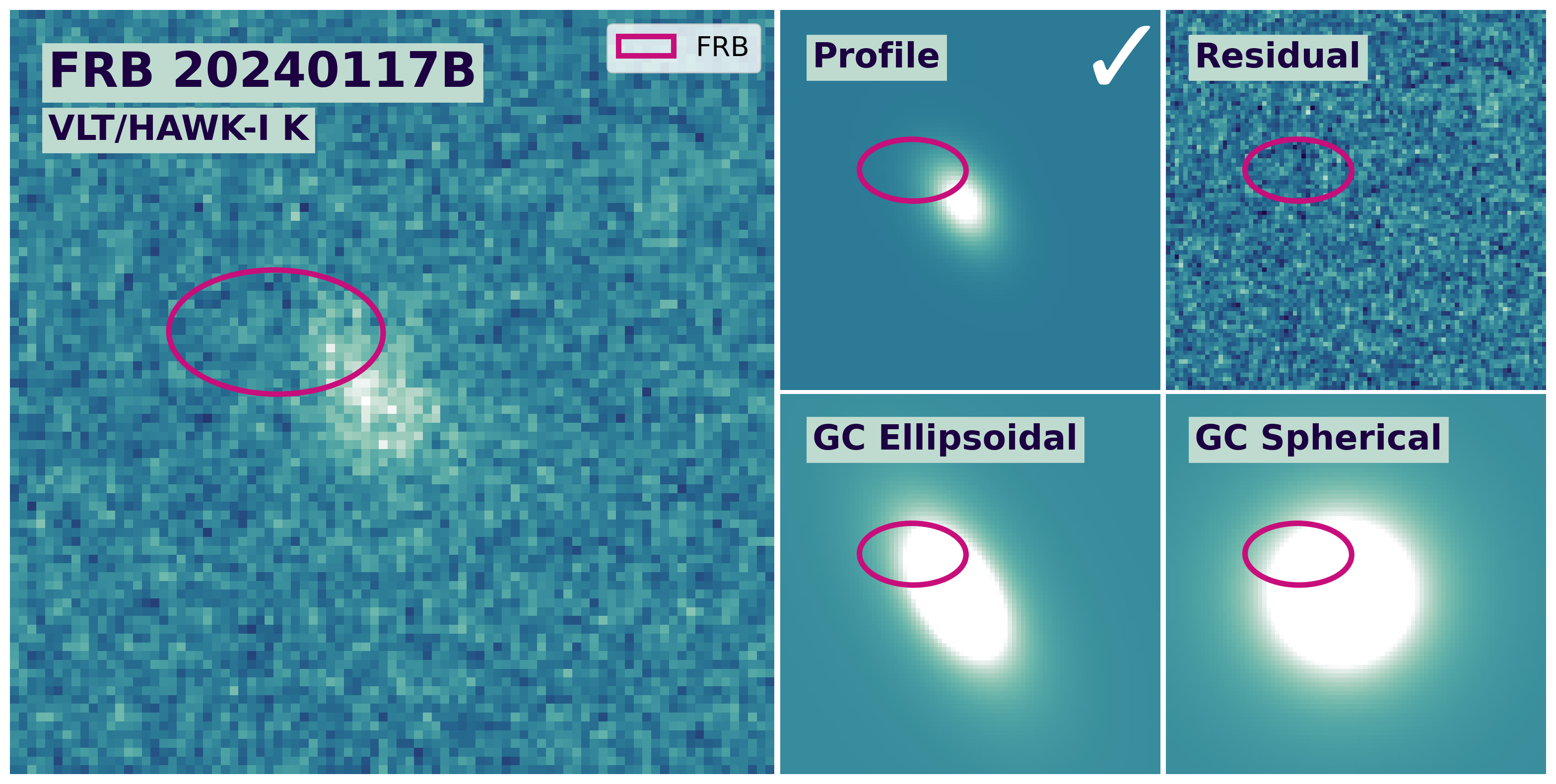}
    \includegraphics[width=0.39\textwidth]{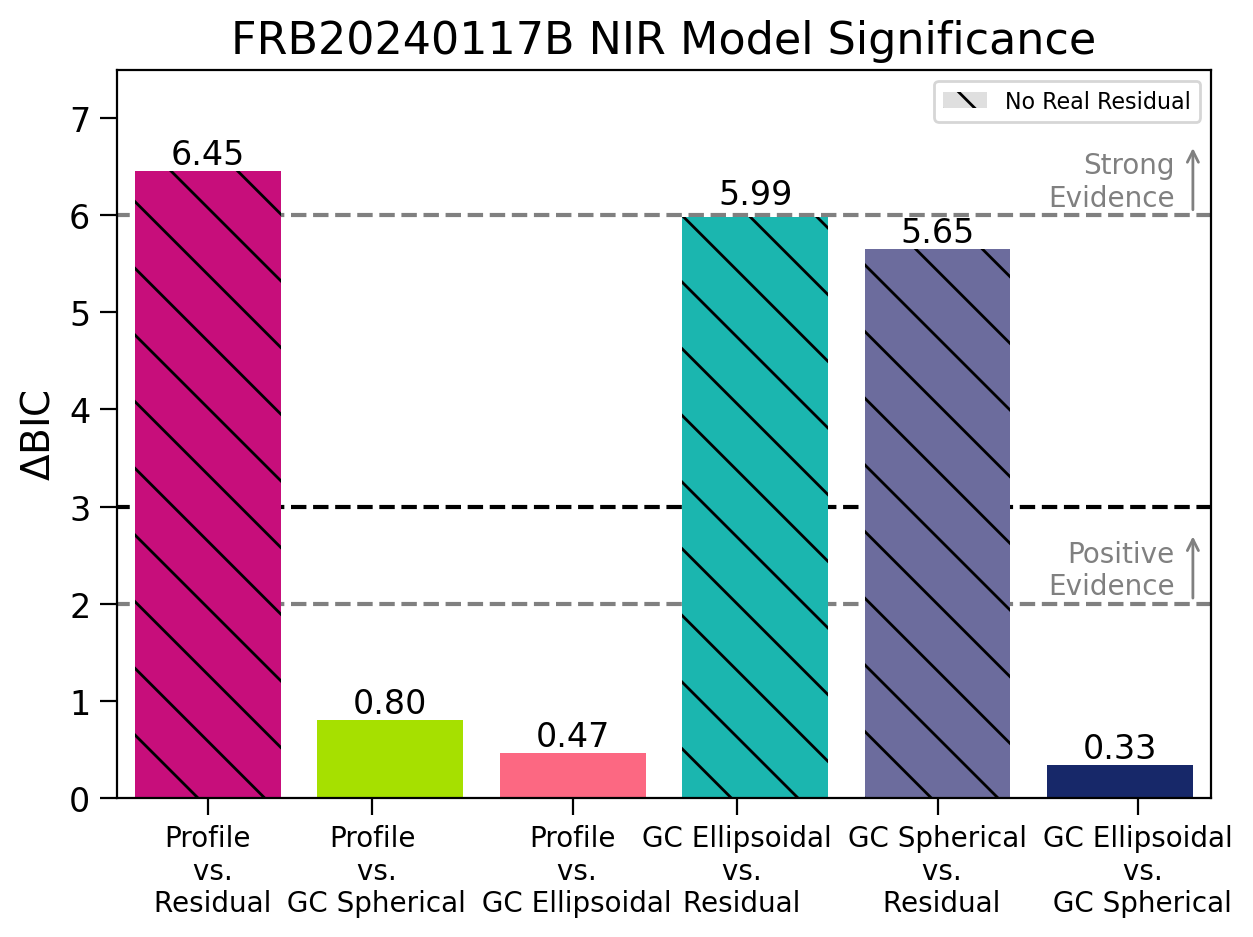}
    \caption{\textbf{(Left:)} Imaging and light profiles for FRB\,20240117B in the NIR. \textbf{(Right:)} Comparison of the $\Delta$BIC between the four models for FRB\,20240117B in the NIR.}
    \label{fig:appendix_240117_NIR}
\end{figure}

\begin{figure}
    \centering
    \includegraphics[width=0.6\textwidth]{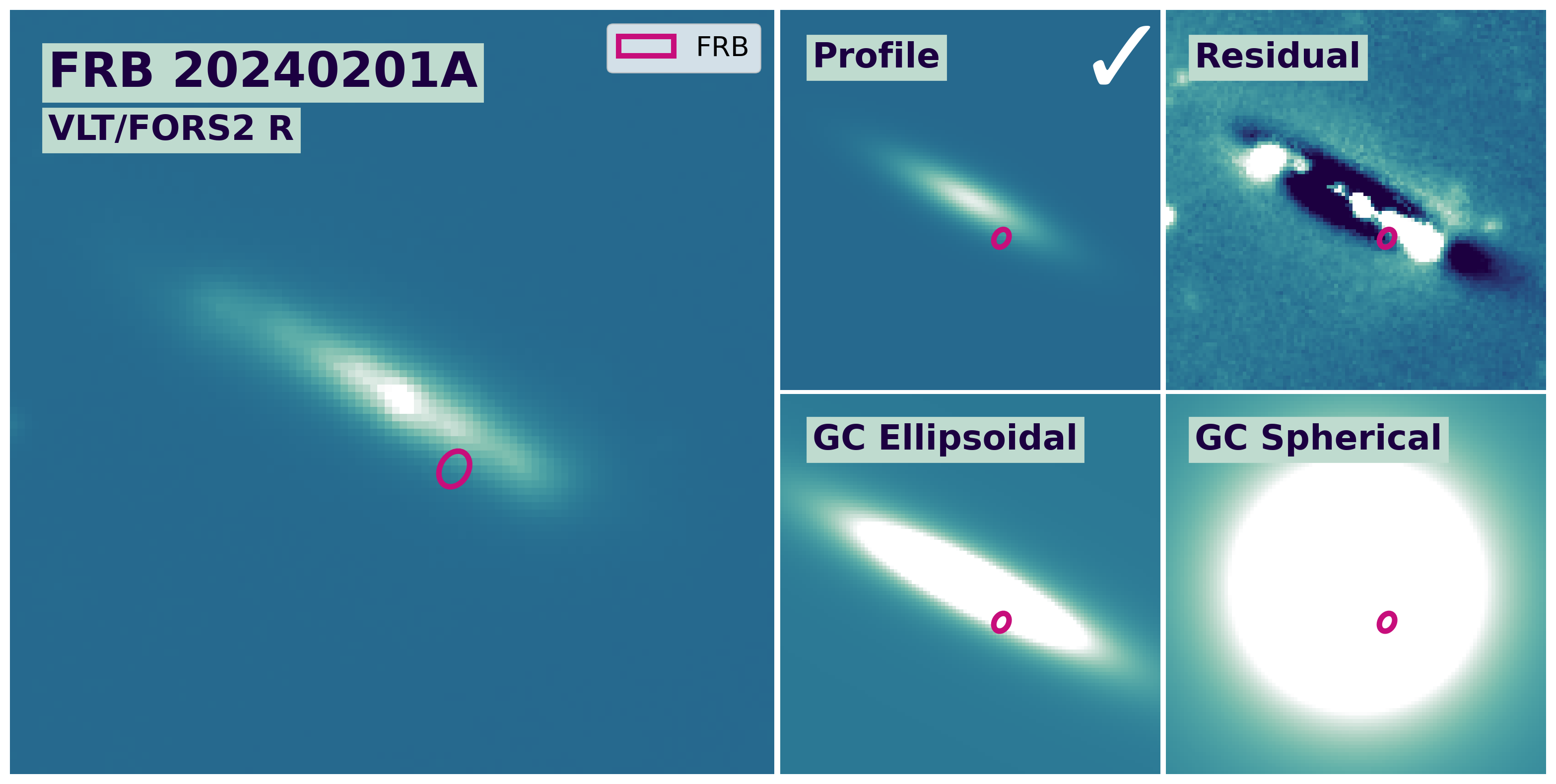}
    \includegraphics[width=0.39\textwidth]{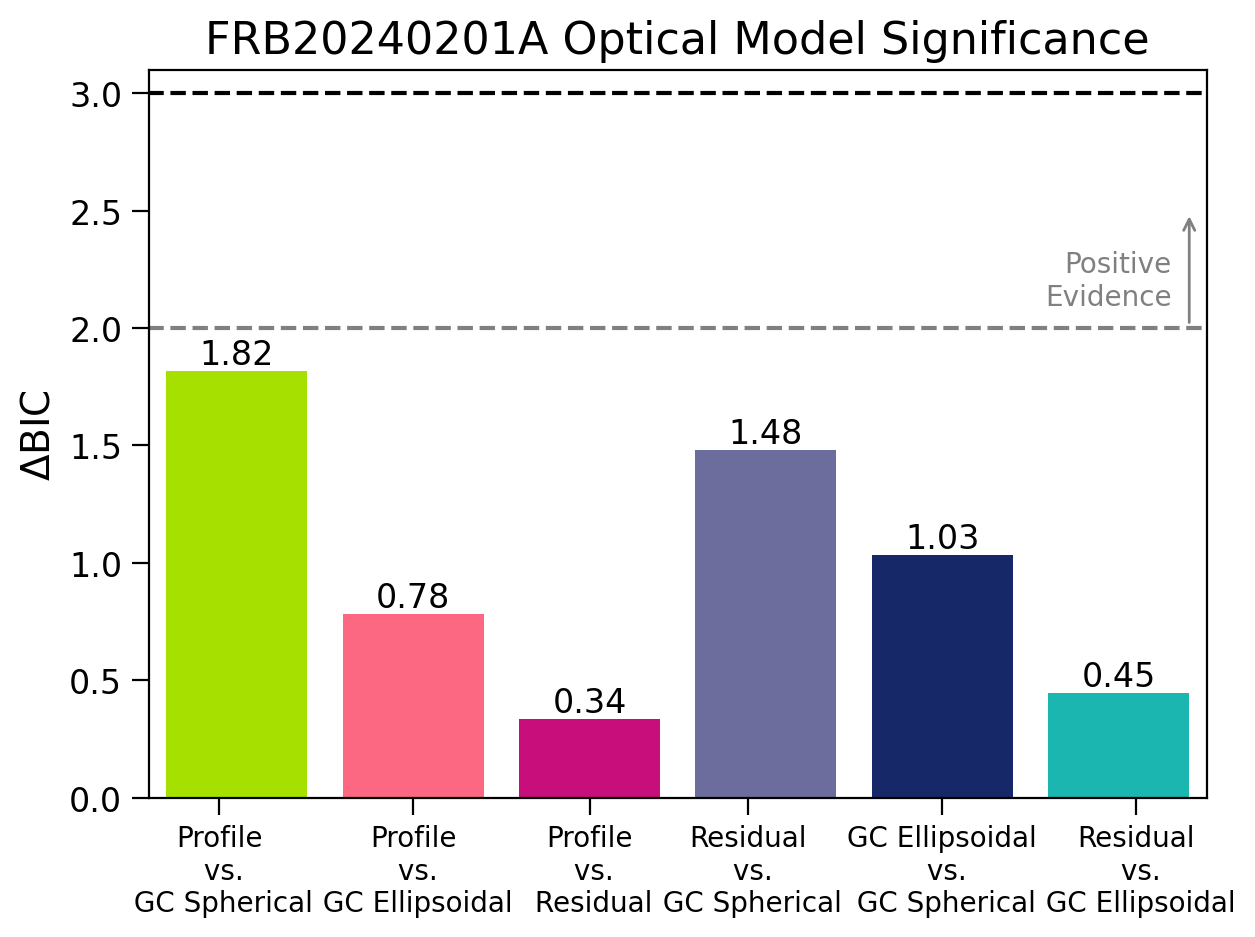}
    \caption{\textbf{(Left:)} Imaging and light profiles for FRB\,20240201A in the optical. \textbf{(Right:)} Comparison of the $\Delta$BIC between the four models for FRB\,20240201A in the optical.}
    \label{fig:appendix_240201_opt}
\end{figure}

\begin{figure}
    \centering
    \includegraphics[width=0.6\textwidth]{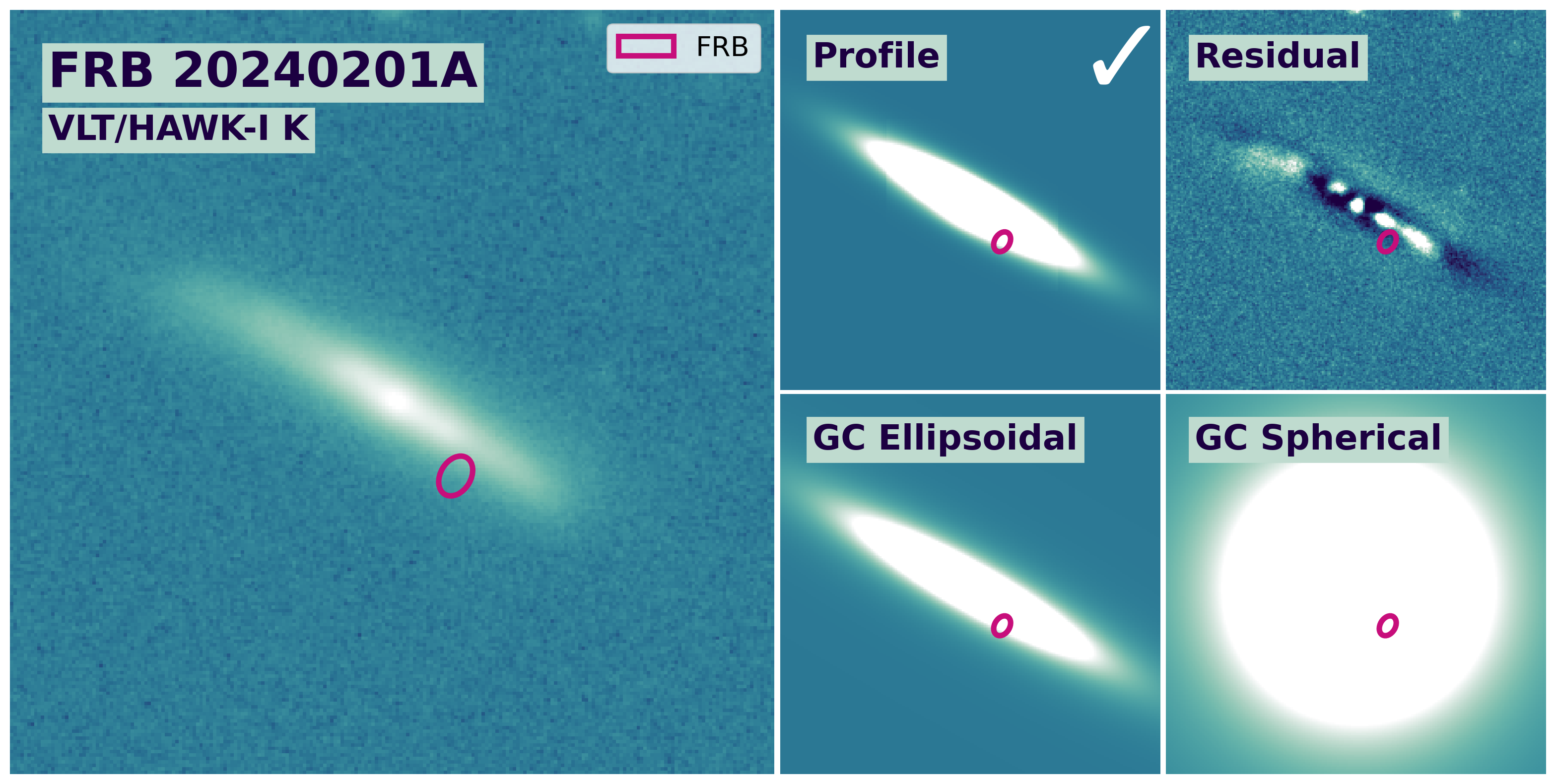}
    \includegraphics[width=0.39\textwidth]{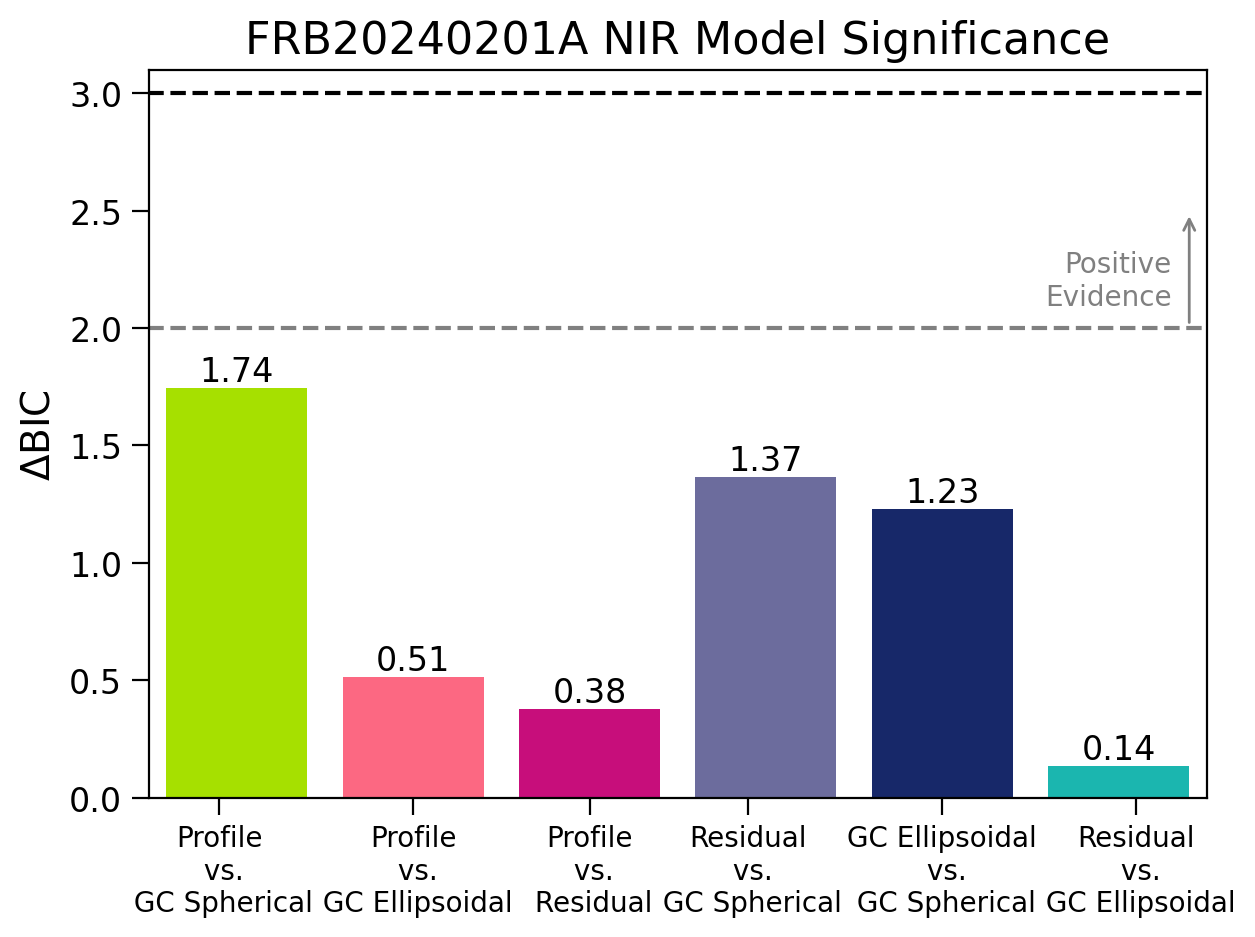}
    \caption{\textbf{(Left:)} Imaging and light profiles for FRB\,20240201A in the NIR. \textbf{(Right:)} Comparison of the $\Delta$BIC between the four models for FRB\,20240201A in the NIR.}
    \label{fig:appendix_240201_NIR}
\end{figure}

\begin{figure}
    \centering
    \includegraphics[width=0.6\textwidth]{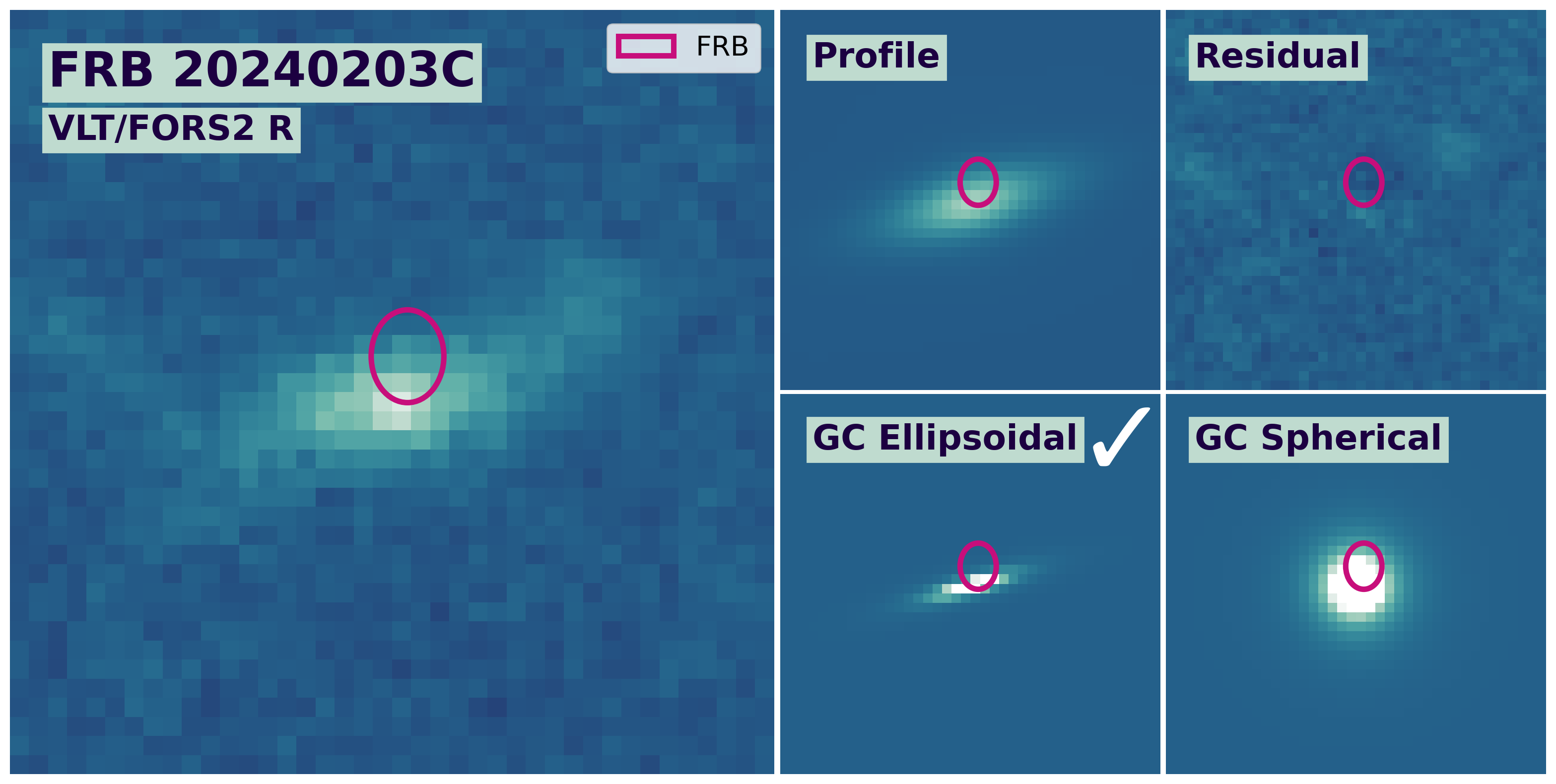}
    \includegraphics[width=0.39\textwidth]{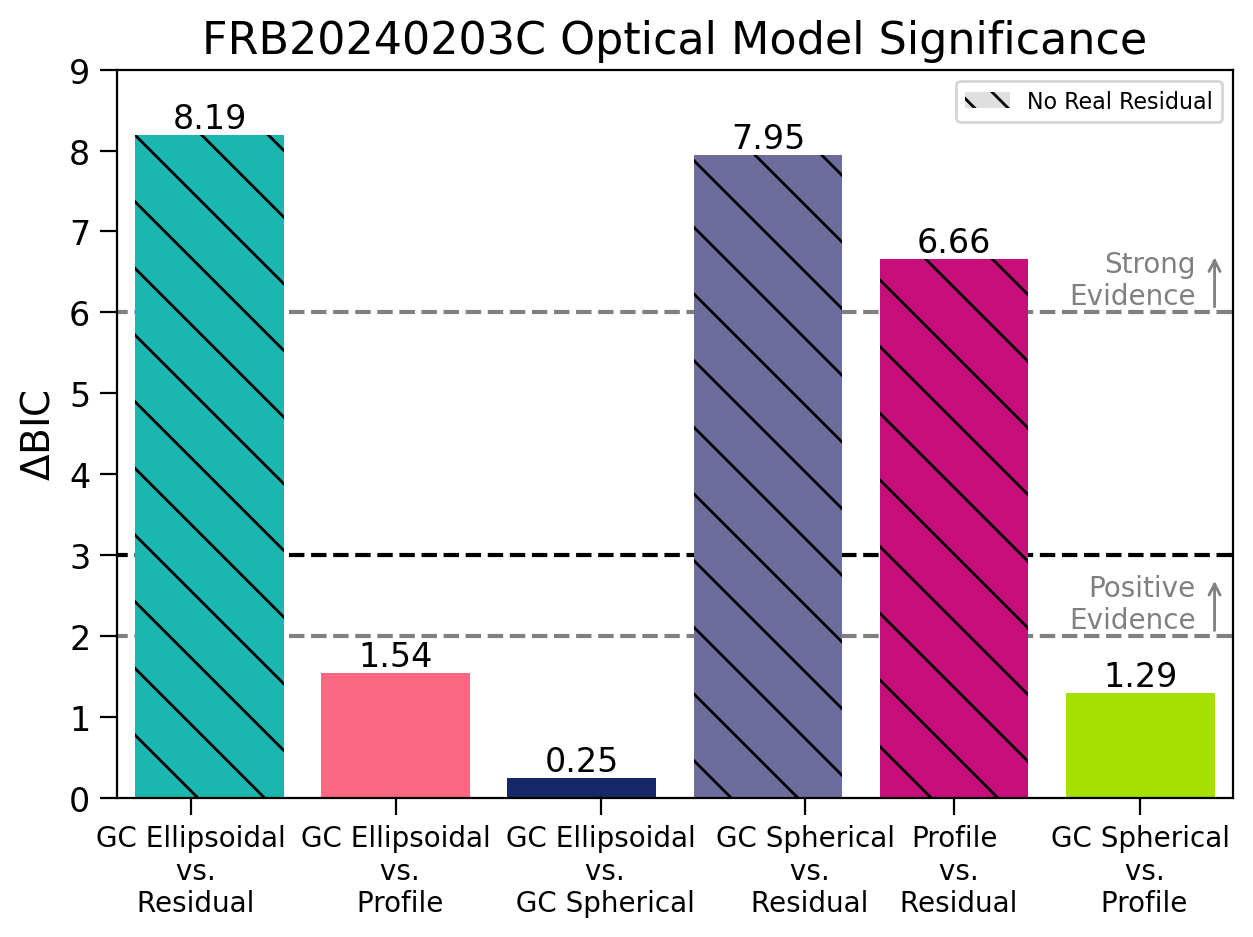}
    \caption{\textbf{(Left:)} Imaging and light profiles for FRB\,20240203C in the optical. \textbf{(Right:)} Comparison of the $\Delta$BIC between the four models for FRB\,20240203C in the optical.}
    \label{fig:appendix_240203_opt}
\end{figure}

\begin{figure}
    \centering
    \includegraphics[width=0.6\textwidth]{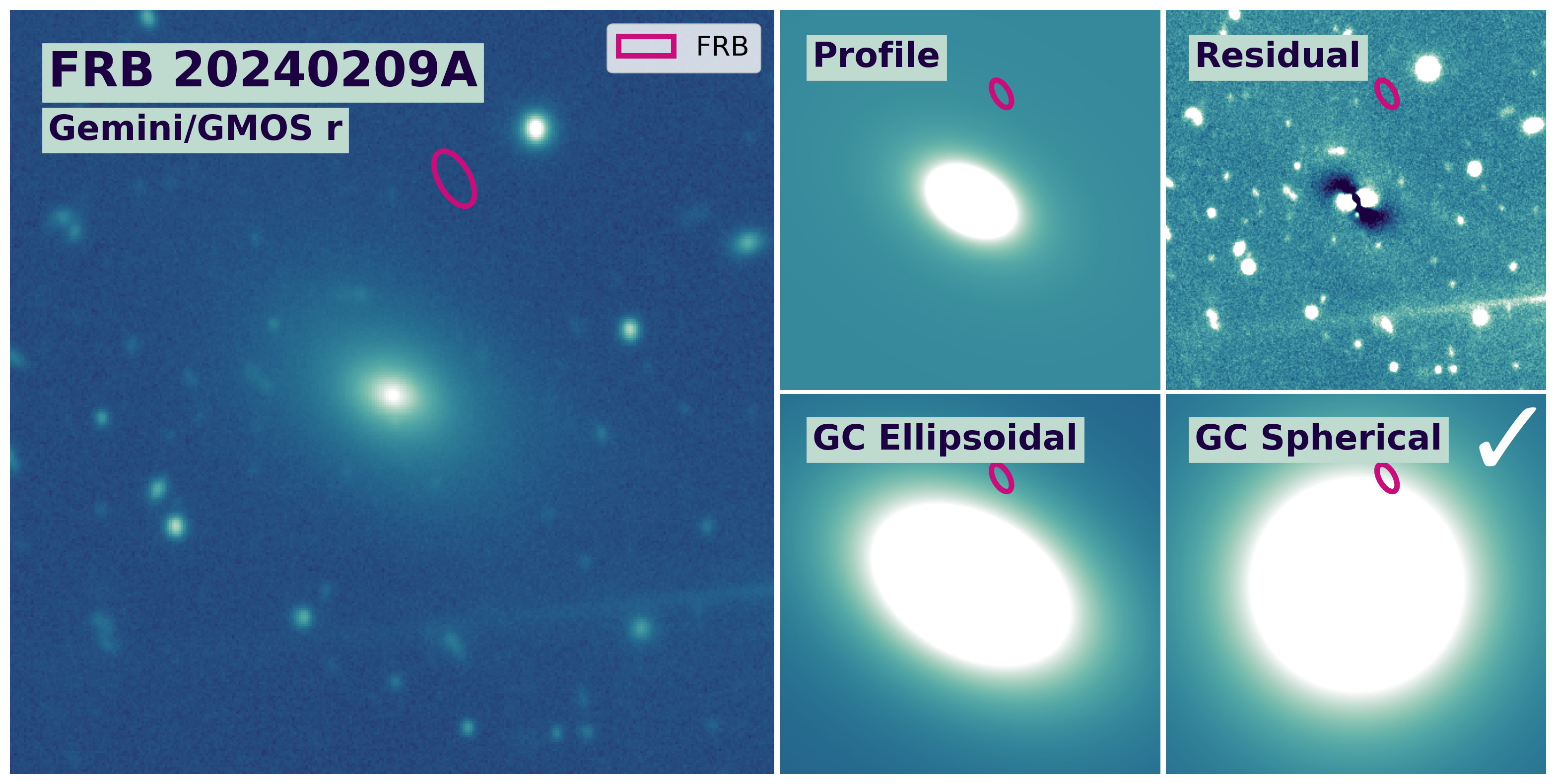}
    \includegraphics[width=0.39\textwidth]{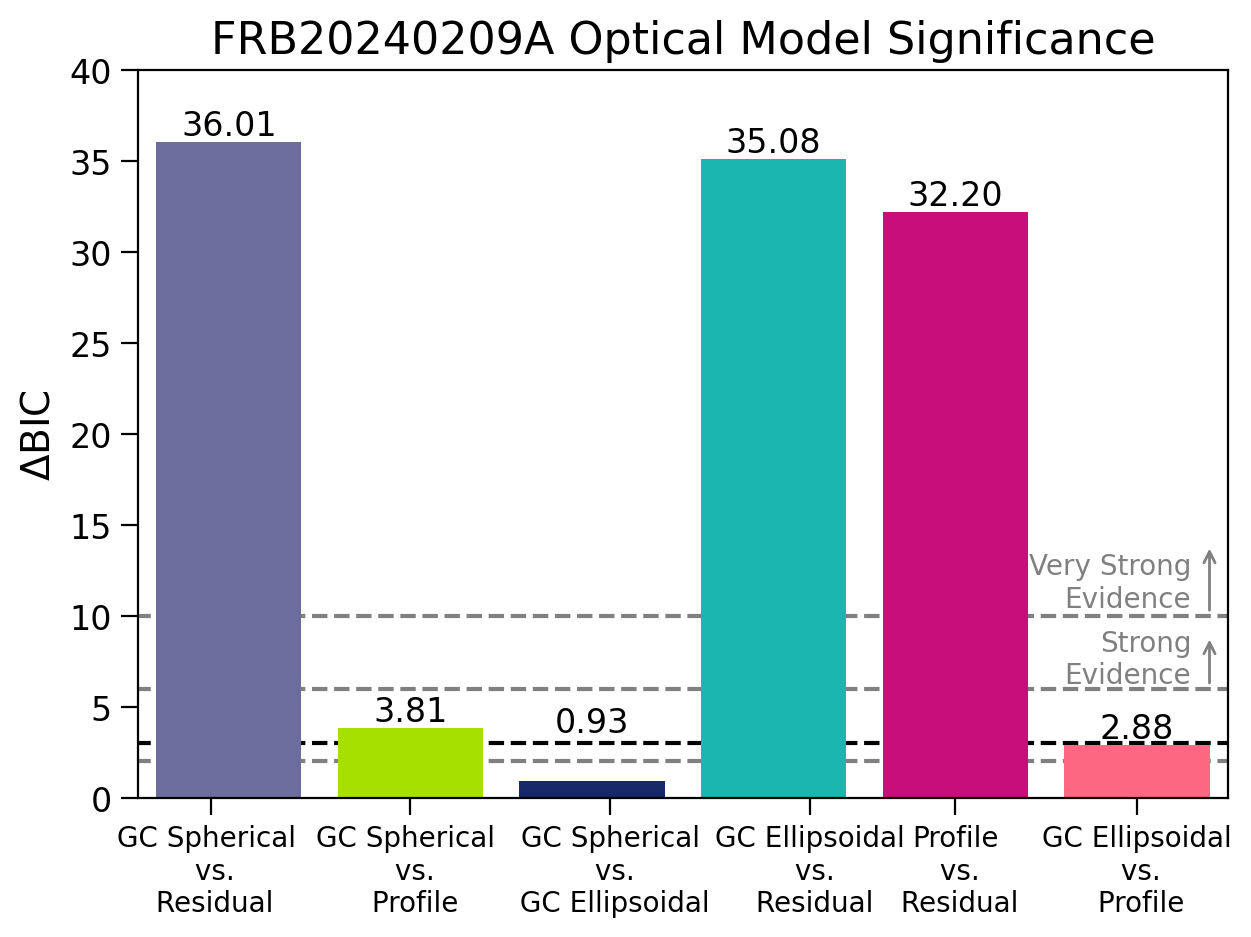}
    \caption{\textbf{(Left:)} Imaging and light profiles for FRB\,20240209A in the optical. \textbf{(Right:)} Comparison of the $\Delta$BIC between the four models for FRB\,20240209A in the optical.}
    \label{fig:appendix_240209_opt}
\end{figure}

\begin{figure}
    \centering
    \includegraphics[width=0.6\textwidth]{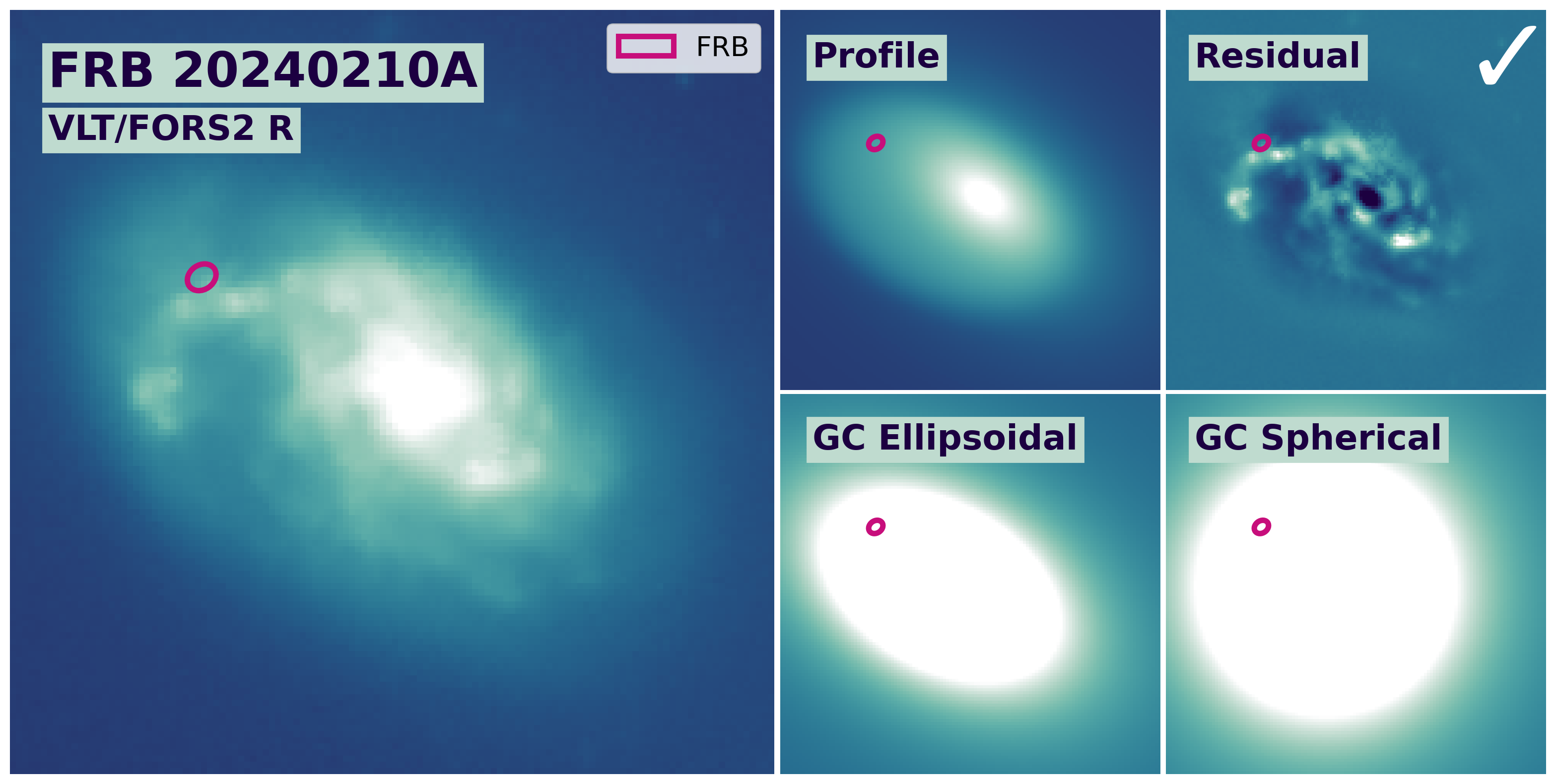}
    \includegraphics[width=0.39\textwidth]{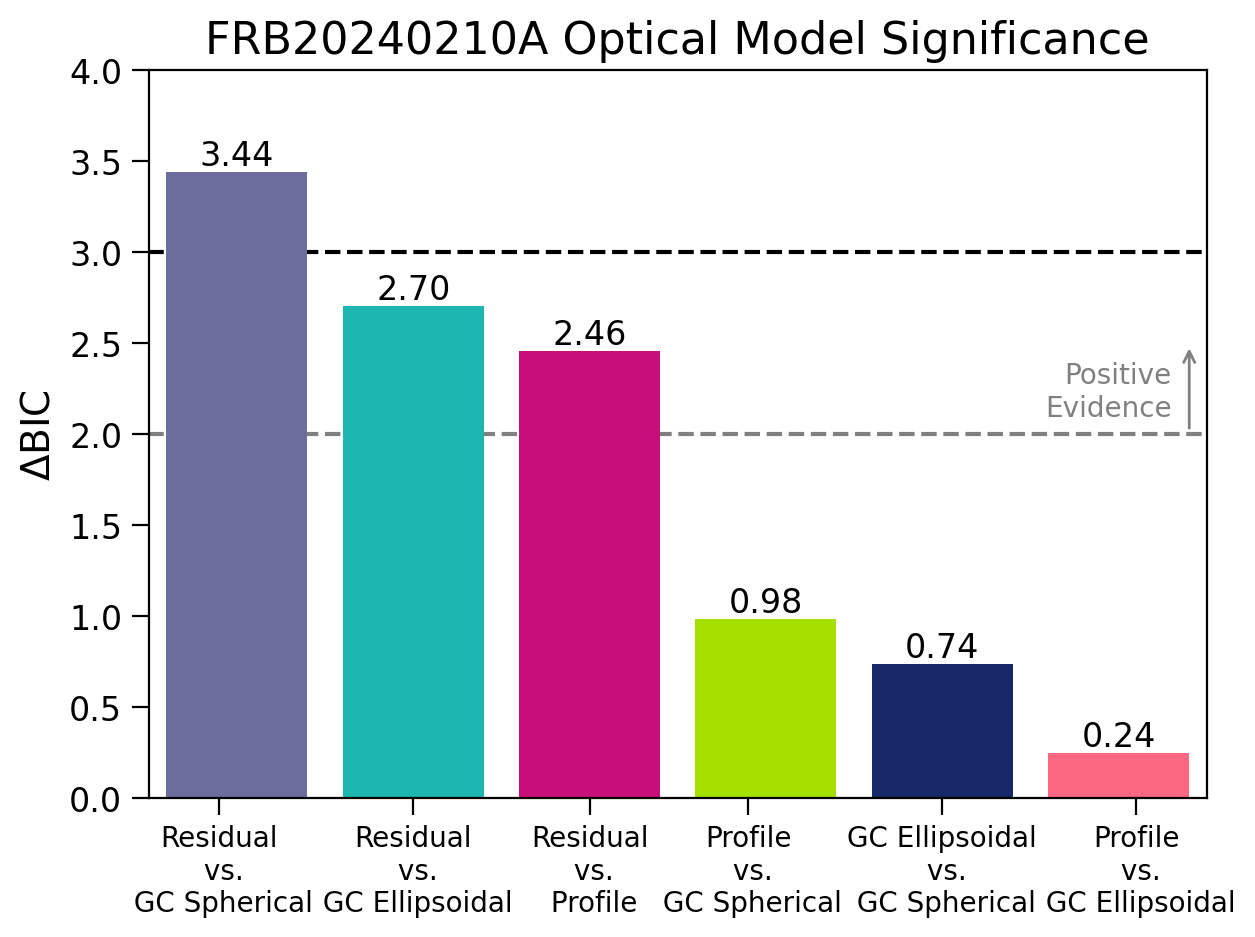}
    \caption{\textbf{(Left:)} Imaging and light profiles for FRB\,20240210A in the optical. \textbf{(Right:)} Comparison of the $\Delta$BIC between the four models for FRB\,20240210A in the optical.}
    \label{fig:appendix_240210_opt}
\end{figure}

\begin{figure}
    \centering
    \includegraphics[width=0.6\textwidth]{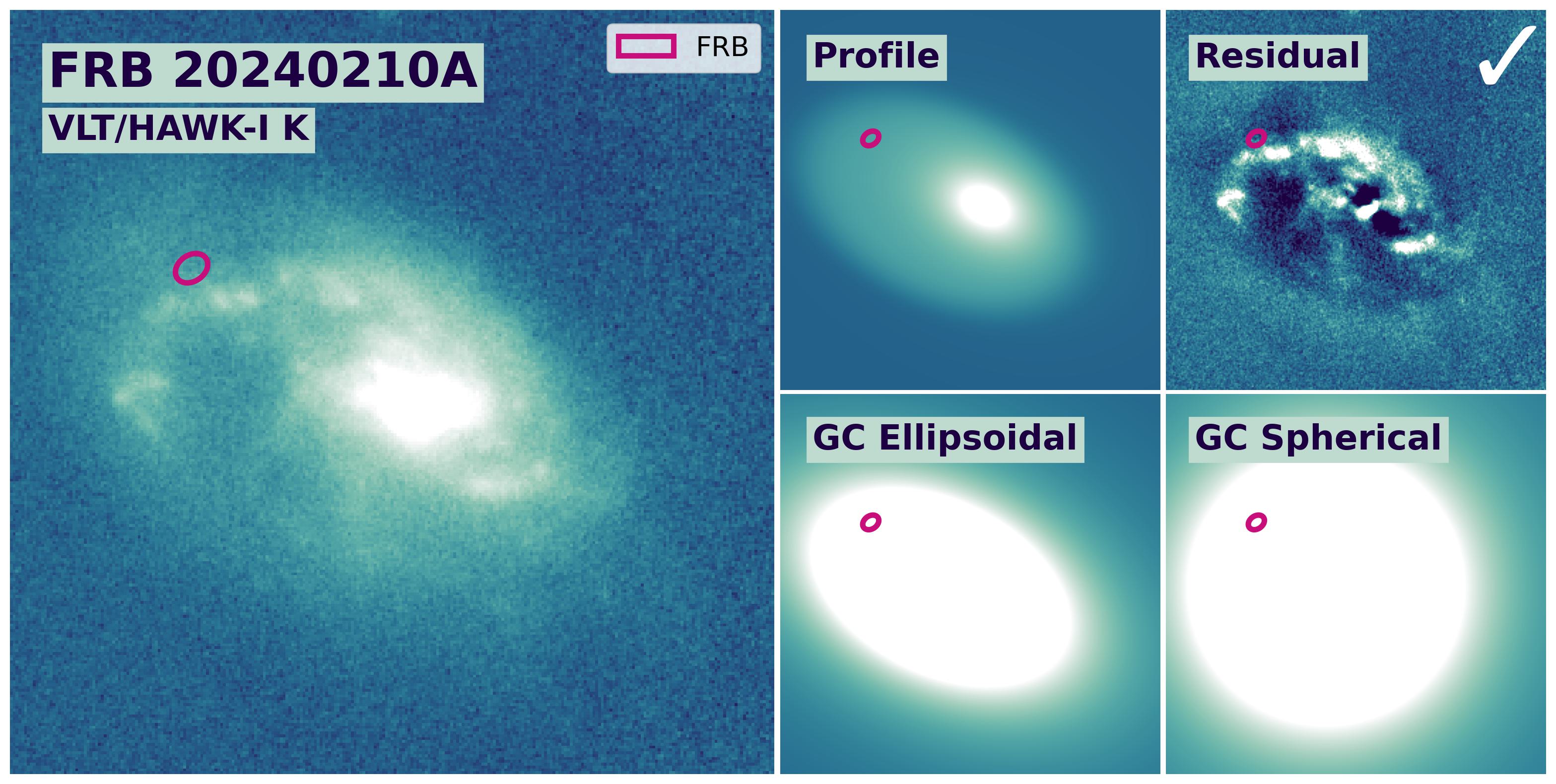}
    \includegraphics[width=0.39\textwidth]{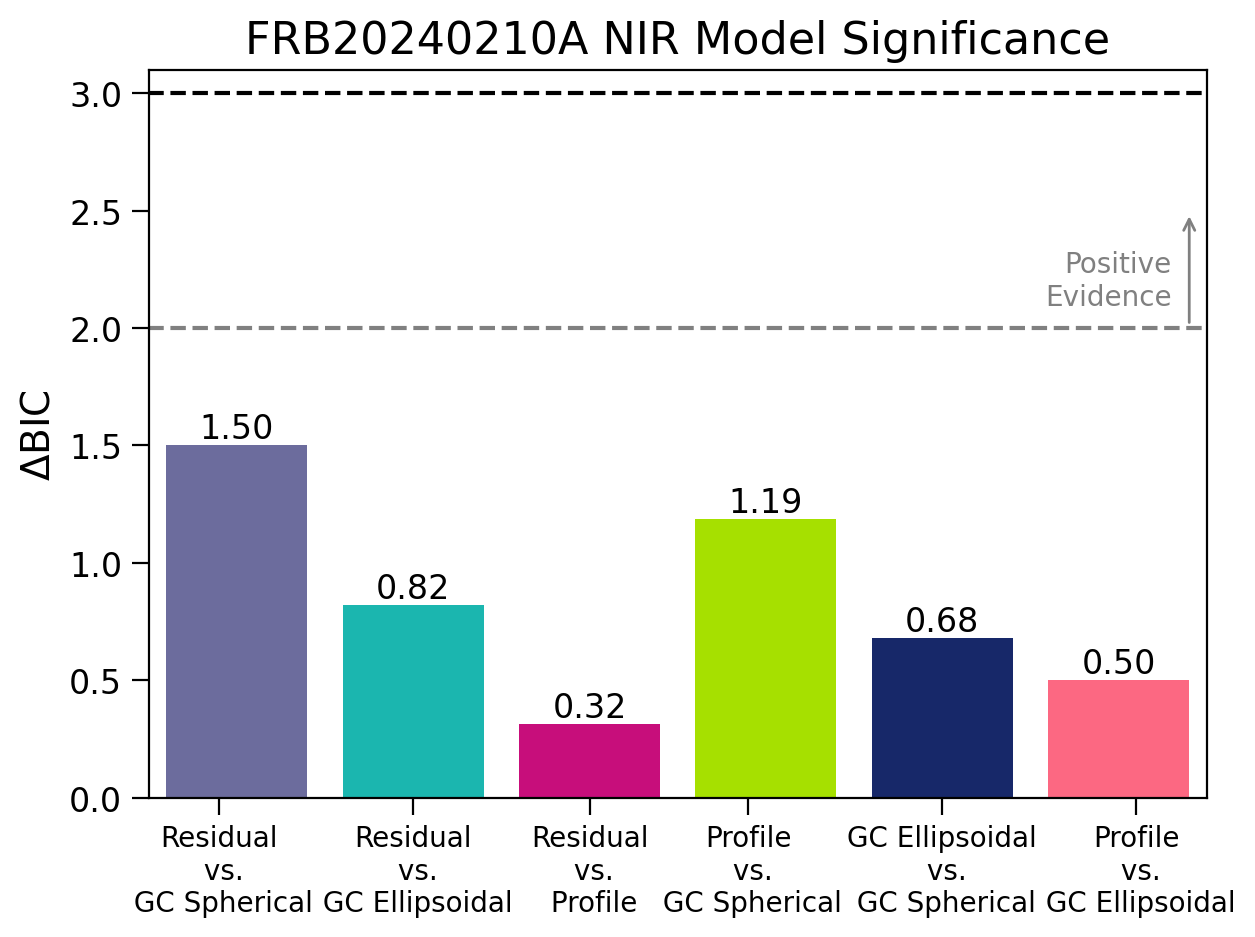}
    \caption{\textbf{(Left:)} Imaging and light profiles for FRB\,20240210A in the NIR. \textbf{(Right:)} Comparison of the $\Delta$BIC between the four models for FRB\,20240210A in the NIR.}
    \label{fig:appendix_240210_NIR}
\end{figure}

\begin{figure}
    \centering
    \includegraphics[width=0.6\textwidth]{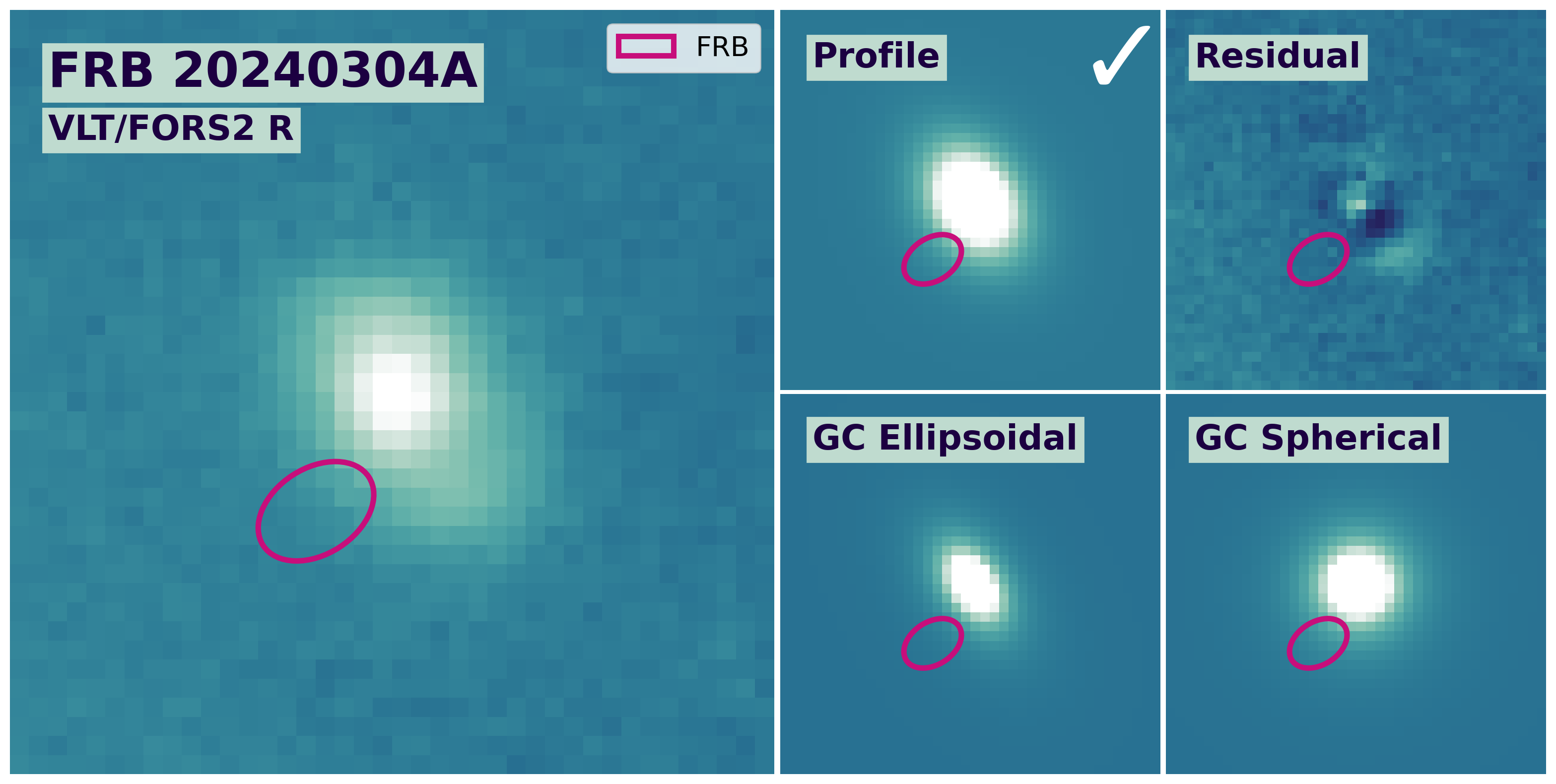}
    \includegraphics[width=0.39\textwidth]{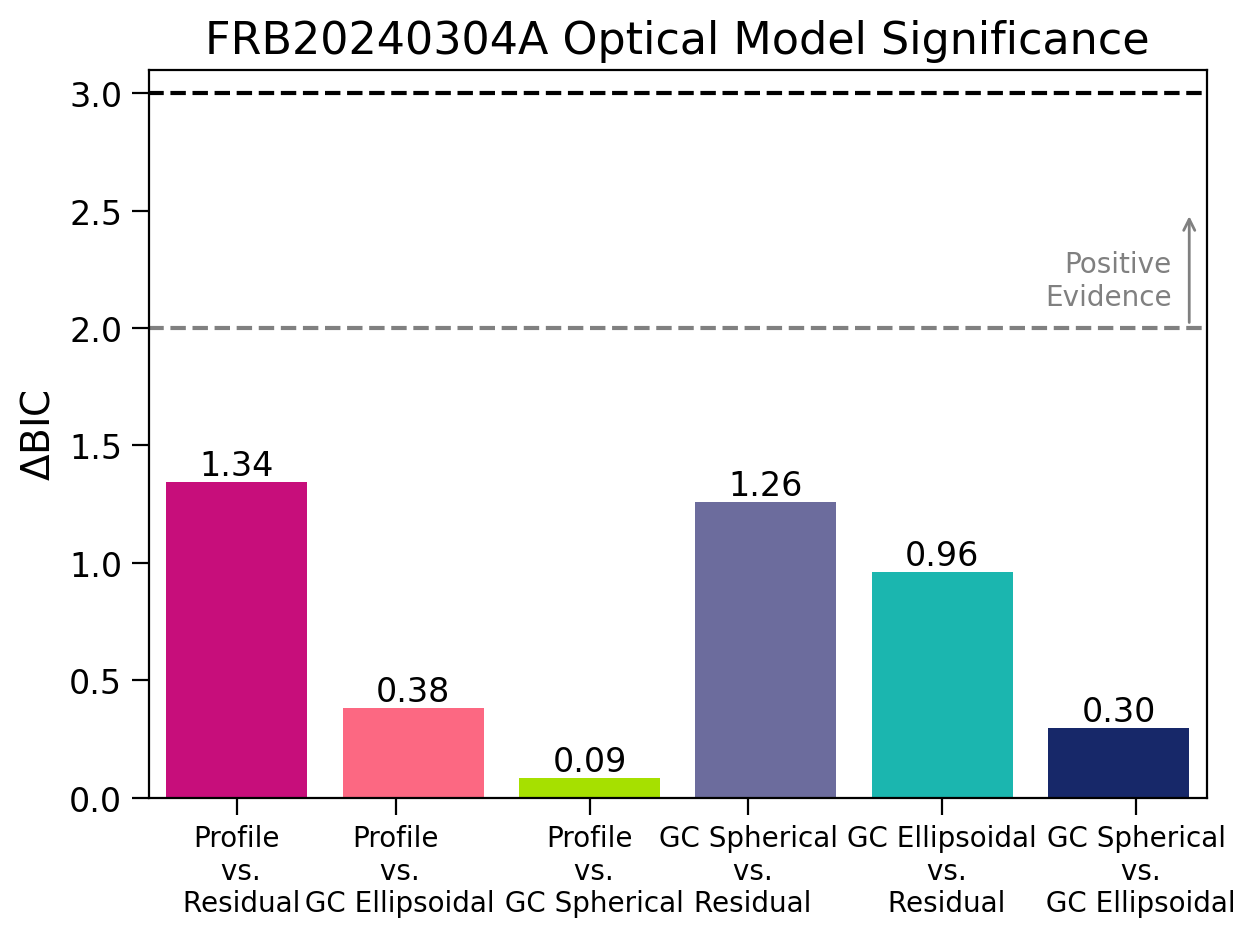}
    \caption{\textbf{(Left:)} Imaging and light profiles for FRB\,20240304A in the optical. \textbf{(Right:)} Comparison of the $\Delta$BIC between the four models for FRB\,20240304A in the optical.}
    \label{fig:appendix_240304_opt}
\end{figure}

\begin{figure}
    \centering
    \includegraphics[width=0.6\textwidth]{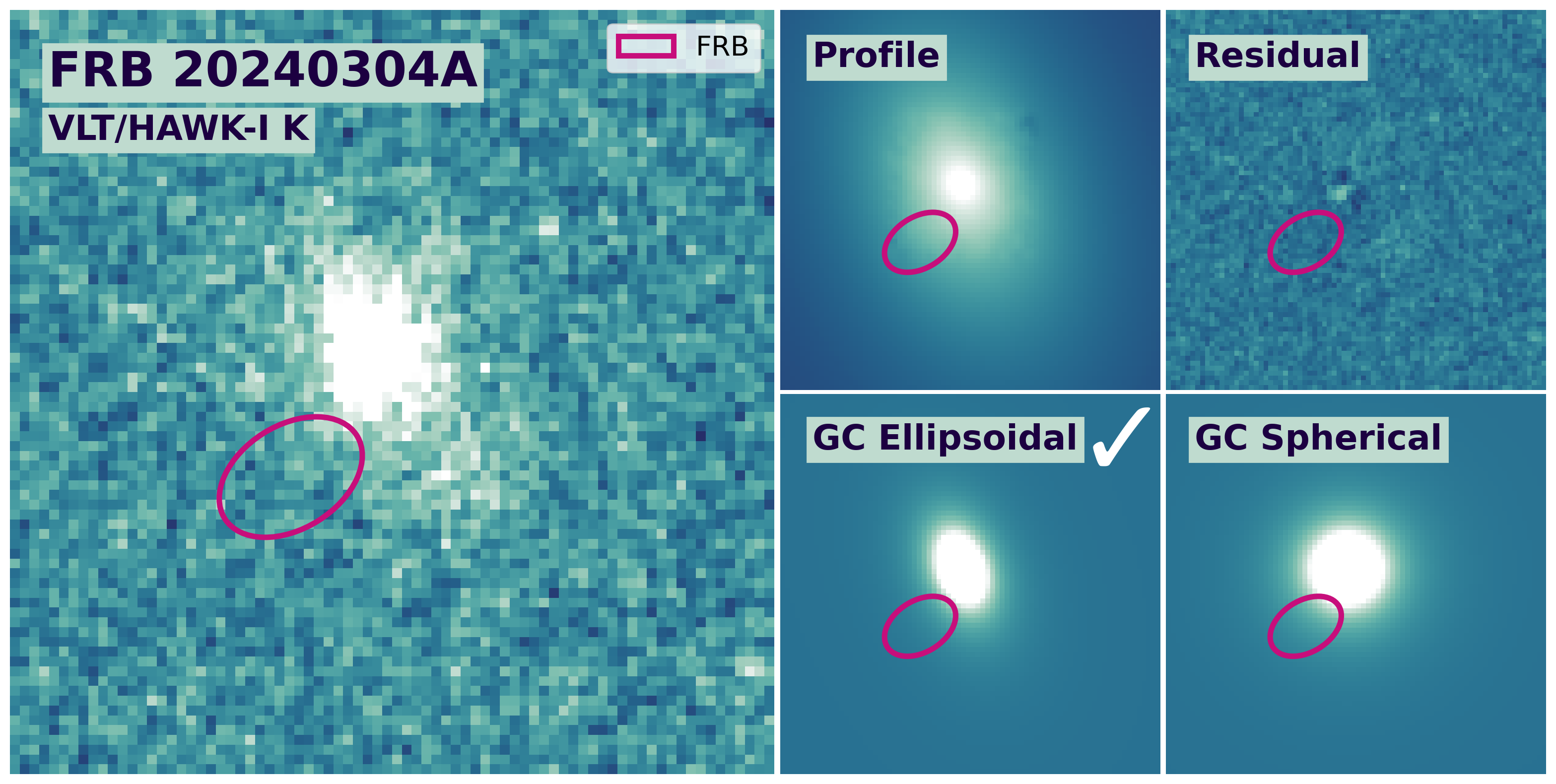}
    \includegraphics[width=0.39\textwidth]{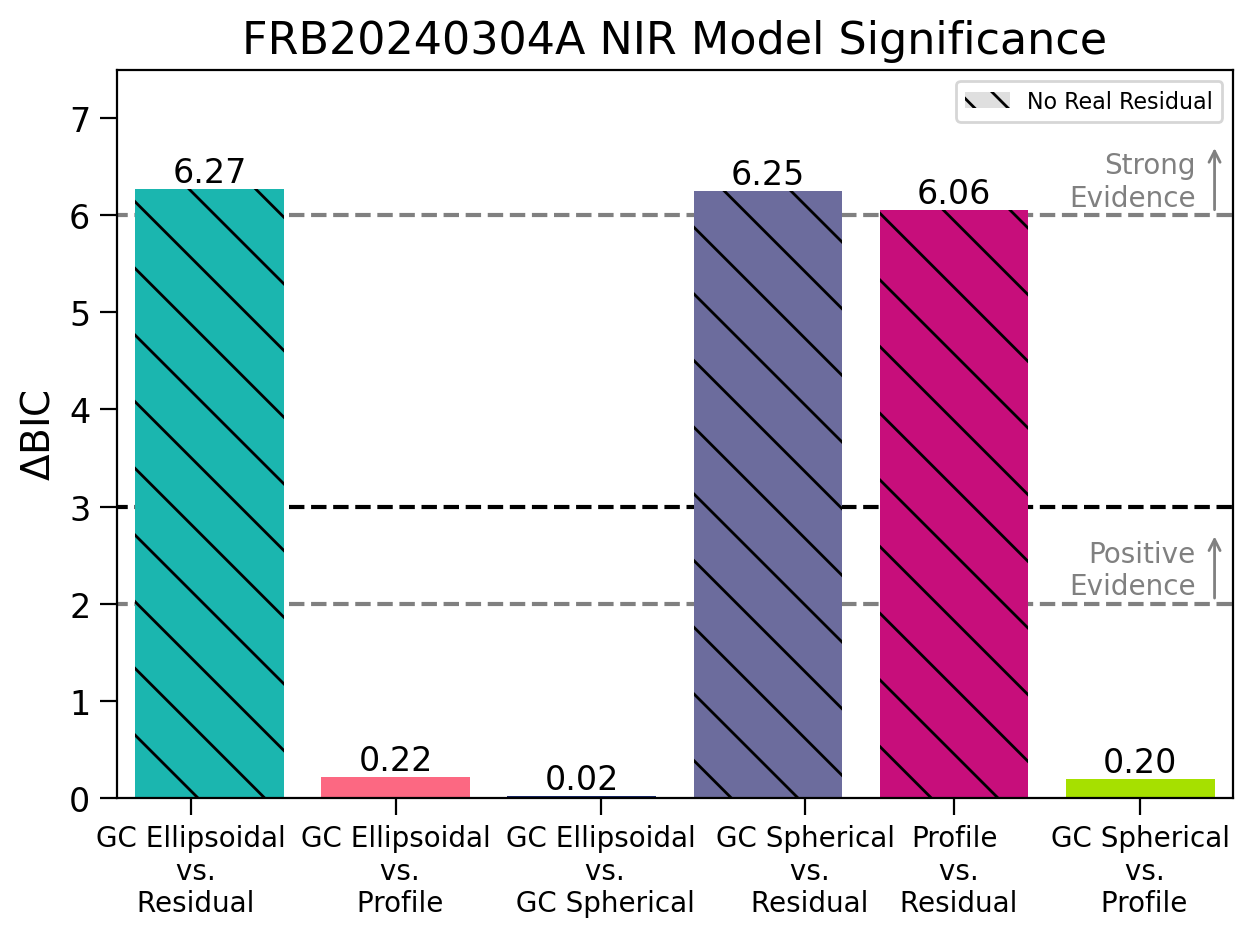}
    \caption{\textbf{(Left:)} Imaging and light profiles for FRB\,20240304A in the NIR. \textbf{(Right:)} Comparison of the $\Delta$BIC between the four models for FRB\,20240304A in the NIR.}
    \label{fig:appendix_240304_NIR}
\end{figure}

\begin{figure}
    \centering
    \includegraphics[width=0.6\textwidth]{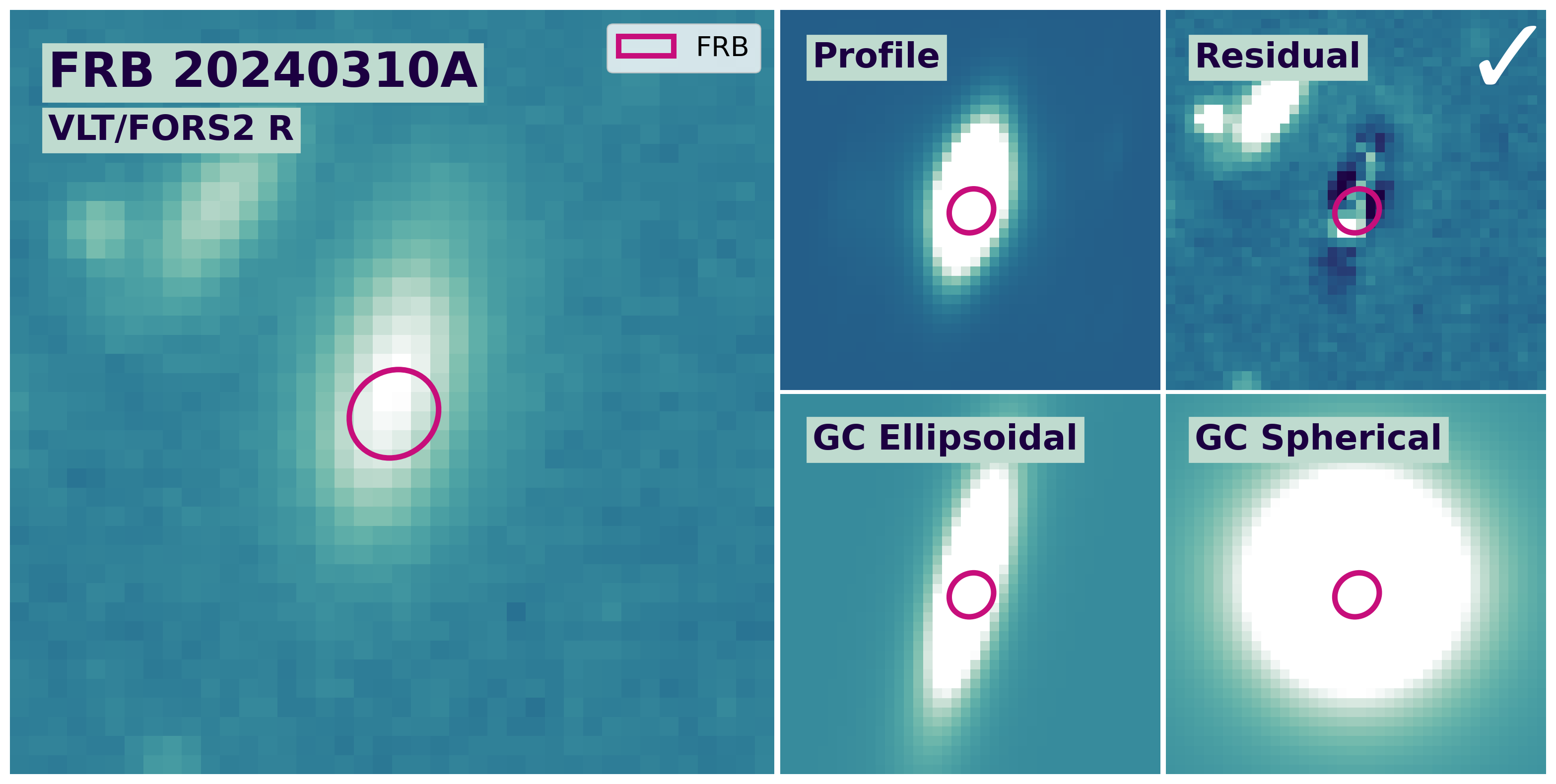}
    \includegraphics[width=0.39\textwidth]{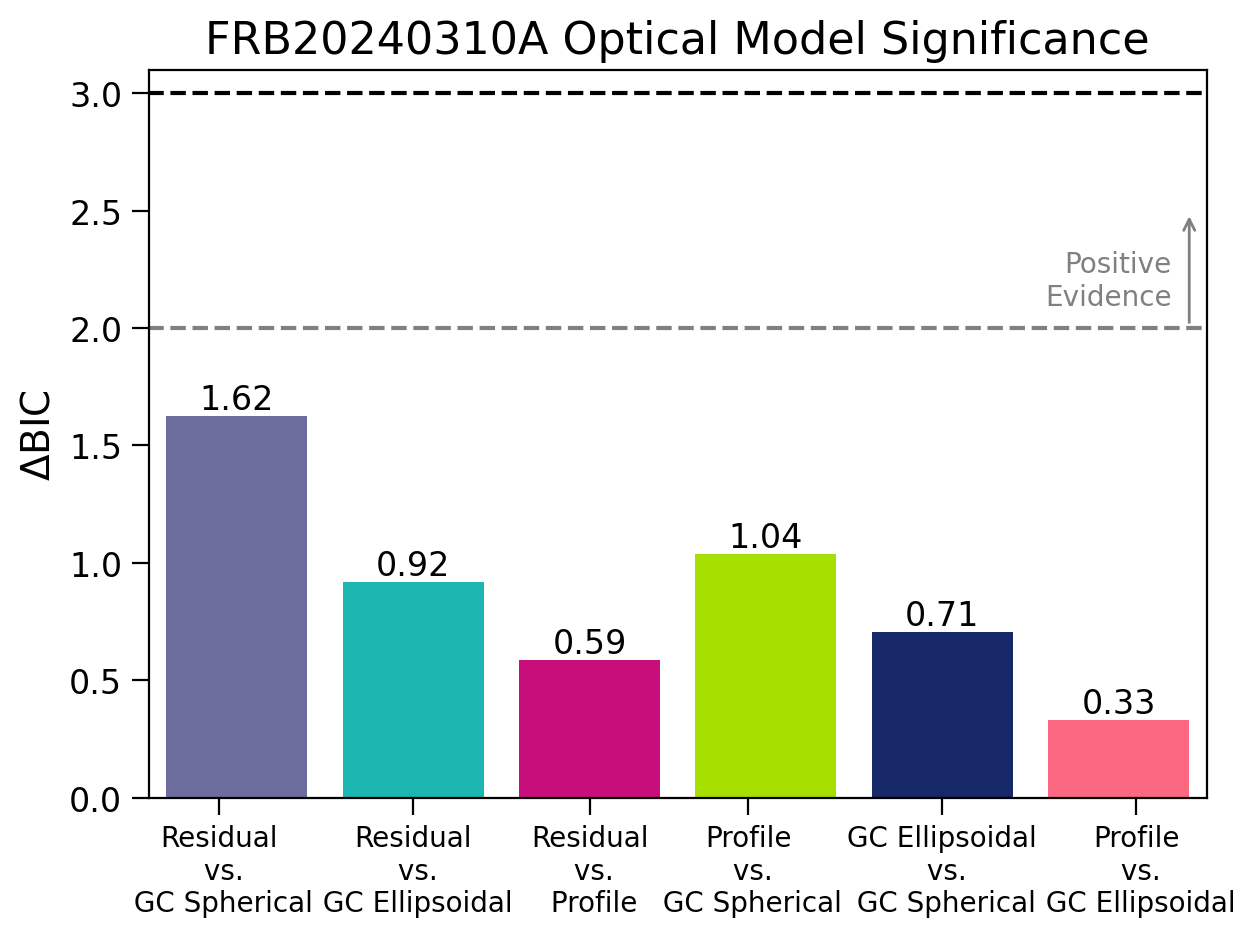}
    \caption{\textbf{(Left:)} Imaging and light profiles for FRB\,20240310A in the optical. \textbf{(Right:)} Comparison of the $\Delta$BIC between the four models for FRB\,20240310A in the optical.}
    \label{fig:appendix_240310_opt}
\end{figure}

\begin{figure}
    \centering
    \includegraphics[width=0.6\textwidth]{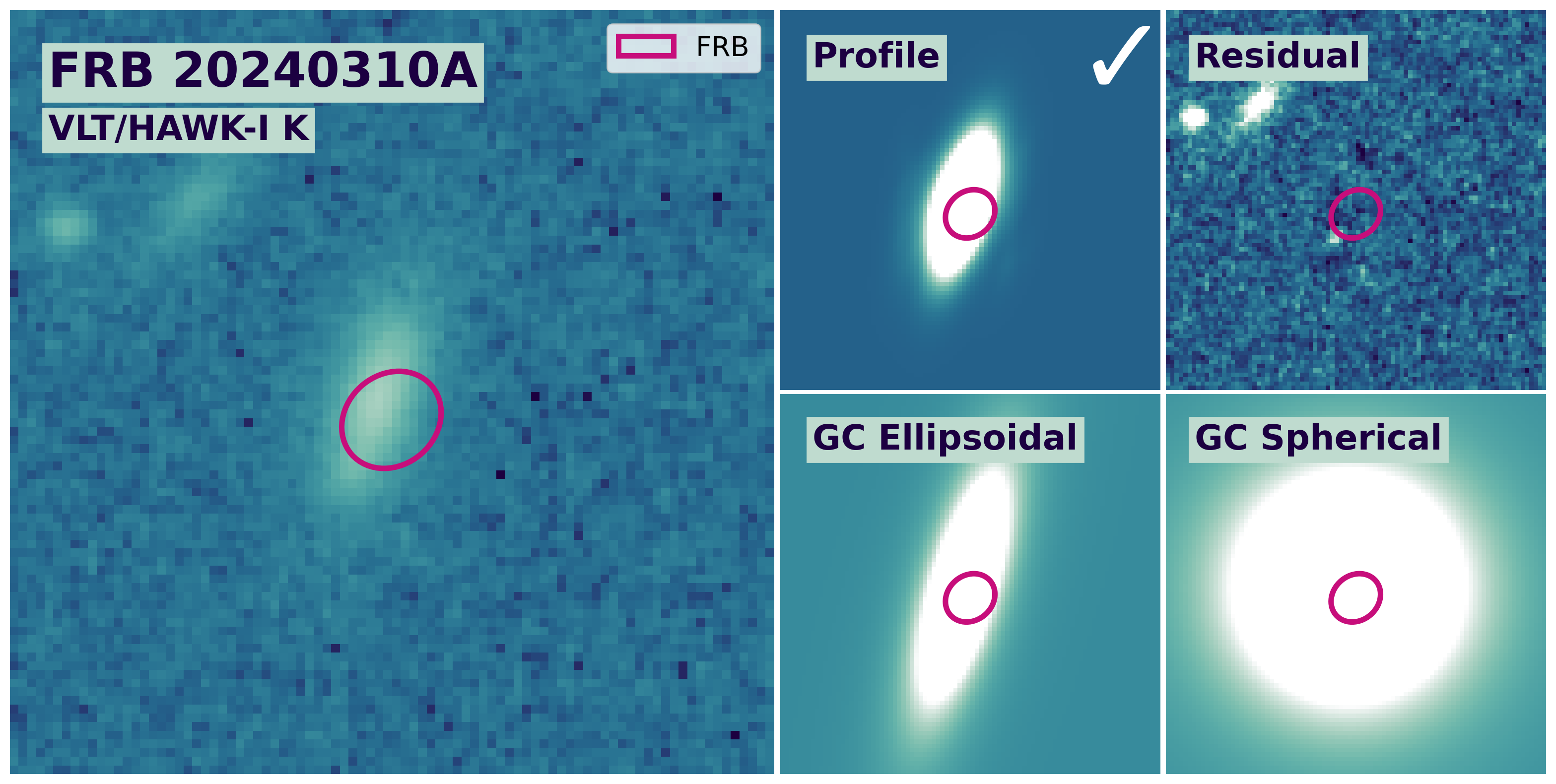}
    \includegraphics[width=0.39\textwidth]{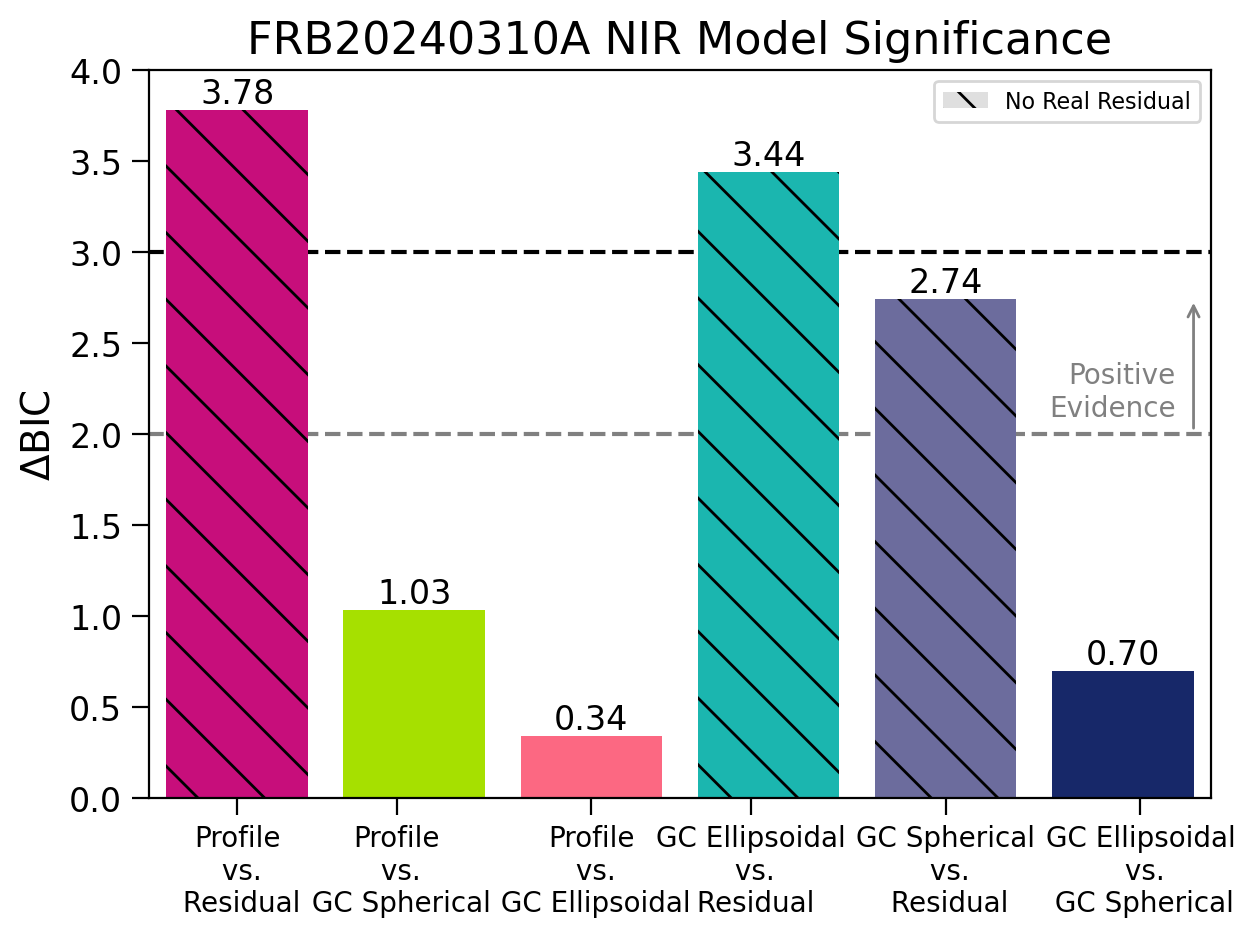}
    \caption{\textbf{(Left:)} Imaging and light profiles for FRB\,20240310A in the NIR. \textbf{(Right:)} Comparison of the $\Delta$BIC between the four models for FRB\,20240310A in the NIR.}
    \label{fig:appendix_240310_NIR}
\end{figure}

\begin{figure}
    \centering
    \includegraphics[width=0.6\textwidth]{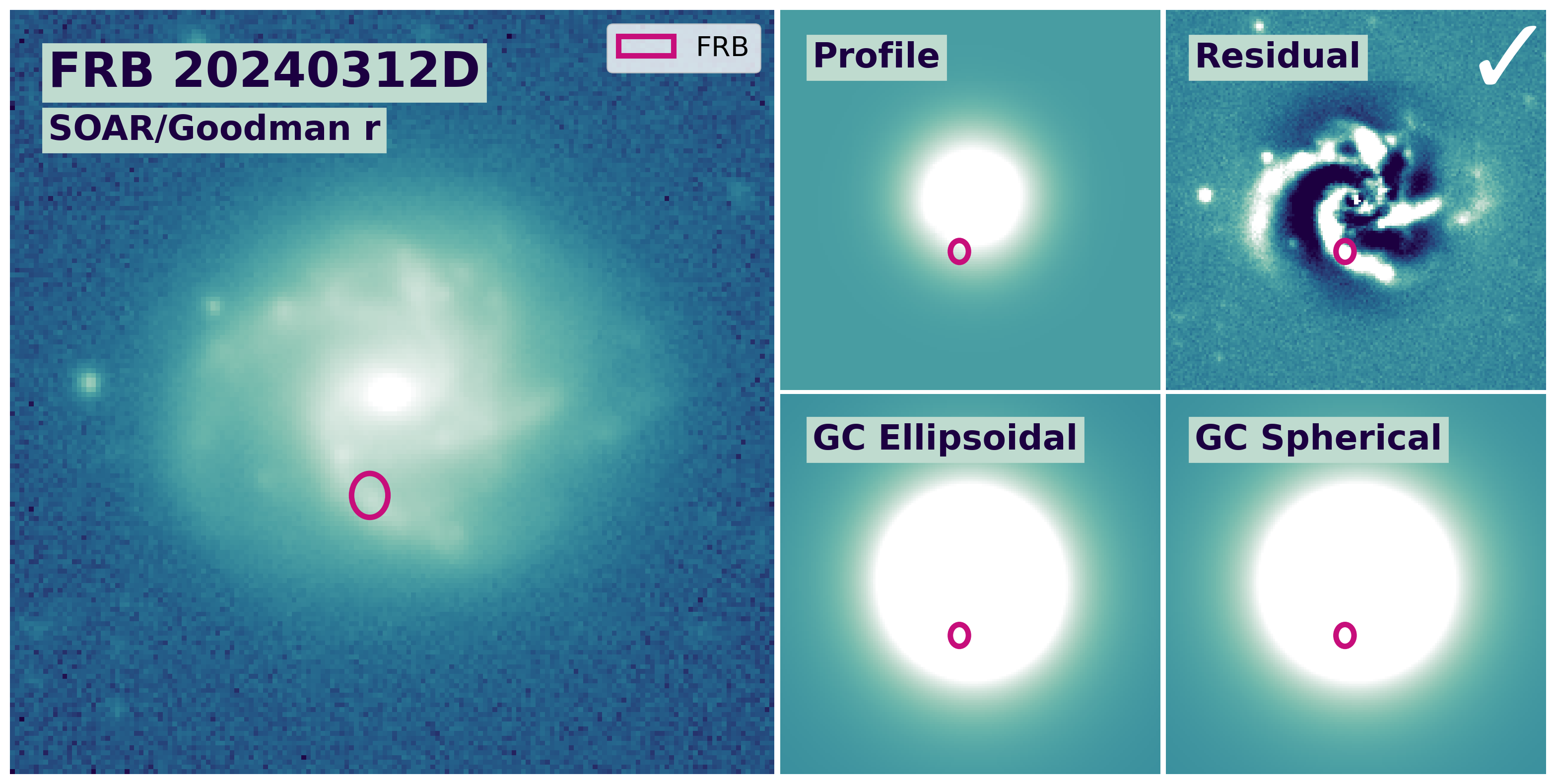}
    \includegraphics[width=0.39\textwidth]{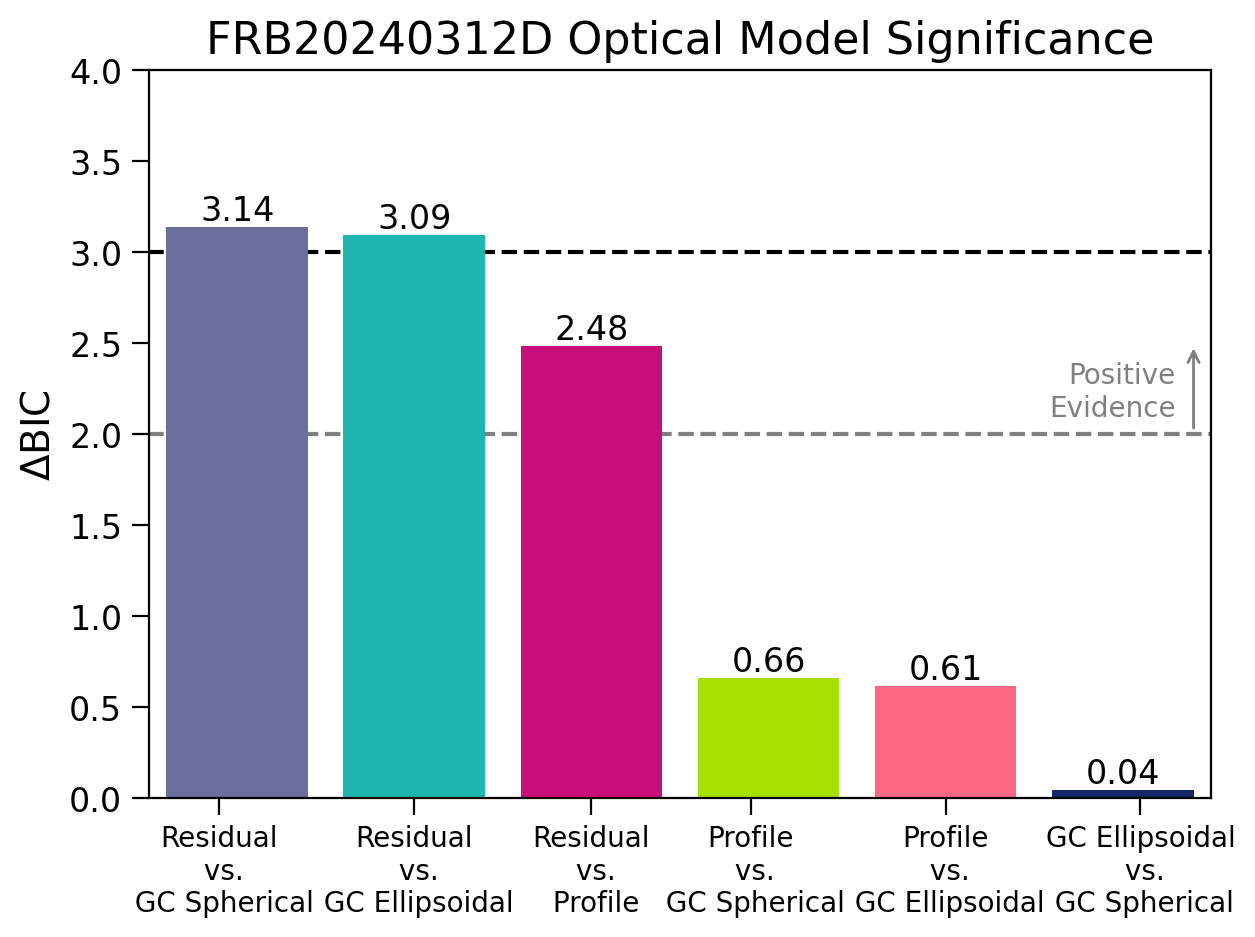}
    \caption{\textbf{(Left:)} Imaging and light profiles for FRB\,20240312D in the optical. \textbf{(Right:)} Comparison of the $\Delta$BIC between the four models for FRB\,20240312D in the optical.}
    \label{fig:appendix_240312_opt}
\end{figure}

\begin{figure}
    \centering
    \includegraphics[width=0.6\textwidth]{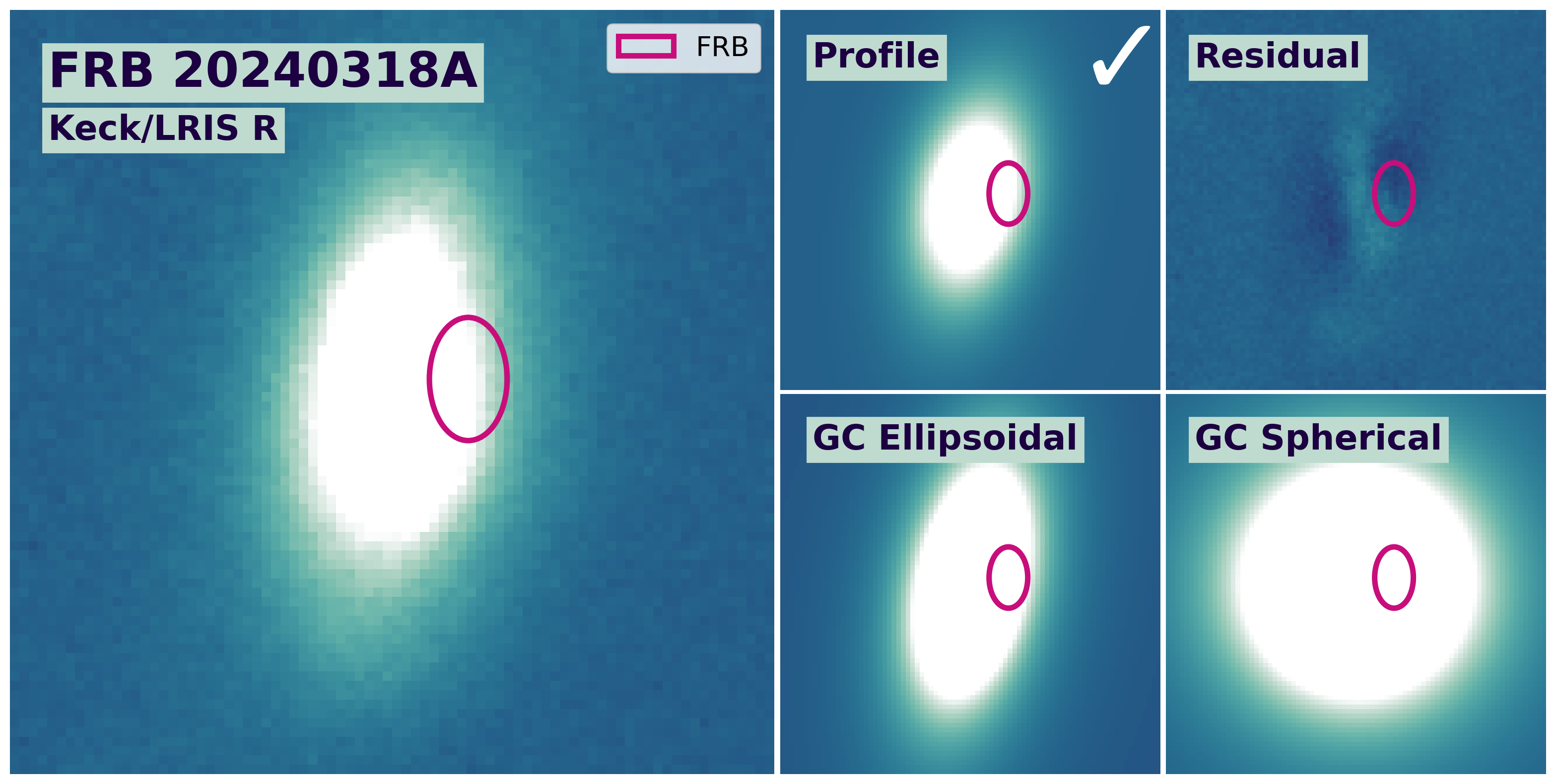}
    \includegraphics[width=0.39\textwidth]{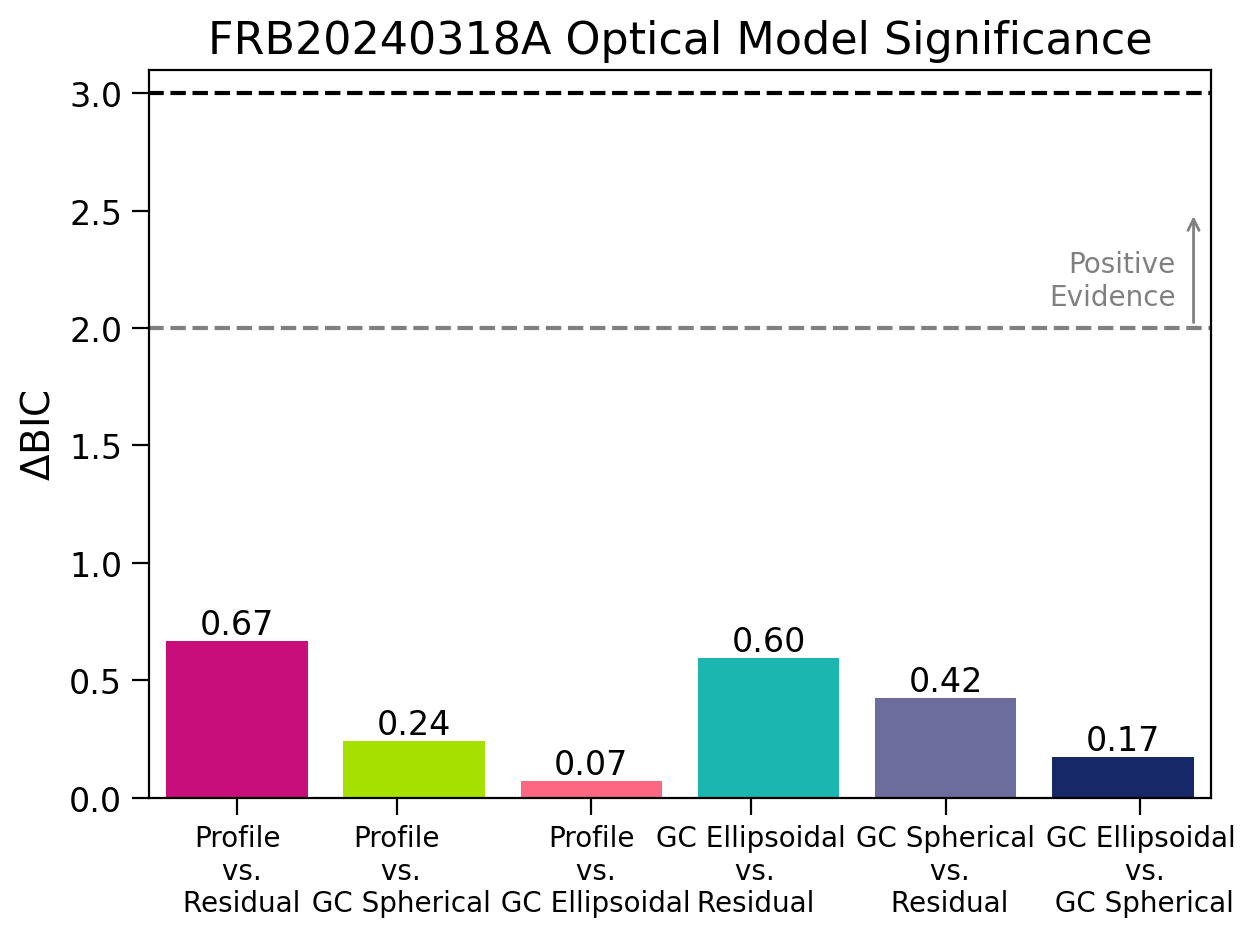}
    \caption{\textbf{(Left:)} Imaging and light profiles for FRB\,20240318A in the optical. \textbf{(Right:)} Comparison of the $\Delta$BIC between the four models for FRB\,20240318A in the optical.}
    \label{fig:appendix_240318_opt}
\end{figure}

\begin{figure}
    \centering
    \includegraphics[width=0.6\textwidth]{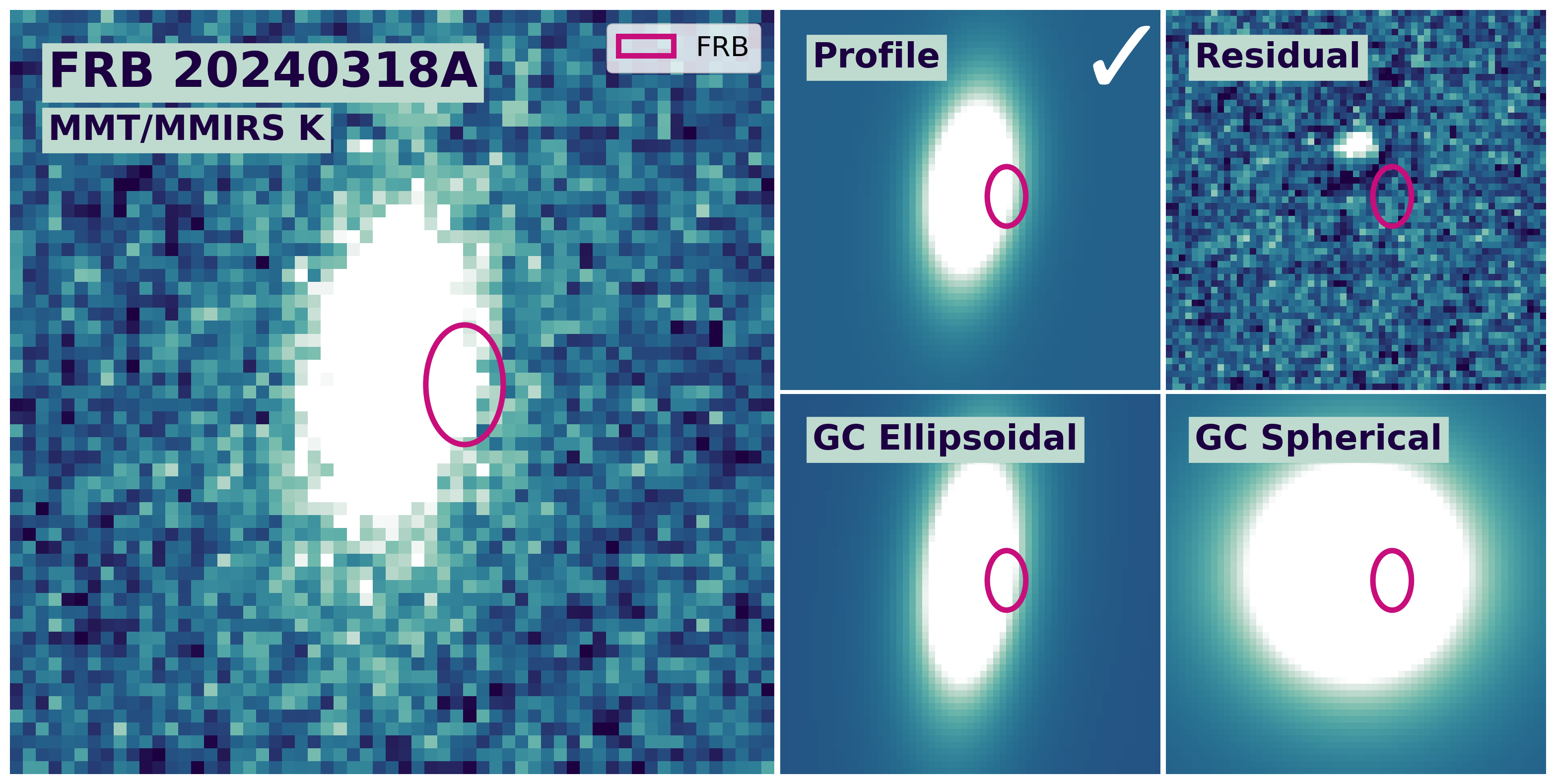}
    \includegraphics[width=0.39\textwidth]{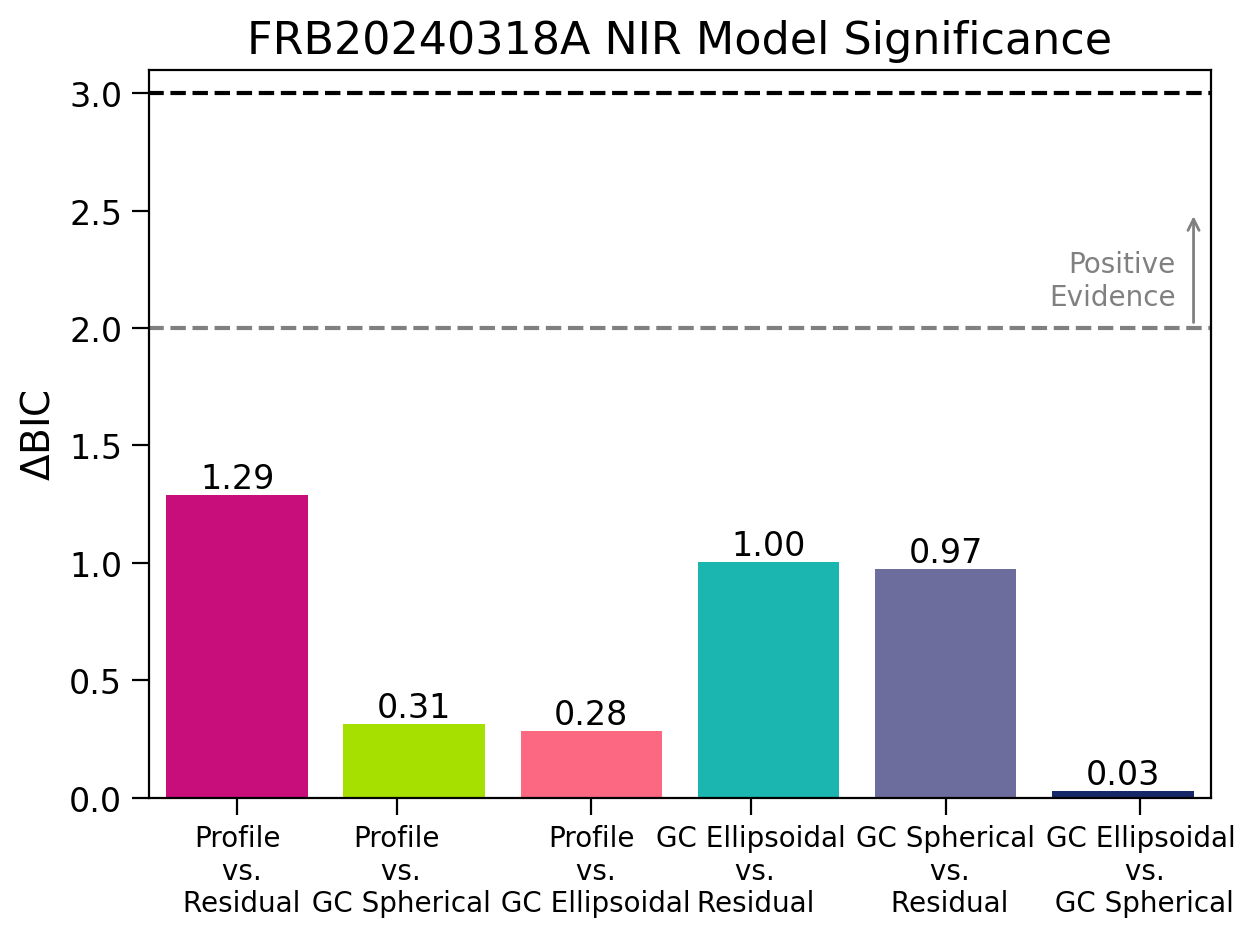}
    \caption{\textbf{(Left:)} Imaging and light profiles for FRB\,20240318A in the NIR. \textbf{(Right:)} Comparison of the $\Delta$BIC between the four models for FRB\,20240318A in the NIR.}
    \label{fig:appendix_240318_NIR}
\end{figure}

\begin{figure}
    \centering
    \includegraphics[width=0.6\textwidth]{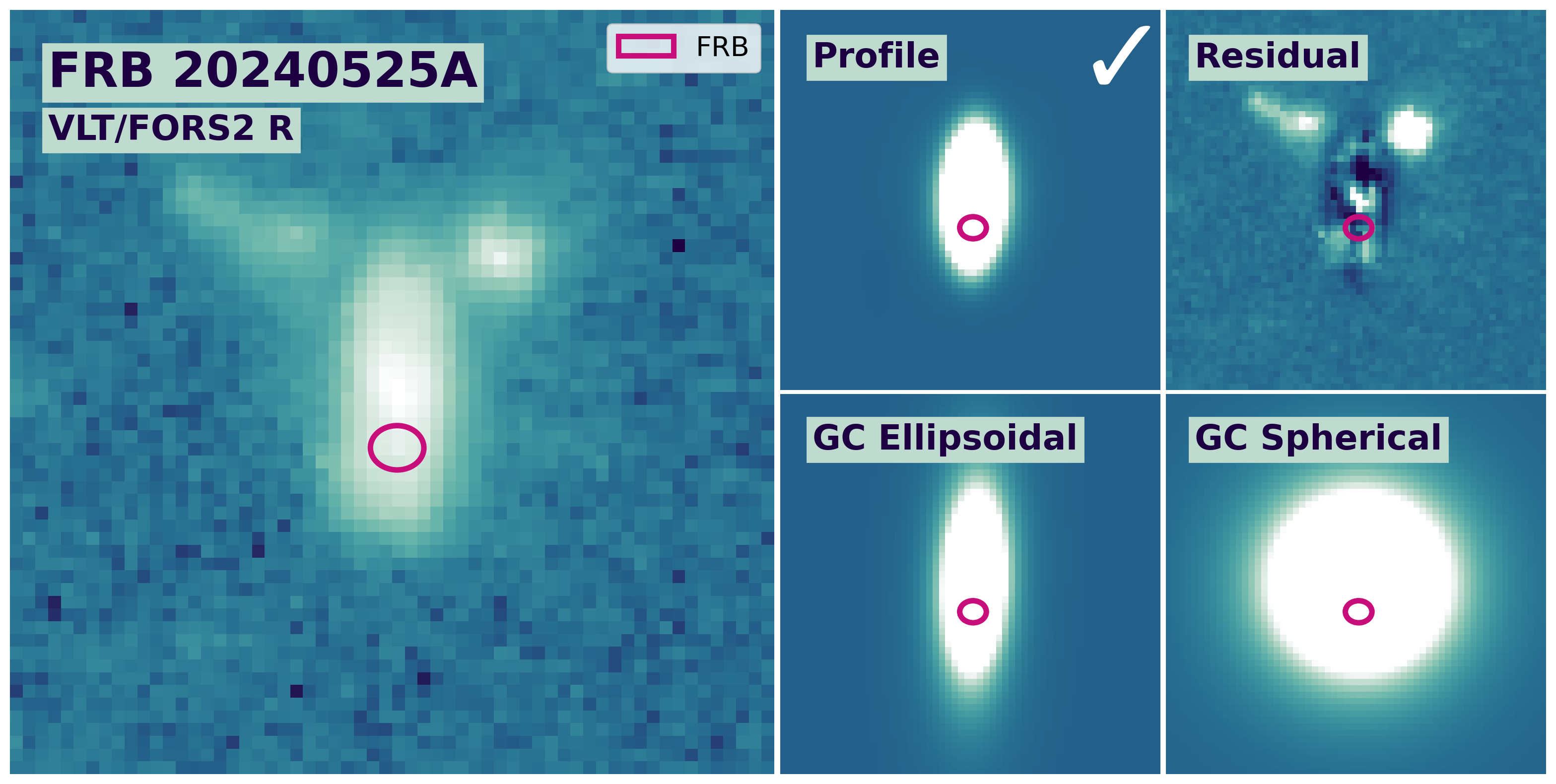}
    \includegraphics[width=0.39\textwidth]{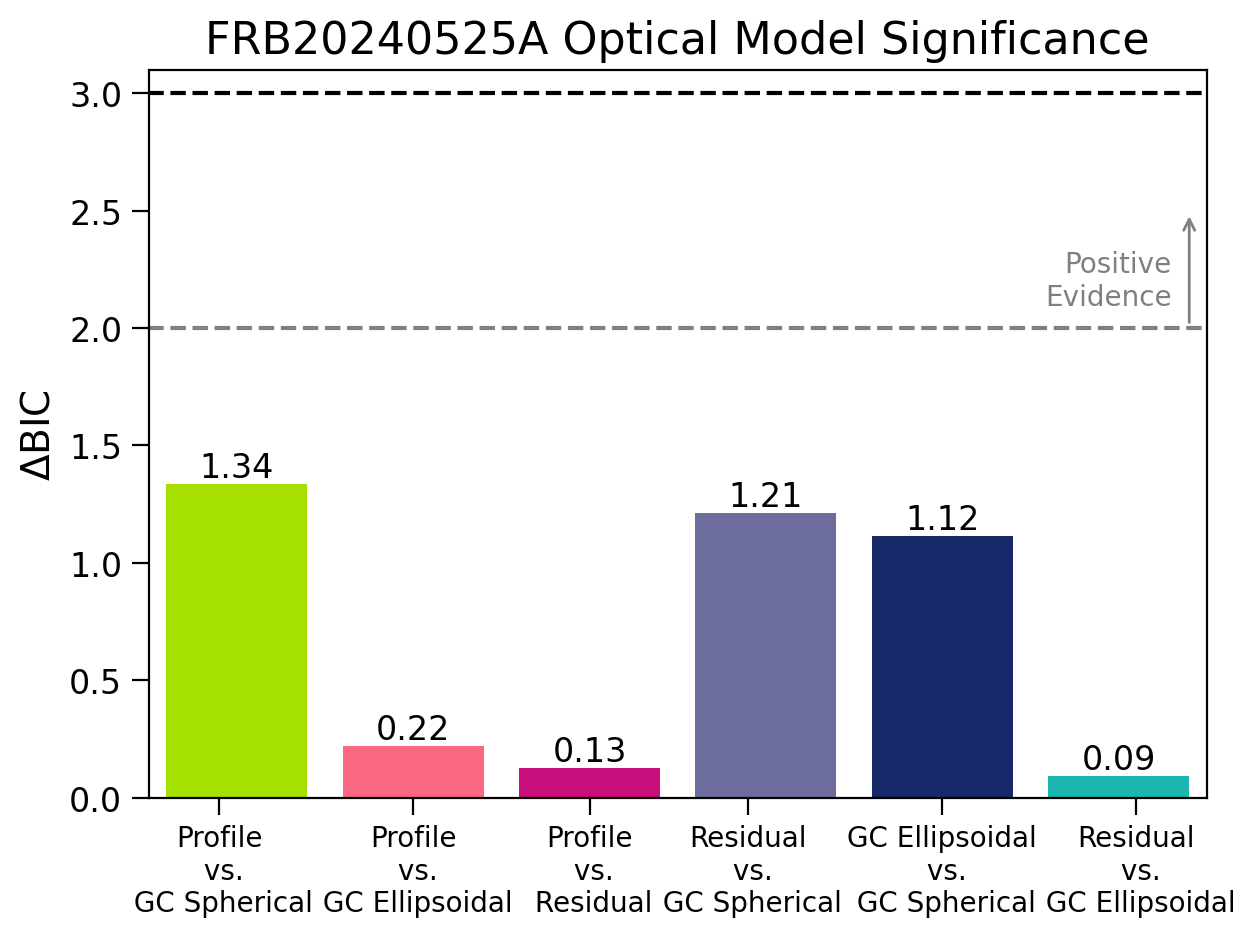}
    \caption{\textbf{(Left:)} Imaging and light profiles for FRB\,20240525A in the optical. \textbf{(Right:)} Comparison of the $\Delta$BIC between the four models for FRB\,20240525A in the optical.}
    \label{fig:appendix_240525_opt}
\end{figure}

\begin{figure}
    \centering
    \includegraphics[width=0.6\textwidth]{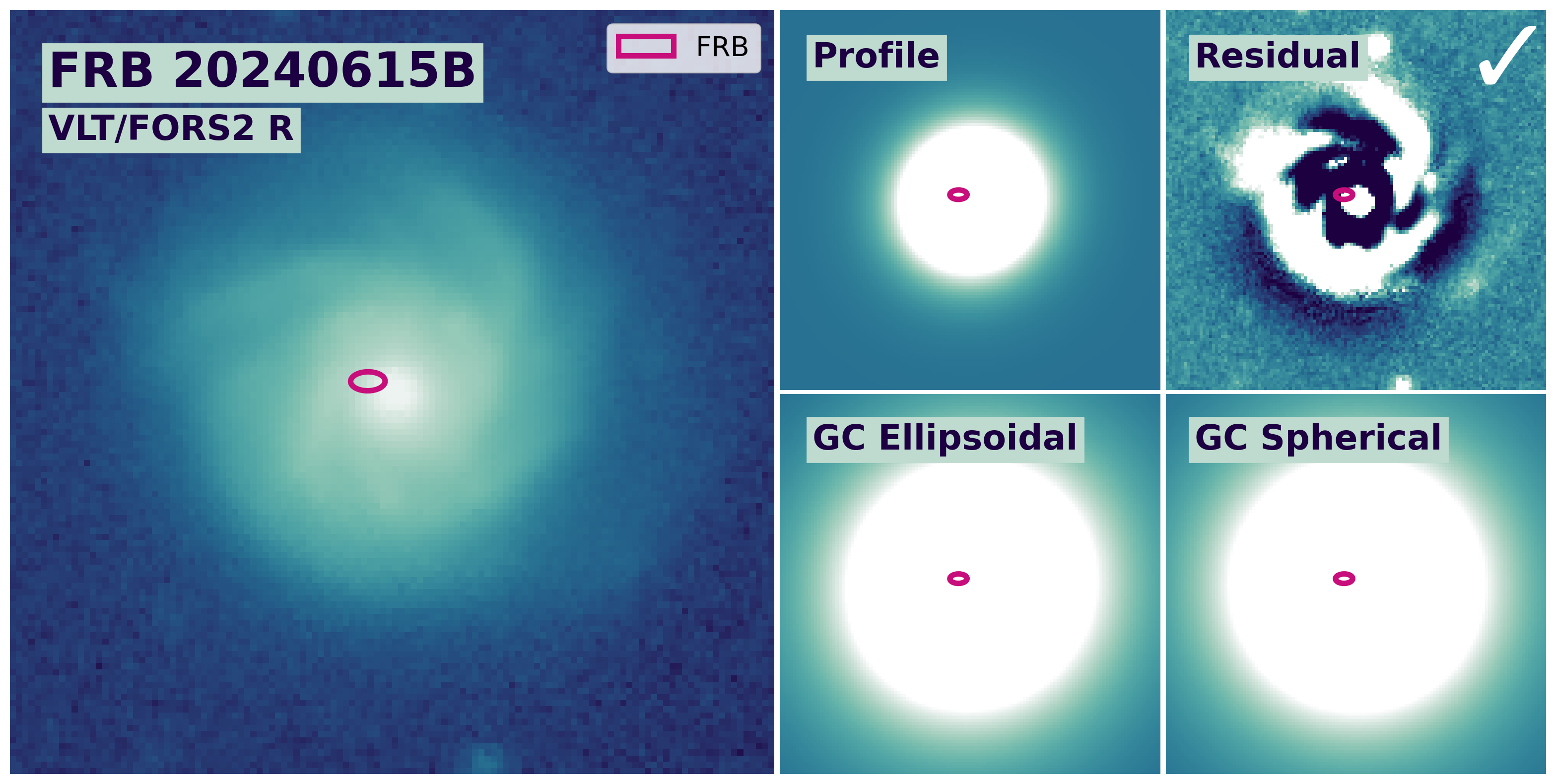}
    \includegraphics[width=0.39\textwidth]{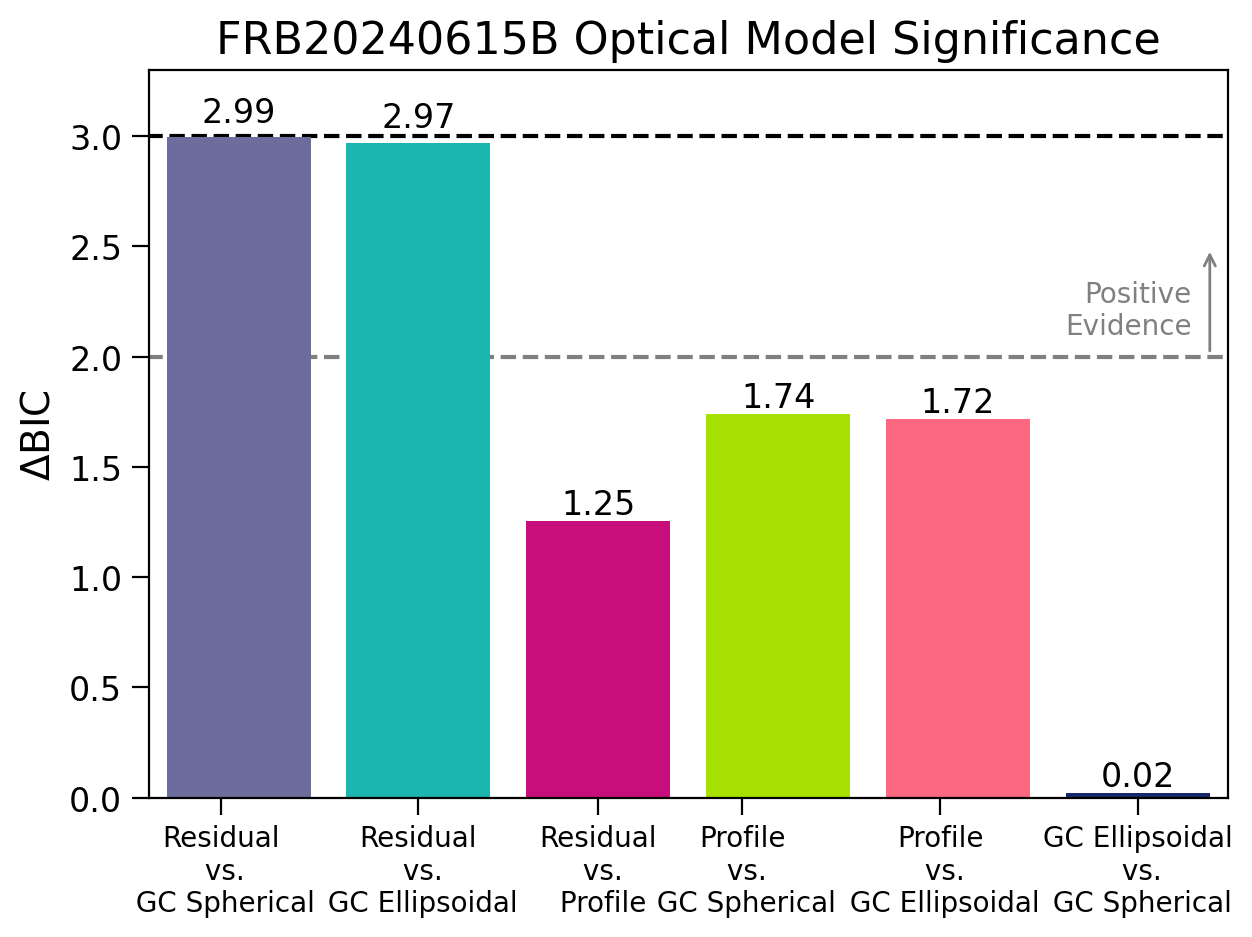}
    \caption{\textbf{(Left:)} Imaging and light profiles for FRB\,20240615B in the optical. \textbf{(Right:)} Comparison of the $\Delta$BIC between the four models for FRB\,20240615B in the optical.}
    \label{fig:appendix_240615_opt}
\end{figure}

\begin{figure}
    \includegraphics[width=0.6\textwidth]{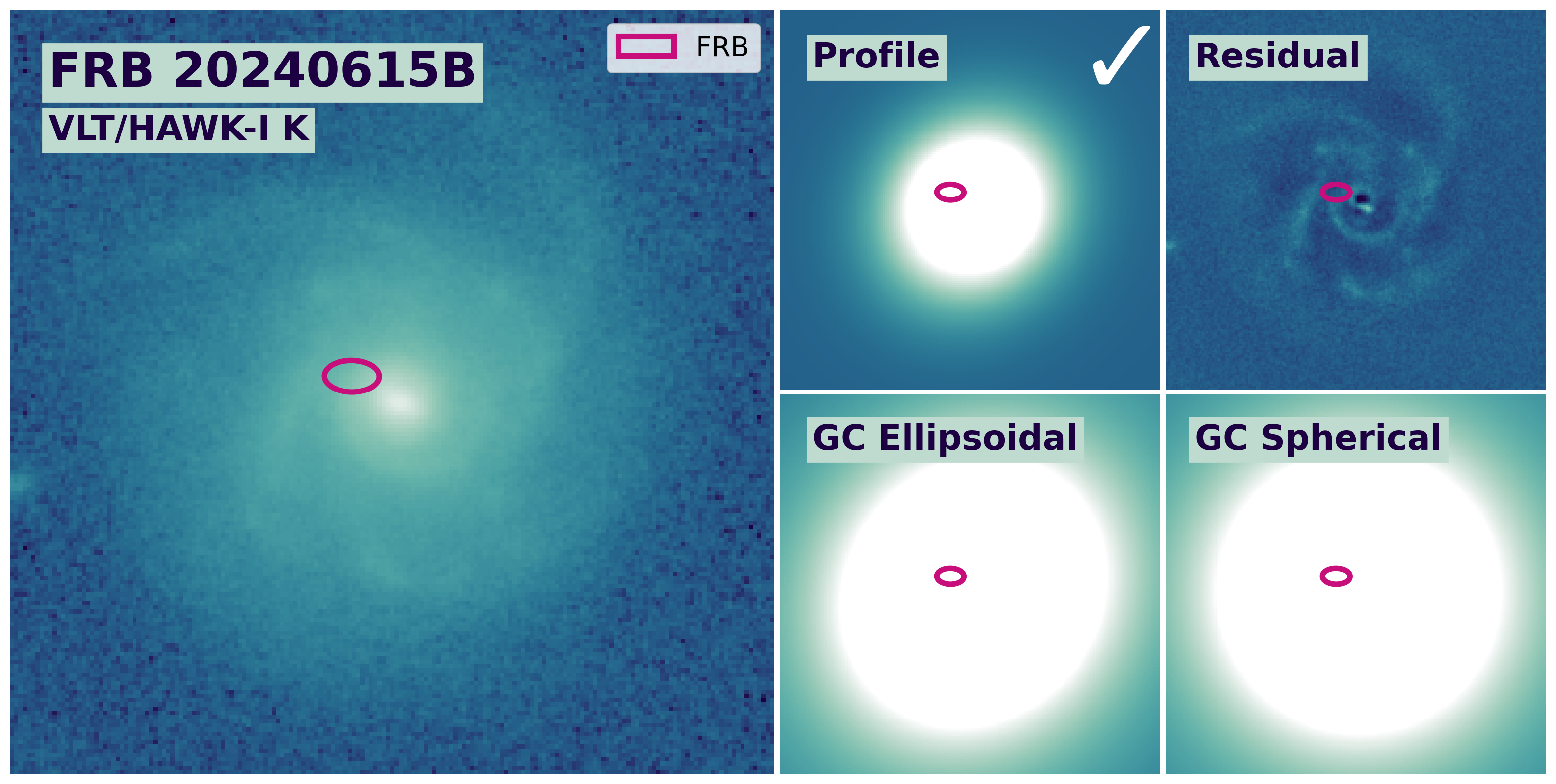}
    \includegraphics[width=0.39\textwidth]{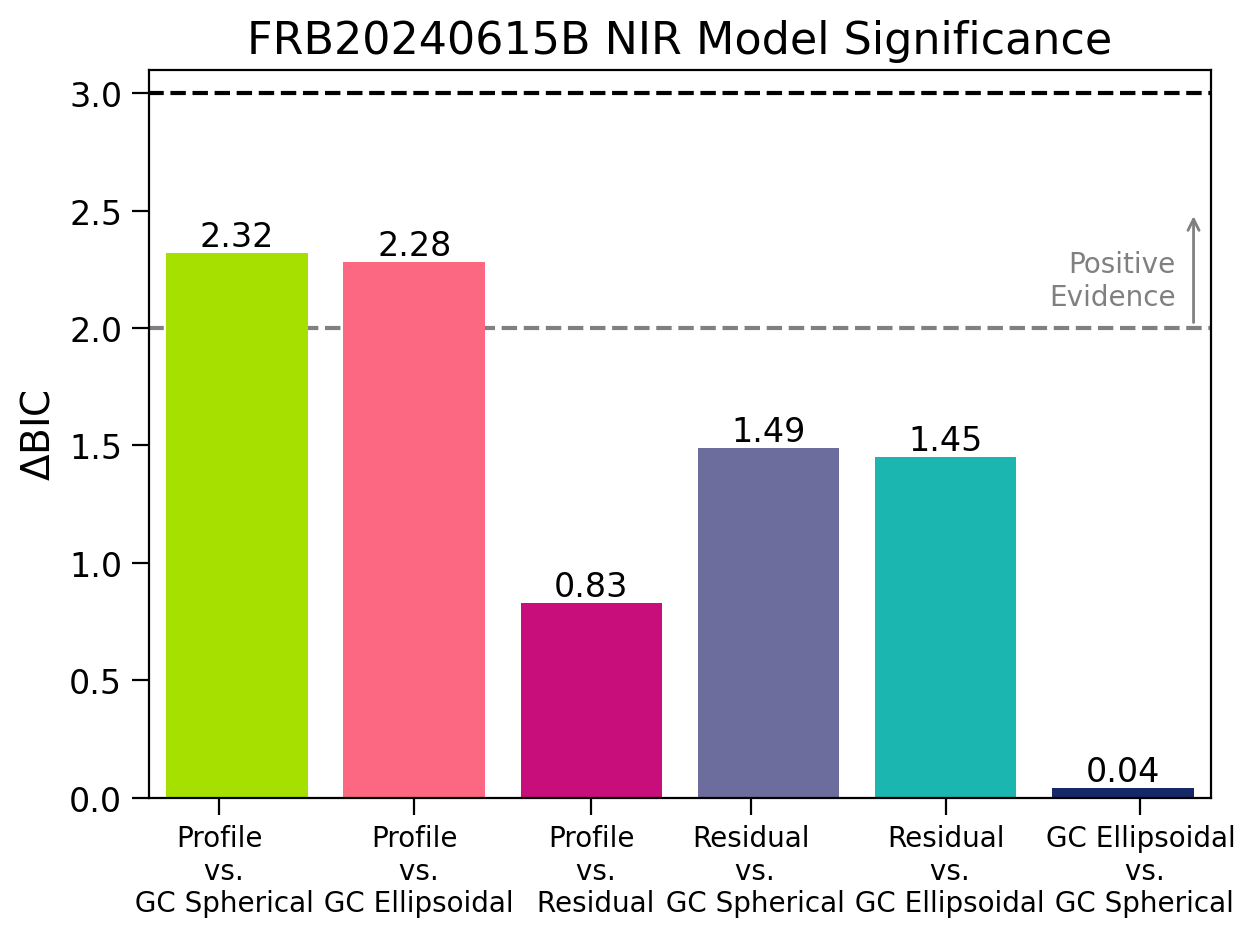}
    \caption{\textbf{(Left:)} Imaging and light profiles for FRB\,20240615B in the NIR. \textbf{(Right:)} Comparison of the $\Delta$BIC between the four models for FRB\,20240615B in the NIR.}
    \label{fig:appendix_240615_NIR}
\end{figure}

\begin{figure}
    \centering
    \includegraphics[width=0.6\textwidth]{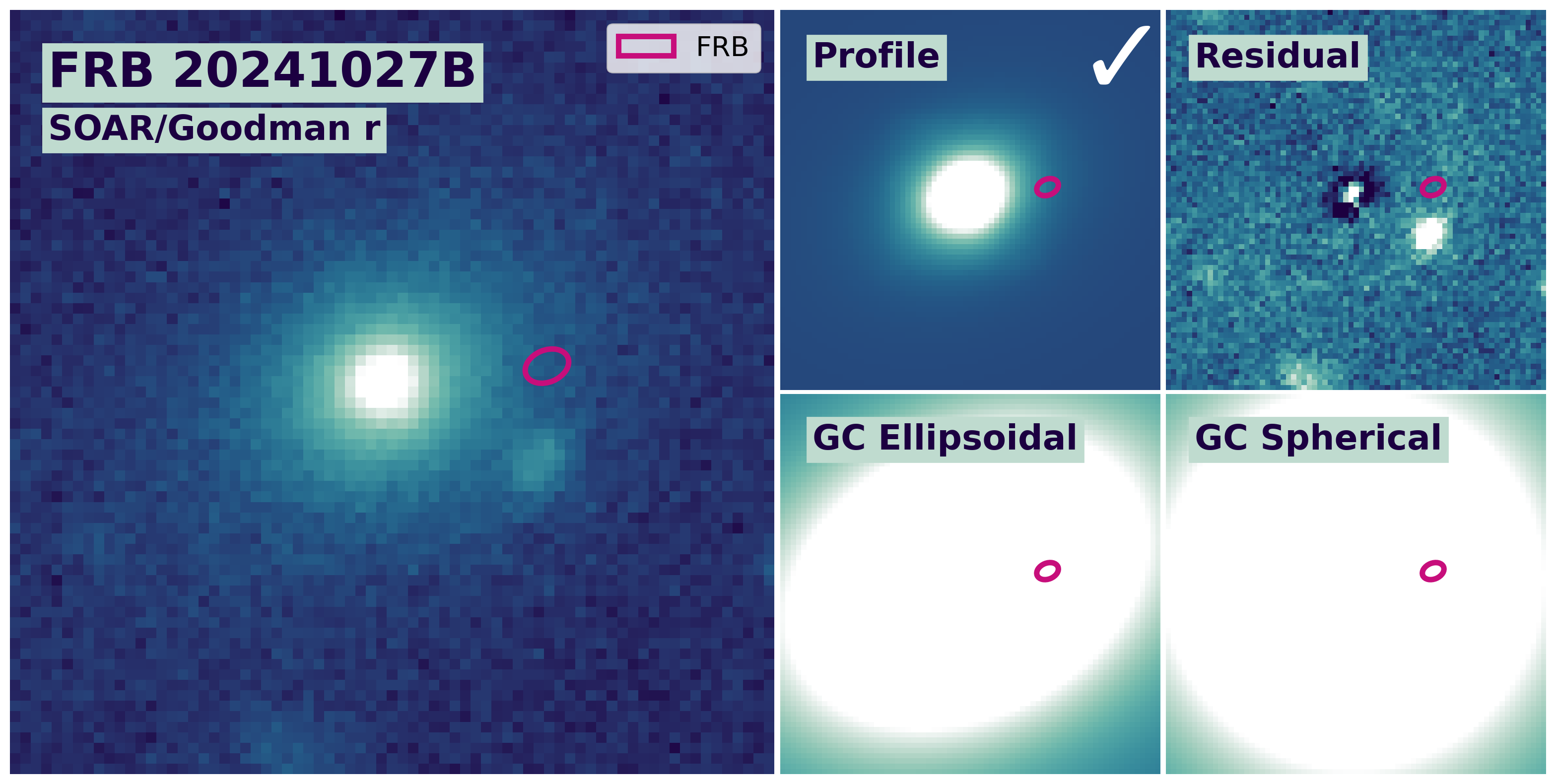}
    \includegraphics[width=0.39\textwidth]{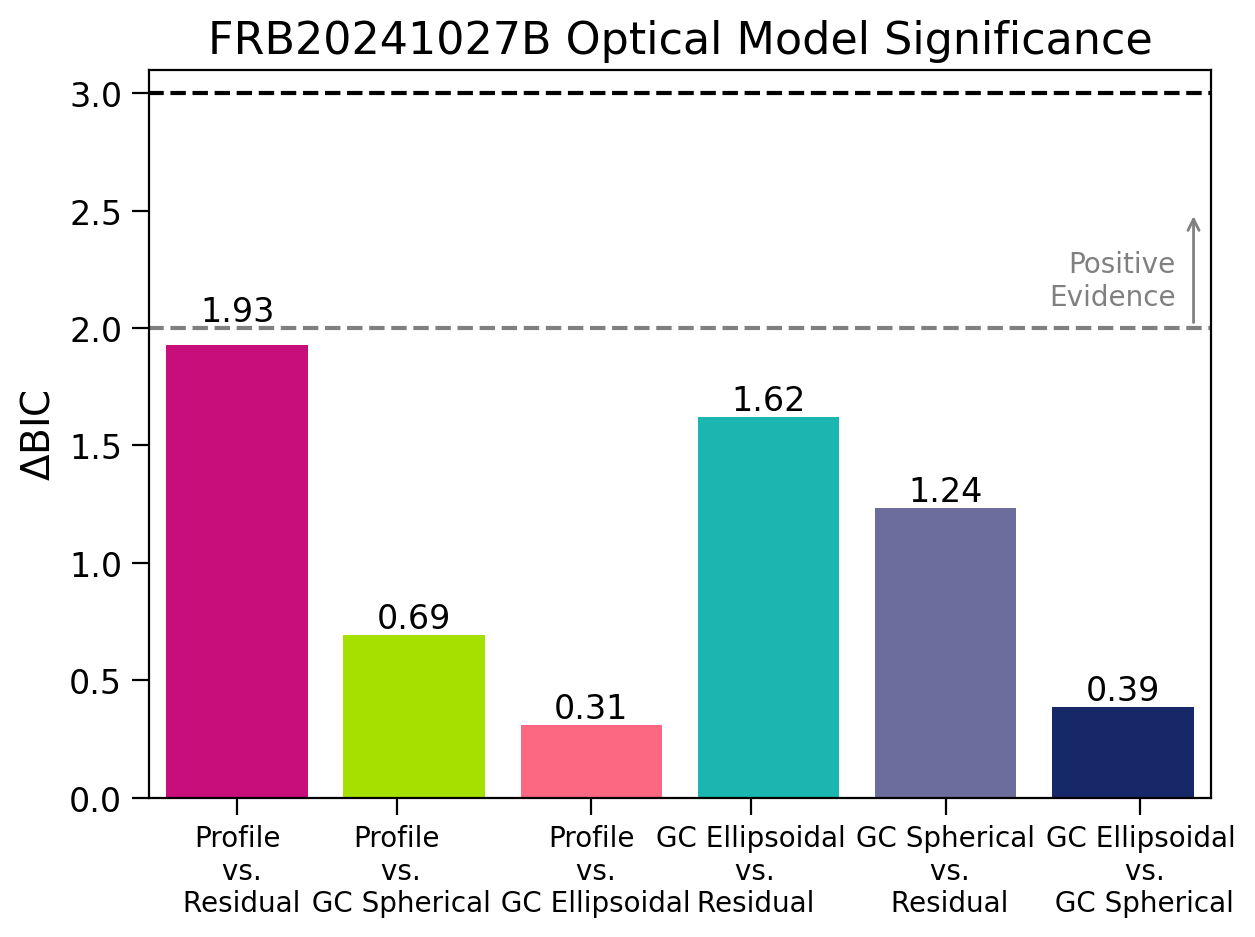}
    \caption{\textbf{(Left:)} Imaging and light profiles for FRB\,20241027B in the optical. \textbf{(Right:)} Comparison of the $\Delta$BIC between the four models for FRB\,20241027B in the optical.}
    \label{fig:appendix_241027_opt}
\end{figure}

\clearpage
\section{FRB and Astrometric Tie Uncertainties} \label{sec:astrometry_appendix}

We tabulate all relevant terms that encompass our total FRB uncertainty ellipses in Table~\ref{tab:astrometry}.

\startlongtable
\begin{deluxetable*}{l|ccccccccccccc}
\tabletypesize{\footnotesize}
\tablewidth{0pc}
\tablecaption{FRB and Astrometric Tie Uncertainties
\label{tab:astrometry}}
\tablehead{
\colhead{FRB} &
\colhead{$a_{\rm FRB}$} &
\colhead{$b_{\rm FRB}$} &
\colhead{$\theta_{\rm FRB}$} &
\colhead{Filter} & 
\colhead{$\sigma_{\rm 1, R.A.}$} &
\colhead{$\sigma_{\rm 1, Decl.}$} &
\colhead{$\sigma_{\rm 2, R.A.}$} &
\colhead{$\sigma_{\rm 2, Decl.}$} &
\colhead{$\sigma_{\rm Host, R.A.}$} &
\colhead{$\sigma_{\rm Host, Decl.}$} &
\colhead{$a_{\rm Total}$} &
\colhead{$b_{\rm Total}$} & \\
\colhead{} &
\colhead{[arcsec]} &
\colhead{[arcsec]} &
\colhead{[deg E of N]} &
\colhead{} & 
\colhead{[mas]} &
\colhead{[mas]} &
\colhead{[mas]} &
\colhead{[mas]} &
\colhead{[mas]} &
\colhead{[mas]} &
\colhead{[arcsec]} &
\colhead{[arcsec]} 
}
\startdata
20180916B  & 2.3e-3 & 2.3e-3 & 0 & F110W & 17 & 23 & - & - & 1.7 & 1.9 & 2.3e-2 & 1.7e-2  \\
20180924B & 0.18 & 0.18 & 0 & F160W & 24 & 23 & - & - & 0.80 & 0.80 & 0.18 & 0.18 \\
20190102C & 0.96 & 0.81 & 0 & F160W & 39 & 66 & - & - & 1.4 & 2.4 & 0.96 & 0.81 \\
20190608B & 0.38 & 0.35 & 90 & F160W & 5.6 & 9.7 & - & - & 0.6 & 0.5 & 0.38 & 0.36 \\
20190714A & 0.55 & 0.39 & 90 & F160W & 28 & 20 & - & - & 0.5 & 1.0 & 0.55 & 0.39 \\
20191001A & 0.46 & 0.43 & 81.9 & F160W & 73 & 86 & - & - & 0.3 & 0.4 & 0.47 & 0.44 \\
20200120E & 0.13 & 0.12 & 0 & $r$ & 110 & 110 & - & - & 2.7 & 3.3 & 0.17 & 0.16 \\
20200430A & 0.46 & 0.43 & 17.24 & $I$ & 50 & 40 & - & - & 7.7 & 9.6 & 0.46 & 0.43 \\
20200906A & 0.51 & 0.47 & 39 & $I$ & 63 & 72 & - & - & 7.4 & 8.8 & 0.52 & 0.47 \\
20201124A & 2.7e-3 & 2.6e-3 & 90 & F475X & 3.6 & 4.2 & - & - & 0.5 & 0.3 & 4.5e-3 & 4.5e-3 \\
 & & & & F160W &  &  & 12 & 15 & 0.1 & 0.1 & 1.5e-2 & 1.2e-2 \\
20210117A & 0.42 & 0.42 & 73.05 & $I$ & 54 & 46 & - & - & 15 & 19 & 0.42 & 0.42 \\
  & & & & $K$ & & & 160 & 150 & & & 0.47 & 0.43 \\
20210320C & 0.47 & 0.45 & 43.73 & $I$ & 42 & 31 & - & - & 31 & 4.5 & 0.47 & 0.45 \\
20210807D & 0.50 & 0.50 & 0 & $I$ & 69 & 150 & - & - & 6.5 & 4.8 & 0.52 & 0.50 \\
 & & & & $K$ & & & 110 & 110 & & & 0.53 & 0.52 \\
20211127I & 0.51 & 0.51 & 65.1 & $I$ & 43 & 51 & - & - & 2.4 & 3.3 & 0.51 & 0.51 \\
 & & & & $K$ & & & 64 & 54 & & & 0.52 & 0.51 \\
20211203C & 0.44 & 0.44 & 80.7 & $R$ & 54 & 39 & - & - & 2.5 & 2.3 & 0.44 & 0.44 \\ 
20211212A & 0.50 & 0.46 & 45.58 & $I$ & 44 & 41 & - & - & 2.2 & 1.9 & 0.50 & 0.46 \\
 & & & & $K$ & & & 53 & 120 & & & 0.51 & 0.46 \\
20220105A & 1.51 & 0.82 & $-$30.45 & $R$ & 100 & 110 & - & - & 4.6 & 4.8 & 1.52 & 0.82 \\
 & & & & $K$ & & & 170 & 170 & & & 1.54 & 0.83 \\
20220725A & 0.46 & 0.44 & $-$56.11 & $R$ & 160 & 110 & - & - & 1.6 & 1.5 & 0.49 & 0.45 \\
 & & & & $K$ & & & 190 & 160 & & & 0.55 & 0.45 \\
20220918A & 0.46 & 0.43 & $-$20.4 & $R$ & 210 & 63 & - & - & 16 & 10 & 0.51 & 0.43 \\
 & & & & $K$ & & & 390 & 130 & & & 0.65 & 0.43 \\
20221106A & 0.60 & 0.53 & $-$58.11 & $R$ & 280 & 250 & - & - & 2.5 & 22.8 & 0.70 & 0.54 \\
 & & & & $K$ & & & 160 & 160 & & & 0.73 & 0.54 \\
20230526A & 0.43 & 0.42 & $-$65.31 & $R$ & 410 & 260 & - & - & 7.2 & 7.4 & 0.59 & 0.50 \\
 & & & & $K$ & & & 230 & 150 & & & 0.63 & 0.52 \\
20230708A & 0.47 & 0.43 & $-$63.24 & $R$ & 110 & 66 & - & - & 12 & 8.9 & 0.48 & 0.44 \\
 & & & & $K$ & & & 270 & 160 & & & 0.55 & 0.47 \\ 
20230902A & 0.69 & 0.55 & $-$72.27 & $R$ & 110 & 78 & - & - & 3.4 & 3.9 & 0.70 & 0.56 \\
 & & & & $K$ & & & 230 & 160 & & & 0.74 & 0.59 \\
20231226A & 0.51 & 0.48 & $-$13.19 & $R$ & 95 & 140 & - & - & 3.6 & 2.9 & 0.53 & 0.49 \\
 & & & & $K$ & & & 160 & 180 & & & 0.56 & 0.51 \\
20231230D & 0.70 & 0.43 & 90 & $R$ & 60 & 66 & - & - & 7.2 & 5.6 & 0.71 & 0.44 \\
 & & & & $K$ & & & 150 & 150 & & & 0.72 & 0.46 \\
20240117B & 1.29 & 0.72 & 90 & $R$ & 73 & 140 & - & - & 7.7 & 7.4 & 1.29 & 0.73 \\
 & & & & $K$ & & & 170 & 180 & & & 1.30 & 0.75 \\
20240201A & 0.52 & 0.46 & $-$30.04 & $R$ & 270 & 250 & - & - & 3.1 & 2.1 & 0.63 & 0.47 \\
 & & & & $K$ & & & 150 & 140 & & & 0.66 & 0.47 \\
20240203C & 0.60 & 0.47 & 0 & $R$ & 74 & 120 & - & - & 25 & 10 & 0.61 & 0.47 \\ 
20240209A & 2.12 & 1.08 & 9.54 & $R$ & 200$^{a}$ & 200$^{a}$ & - & - & 4.2 & 3.4 & 2.13 & 1.09 \\
20240210A & 0.55 & 0.49 & $-$54.14 & $R$ & 190 & 200 & - & - & 1.8 & 1.5 & 0.61 & 0.49 \\
 & & & & $K$ & & & 160 & 180 & & & 0.66 & 0.49 \\
20240304A & 0.81 & 0.56 & $-$57.04 & $R$ & 69 & 110 & - & - & 9.8 & 8.8 & 0.82 & 0.56 \\
 & & & & $K$ & & & 150 & 160 & & & 0.85 & 0.56 \\
20240310A & 0.59 & 0.55 & $-$48.6 & $R$ & 160 & 79 & - & - & 2.0 & 2.7 & 0.61 & 0.55 \\ 
 & & & & $K$ & & & 180 & 120 & & & 0.65 & 0.56 \\
20240312D & 1.29 & 1.10 & 0 & $r$ & 480 & 310 & - & - & 2.4 & 2.2 & 1.38 & 1.14 \\ 
20240318A & 0.79 & 0.54 & 0 & $R$ & 120 & 380 & - & - & 160 & 100 & 0.89 & 0.56 \\ 
 & & & & $K$ & & & 270 & 220 & & & 0.93 & 0.60 \\
20240525A & 0.48 & 0.42 & 90 & $R$ & 210 & 110 & - & - & 2.8 & 4.0 & 0.52 & 0.43 \\ 
20240615B & 0.66 & 0.32 & 90 & $R$ & 190 & 200 & - & - & 1.2 & 1.2 & 0.68 & 0.37 \\ 
 & & & & $K$ & & & 130 & 140 & & & 0.69 & 0.40 \\
20241027B & 0.64 & 0.45 & $-$67.27 & $r$ & 110 & 49 & - & - & 16 & 12 & 0.65 & 0.46 \\
\enddata
\tablecomments{Details of the uncertainties associated with the FRBs and images. $a_{\rm FRB}$ and $b_{\rm FRB}$ are the combined statistical and systematic uncertainties on the FRB, represented as semi-major and semi-minor axes of the localization ellipse. $\theta_{\rm FRB}$ is the position angle of the ellipse, oriented in degrees East of North. References for these values can be found in Table~\ref{tab:sample}. $\sigma_{\rm 1}$ denotes the astrometric tie uncertainty of the image. When both optical and infrared images are available, we tie the infrared image to the optical image; $\sigma_{\rm 2}$ refers to the uncertainties associated with this process. $a_{\rm Total}$ and $b_{\rm Total}$ are the quadrature sums of all relevant uncertainties (see Section~\ref{sec:astrom} for details); the position angle $\theta_{\rm FRB}$ is not changed.}
$^{a}$ Assumed fiducial values. \\
\end{deluxetable*}

\section{Details of New Spectroscopy} \label{sec:spec_appendix}

We tabulate all detected emission and absorption features in the host galaxy spectroscopy newly presented in this work in Table~\ref{tab:spec}.

\begin{deluxetable*}{l|ccc}[b]
\tablewidth{0pc}
\tablecaption{Observed Spectroscopic Features in the Hosts of Localized FRBs
\label{tab:spec}}
\tablehead{
\colhead{FRB} &
\colhead{$z$} &
\colhead{Emission Lines} &
\colhead{Absorption Lines}
}
\startdata
20240203C & 0.2426 & [O\,{\sc ii}]$\lambda\lambda 3726, 3729$, [Ne\,{\sc iii}]$\lambda 3869$, H$\delta$, H$\gamma$, H$\beta$,  & None \\
          &         & [O\,{\sc iii}]$\lambda\lambda 4959, 5007$, [N\,{\sc ii}]$\lambda\lambda 6548, 6583$, H$\alpha$, [S\,{\sc ii}]$\lambda\lambda 6716, 6731$ & \\
\hline
20240312D & 0.0495 & [O\,{\sc ii}]$\lambda\lambda 3726, 3729$, H$\beta$, H$\alpha$, [N\,{\sc ii}]$\lambda 6584$, [S\,{\sc ii}]$\lambda 6716$ & None \\
\hline
20240318A & 0.1120 & [O\,{\sc ii}]$\lambda\lambda 3726, 3729$, H$\beta$, [O\,{\sc iii}]$\lambda 5007$, [N\,{\sc ii}]$\lambda\lambda 6548, 6584$, & Ca {\sc ii} H\&K\\
          &         & H$\alpha$, [S\,{\sc ii}]$\lambda\lambda 6716, 6731$ \\
\hline
20240525A & 0.3266 & [O\,{\sc ii}]$\lambda\lambda 3726, 3729$, H$\beta$, [O\,{\sc iii}]$\lambda\lambda 4959, 5007$ & None \\
          &        & H$\alpha$, [N\,{\sc ii}]$\lambda 6584$, [S\,{\sc ii}]$\lambda 6716, 6731$ \\ 
\hline
20240615B & 0.073 & [O\,{\sc ii}]$\lambda\lambda 3726, 3729$, H$\delta$, H$\gamma$, H$\beta$, \\
          &        & [N\,{\sc ii}]$\lambda\lambda 6548, 6583$, H$\alpha$, [S\,{\sc ii}]$\lambda\lambda 6716, 6731$ & None \\
\hline
20241027B & 0.336 & None & Ca {\sc ii} H\&K, G-band, H$\beta$, Mg I, Na I \\
\enddata
\end{deluxetable*}


\bibliography{main}{}
\bibliographystyle{aasjournal}



\end{CJK*}
\end{document}